\documentclass[11pt,a4paper]{article}                           

\usepackage{a4}          

\usepackage{amsmath}
\usepackage{amssymb}
\usepackage[usenames]{color} 
\usepackage{multirow}
\usepackage{graphicx}
\usepackage{epstopdf}
\usepackage{array}
\usepackage{hhline}
\usepackage{cite}
\usepackage{microtype}
\usepackage[bf]{caption}
\usepackage{color}
\usepackage[usenames,dvipsnames]{xcolor}
\usepackage{subcaption}
\usepackage{pstool}
\usepackage{tcolorbox}

\usepackage{ifpdf}
\ifpdf
  \usepackage[pdftex,
    pdftitle={},
    pdfauthor={},
    pdfsubject={},
    bookmarksopen, bookmarksnumbered, bookmarksopenlevel=2]{hyperref}
\fi
\def\hybrid{\topmargin -20pt    \oddsidemargin 0pt
        \headheight 0pt \headsep 0pt
        \textwidth 6.25in       
        \textheight 9 in       
        \marginparwidth .875in
        \parskip 5pt plus 1pt 
          \jot = 1.5ex
   }
\hybrid
\numberwithin{equation}{section}
\numberwithin{table}{section}

\newcommand{\beq}{\begin{equation}}  \newcommand{\eeq}{\end{equation}}
\newcommand{\bal}{\begin{aligned}}   \newcommand{\eal}{\end{aligned}}
\newcommand{\bea}{\begin{eqnarray}}  \newcommand{\eea}{\end{eqnarray}}

\newcommand{\bmat}{\left(\begin{array}}
\newcommand{\emat}{\end{array}\right)}

\newcommand{\nn}{\nonumber}

\newcommand{\cO}{\mathcal{O}}

\newcommand{\cC}{\mathcal{C}}
\newcommand{\cD}{\mathcal{D}}
\newcommand{\cL}{\mathcal{L}}
\newcommand{\cS}{\mathcal{S}}

\newcommand{\cH}{\mathcal{H}}

\newcommand{\cF}{\mathcal{F}}
\newcommand{\cI}{\mathcal{I}}
\newcommand{\cJ}{\mathcal{J}}
\newcommand{\cR}{\mathcal{R}}

\newcommand{\cV}{\mathcal{V}}

\newcommand{\cM}{\mathcal M}
\newcommand{\cQ}{\mathcal Q}

\usepackage{cancel}

\newcommand{\be}{\begin{equation}}
\newcommand{\ee}{\end{equation}}

\newcommand{\half}{\frac{1}{2}}

\begin{document}

\baselineskip=14pt
\parskip 5pt plus 1pt

\vspace*{-1.5cm}
\begin{flushright}    
  {\small MPP-2019-53
  }
\end{flushright}

\vspace{2cm}
\begin{center}        
  {\LARGE The Swampland: Introduction and Review}
\end{center}

\vspace{0.5cm}
\begin{center}        
{\large Eran Palti}
\end{center}

\vspace{0.15cm}
\begin{center}        
  \emph{Max-Planck-Institut f\"ur Physik (Werner-Heisenberg-Institut), \\
Fohringer Ring 6, 80805 Munchen, Germany}
             \\[0.15cm]
 
\end{center}

\vspace{2cm}


\begin{abstract}
The Swampland program aims to distinguish effective theories which can be completed into quantum gravity in the ultraviolet from those which cannot. This article forms an introduction to the field, assuming only a knowledge of quantum field theory and general relativity. It also forms a comprehensive review, covering the range of ideas that are part of the field, from the Weak Gravity Conjecture, through compactifications of String Theory, to the de Sitter conjecture. 
\end{abstract}

\thispagestyle{empty}
\clearpage

\setcounter{page}{1}


\newpage

\tableofcontents

\section{Introduction}
\label{sec:intro}

Self-consistency can be an important tool for understanding the structure of physical theories. Particularly so since it typically becomes more powerful at increasing energy scales, where empirical constraints become less accessible. Within the framework of Quantum Field Theory the prototypical examples are anomalies, which restrict the particle spectra in the theory. But at higher energy scales, such as those relevant to quantum gravity physics, it appears that self-consistency may be strong enough to almost uniquely fix the theory. String theory, in particular, manifests such characteristics. Indeed, already at the string scale perturbative string theories are almost uniquely fixed including, remarkably, the dimension of space-time and any possible constant parameters. 

\begin{figure}[t]
\centering
 \includegraphics[width=0.9\textwidth]{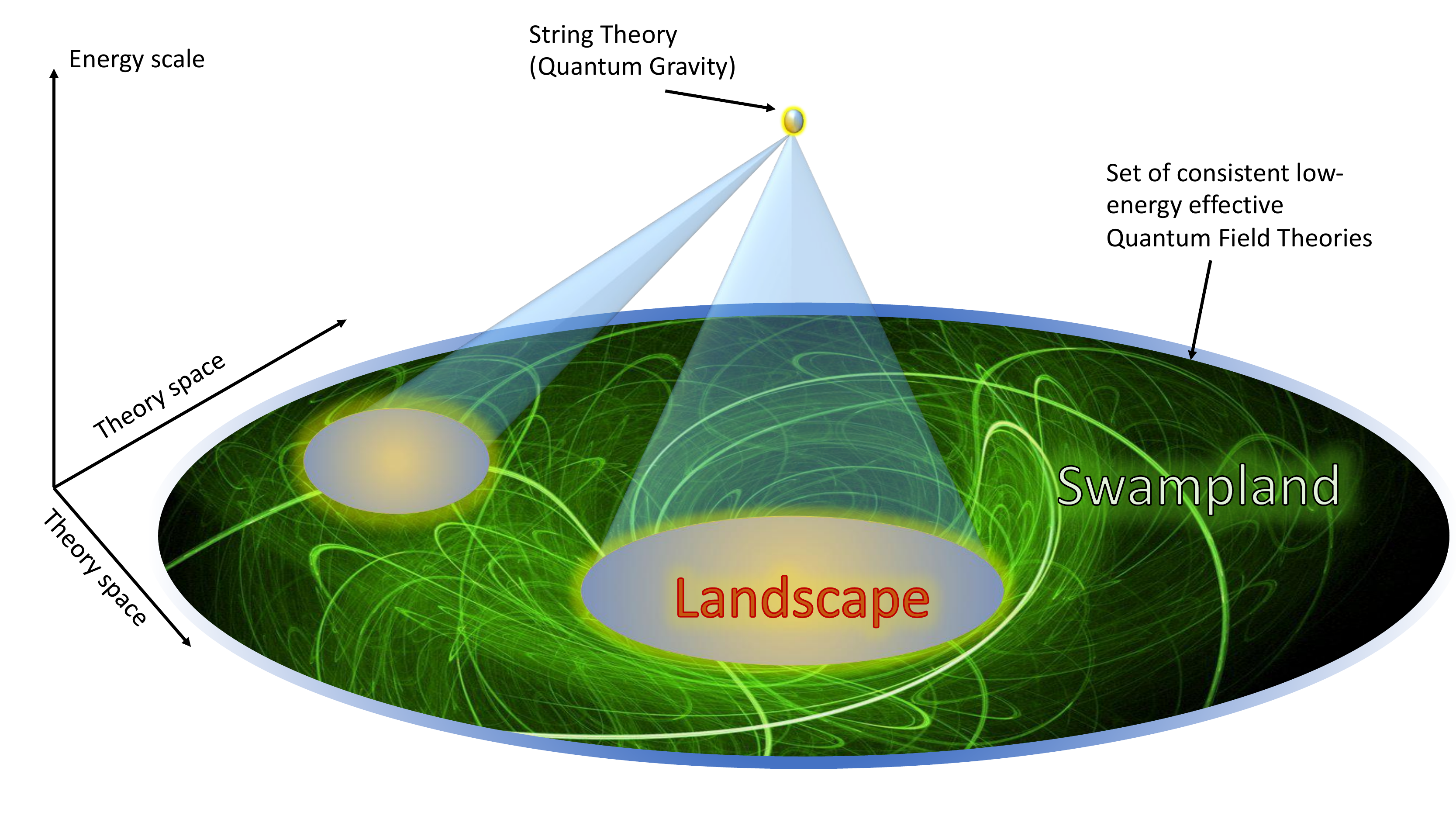}
\caption{A schematic illustration of the space of (apparently) self-consistent effective quantum field theories. The sub-set which could arise from string theory is the string Landscape, while all the other theories are in the string Swampland.}
\label{fig:sw}
\end{figure}

Self-consistency becomes much less powerful at low energies, even for theories which include gravity. In string theory this manifests as the existence, within our current understanding, of a huge number of resulting low-energy effective theories. Each such theory is constructed about a different vacuum of string theory, and the rich vacuum structure of the theory, the so-called String Theory {\it Landscape}, then translate into a large spectrum of effective theories. However, it is important to not confuse this richness of structure with a complete absence of constraints. The resulting set of theories still picks out only a subset of all the possible apparently self-consistent effective theories. The use of apparently here means that there is nothing manifestly wrong with the effective theory, but an inconsistency would manifest should one try to complete it in the ultraviolet. The idea of the String Theory {\it Swampland} was introduced in \cite{Vafa:2005ui} as a way to quantify and refer to these residual low-energy constraints.\footnote{See also \cite{DouglasStrings2005} for similar ideas at the same time.} More precisely:
\begin{center}
The {\bf Swampland} can be defined as the set of (apparently) consistent effective field theories \\ that cannot be completed into quantum gravity in the ultraviolet.
\end{center}
So string theory might lead to a large Landscape of effective low-energy theories, but there is an even larger Swampland of effective theories that cannot come from string theory. This is illustrated in figure \ref{fig:sw}. 
Note that we phrased the definition of the Swampland using a general notion of quantum gravity, rather than specifically string theory. For simplicity of notation, we will rarely distinguish between such a general quantum theory of gravity and string theory, but it is natural to define the Swampland in this more general sense. 

Of course, the abstract concept of the Swampland has no useful meaning unless we understand how to distinguish between effective field theories that are in the Landscape from those in the Swampland. The criteria for not being in the Swampland should further be formulated purely in terms of properties of the low-energy effective theory itself, so without reference to a particular ultraviolet origin. Determining such criteria is a difficult task, and not all proposed Swampland criteria are on the same footing. They vary in the amount of evidence supporting them and in the severity of their implications. The criteria, perhaps with one or two exceptions, are not proven from a microscopic physics perspective and so are usually stated as conjectures. Nonetheless, as we will review, some of these conjectures are backed by significant, and sometimes remarkable, evidence. 

\begin{figure}[t]
\centering
 \includegraphics[width=0.9\textwidth]{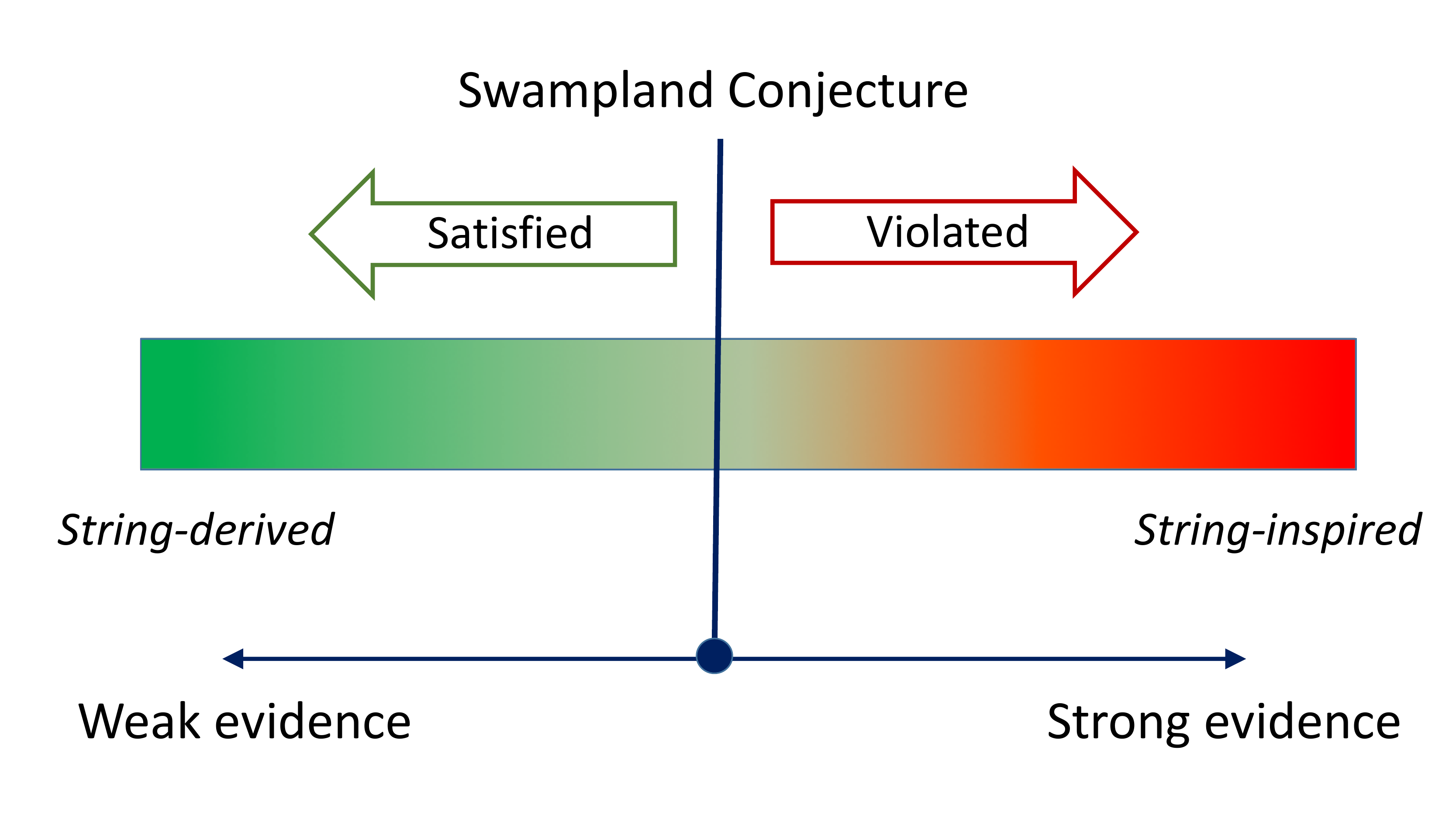}
\caption{A schematic illustration of the spectrum of vacuum constructions in string theory. The most rigorously understood, string-derived, vacua are on the left while the more loosely connected, string-inspired, vacua are on the right. A Swampland conjecture can be placed on the spectrum such that all known vacua of increasing rigour satisfy it. A conjecture placed to the left of the spectrum is said to have weaker evidence for it than one placed to the right.}
\label{fig:ss}
\end{figure}

One approach towards providing evidence for a Swampland conjecture is by utilising known vacua of string theory as a type of experimental data. So if all the known vacua satisfy a conjecture then string theory data supports it. This is the approach that is typically adopted, but it suffers from an inherit difficulty which is that known vacua of string theory themselves have varying levels of rigour. The best understood are vacua where we have a full string world-sheet description, these are usually supersymmetric and have relatively simple geometries for the extra dimensions, typically some orbifold. We will refer to these as {\it string-derived}. On the other end of the spectrum are proposed vacua of string theory which involve a large number of assumptions. Eventually, such vacua are better thought of as constructions in quantum field theory that are motivated by the type of structures one finds in string theory, we will refer to these as {\it string-inspired}. A given Swampland conjecture would typically be of the form that at some point on this spectrum of vacua, all vacua of increasing rigour satisfy the conjecture. The question of whether a Swampland conjecture is proven or disproven in string theory is therefore somewhat subjective. If we insist on only counting string-derived vacua then many conjectures are proven to hold for all known vacua. While a counter-argument would be that this is due to technical limitations and there are many other vacua of string theory which violate the conjecture but we simply are not able to rigorously construct them. This debate is captured by stating that string theory provides evidence, of varying levels of strengths, for a given conjecture. This situation is illustrated in figure \ref{fig:ss}.

Another approach that is commonly adopted to study Swampland criteria is utilising quantum gravity arguments directly in the low-energy effective theory. There are low-energy aspects of quantum gravity that are expected to be universal, in particular relating to black holes and to the holographic nature of gravity. Relative to utilising string theory constructions, this approach has the advantage of broader generality. On the other hand, the arguments often lack the explicit details and concreteness of string theory based evidence. 

The ideal approach for establishing Swampland criteria is a direct derivation from microscopic physics. Of course, no physics theory is assumption free, but it may be that one could uncover some ultraviolet principle of string theory that we had missed, or had not sufficiently appreciated, that will then lead directly to the Swampland criteria. This is as opposed to the more indirect arguments based on data from string theory vacua. In many ways, uncovering such new underlying microscopic physics can be considered the big-picture aim of the Swampland program. The different string vacua, and general quantum gravity arguments, can then be thought of as supplying the experimental data towards the construction of a new theory. Indeed, there are some signs that we are on the right track towards such a result. As we will review, various Swampland criteria, while initially seemingly unrelated, have been increasingly found to be part of an interconnected structure. Further, much of this structure can be shown to be deeply related to emergent dynamics in the infrared, and we will dedicate section \ref{sec:emergence} in this review to developing this idea. 

The Swampland program builds on all three of the approaches, string theory constructions, quantum gravity arguments, and microscopic physics, in order to establish evidence for constraints. It is encouraging that, as we will review in detail, the three perspectives appear remarkably in sync, and lead to results consistent with each other. It is this consistency, and the multiple sources of evidence, which suggest that at least at the general level, the Swampland program is on the right track.

\begin{figure}[t]
\centering
 \includegraphics[width=0.8\textwidth]{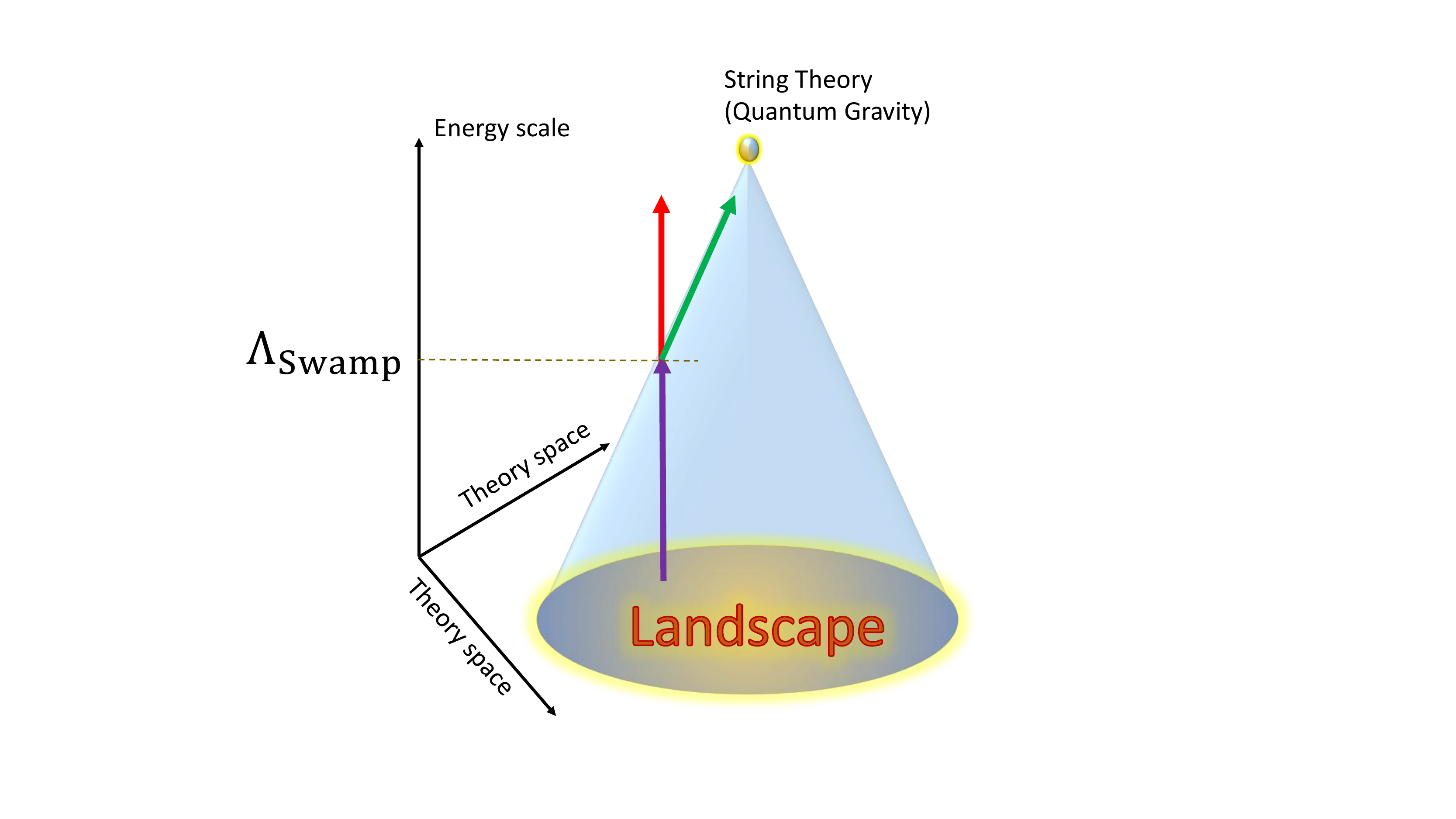}
\caption{Figure showing a schematic interpretation of the scale $\Lambda_{\mathrm{Swamp}}$ at which an effective theory must be modified if it is to be able to complete into quantum gravity in the ultraviolet.}
\label{fig:swcone}
\end{figure}

A Swampland criterium typically has some universal structure. One considers a class of effective QFTs that are self-consistent up to a cutoff scale $\Lambda_{\mathrm{QFT}}$. These theories are then coupled to gravity, the coupling being associated to a finite gravitational strength, so a finite value of the Planck mass $M_p$. The resulting theory will then (be conjectured to) have a new energy scale $\Lambda_{\mathrm{Swamp}}$ above which it becomes inconsistent if left unmodified. The required modification may be mild, say the introduction of a new particle into the theory, or it may be very substantial. We can think of the scale $\Lambda_{\mathrm{Swamp}}$ as the point where the theory must start to `plan ahead' if it is to reach a consistent quantum gravity theory in the ultraviolet. So it is not necessarily that something would go wrong with the theory at $\Lambda_{\mathrm{Swamp}}$, but rather if nothing changes then the theory would not be completable in the ultraviolet. This is illustrated in figure \ref{fig:swcone}. 

\begin{figure}[t]
\centering
 \includegraphics[width=0.8\textwidth]{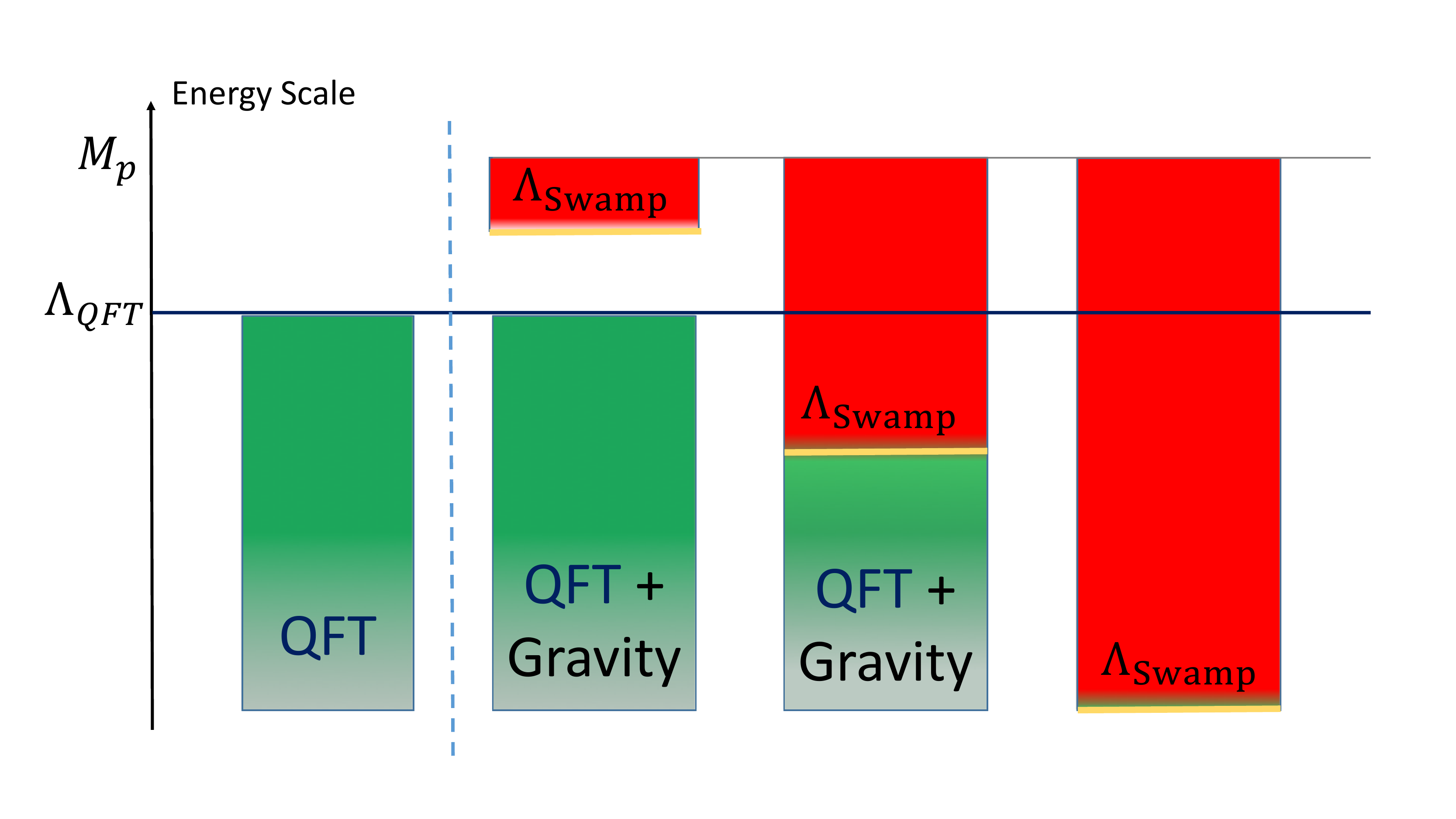}
\caption{Figure showing various cutoff scales on effective theories. The first case is a pure QFT with a cutoff $\Lambda_{\mathrm{QFT}}$. The second is a QFT coupled to Gravity, with parameter values such that the new Swampland cutoff $\Lambda_{\mathrm{Swamp}}$ lies above the QFT cutoff. Varying the parameters one may reach the third case where the Swampland cutoff is far below the QFT one $\Lambda_{\mathrm{Swamp}} \ll \Lambda_{\mathrm{QFT}}$, leading to a strong constraint on the effective theory due to the presence of gravity. Finally, the fourth case illustrates that it may be that for certain parameter ranges the theory may be inconsistent already at the lowest non-trivial energy scale in the theory.}
\label{fig:swgen}
\end{figure}

Depending on the parameter values of the effective QFT, the scale $\Lambda_{\mathrm{Swamp}}$ may be above the scale $\Lambda_{\mathrm{QFT}}$, in which case there is no effect on the effective theory, or it may be below it, in which case the effective theory with the original cutoff becomes inconsistent due to coupling to gravity. The scale $\Lambda_{\mathrm{Swamp}}$ will typically behave such that $\Lambda_{\mathrm{Swamp}} \rightarrow \infty$ as $M_p \rightarrow \infty$, as expected since it was tied to the coupling of gravity. However, depending on the parameters of the QFT, one may have $\Lambda_{\mathrm{Swamp}} \ll \Lambda_{\mathrm{QFT}} \ll M_p$, strongly affecting the low-energy theory. Of course, it is such setups which are of most interest. The most extreme situation is one where $\Lambda_{\mathrm{Swamp}}$ is lower than any non-trivial energy scale in the effective theory, and so essentially the whole theory is in the Swampland. The various possibilities are illustrated in figure \ref{fig:swgen}. 

So far we have not explicitly distinguished between string theory and quantum gravity. However, some of the motivations for Swampland criteria can be understood without any reference to string theory. We will try to highlight such qualities when they arise. Practically, almost all of the Swampland program is guided one way or another by string theory. This need not mean that it is tied to string theory necessarily, for example the modern form of holography in anti-de Sitter space was discovered through string theory but can be formulated in a more general quantum gravitational sense. In other words, string theory certainly seems to be {\it a} quantum theory of gravity, and can be utilised to study such physics irrespective of whether it is {\it the} quantum theory of gravity relevant for our universe.

This introduction and review article has multiple aims. Foremost, it aims at forming a practical and coherent starting point for those interested in studying the Swampland. The hope is that much of it will be accessible even to readers with little experience in string theory. And that the review can form a base from which cutting edge research is made accessible. While string theory is a mature field, the Swampland aspect of it can be considered to be at a relatively early stage, with much of the work being about establishing what are the candidates for the main underlying ideas and physics, and how can these be tested. It is therefore not a field that can be reviewed as a list of derivable, provable results starting from underlying axioms. Indeed, one of the main research directions is about establishing what the microscopic physics underlying the Swampland results actually could be. Instead, it is currently very much a collection of arguments, from a variety of sources, for certain proposals about the Swampland. In presenting an introduction and review of the field, emphasis will therefore be placed on explaining as many as possible of the existing arguments for proposed criteria of the Swampland, so that the reader may judge for themselves the strength of the evidence. Similarly, the focus in introducing string theory tests of the criteria will be on a detailed and pedagogical introduction to simple examples, that will establish the key ideas and rules, and which will form the solid starting point from which to explore more complicated scenarios. Nonetheless, the review will be as comprehensive as possible in touching on as many of the ideas in the field as possible, at least in some way. 

Note that in \cite{Brennan:2017rbf} another review of the Swampland was presented. This is recommended as offering a different perspective and approach to reviewing the field, and while this review will be more comprehensive, there are many interesting ideas and perspectives in \cite{Brennan:2017rbf} which will not be discussed here. 

It is difficult to fully understand the ideas of the field without knowing some string theory. Therefore, in section \ref{sec:first} we start with a quick focused introduction to string theory, before presenting a first encounter with simple examples of Swampland criteria. In section \ref{sec:map} we present a general overview of the Swampland, including more general, typically black hole based, arguments for some of the conjectures. In section \ref{sec:stringcomp} we discuss tests of the conjectures in string theory. In section \ref{sec:emergence} we present a proposal, based on emergence, for the microscopic physics underlying some of the primary Swampland criteria. In section \ref{sec:moreswamp} we discuss more proposals for different Swampland criteria. In section \ref{sec:dsconj} we discuss the de Sitter and refined de Sitter conjectures. We present a summary and outlook in section \ref{sec:summary}.

In this review we will work with the following conventions. We set $\hbar=c=1$ and work in mostly-plus signature for the metric, so the flat-space metric is
\be
\eta_{\mu\nu} = \mathrm{diag\;} \left( -,+,+,...,+\right) \;.
\ee
We define the reduced Planck mass in $d$-dimensions $M_p^d$ as the coefficient in front of the $d$-dimensional Ricci scalar $R^d$ in the action
\be
S_d = \int d^dX \sqrt{-G} \left[ \frac{\left(M^d_p\right)^{d-2}}{2} R^d+ ... \right] \;.
\ee 
Our units are chosen such that, in a $d$-dimensional theory, the $d$-dimensional reduced Planck mass $M_p^d$ is set to one $M^d_p=1$. We will sometimes reinstate it for clarity purposes, in which case the four-dimensional value is 
\be
M_p = \sqrt{\frac{1}{8 \pi G_N}} \simeq 2.4 \times 10^{18} \;\mathrm{GeV} \;,
\ee
where $G_N$ is Newton's constant.

\section{First encounters with the Swampland}
\label{sec:first}

This section provides a first encounter with some of the Swampland criteria. The first part, section \ref{sec:focst}, is aimed at those who have little experience with string theory and so will be a quick and focused introduction to some of the key concepts of string theory. Readers familiar with string theory can safely skip this and only use it as a reference to the notation and conventions of the paper. The introduction to string theory is essential to convey how Swampland criteria have deeply string theoretic, or more generally quantum gravitational, physics built into them. Following this introduction, in section \ref{sec:fcsdc}, we will consider dimensional reduction of the bosonic string on a circle. This will provide a first encounter with the Swampland distance conjecture \cite{Ooguri:2006in}. In section \ref{sec:fewgcst} we then go on to show how it also manifests the physics of the Weak Gravity Conjecture \cite{ArkaniHamed:2006dz}. Finally, in section \ref{sec:fespec} we will introduce the idea of the Species scale. 

\subsection{A focused introduction to String Theory}
\label{sec:focst}

This subsection gives a very quick introduction to string theory with a focus on the basics necessary to understand the first encounter with the Swampland. It is based on very familiar material covered by any standard textbook on string theory. See for example \cite{Blumenhagen:2013fgp}.

\subsubsection{The Nambu-Goto action for a particle}

\begin{figure}[t]
\centering
 \includegraphics[width=0.8\textwidth]{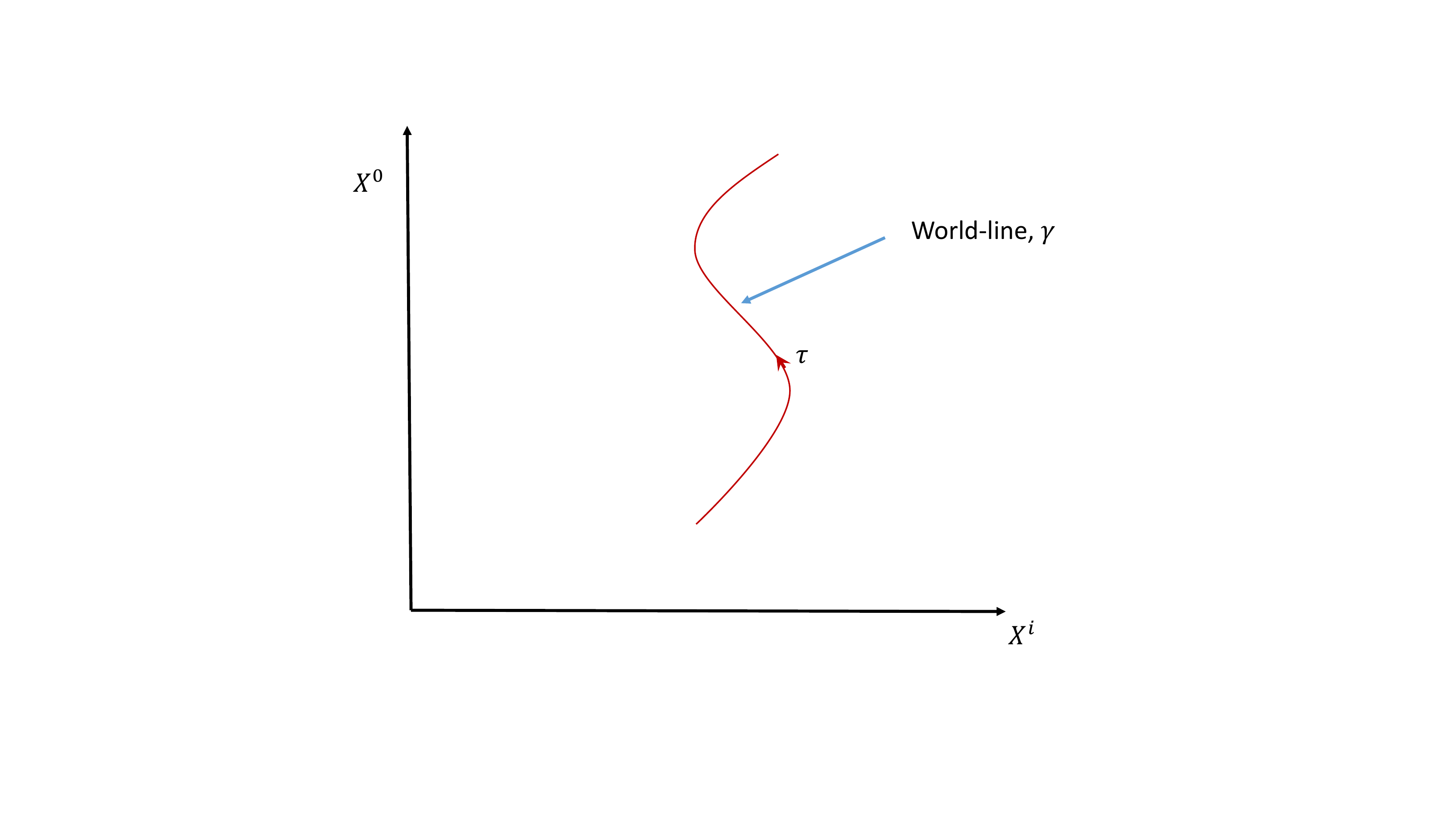}
\caption{Figure illustrating the world-line of a particle.}
\label{fig:pwl}
\end{figure}

To begin let us consider a point particle. We let it propagate in $D$-dimensional spacetime, so on $\mathbb{R}^{1,D-1}$. To describe the motion of the particle we can split the coordinates into $X^0=t$ and $X^i$, where $i=1,...,D-1$. Then its motion is associated to a world-line $\gamma$ which specifies the $X^i\left(X^0\right)$ as a function of $X^0$. We can also describe the particle world-line in relativistic coordinates by using a world-line parameter $\tau$ so that $\gamma$ specifies $X^{\mu}\left(\tau\right)$, where $\mu=0,1,...,D-1$,
\be
\gamma \;:\; \tau \hookrightarrow X^{\mu}\left(\tau\right) \in \mathbb{R}^{1,D-1} \;.
\ee
These ways of describing the particle motion are illustrated in figure \ref{fig:pwl}. 

In Minkowski space the line element is taken as
\be
ds^2 = \eta_{\mu\nu} dX^{\mu} dX^{\nu} \;.
\ee
We can then write an action for the particle, called the Nambu-Goto action, which is just the length of its (time-like) worldline
\be
S_{\mathrm{NG}} = - m \int_{\gamma} \sqrt{-ds^2} = - m  \int_{\gamma}\left(-\dot X^2 \right)^{\frac12} d\tau \;,
\ee
where $\dot X^2 = \eta_{\mu\nu} \dot X^{\mu} \dot X^{\nu}$ and $\dot X^{\mu} = \frac{\partial X^{\mu}}{\partial \tau}$. The constant parameter $m$ will be related to the mass of the particle.

To see that this action correctly describes the motion of the particle we can associate to it a Lagrangian
\be
S_{\mathrm{NG}} = - m \int_{\gamma} d \tau L\left( \tau \right) \;.
\ee
The canonical momentum is then
\be
p_{\mu} = \frac{\partial L}{\partial \dot X^{\mu}} = \frac{m \dot X_{\mu}}{\left(-\dot X^2 \right)^{\half}} \;.
\ee
We therefore arrive at the constraint
\be
p^2 + m^2 = 0 \;,
\ee
which shows that $m$ is indeed the mass of the particle. The equation of motion for $X^{\mu}$ in turn gives
\be
m \ddot X^{\mu} = 0 \;,
\ee
so the particle is freely propagating.

\subsubsection{The Polyakov action for a particle}

We can also consider a different action for the particle, called the Polyakov action. It is defined as
\be
S_{P} = \frac12 \int_{\gamma} d \tau e\left(\tau\right) \left[ \frac{1}{e\left(\tau\right)^2} \dot X^2 - m^2 \right] \;.
\ee
The degree of freedom $e\left(\tau\right)$ is called the world-line metric. Its equation of motion gives 
\be
\dot X^2 + m^2 e\left(\tau\right)^2 = 0 \;.
\ee
Since this is an algebraic constraint we can use it to eliminate the world-line metric in the Polyakov action which gives
\be
S_P = \frac12 \int_{\gamma} d \tau e\left(\tau\right) \left[ -2m^2 \right] = - m  \int_{\gamma}\left(-\dot X^2 \right)^{\frac12} d\tau = S_{\mathrm{NG}} \;.
\ee
Therefore, the Polyakov and Nambu-Goto actions are classically equivalent. However, the utility of the Polyakov action is that it is much easier to quantize.

\subsubsection{The String Worldsheet}

\begin{figure}[t]
\centering
 \includegraphics[width=0.8\textwidth]{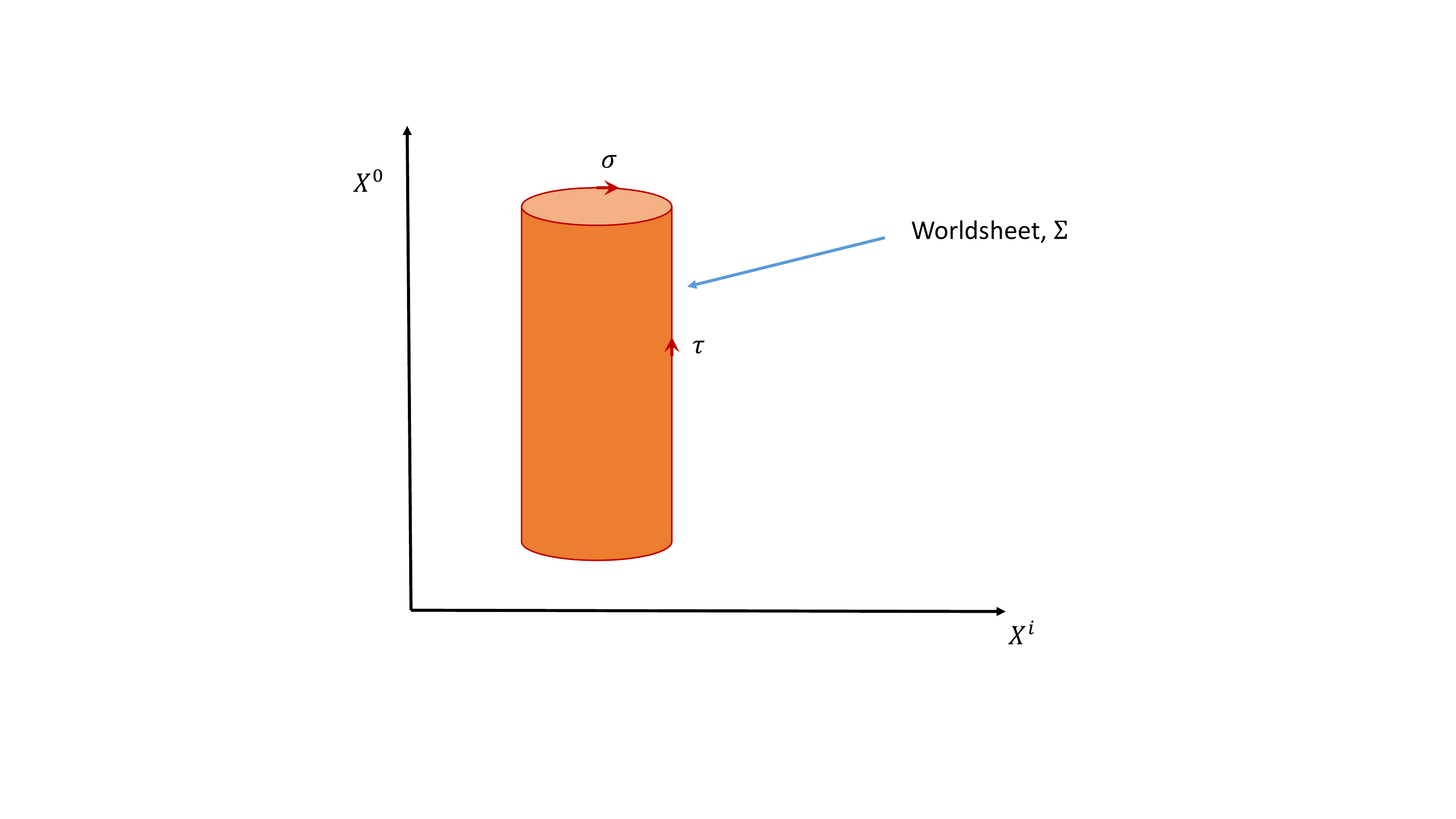}
\caption{Figure illustrating the worldsheet of a string.}
\label{fig:sws}
\end{figure}

We now go through the same process we did for the particle but for a string. A string sweeps out a two-dimensional worldsheet $\Sigma$ parameterised by two coordinates $\left(\sigma,\tau\right)$. So we have
\be
\Sigma \;:\; \left(\sigma,\tau\right) \hookrightarrow X^{\mu}\left(\sigma,\tau\right) \in \mathbb{R}^{1,D-1} \;.
\ee
We take the coordinates to have the ranges
\be
0 \leq \sigma \leq 2\pi \;,\;\; \tau \in \mathbb{R} \;,
\ee
We will mostly be concerned with closed, rather than open, strings. We therefore identify
\be
\sigma \simeq \sigma + 2 \pi \;.
\ee
The coordinates therefore parameterise the string worldsheet as shown in figure \ref{fig:sws}. We will often denote 
\be
\left\{ \sigma, \tau \right\} \equiv \xi^a \;,\;\; a=0,1 \;.
\ee

We want to describe the dynamics of the string through an action. The associated Polyakov action is
\be
S_P = -\frac{T}{2} \int_{\Sigma} d^2 \xi \left( - \mathrm{det}\; h \right)^{\frac12} h^{ab}\left( \xi \right) \partial_a X^{\mu}\left(\xi \right) \partial_b X^{\nu}\left(\xi \right) \eta_{\mu\nu} \;.
\label{Paction}
\ee
Here $h_{ab}\left(\xi \right)$ is the worldsheet metric. $T$ is the string tension, which is often denoted in terms of a parameter $\alpha'$ as
\be
T \equiv \frac{1}{2 \pi \alpha'} \;.
\ee
Note that other similar scales are the string length $l_s$ and the string scale $M_s$, defined as
\be
l_s \equiv \sqrt{\alpha'} \;,\;\; M_s \equiv \frac{1}{2\pi\sqrt{\alpha'}} \;.
\ee
The mass dimensions of the coordinates are $\left[X^{\mu}\right]=-1$ and $\left[\xi^a\right]=0$.

We should think of the worldsheet action (\ref{Paction}) as specifying a two-dimensional theory with scalar fields $X^{\mu}\left(\xi\right)$. Such theories are called sigma models. The space-time in which the string propagates, parameterised by the $X^{\mu}$, is known as the Target space of the worldsheet theory.  The metric on that spacetime, here $\eta_{\mu\nu}$, is the metric on the field space of the scalar fields $X^{\mu}$. So strings propagating in different target spaces have different metrics on the scalar field spaces. 

\subsubsection{The worldsheet symmetries}

The worldsheet theory (\ref{Paction}) is invariant under local diffeomorphisms
\be
\xi^a \rightarrow \tilde{\xi}^a\left(\xi\right) \;.
\label{WSdiff}
\ee
It is also invariant under Weyl transformations, which are defined as
\be
\delta X^{\mu} = 0 \;,\;\; h_{ab} \rightarrow \tilde{h}_{ab} = e^{2 \Lambda\left(\xi\right)} h_{ab} \;.
\label{wet}
\ee
To see this directly note that under (\ref{wet}) we have $\sqrt{-\mathrm{det}\;h} \rightarrow e^{2 \Lambda\left(\xi\right)} \sqrt{-\mathrm{det}\;h}$. 

The worldsheet symmetries can be used to completely fix the worldsheet metric $h_{ab}$. It is worth looking at this generally. For a $D$-dimensional theory, we can count the number of degrees of freedom in the metric, a symmetric tensor, and in the diffeomorphism and Weyl symmetries, these are shown in table \ref{tab:dof}. 
\begin{table}
\centering
\begin{tabular}{|c|c|}
\hline
Object & Degrees of freedom \\
\hline
$h_{ab}$ & $\frac12 D\left(D+1\right)$ \\
\hline
Diffeomorphisms &$D$ \\
\hline
Weyl & 1\\
\hline
\end{tabular}
\caption{Table showing the degrees of freedom in the metric and symmetries in $D$-dimensions.} 
\label{tab:dof}
\end{table}
We see that for $D=2$, so a string, the number of symmetry parameters is the same as the degrees of freedom of the metric. Using the symmetries we can therefore set
\be
h_{ab} = \eta_{ab} = \left( \begin{array}{cc} -1 & 0 \\ 0 & 1 \end{array} \right) \;.
\ee
This is called flat gauge.

It is important to note that even though we can gauge away the metric we must still impose its equations of motion. The equations of motion for the metric correspond to the vanishing of the energy momentum tensor of the theory
\be
T_{ab} = 0 \;,
\label{vcon}
\ee
where
\be
T_{ab} \equiv \frac{4\pi}{\sqrt{-\mathrm{det}\;h}} \frac{\delta S_P}{\delta h^{ab}} \;.
\ee
The resulting constraint (\ref{vcon}) is called a Virasoro constraint, and it will play an important role when we quantize the string.

From table \ref{tab:dof} we see that strings are special extended objects. For $D>2$ we cannot use the symmetries of the Polyakov action to remove the metric degrees of freedom. This makes the extended objects with $D>2$ much more difficult to quantize. From a modern perspective, we do not think of strings as any more `fundamental' than other extended objects, but what makes them special is that they can be readily quantized. 

In flat gauge the Polyakov action reduces to the action of a set of free scalar fields
\be
S_P = \frac{T}{2} \int_{\Sigma} d\sigma d\tau \left[ \left(\partial_{\tau} X \right)^2 - \left(\partial_{\sigma} X \right)^2 \right] \;.
\ee
It is convenient to go to so-called light-cone coordinates
\be
\xi^{\pm} \equiv \tau \pm \sigma \;,\;\; \partial_{\pm} \equiv \frac12 \left(\partial_{\tau}  \pm \partial_{\sigma}  \right)\;.
\ee
In light-cone coordinates the Polyakov action reads
\be
S_P = T  \int_{\Sigma} d\xi^{+}d\xi^- \partial_+ X \partial_- X \;.
\ee
The equations of motion for the $X^{\mu}$ are readily obtained
\be
\partial_+ \partial_- X^{\mu} = 0 \;.
\ee
We can therefore write $X^{\mu}$ as a sum of left-moving and right-moving waves along the string
\be
X^{\mu} = X^{\mu}_L\left(\xi^+\right) + X^{\mu}_R\left(\xi^-\right) \;.
\ee
And we must impose periodic boundary conditions $X^{\mu} \left(\tau,\sigma=0 \right) = X^{\mu} \left(\tau,\sigma=2\pi\right)$. The most general solution is 
\bea
X_R^{\mu} \left(\xi^- \right) &=& \frac12 \left(x^{\mu}+c^{\mu}\right) + \frac12 \alpha' p_R^{\mu}\xi^- + i \sqrt{\frac{\alpha'}{2}} \sum_{n \in \mathbb{Z}\;,\; n \neq 0} \frac{1}{n} \alpha_n^{\mu} e^{-i n \xi^-} \;, \nn \\
X_L^{\mu} \left(\xi^+ \right) &=& \frac12 \left(x^{\mu}-c^{\mu}\right) + \frac12 \alpha' p_L^{\mu}\xi^+ + i \sqrt{\frac{\alpha'}{2}} \sum_{n \in \mathbb{Z}\;,\; n \neq 0} \frac{1}{n} \tilde{\alpha}_n^{\mu} e^{-i n \xi^+} \;.
\label{gemodex}
\eea
Here $x^{\mu}$, $c^{\mu}$, $p_L^{\mu}$, $p_R^{\mu}$, $\alpha_n^{\mu}$ and $\tilde{\alpha}_n^{\mu}$ are constants. Periodicity in $\sigma$ implies
\be
p_L^{\mu} = p_R^{\mu} \equiv p^{\mu} \;. 
\ee 
If we average $X^{\mu}$ over the string we have
\be
q^{\mu} \equiv \frac{1}{2\pi} \int_0^{2\pi} d \sigma X^{\mu} = x^{\mu} + \alpha' p^{\mu} \tau \;.
\ee
So $x^{\mu}$ is the centre of mass position, and $p^{\mu}$ is the target space momentum. 

It will be important for later to note that even after the gauge fixing the worldsheet metric, there are still residual symmetries 
\be
\xi^{\pm} \rightarrow \tilde{\xi}^{\pm}\left(\xi^{\pm}\right) \;.
\label{rckvs}
\ee
These are associated to so-called conformal Killing vectors.\footnote{Note that these are an infinitesimal subset of diffeomorphisms because the transformations restrict to only one coordinate. This is consistent with the earlier counting argument in table \ref{tab:dof}.}

\subsubsection{The string spectrum}
\label{sec:2sp}

So far we have considered the string in a classical sense, but in order to study the spectrum of excitations we need to quantize it. We will do this using so-called light-cone quantization. The starting point is to introduce target-space light-cone coordinates
\be
X^{\pm} \equiv \frac{1}{\sqrt{2}} \left(X^0 \pm X^{D-1} \right) \;,\;\; X^i \;,\;\; i=1,...,D-2 \;.
\ee
The target-space metric then becomes
\be
\eta_{+-} = \eta_{-+} = -1 \;, \;\; \eta_{ij} = \delta_{ij} \;.
\ee
And this gives an inner product
\be
X^2 = -2 X^+ X^- + \dot X^i \dot X^i \;.
\ee

Consider now the expansion for $X^+$, this reads
\bea
X^+\left(\tau,\sigma\right) &=& x^+ + \alpha' p^+ \tau + i \sqrt{\frac{\alpha'}{2}} \sum_{n \in \mathbb{Z} \;,\; n \neq 0} \frac{1}{n} \alpha^+_n e^{-in \xi^-} + i \sqrt{\frac{\alpha'}{2}} \sum_{n \in \mathbb{Z} \;,\; n \neq 0} \frac{1}{n} \tilde{\alpha}^+_n e^{-in \xi^+} \;.
\eea
Recall that we have a residual infinite dimensional symmetry (\ref{rckvs}) after going to light-cone gauge. We can use this to set all the oscillator modes of the $X^+$ to zero. In that gauge we then have
\be
X^+\left(\tau,\sigma\right)  = x^+ + \alpha' p^+ \tau \;.
\ee
Now recall that we must impose the Virasoro constraints (\ref{vcon}) on the theory. It can be shown that these imply
\be
\partial_{\pm} X^- = \frac{1}{\alpha' p^+} \left(\partial_{\pm}X^i \right)^2 \;.
\label{vconxm}
\ee
Therefore, we see that also the $X^-$ oscillators are given in terms of the transverse oscillators in $X^i$. So only the transverse oscillators are independent degrees of freedom. 

The usefulness of the target-space light-cone gauge is therefore that only the $X^i$ contain physically independent oscillators. This is useful because it automatically projects out two polarizations of the string which are unphysical. This is completely analogous to how a Maxwell field in four dimensions only has two physical polarizations. 

The action in light-cone gauge reads
\bea
S_{LC} &=& \frac{1}{4\pi\alpha'} \int_{\Sigma} d\tau d\sigma \left[ \left(\partial_{\tau} X^i \right)^2 - \left(\partial_{\sigma} X^i \right)^2 + 2\left( - \partial_{\tau} X^+ \partial_\tau X^- + \partial_{\sigma} X^+ \partial_{\sigma} X^- \right)\right] \nn \\
&=& \frac{1}{4\pi\alpha'} \int_{\Sigma} d\tau d\sigma \left[ \left(\partial_{\tau} X^i \right)^2 - \left(\partial_{\sigma} X^i \right)^2 \right] - \int d\tau p^+ \partial_{\tau} q^- \nn \\
&\equiv& \int d\tau L \;,
\eea
where we define
\be
q^- \equiv \frac{1}{2\pi} \int_0^{2\pi} d \sigma X^- \;.
\ee
From this Lagrangian we can define canonical momenta
\be
p_- \equiv \frac{\partial L}{\partial \dot q^-} = - p^+ \;,\;\; \Pi_i \equiv \frac{\partial L}{\partial \dot X^i}  = \frac{\dot X_i}{2 \pi \alpha'} \;.
\ee
We then quantize the theory by introducing the canonical commutation relations
\be
\left[ X^{\mu}\left(\tau,\sigma\right),\Pi^{\mu}\left(\tau,\sigma'\right)\right] = i \eta^{\mu\nu} \delta\left(\sigma -\sigma' \right) \;,
\ee
which give
\bea
\left[x^i,p^i \right] &=& i \delta_{ij} \;, \nn \\
\left[ p^+,q^-\right] &=& i \;, \nn \\
\left[\alpha_m^i,\alpha_n^j \right] &=& m \delta_{n+m,0}\delta_{ij} \;, \nn \\
\left[\tilde{\alpha}_m^i,\tilde{\alpha}_n^j \right] &=& m \delta_{n+m,0}\delta_{ij} \;.
\eea
We therefore follow the usual procedure for quantization, as in quantum field theory, by promoting the oscillator modes to operators acting on a Hilbert space. The $\alpha^i_{-n}$ with $n>0$ are creation operators acting on a vacuum state $\left|0,p\right>$. While the $\alpha^i_{n}$ with $n>0$ are annihilation operators.

Recall that there are no oscillators to quantize for $X^+$, while the $X^-$ oscillators are given in terms of the $X^i$. Explicitly this reads
\be
\alpha^-_n = \frac{1}{2\sqrt{2 \alpha'}p^+} \sum_{m=-\infty}^{m=\infty} \alpha^i_{n-m}\alpha^i_m \;.
\label{lcosc}
\ee
When we quantize the theory the ordering of the $\alpha$'s matters, and so we should write things in terms of normal ordered products and a normal ordering constant $a$ which we need to determine
\be
\alpha^-_n = \frac{1}{2\sqrt{2 \alpha'}p^+} \left( \sum_{m=-\infty}^{m=\infty}: \alpha^i_{n-m}\alpha^i_m: - a \delta_{n,0}\right) \;,
\ee
where
\bea
: \alpha^i_{m}\alpha^i_n:  \equiv \left\{ \begin{array}{cc} \alpha^i_{m}\alpha^i_n & \mathrm{for\;} m \leq n \\ \alpha^i_{n}\alpha^i_m & \mathrm{for\;} n < m \end{array} \right. \;.
\eea
This is the canonical quantization procedure. 

\subsubsection{Criticality and Lorentz Invariance}

The quantization of the theory was performed in special target-space light-cone coordinates. It is therefore not clear that the quantum theory respects Lorentz invariance. Indeed, we will see that requiring the preservation of target-space Lorentz invariance also in the quantum theory will place rather stringent constraints on the theory. 

In general, the generators of Lorentz transformations are
\be
J^{\mu\nu} = \int_0^{2\pi} d\sigma \left( X^{\mu} \Pi^{\nu} - X^{\nu}\Pi^{\mu}\right) \equiv l^{\mu\nu} + E^{\mu\nu} + \tilde{E}^{\mu\nu} \;,
\label{amg}
\ee
where
\bea
l^{\mu\nu} &=& x^{\mu} p^{\nu} - x^{\nu}p^{\mu} \;, \nn \\
E^{\mu\nu} &=& -i \sum_{n=1}^{\infty} \frac{1}{n} \left( \alpha_{-n}^{\mu} \alpha_n^{\nu} -  \alpha_{-n}^{\nu} \alpha_n^{\mu}\right) \;, \nn \\
\tilde{E}^{\mu\nu} &=& -i \sum_{n=1}^{\infty} \frac{1}{n} \left( \tilde{\alpha}_{-n}^{\mu} \tilde{\alpha}_n^{\nu} -  \tilde{\alpha}_{-n}^{\nu} \tilde{\alpha}_n^{\mu}\right) \;.
\eea
Now the Lorentz algebra reads
\be
\left[J^{\mu\nu},J^{\rho\sigma} \right] = i \eta^{\mu\rho}J^{\nu\sigma} + i \eta^{\nu \sigma} J^{\mu \rho} - i \eta^{\mu\sigma} J^{\nu \rho} -i \eta^{\nu\rho} J^{\mu\sigma}\;.
\ee
In particular,
\be
\left[J^{-i},J^{-j} \right] = i \eta^{--}J^{ij} + i \eta^{ij} J^{--} - i \eta^{-j} J^{i-} -i \eta^{i-} J^{-j} = 0 \;.
\ee
However, an explicit calculation yields
\be
\left[J^{-i},J^{-j} \right]  = -\frac{1}{\left(p^+\right)^2} \sum_{m=1}^{\infty} \Delta_m \left( \alpha_{-m}^i \alpha^j_{m} - \alpha_{-m}^j \alpha_{m}^i\right) + \tilde{\left(...\right)} \;,
\label{conan}
\ee
where
\be
\Delta_m \equiv m \frac{26-D}{12} + \frac{1}{m}\left[ \frac{D-26}{12} + 2\left(1-a\right)\right]\;.
\ee
Therefore, we find that maintaining Lorentz invariance at the quantum level requires
\be
D=26 \;,\;\; a=1 \;.
\label{crit}
\ee
This is a rather remarkable result. It shows that while the classical string is consistent in any number of dimensions, the quantum bosonic string is only consistent in 26 dimensions.\footnote{This is not quite the full picture, it is possible to consider non-critical strings, so with a different target-space dimensionality. For $D<26$ this is achieved by breaking Lorentz invariance through a linear dilaton background. For $D>26$ this leads to a potential for the dilaton. Such theories however are as yet not well-understood and so we will not discuss them here.} The restriction on the number of dimensions is called criticality.  In fact, there are many different ways to arrive at this criticality result. Note that for the Superstring, so a string theory incorporating supersymmetry, the critical number of dimensions changes to 10.

\subsubsection{The quantum string spectrum}

Having quantized the string we can now examine its spectrum. The classical Hamiltonian is given by
\bea
H &=& p_- \dot q^- + \int_0^{2\pi} d\sigma \Pi_i \dot X^i - L \nn \\
 &=& \frac{1}{4\pi\alpha'}\int_0^{2\pi} d\sigma \left[ \left( \partial_{\tau}X^i\right)^2 + \left( \partial_{\sigma}X^i\right)^2 \right] \nn \\
 & =& \frac{\alpha'}{2} p^i p^i + \frac12 \sum_{n=-\infty}^{\infty} \left( \alpha^i_{-n} \alpha^i_n + \tilde{\alpha}^i_{-n} \tilde{\alpha}^i_n\right) \;.
\label{hami}
\eea
This is related to $X^-$ through the Virasoro constraint (\ref{vconxm}) 
\be
\partial_{\tau} X^- = \frac{1}{2p^+\alpha'}\left[ \left( \partial_{\tau}X^i\right)^2 + \left( \partial_{\sigma}X^i\right)^2 \right] \;,
\ee
so that
\be
p^- = \frac{1}{2\pi\alpha'} \int_0^{2\pi} d\sigma \partial_{\tau}X^- = \frac{H}{\alpha' p^+} \;.
\label{pppmma}
\ee
In quantizing we need to normal order the $\alpha$, so we have 
\be
p^+ p^- = \frac{1}{\alpha'} \left[ N_{\perp} + \tilde{N}_{\perp} - 2a + \frac{\alpha'}{2} p^i p^i\right] \;,
\ee
where we define
\be
N_{\perp} \equiv \sum_{n=1}^{\infty} : \alpha_{-n}^i \alpha_n^i : \;,\;\; \tilde{N}_{\perp} \equiv \sum_{n=1}^{\infty} : \tilde{\alpha}_{-n}^i \tilde{\alpha}_n^i :  \;.
\ee

The mass in the target-space is given by
\be
M^2 = -p^2 = 2 p^+p^- - p^i p^i = \frac{2}{\alpha'} \left( N_{\perp} + \tilde{N}_{\perp} - 2a\right) \;.
\ee
Finally, we note that for the closed string we have a symmetry of translations along $\sigma$, and this can be shown to imply the level matching condition
\be
N_{\perp}  = \tilde{N}_{\perp} \;,
\label{lvlmatch}
\ee
so the expression for the mass becomes
\be
M^2 = \frac{4}{\alpha'} \left( N_{\perp}-1 \right) \;.
\ee
Here we used the fact that the normal ordering constant is fixed by criticality to $a=1$ (\ref{crit}).

Now we can examine the spectrum on the string according to how many oscillators are present:

\subsubsection*{$N_{\perp}=0$}

Here we have
\be
M^2 = -\frac{4}{\alpha'} \;.
\ee
This is a tachyonic mode, which means that it is signaling an instability in the bosonic string. For superstrings this mode will be absent, and so such strings are stable. The tachyon will not play an important role for us and so we will not discuss it further.

\subsubsection*{$N_{\perp}=1$}

In this case we have 
\be
M^2 = 0 \;.
\ee
This is therefore the massless spectrum of the quantum bosonic string. It is given by 
\be
\xi_{ij} \tilde{\alpha}^i_{-1} \alpha^j_{-1} \left|0,p\right> \;,\;\; i,j=1,...,24 \;.
\ee
We can decompose the tensor $\xi_{ij}$ into irreducible representations of $SO(24)$ as
\be
\xi_{ij} = g_{\left(ij\right)} + B_{\left[ij\right]} + \Phi \;, 
\label{decomp}
\ee
where $g_{\left(ij\right)}$ is traceless symmetric, $ B_{\left[ij\right]} $ is anti-symmetric and $\Phi$ is a scalar (corresponding to the trace).

We therefore find a massless, transversely polarized, symmetric tensor field $g_{ij}$. This is a graviton! Indeed, it can be shown that it has spin 2. So the quantum bosonic string contains gravitational modes in its spectrum. 

We also find an anti-symmetric massless tensor $ B_{\left[ij\right]} $ termed the Kalb-Ramond field. We will return to this field later when we discuss compactifications. 

The massless scalar $\Phi$ is called the dilaton. It actually determines the coupling constant for string interactions. 

\subsubsection*{$N_{\perp}>1$}

These are massive oscillator string modes, with a mass starting at $M_s$. They are crucial for showing the finiteness of string theory scattering amplitudes. But we will not study them further.

\subsubsection{The low-energy effective action}

We see that the string has a set of massless fields and some very massive fields. 
We can consider non-trivial vacuum configurations of the massless fields $G_{\mu\nu}$, $B_{[\mu\nu]}$ and $\Phi$. These can be thought of as coherent states of the string excitation modes. In such a non-trivial background the string worldsheet theory is modified to
\bea
\label{stract}
S_P &=& -\frac{T}{2} \int_{\Sigma} d^2 \xi \left( - \mathrm{det}\; h \right)^{\frac12} h^{ab}\left( \xi \right) \partial_a X^{\mu}\left(\xi \right) \partial_b X^{\nu}\left(\xi \right) \eta_{\mu\nu} \nn \\
&\rightarrow&  -\frac{T}{2} \int_{\Sigma} d^2 \xi \left( - \mathrm{det}\; h \right)^{\frac12} \left[ h^{ab}\left( \xi \right) \partial_a X^{\mu}\left(\xi \right) \partial_b X^{\nu}\left(\xi \right) G_{\mu\nu} \right. \nn \\ 
& & + \left. i B_{[\mu\nu]} \partial_a X^{\mu}\left(\xi \right) \partial_b X^{\nu}\ \epsilon^{ab} + \alpha' \Phi R^{(2)}\right] \;.
\label{pagen}
\eea
Here $\epsilon^{ab}$ is the anti-symmetric unit tensor, and $R^{(2)}$ is the two-dimensional Ricci scalar. The action (\ref{stract}) is not, in general, conformally invariant at the quantum level. We can think of the fields $G_{\mu\nu}$, $B_{[\mu\nu]}$ and $\Phi$ as coupling constants in the action which will run. The worldsheet theory is difficult to quantize, but we can study it for small curvature backgrounds by expanding about flat space (as well as some specific other backgrounds for which we can solve the worldsheet theory). This expansion is controlled by $\alpha'$ since each derivative, associated to higher curvature, will by dimensional analysis be accompanied with a power of $\alpha'$.

Their expanded $\beta$-functions take the form
\bea
\label{betafun}
\beta^G_{\mu\nu} &=& \alpha' R_{\mu\nu} + 2 \alpha' \nabla_{\mu} \nabla_{\nu} \Phi - \frac{\alpha'}{4} H_{\mu \lambda \kappa} H_{\nu}^{\lambda\kappa} + {\cal O}\left(\left(\alpha' \right)^2 \right)  \;, \nn \\
\beta^B_{\mu\nu} &=& -\frac{\alpha'}{2} \nabla^{\gamma}H_{\gamma\mu\nu} + \alpha' \left(\nabla^{\gamma}\Phi \right) H_{\gamma\mu\nu} + {\cal O}\left(\left(\alpha' \right)^2 \right)  \;, \nn \\
\beta^{\Phi} &=& -\frac{\alpha'}{2} \nabla^2 \Phi + \alpha' \nabla_{\gamma} \Phi \nabla^{\gamma} \Phi - \frac{\alpha'}{24} H_{\mu\nu\gamma}H^{\mu\nu\gamma} + {\cal O}\left(\left(\alpha' \right)^2 \right)  \;.
\eea 
The $H$ is defined as
\be
H_{\mu\nu\rho} \equiv \partial_{[\mu}B_{\nu\rho]} \;,
\ee
where the square brackets denote anti-symmetrization of the indices. 
Consistency of string theory requires that the world-sheet theory be conformally invariant. This means that the $\beta$-functions (\ref{betafun}) should vanish. The resulting constraints can be written as the equations of motion of an action. This action is then the low-energy effective theory describing the massless fields of the string. The effective action for this theory is given by
\be
S_D = 2\pi M_s^{D-2} \int d^D X \sqrt{-G} e^{-2\Phi} \left( R - \frac{1}{12} H_{\mu\nu\rho}H^{\mu\nu\rho} + 4 \partial_{\mu} \Phi \partial^{\mu} \Phi \right) \;.
\label{steff}
\ee
We have written the action for general dimension $D$ so that we can treat both the bosonic string $D=26$ and the superstring $D=10$, since both contain this massless spectrum. The action is composed purely of the kinetic terms for the fields, as expected since they are massless.

Note that in the case of the bosonic string there is also a tachyon mode, which we neglect here. We are only studying the dynamics of the massless modes. This is not really consistent, but can be done completely consistently for the superstring with $D=10$.

\subsection{First encounter with the Swampland Distance Conjecture}
\label{sec:fcsdc}

We have seen that the bosonic string lives in 26 dimensions. The superstring lives in 10 dimensions. These both seem to be directly incompatible with the observed universe. However, this need not be the case. The point is that the additional dimensions may be compact and small, so that they have yet to be observed. This naturally leads to thinking about string theory in a space-time which has a compact direction. The simplest such setting is the case where one of the dimensions is in the shape of a circle. We will study this in this section and this will lead to our first encounter with a Swampland criterium: the distance conjecture.  

\subsubsection{Compactification of field theory on a circle}
\label{sec:compft}

We consider $D=d+1$ dimensional space-time. The spatial direction $X^d$ is taken to be compact in the shape of a circle so is periodically identified
\be
X^d \simeq X^d + 1 \;.
\ee
We are interested in looking at the effective theory in the $d$ non-compact dimensions. 

First, recall that we are working in Planck units, which in this case we therefore set as $M_p^d=1$, where $M_p^d$ denotes the $d$-dimensional Planck mass. The periodicity of $X^d$ is set to one in those units.  

We can write the metric on the $D$-dimensional space as
\be
ds^2 \equiv G_{MN} dX^M dX^N = e^{2\alpha\phi}g_{\mu\nu} dX^{\mu} dX^{\nu} + e^{2 \beta\phi} \left( dX^d\right)^2 \;.
\label{metan}
\ee
So here we have introduced the coordinates $X^M$ which are $D$-dimensional, so $M=0,...,d$, while $\mu=0,...,d-1$. The $D$-dimensional metric is $G_{MN}$ and we take it as a product metric. The $d$-dimensional metric is $g_{\mu\nu}$. In practice we will take this to be $\eta_{\mu\nu}$ but we keep it general for now. The metric has a parameter $\phi$ which can be regarded as a $d$-dimensional scalar field. 
The constants $\alpha$ and $\beta$ are chosen to be
\be
\alpha^2 = \frac{1}{2\left(d-1\right)\left(d-2\right)} \;,\;\; \beta = -\left(d-2\right)\alpha \;.
\ee

Let us look at the circumference of the circle, denoted $2\pi R$, it is given by
\be
2\pi R \equiv \int_0^{1} \sqrt{G_{dd}} dX^d = e^{\beta \phi} \;.
\label{Rphi} 
\ee
We see that the radius of the circle is a dynamical field in $d$-dimensions. We will be interested in the behaviour of the $d$-dimensional theory under variations of the expectation value of the field $\phi$, which amounts to variations of the size of the circle.

The first thing we want to do is decompose the $D$-dimensional Ricci scalar $R^D$ for the metric (\ref{metan}). We have
\be
\int d^DX \sqrt{-G}R^D = \int d^dX \sqrt{-g} \left[R^d - \frac12\left(\partial \phi \right)^2 \right] \;.
\label{ddim} 
\ee
We observe that indeed $\phi$ picks up dynamics, and that it is canonically normalized. 

Now consider introducing a massless $D$-dimensional scalar field $\Psi$. Since the $d^{\mathrm{th}}$ dimension is periodic so must $\Psi$ be, therefore we can decompose it as
\be
\Psi\left(X^M\right) = \sum_{n=-\infty}^{\infty} \psi_n\left(X^{\mu}\right)e^{2\pi i n X^d} \;.
\label{KKexp}
\ee
The modes $\psi_n$ are $d$-dimensional scalar fields. The mode $\psi_0$ is called the zero-mode of $\Psi$, while the $\psi_n$ are called Kaluza-Klein (KK) modes. Note that the momentum is quantized along the compact direction
\be
-i\frac{\partial}{\partial X^d} \Psi = 2 \pi n \Psi \;.
\ee
For simplicity we now restrict to $g_{\mu\nu}=\eta_{\mu\nu}$. Since $\Psi$ is massless in $D$-dimensions, its equation of motion is
\be
\partial^M\partial_M \Psi = \left(e^{-2\alpha \phi} \partial^{\mu}\partial_{\mu} + e^{-2\beta\phi} \partial^2_{X^d}\right) \Psi=0 \;.
\ee
This gives the equations of motion for the $\psi_n$ modes
\be
\left[\partial^{\mu} \partial_{\mu} - \left(\frac{1}{2\pi R}\right)^{2} \left(\frac{1}{2\pi R}\right)^{\frac{2}{d-2}} \left(2 \pi n \right)^2 \right] \psi_n = 0 \;.
\ee
We can therefore read off the mass of the KK modes as 
\be
M_n^2 = \left(\frac{n}{R}\right)^2  \left(\frac{1}{2\pi R}\right)^{\frac{2}{d-2}} \;.
\label{kkmasf}
\ee
So in the $d$-dimensional theory the KK modes are a massive tower of states with increasing masses as in (\ref{kkmasf}).

\subsubsection{Compactification of string theory on a circle}

Now let us repeat this exercise in string theory by considering strings on a circle of radius $R$.  
We would like to connect with our results in section \ref{sec:focst}, but those were performed for a metric 
\be
ds^2=\eta_{MN} dX_{(s)}^M dX_{(s)}^N \;,
\label{sfmet}
\ee
rather than (\ref{metan}). The subscripts on $X_{(s)}^M$ are to remind us that we are working with the metric (\ref{sfmet}). For now we will proceed with the metric (\ref{sfmet}) and take the $X_{(s)}^d$ direction as $R$-periodic
\be
X_{(s)}^d \simeq X_{(s)}^d + 2 \pi R \;.
\ee
We will reconnect to the metric (\ref{metan}) later.

\begin{figure}[t]
\centering
 \includegraphics[width=0.8\textwidth]{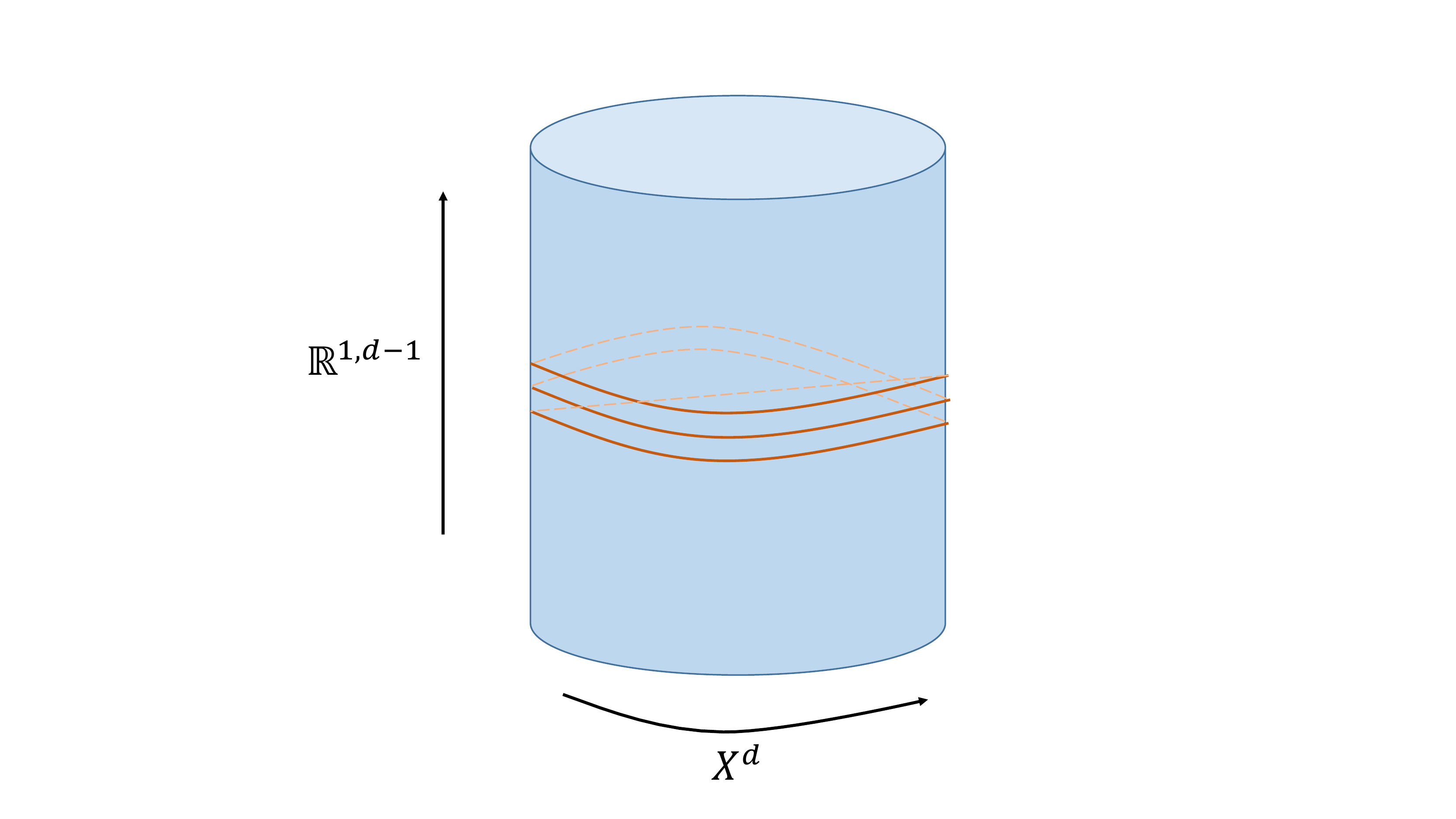}
\caption{Figure showing a string winding around a circler dimension 3 times.}
\label{fig:wi}
\end{figure}

We consider the bosonic mode expansion, as in (\ref{gemodex}), but now we will not impose yet $X_{(s)}^{M}\left(\sigma+2\pi,\tau \right) = X_{(s)}^{M}\left(\sigma,\tau \right)$ on the linear terms in $\sigma$. So we have
\be
X_{(s)}^{M}\left(\tau,\sigma\right) = x^{\mu} +\alpha' p^{M} \tau + \frac{\alpha'}{2}\left(p^{M}_L - p^{M}_R \right) \sigma + \mathrm{oscillators} \;.
\ee
We have allowed here for independent left-moving and right-moving momenta, and the overall momentum of the string is half their sum
\be
p^{M} = \frac12 \left( p^{M}_R + p^{M}_L \right)\;.
\ee
Recall that because the $X^d$ direction is compact this is quantized. The appropriate quantization, as we will soon see, is
\be
p^d = \frac{n}{R} \;.
\ee
In the non-compact space we imposed $X_{(s)}^{\mu}\left(\sigma+2\pi,\tau \right) = X_{(s)}^{\mu}\left(\sigma,\tau \right)$ which lead to $p^{\mu}_R=p^{\mu}_L$, but for a circle we may have a winding string
\be
X_{(s)}^{d}\left(\sigma+2\pi,\tau \right) = X_{(s)}^{d}\left(\sigma,\tau \right) + w 2 \pi R \;,
\ee
with $w \in \mathbb{Z}$. The string is wrapping around the circle $w$ times, as illustrated in figure \ref{fig:wi}. 
For such a winding string we therefore have
\be
\frac{\alpha'}{2}\left(p_L^d - p_R^d\right) = w R \;.
\ee

Now consider the mass spectrum for the string on such a background. We again go to target-space light-cone gauge. The Hamiltonian (\ref{hami}) now reads
\be
H = \frac{\alpha'}{2} \left[ \frac14 \left(p_L^d - p_R^d\right)^2 + p^{\alpha} p^{\alpha} + \left( p^d\right)^2 \right] + \left(N_{\perp} + \tilde{N}_{\perp} - 2 \right)  \;,
\ee
where we split the index $i =\left\{\alpha,d\right\}$.
Note that we no longer have the level matching condition (\ref{lvlmatch}), but instead have
\be
N_{\perp} - \tilde{N}_{\perp} = n w\;.
\ee
Then the $d$-dimensional mass is given by $-p_{\mu} p^{\mu} = 2p^+ p^- -p^{\alpha} p^{\alpha}$ which, for states with no oscillators excited, leads to 
\be
\left(M^{(s)}_{n,w}\right)^2 = \left(\frac{n}{R}\right)^2 + \left(\frac{w R}{\alpha'}\right)^2 \;.
\label{smasc}
\ee

We would like to connect this result with the effective action (\ref{ddim}). However, to do that we need to change from the metric (\ref{sfmet}) to the metric (\ref{metan}). This is called going from the string frame to the Einstein frame. The difference is the factor of $e^{2 \alpha \phi}$ multiplying the $g_{\mu\nu}$ directions. To get the Einstein frame mass for the states we simply note that $p^{\mu}p_{\mu}$ has one inverse factor of the metric and so we need to multiply masses by a factor of $e^{2 \alpha \phi} = \left(\frac{1}{2\pi R}\right)^{\frac{2}{d-2}}$. 

This does not quite do the full job. If we look at the effective action coming from string theory (\ref{steff}) we see that there is an overall factor of the exponential of the dilaton $e^{-2\Phi}$. An important object is the $d$-dimensional dilaton $\Phi^d$ defined as
\be
\Phi^d \equiv \Phi - \frac12 \log \left(2 \pi R M_s \right)\;.
\ee
We would like to look at variations of $R$ which keep $\Phi^d$ fixed. This means that we must vary 
\be
e^{-2\Phi} \sim \frac{1}{2 \pi R M_s} \;.
\ee
But if we consider the definition of the $d$-dimensional Planck mass $M_p^d$ coming from the string effective action (\ref{steff}) 
\be
\label{mpmsdire}
\frac{\left(M_p^d\right)^{d-2}}{2} \equiv 2\pi M_s^{D-2} e^{-2\Phi} \;,
\ee
we see that in order to stay in the Einstein frame $M_p^d=1$ we have to choose our units such that 
\be
M_s \sim \left( 2 \pi R \right)^{\frac{1}{d-2}} \;.
\label{MsR}
\ee
This will then affect the mass of the winding modes in (\ref{smasc}) because of the factor of $\alpha'$.

Performing the change of frames then finally leads to the Einstein frame mass
\be
\left(M_{n,w}\right)^2 = \left(\frac{1}{2\pi R}\right)^{\frac{2}{d-2}} \left(\frac{n}{R}\right)^2  + \left(2\pi R\right)^{\frac{2}{d-2}} \left(\frac{w R}{\alpha'_0}\right)^2 \;,
\label{essmasc}
\ee
where the subscript on $\alpha'_0$ denotes that the $R$ scaling has been taken out.
We see that this indeed matches the simple field theory calculation for the KK masses (\ref{kkmasf}). 

\subsubsection{The Swampland Distance Conjecture}

\begin{figure}[t]
\centering
 \includegraphics[width=0.8\textwidth]{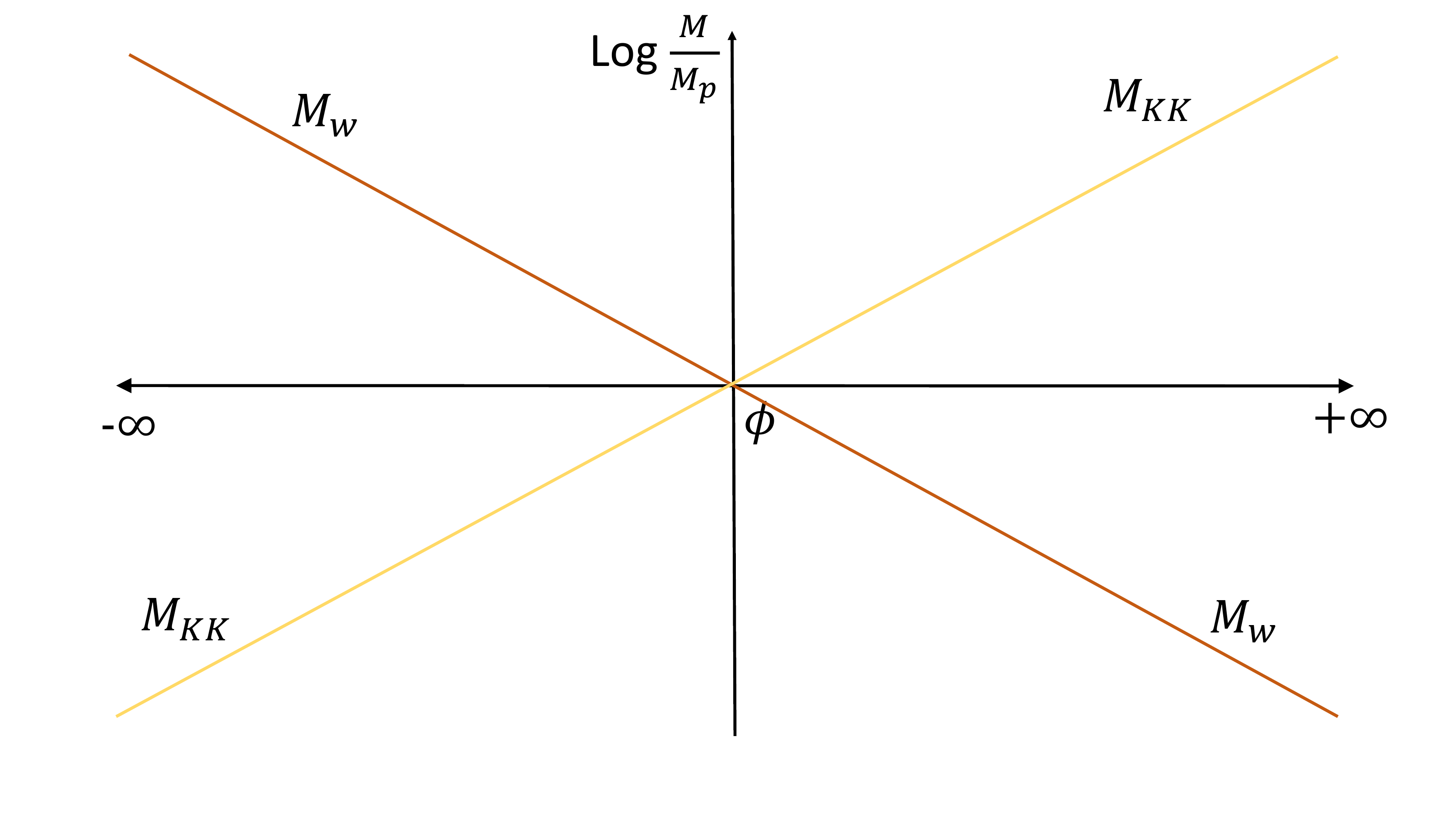}
\caption{Figure showing the mass scale, on a log plot, for the KK and winding towers as a function of the scalar field $\phi$ expectation value. The gradient of the slope is the exponent $\alpha$. The $\mathbb{Z}_2$ symmetry in the figure is due to T-duality.}
\label{fig:swtd}
\end{figure}

We can now study the $d$-dimensional effective theory. The action is given in (\ref{ddim}), and this must be supplemented by the spectrum (\ref{essmasc}). We are particularly interested in how the spectrum of states behaves under variations of the expectation value of the field $\phi$. This is easy to determine from the simple relation (\ref{Rphi}). The possible expectation values of the field $\phi$ define a field space ${\cal M}_{\phi}$, which in this case has one infinite real dimension. So we can consider
\be
{\cal M}_{\phi} \;: \: -\infty < \phi < \infty \;.
\ee
Let us define a variation of $\phi$ from some initial value $\phi_i$ to some final value $\phi_f$ as
\be
\Delta \phi = \phi_f - \phi_i \;.
\ee 
We now note that there are two infinite towers of massive states in this theory. The tower of KK modes, with masses given by $M_{n,0}$ in (\ref{essmasc}), and a tower of winding modes given by $M_{0,m}$. We can associate to each tower a mass scale, which is the universal factor multiplying the integers $n$ and $m$. Using (\ref{Rphi}) we can write these mass scales as
\be
\label{swcircma}
M_{KK} \sim e^{\alpha \phi} \;,\;\; M_{w} \sim e^{-\alpha \phi} \;,
\ee
where 
\be
\alpha = \sqrt{2} \left(\frac{d-1}{d-2} \right)^{\frac12} > 0 \;.
\ee
We therefore can make the following observation. For any $\Delta \phi$ there exists an infinite tower of states, with some associated mass scale $M$, which becomes light at an exponential rate in $\Delta \phi$
\be
M\left(\phi_i + \Delta \phi\right) \sim M\left(\phi_i\right) e^{-\alpha \left|\Delta \phi\right|} \;.
\label{exsc}
\ee
This is illustrated in figure \ref{fig:swtd}. 
There are some important things to note about this observation
\begin{itemize}
\item The tower of states which becomes light is the KK tower if $\Delta \phi < 0$ while it is the winding tower if  $\Delta \phi > 0$. So some tower of states always become light no matter what the sign of $\Delta \phi$ is. 
\item The behaviour (\ref{exsc}) is deeply string theoretic. It is not true in quantum field theory because one set of states are winding states which are absent in field theory.
\item The product of the mass scales of the two towers is independent of $\phi$. 
\item The exponent $\alpha$ in the mass is a constant of order one. 
\item The field $\phi$ is canonically normalized, so the behaviour of the mass scales is exponential in the proper distance in $\phi$ field space. 
\item If $\left|\Delta \phi \right| \rightarrow \infty$ then an infinite number of states become massless, which means that there is no description of that locus in a $d$-dimensional quantum field theory.\footnote{We can describe it as a $D$-dimensional theory. Remarkably, this is true for either a very large or very small radius of the circle.}
\end{itemize}
The last point has a continuous analogue. If we consider an effective field theory which has a cutoff $\Lambda$ below the mass scale of an infinite tower of states, then this field theory can only hold for a finite range of expectation values of $\phi$. 

The behaviour (\ref{exsc}) is very interesting and it is natural to wonder if there is a deep reason behind it, and if so, then if it is a general property of string theory. There is good reason to expect that the answer is positive to both of these questions. One clue is in the origin of the two towers, the KK and winding modes. These towers a deeply related, indeed there is a ${\mathbb Z}_2$ symmetry which interchanges them. This is called T-duality, and it is most directly seen in the string frame where we observe that the mass spectrum (\ref{smasc}) is invariant under the action
\be
\mathrm{T-duality}\;:\; R \leftrightarrow \frac{\sqrt{\alpha'}}{R} \;.
\ee
It can be shown that this is not only a symmetry of the mass spectrum, but of the full string theory. In fact it can be embedded into a gauge symmetry which becomes manifest at the self-dual radius $R =  \frac{\sqrt{\alpha'}}{R}$. Duality is a very deep property of string theory. There are many more dualities than T-duality. Indeed, all known string theories are themselves related by dualities. It is then natural to expect that there are many different towers of states which are dual, and this duality is such that as one moves in the parameter space of the theory, which in string theory means in the scalar field space, the product of the mass scale of the dual towers stays constant and so one must become light in any direction. As we move an infinite distance in parameter space the tower must become massless.

This kind of reasoning, and various simple examples in string theory, let to the proposal of the Swampland Distance Conjecture (SDC) in \cite{Ooguri:2006in}. The conjecture is at heart analogous to (\ref{exsc}) but can be phrased more generally and precisely as follows. 

\begin{tcolorbox}
{\bf Swampland Distance Conjecture } \;\cite{Ooguri:2006in}
{\it 
\begin{itemize}
\item Consider a theory, coupled to gravity, with a moduli space $\cM$ which is parametrized by the expectation values of some field $\phi^i$ which have no potential. Starting from any point $P \in \cM$ there exists another point $Q \in \cM$ such that the geodesic distance between $P$ and $Q$, denoted $d\left(P,Q\right)$, is infinite. 
\item There exists an infinite tower of states, with an associated mass scale $M$, such that 
\be
\label{sdc}
M\left(Q\right) \sim M\left(P\right) e^{-\alpha d\left(P,Q\right) } \;,
\ee
where $\alpha$ is some positive constant.
\end{itemize}
}
\end{tcolorbox}

\begin{figure}[t]
\centering
 \includegraphics[width=0.8\textwidth]{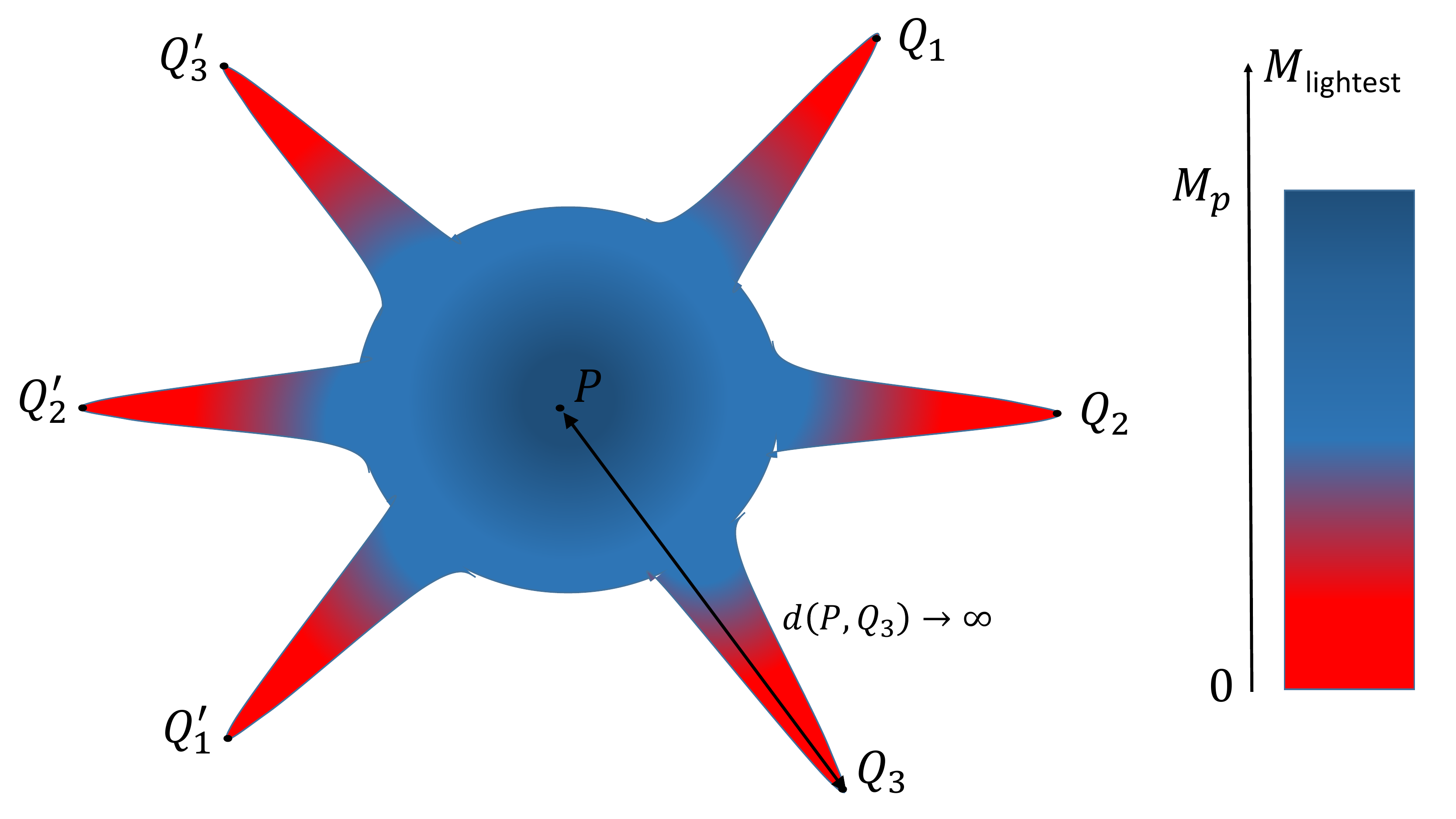}
\caption{Figure showing a schematic illustration of string moduli space. The distance from any point $P$ in the bulk of moduli space to any point $Q_i$, or $Q'_i$, is infinite. $M_{\mathrm{lightest}}$ denotes the mass scale of the lightest tower of states in the theory, and this goes to zero at any point $Q_i$. Points $Q_i$ and $Q'_i$ are related by duality and the light towers of states are interchanged between them. }
\label{fig:sdcg}
\end{figure}

Note that because this is an asymptotic statement about infinite distance $d\left(P,Q\right)\rightarrow \infty$ the mass scale value at $P$ is not so important. The behaviour of the conjecture is illustrated schematically in figure \ref{fig:sdcg}. We will discuss the Swampland Distance Conjecture is more detail in sections \ref{sec:sdc} and \ref{sec:stringcomp}. 

This is our first encounter with a Swampland conjecture. It typifies many of the general properties of conjectures about the Swampland. 
\begin{itemize}
\item It is supported by explicit constructions in string theory.
\item There are some general, but imprecise, arguments for why it may be expected to hold generally.
\item It goes beyond quantum field theory and general relativity.
\end{itemize}

It is worth thinking generally about the Swampland Distance Conjecture. From the perspective of quantum field theory it is quite surprising. We have seen that the conjecture can only hold due to string theory, or more generally, quantum gravity. Typically, the mass scale associated to such physics is $M_p$, and one might expect that working at energy scales far below the Planck mass would mean that we lose sensitivity to such physics. But the conjecture says that if in the bulk of moduli space, the blue region in figure \ref{fig:sdcg}, the tower of states has a mass scale around the Planck mass $M_p$, then at large field expectation values this mass scale is exponentially lower than $M_p$. Therefore, it claims that the naive application of decoupling in effective quantum field theory breaks down at an exponentially lower energy scale than expected whenever a field develops a large expectation value. We will go into much more detail regarding aspects of this conjecture in the review, but for now let us proceed to a first encounter with a related conjecture. 

\subsection{First encounter with the Weak Gravity Conjecture}
\label{sec:fewgcst}

In the previous subsection we studied compactification of string theory on a circle, but have neglected a whole sector of physics within such compactifications associated to gauge fields. In this section we will study this physics and encounter the Weak Gravity Conjecture \cite{ArkaniHamed:2006dz}. 

\subsubsection{Compactification with gauge fields}
\label{sec:fecompga}

In section \ref{sec:fcsdc} we considered the reduction of string theory on a metric (\ref{metan}). The metric encoded the parameter $\phi$ which became a dynamical field in the lower $d$-dimensional theory. There is another degree of freedom in the metric which becomes a $d$-dimensional gauge, rather than scalar, field $A_{\mu}$. It is associated to mixed terms as
\be
ds^2 = e^{2 \alpha \phi} g_{\mu\nu} dX^{\mu} dX^{\nu} + e^{2 \beta \phi} \left(dX^d + A_{\mu} dX^{\mu} \right)^2 \;.
\label{metana}
\ee 
Dimensionally reducing the Ricci scalar now gives
\be
\int d^DX \sqrt{-G}R^D = \int d^dX \sqrt{-g} \left[R^d - \frac12\left(\partial \phi \right)^2  - \frac14 e^{-2\left(d-1\right)\alpha \phi}F_{(A),\mu\nu}F_{(A)}^{\mu\nu}\right] \;,
\label{ddimA}
\ee
where $F_{(A),\mu\nu}=\frac12\partial_{[\mu}A_{\nu]}$ is the gauge field kinetic term. We therefore see that the lower dimensional theory has a propagating $U(1)$ gauge field with gauge coupling
\be
g_{(A)} = e^{\left(d-1\right)\alpha \phi} = \frac{1}{2 \pi R} \left(\frac{1}{2\pi R}\right)^{\frac{1}{d-2}}  \;.
\label{zmgc}
\ee 
The gauge symmetry associated to the gauge field, with a local gauge parameter $\lambda\left(X^{\nu}\right)$, is coming from the circle isometry
\be
A_{\mu} \rightarrow A_{\mu} - \partial_{\mu} \lambda\left(X^{\nu}\right) \;,\;\; X^d \rightarrow X^d + \lambda\left(X^{\nu}\right)\;.
\label{gaugesymm}
\ee

Recall the KK expansion for the a $D$-dimensional field (\ref{KKexp}). From the gauge symmetry transformation (\ref{gaugesymm}) we therefore see that the KK modes $\psi_n$ are charged under the KK $U(1)$ gauge field $A_{\mu}$. Their charge is
\be
q^{(A)}_n = 2 \pi n \;,
\ee
which, as expected, is quantized. We now note that there is a relation between the charge and mass of the KK modes
\be
g_{(A)} q^{(A)}_n  = M_{n,0} \;,
\label{eWGCe}
\ee
where the KK mass is as in (\ref{essmasc}).

Let us now consider the String theory effective action (\ref{steff}). If we compactify this action on a circle we will obtain a gauge field coming from the gravitational sector as in (\ref{ddimA}). However, we will also obtain a second gauge field $V_{\mu}$ coming from the Kalb-Ramond $B$-field with one index along the $X^d$ direction
\be
V_{\mu} \equiv B_{[\mu d]} \;.
\ee
The kinetic terms for $V_{\mu}$ come from dimensional reduction of the kinetic terms of the Kalb-Ramond field. This gives
\be
\int d^dX \sqrt{-g} \left[R^d - \frac14 e^{-2\left(\alpha+\beta\right) \phi}F_{{(V)},\mu\nu}F_{(V)}^{\mu\nu}\right] \;.
\ee
The factor in front of kinetic terms comes from reducing $\sqrt{-G} H_{\mu\nu d}H^{\mu\nu d}$ so that 
\be
e^{-2\left(\alpha+\beta\right) \phi} = \underbrace{e^{2\alpha \phi}}_{\sqrt{-G}} \; \underbrace{e^{-4\alpha \phi}}_{\left(G^{\mu\nu}\right)^2}\; \underbrace{e^{-2\beta \phi}}_{G^{dd}} \;.
\ee
We therefore find a gauge coupling of 
\be
g_{(V)} = e^{\left(\alpha+\beta\right)\phi} = 2 \pi R \left(\frac{1}{2\pi R}\right)^{\frac{1}{d-2}}  \;.
\ee
The states that are charged under $V_{\mu}$ are the winding modes of the string. To see this we can evaluate directly the Polyakov action (\ref{pagen}) for a string wrapping the $X^d$ direction $w$ times (in the Einstein frame). Because the string wraps $X^d$ we can set $\sigma=\frac{2 \pi}{w} X^d$ and so
\bea
S_P &=& -\frac{T}{2} \int_{\Sigma} d \tau d \sigma \left[ 2 i V_{\mu} \partial_{\tau} X^{\mu} \partial_{\sigma} \left(\frac{ w \sigma}{2\pi} \right)\right] \nn \\
&=& -i \frac{w}{2\pi \alpha'}\int_{\gamma} d\tau \left(\partial_{\tau} X^{\mu}\right) V_{\mu}
\label{cpwl} \;.
\eea
This is just the world-line action of a particle with charge
\be
q^{(V)}_w = \frac{w}{2\pi \alpha'} \left(2 \pi R \right)^{\frac{2}{d-2}}\;,
\ee
where we have included the extra factor of $\left(2 \pi R \right)^{\frac{2}{d-2}}$ coming from (\ref{MsR}). 
We therefore have
\be
g_{(V)}q^{(V)}_w  = M_{0,w} \;,
\label{gvmw}
\ee
where the winding mass is as in (\ref{essmasc}).

We therefore find that there are two gauge fields in $d$-dimensions $A_{\mu}$ and $V_{\mu}$ and there are states charged under them, KK modes and winding modes respectively. Further, the charged states have interesting relations between their charges and masses (\ref{gvmw}) and (\ref{eWGCe}). We will investigate these properties in the context of the Swampland below. But before that we note that there is a symmetry in the theory where we exchange the gauge fields $A_{\mu}$ and $V_{\mu}$ and the KK and winding modes. This is in fact just the T-duality symmetry we encountered already. 

\subsubsection{The Weak Gravity Conjecture}
\label{sec:fewgc}

We have seen that string compactifications on a circle lead to two gauge fields, $A_{\mu}$ and $V_{\mu}$, whose gauge couplings are inverse to each other. The $d$-dimensional effective theory had states charged under the two gauge fields whose charge mass were related as $m = g q$. Further, these states were actually the first in an infinite tower of such states with an associated mass scale $m=g$. This tower of states was associated with a new description of the effective theory, in the case of the KK tower, this description was the $D$-dimensional theory. We may then consider this mass scale as a cutoff on the $d$-dimensional effective theory. 

These properties are captured by the Weak Gravity Conjecture which can be stated as follows.\footnote{The reader might wonder how the factor of $\sqrt{\frac{d-2}{d-3}}$ fits in with (\ref{eWGCe}). We will explain this in section \ref{sec:wgcsca}.}

\begin{tcolorbox}
{\bf Weak Gravity Conjecture ($d$-dimensions)}\footnote{The conjecture in \cite{ArkaniHamed:2006dz} was presented only for four dimensions, and is presented in (\ref{ewgc}) and (\ref{mwgc}). The logic behind the electric WGC can be simply generalized to $d$-dimensions, see for example \cite{Heidenreich:2015nta}. The Magnetic WGC for $d$-dimensions here is deduced naturally from the known examples.} \;\cite{ArkaniHamed:2006dz}
{\it 
\newline
\newline
Consider a theory, coupled to gravity, with a $U(1)$ gauge symmetry with gauge coupling $g$
\be
S = \int d^dX \sqrt{-g} \left[ \left(M_p^d\right)^{d-2} \frac{R^d}{2} - \frac{1}{4g^2} F^2 + ... \right] \;.
\ee
\begin{itemize}
\item (Electric WGC) There exists a particle in the theory with mass $m$ and charge $q$ satisfying the inequality
\be
\label{ewgcd}
m \leq \sqrt{\frac{d-2}{d-3}} g q \left(M_p^d\right)^{\frac{d-2}{2}} \;.
\ee
\item (Magnetic WGC) The cutoff scale $\Lambda$ of the effective theory is bounded from above approximately by the gauge coupling
\be
\label{mwgcd}
\Lambda \lesssim g \left(M_p^d\right)^{\frac{d-2}{2}}  \;.
\ee
\end{itemize}
}
\end{tcolorbox}

We see that the cutoff scale $\Lambda$ of the magnetic WGC, in the case of string theory compactifications on a circle, is associated to the mass scale of an infinite tower of states. We will present arguments that this may be general. Note that the WGC presents only an upper bound in this cutoff, due to quantum gravity physics. As is the case with general Swampland criteria, as illustrated in figure \ref{fig:swgen}, there may be a lower cutoff scale above which the effective theory changes in character such that the WGC need not apply.  

We will discuss the Weak Gravity Conjecture in much more detail in this review, but already at this stage it is worth thinking about connections between the Weak Gravity Conjecture and the Swampland Distance Conjecture. From (\ref{eWGCe}) and (\ref{swcircma}) we see that the mass of the KK tower behaves as $M_{KK} \sim g^{(A)} \sim e^{\alpha \phi}$. The winding mode tower satisfies a similar relation $M_{w} \sim g_{(V)} \sim e^{-\alpha \phi}$, and the two are related by T-duality. These relations identify the tower of states responsible for the cutoff $\Lambda$ in the Weak Gravity Conjecture, with the tower of states of the Swampland Distance Conjecture. This identification was first proposed in general in \cite{Heidenreich:2016aqi,Klaewer:2016kiy}, and subsequently found to hold in a wide range of string theory settings. 

%
%
%
%
%
%

\subsection{Strong coupling and the Species scale}
\label{sec:fespec}

We have seen that both the Weak Gravity Conjecture and the Swampland Distance Conjecture can be associated with the mass scale of an infinite tower of states. This mass scale acts as a type of soft cutoff on the effective theory. One could imagine including a few of the KK modes still within the lower, $d$-dimensional, theory. Let us consider how many such modes we can include in an effective theory. Since in some sense we know the ultraviolet completion of the $d$-dimensional theory, which is a $D$-dimensional theory, we know that the universal quantum gravity bound due to strong coupling should be the $D$-dimensional Planck mass $M_p^D$. We can therefore count how many KK modes we can include before reaching this scale. 

The counting can be done most straightforwardly without going to the $d$-dimensional Einstein frame. So we directly dimensionally reduce with a metric
\be
ds^2 = g_{\mu\nu} dX^{\mu} dX^{\nu} + \left(2\pi R \right)^2 \left(dX^d\right)^2 \;,
\ee
which gives a relation between the Planck masses
\be
\left(M_p^d\right)^{d-2} = \left(M^D_p\right)^{D-2} 2\pi R \;.
\ee
So, in the $d$-dimensional Planck units we are using, we have $M^D_p \sim R^{\frac{1}{2-D}}$. The KK scale in this frame is $M_{KK} \sim \frac{1}{R}$, and so we find that the number of states $N_s$ we can include before the true quantum gravity cutoff scale, $M_p^D$, is
\be
N_s \sim \frac{M_p^D}{M_{KK}} \sim R^{\frac{d-2}{D-2}} \;.
\ee
We can then rewrite the quantum gravity cutoff scale in terms of $N_s$ as
\be
\label{Ddimspecies}
M_p^D \sim \frac{1}{N_s^{\frac{1}{d-2}}} \;.
\ee
This is the scale where gravity becomes strongly coupled in the higher dimensional theory which we are reaching by including the KK modes. 

The scale $M_p^D$ is, for $R \gg 1$, much lower than the $d$-dimensional Planck mass $M_p^d$ (which we have set to unity). So gravity becomes strongly coupled at a parametrically lower scale than one might have naively guessed. We can understand this microscopically in terms of the fact that the gravitational force is really $D$-dimensional rather than $d$-dimensional, and so is diluted by more than we might have guessed. At energy scales below the compactification radius this dilution is manifest as the $d$-dimensional force appearing weaker than it actually is, leading to an incorrect estimation of the scale at which it becomes strongly coupled. This mechanism is famously exploited as a proposed solution to the hierarchy problem by postulating that for sufficiently large extra dimensions the strong coupling scale of gravity may actually be close to the electroweak scale \cite{ArkaniHamed:1998rs}. 

The premature strong coupling scale of gravity can be written in terms of the number of states below it in the form (\ref{Ddimspecies}). In this case, $N_s$ is counting KK modes, but it is actually conjectured that the form (\ref{Ddimspecies}) is generally true. One can formulate this by defining the Species scale $\Lambda_s$ which sets the scale of strong coupling for gravity. 

\begin{tcolorbox}
{\bf The Species Scale (Conjecture)}
\newline
\newline
{\it 
Consider a $d$-dimensional theory, coupled to gravity with a Planck mass $M_p^d$, which has $N_s$ particle states below a cutoff scale $\Lambda$. Then within any weakly-coupled gravitational regime there is a bound
\be
\label{spscb}
\Lambda < \Lambda_s \equiv \frac{M^d_p}{N_s^{\frac{1}{d-2}}} \;.
\ee
}
\end{tcolorbox}

The species scale has a long history associated to various inconsistencies that could arise for a large number of species. An incomplete list of studies of this in the context of holography is \cite{PhysRevD.34.373,Bekenstein:1993dz,Jacobson:1994iw,Bekenstein:2000sw,Veneziano:2001ah}. The formulation of the scale presented above, is more closely related to \cite{Dvali:2001gx,ArkaniHamed:2005yv,Distler:2005hi,Dimopoulos:2005ac,Dvali:2007hz,Dvali:2009ks}, though we will see in section \ref{sec:cebss} that the two approaches are related.

As in the other Swampland conjectures, our first encounter with the species scale is in the rather specific context of KK reduction. However, we will present increasingly general arguments for its validity as the review progresses.

\section{Charting the Swampland}
\label{sec:map}

The study of what criteria differentiate a theory in the Landscape from one in the Swampland is a substantial and somewhat complex field. The evidence for the proposed criteria typically stems from a number of sources. In section \ref{sec:first} we studied evidence from rather simple string theory constructions. We will develop the evidence from string theory in more detail in section \ref{sec:stringcomp}. Evidence from string theory constructions can be thought of as indirect. We may consider string theory vacua as providing experimental input for proposed theories. A more direct approach would be to start from some general property of the ultraviolet physics and directly derive Swampland criteria. Such an approach is still in its early stages, but we will dedicate section \ref{sec:emergence} to studying a proposal for how such a direct picture could work. 

In this section we will discuss an approach to gathering evidence which is somewhere in the middle between examples in string theory and a direct microscopic picture. The idea is to try and motivate general properties of quantum gravity that would manifest in the infrared, and use these to arrive at Swampland criteria. This approach is also indirect, it assumes that the ultraviolet physics would be structured in such a way that certain infrared properties of gravity would be guaranteed, rather than working directly from the ultraviolet. These assumed infrared properties should not be thought of as proven, in general they are conjectural and some may be incorrect. To make this clear we will present the arguments for each proposal, or conjecture, and the reader may consider how seriously this evidence should be regarded. The idea is not necessarily to prove Swampland criteria, though some arguments may be considered by some as close to proofs, but rather to build a network of {\it signposts} pointing us towards what kind of criteria could be part of the Swampland, and how they may be connected. We are seeking to build a framework in which to place new and old ideas about the Swampland. One way to motivate this is as generating hypothesis that can then be tested in string theory. After all, much of the Swampland is uncharted, and we need to first establish what are the candidate Swampland criteria. Another is that the signposts may help us uncover the underlying microscopic physics. Yet another is that finding a large number of independent indirect arguments towards the same result may still be regarded as strong evidence. Finally, it is not inconceivable that some criteria can really be proven from infrared properties of gravity, especially utilizing holography.


\subsection{Global symmetries in quantum gravity}

Many of the Swampland criteria can be thought of as attempts at extending the idea that quantum gravity should have no global symmetries. In some sense, the absence of globals symmetries is arguably the first proposed Swampland criterion. It is an old idea with a number of motivating arguments. A good account of this subject can be found in \cite{Banks:2010zn}. One motivation is that in perturbative string theory a global symmetry on the worldsheet is always gauged in the target space \cite{Banks:1988yz}. In anti-de Sitter space it is possible to utilize holography as powerful tool to study this question, and recently a proof of the absence of global symmetries was presented in \cite{Harlow:2018tng,Harlow:2018jwu}. We can then formulate this as a conjecture.

\begin{tcolorbox}
{\bf No Global Symmetries Conjecture} \;\cite{Banks:1988yz,Banks:2010zn}
\newline
\newline
{\it 
A theory with a finite number of states, coupled to gravity, can have no exact global symmetries. 
}
\end{tcolorbox}

An important argument against global symmetries comes from semi-classical properties of black holes.\footnote{While black holes are the natural objects to utilize for global $U(1)$ symmetries, for shift symmetries the appropriate objects are more closely related to Euclidean wormholes, see \cite{Hebecker:2018ofv} for a recent review.}  Let us first recall some basic properties. The four-dimensional Schwarzschild black hole solution takes the form
\be
\label{schbh}
ds^2 = -f\left(r\right) dt^2 + f\left(r\right)^{-1}dr^2 + r^2 d\Omega^2\;,
\ee
where
\be
f\left(r\right) = 1- \frac{2 M_{\mathrm{ADM}}}{r} \;.
\ee
Here $M_{\mathrm{ADM}}$ is the (ADM) Black Hole mass (measured relative to the asymptotic flat metric). We have used spherical coordinates, with time $t$, radial coordinate $r$, and area element $d\Omega$. The No Hair Theorem \cite{PhysRev.164.1776} states that the metric (\ref{schbh}) is unique for uncharged stationary black holes. In particular, for our purposes, if the black hole was formed by throwing in matter which is charged say under a putative exact $U(1)$ global symmetry, this charge is not reflected in properties of the black hole horizon. Classically, this is not necessarily a problem, but semi-classically the black hole will emit Hawking radiation and thereby lose mass. However, there is no way in which it can lose any global charge because the horizon is not sensitive to such a charge, or in other words, there is nothing which breaks the symmetry between positive and negative charges at the horizon. This means that in this theory which has a global charge, given a black hole of a certain mass we cannot determine what the global charge of the black hole is. There is an infinite uncertainty to an observer outside the black hole, and this uncertainty may be associated with an infinite entropy. However, the entropy of a black hole is expected to be finite and given by the Bekenstein-Hawking entropy which, up to numerical factors, goes as $M_{\mathrm{ADM}}^2$. This is an inconsistency, which leads to the conclusion that quantum gravity should have no global symmetries. 

\begin{figure}[t]
\centering
 \includegraphics[width=0.8\textwidth]{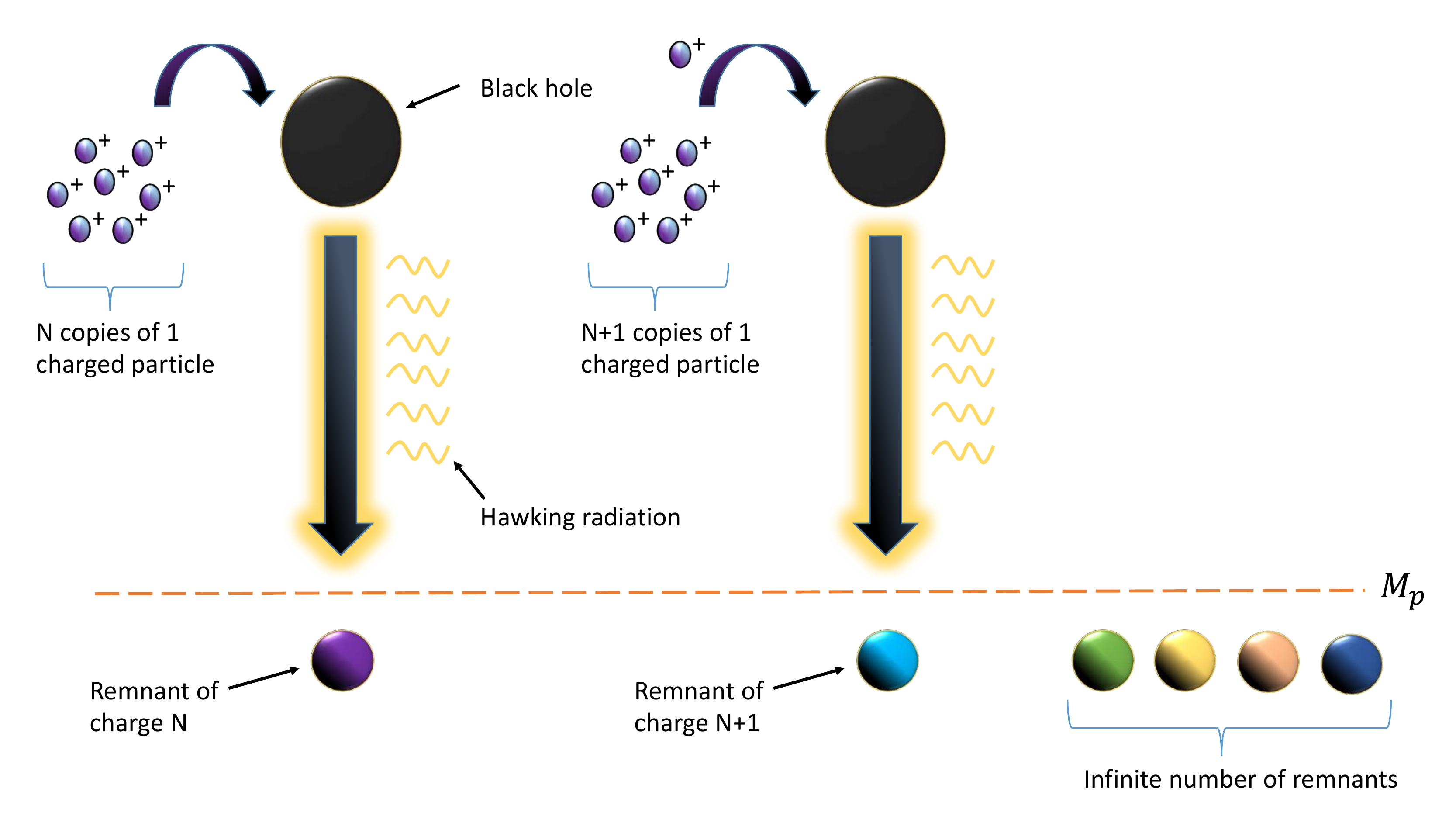}
\caption{Figure illustrating how in the presence of a $U(1)$ global symmetry, black hole formation and decay would lead to an infinite number of stable remnant states below a fixed mass scale.}
\label{fig:bhgr}
\end{figure}

The argument can be formulated in a different way in terms of so-called {\it remnants}. This is just a restriction of the mass of the black hole in the above argument to the smallest mass, so whatever the end point of the Hawking evaporation is. This perspective naturally has an uncertainty associated to it since the end point of Hawking radiation is presumably some deeply quantum gravitational physics, where the horizon no longer has a semi-classical description. However, one can argue that still whatever that regime is, it is not possible for the black hole to shed its global charge. The reason is that if we formed the black hole by throwing in some large number of charged particles, by the time it has shed its mass to reach the Planck scale quantum regime it no longer has enough mass to be able to emit enough of the charged particles to shed its global charge. The result is therefore an object which is stable due to its global charge, but which may be relatively light, so of order the Planck mass. This is a remnant. A quantum theory of gravity with a global symmetry would therefore have an infinite number of remnants. This is illustrated in figure \ref{fig:bhgr}. 

So in the presence of a global symmetry black holes can turn one charged particle below $M_p$ into an infinite number of charged particles below $M_p$. Having an infinite number of states below a fixed mass scale can be reasonably considered to be inconsistent \cite{Susskind:1995da}. However, it is not simple to really prove that something goes wrong. An important point is that the remnants may lie at energy scales where gravity is strongly coupled, say around $M_p$, and so a microscopic derivation of inconsistencies is difficult to reach. One may argue that the remnants can form a violation of an appropriate entropy bound \cite{Banks:2010zn}, but we will see in the following that this is not simple to show for similar reasons. 

This is actually a fundamental obstacle to the approach of attempting to derive Swampland criteria from a semi-classical black hole starting point. The Swampland criteria are typically formulated as statements about states in the theory which lie below the Planck mass, so things that are particle-like in nature. On the other hand, black holes are extended solutions, and are therefore only well understood for masses far above the Planck mass. In between these two regimes lies the quantum gravity regime and any microscopic type argument would have to cross this boundary. Typically, this is only possible to do in a controlled setting with sufficient supersymmetry in a string theory. The notion of black hole entropy forms a bridge between these two regimes allowing us to somewhat by-pass this issue, at least with an appropriate microstates counting interpretation of the entropy, and so the entropy argument above is in this sense stronger than the remnants argument. Nonetheless, there is a clear correlation between the two infinities associated to global symmetries, and since in any case these arguments will form indirect signposts towards the Swampland criteria, we may try to utilize the logic of problems with remnants as a guiding principle. 

\subsection{Completeness in quantum gravity}
\label{sec:complconj}

Unlike global symmetries, gauge symmetries are of course perfectly fine within quantum gravity. However, we may relate gauge symmetries to global symmetries in certain ways and this way attempt to constrain properties of theories with gauge symmetries in quantum gravity. In section \ref{sec:wgccbhs} we will discuss one such relation, which is naturally related to taking the gauge coupling $g$ of a gauge symmetry to zero $g \rightarrow 0$. There is another interesting relation which forms a motivation for the ideas of this section. 

It is most simply illustrated with a $U(1)$ gauge symmetry. Consider a pure Einstein-Maxwell theory with just a $U(1)$ gauge field with gauge coupling $g$,
\be
\label{ccem}
S = \int d^4X \sqrt{-g} \left[ \frac{M_p^{2}}{2} R - \frac{1}{4g^2} F^2 \right] \;.
\ee
The local gauge transformation acts with a scalar parameter $\lambda\left(x\right)$ as $A_{\mu} \rightarrow A_{\mu} + \partial_{\mu} \lambda$. However, we may also consider a transformation $A_{\mu} \rightarrow A_{\mu} + \sigma_{\mu}$, with
\be
\partial_{[\nu} \sigma_{\mu]}=0 \;.
\label{sigclgg}
\ee 
Locally in space-time there is no difference between $\partial_{\mu}\lambda$ and $\sigma_{\mu}$, but physics which is sensitive to more global aspects of the space can tell the difference.\footnote{In terms of de Rham cohomology, the difference between $\partial_{\mu}\lambda$ and $\sigma_{\mu}$ is that the former is exact while the latter is closed.} For example, if we consider the spacetime to be topologically $\mathbb{R}^{1,2}\times S^1$, then Wilson loops wrapping the circle $e^{\int_{S^1} A_{\mu} dx^{\mu}}$, transform under $\sigma_{\mu}$ but not under $\partial_{\mu} \lambda$. Gauging the symmetry $\sigma_{\mu}$ amounts to dropping the condition (\ref{sigclgg}), which means that we would require introducing some anti-symmetric two-tensor (a two-form) to transform as $\partial_{[\nu} \sigma_{\mu]}$ and thereby retain gauge invariance. If we maintain (\ref{sigclgg}), then this is an example of a generalized global symmetry \cite{Gaiotto:2014kfa}. Within quantum gravity we can expect an obstruction to the existence of this global symmetry. 

A simple way to break the generalized global symmetry is the introduction of charged matter. The gauge covariant derivative for a charged field is not invariant under the $\sigma_{\mu}$ transformation but only under $\partial_{\mu} \lambda$. Therefore, we may motivate an idea that in quantum gravity we should not expect a $U(1)$ gauge symmetry with no matter charged under it. This is a mild version of a more general conjecture about quantum gravity called the Completeness Conjecture \cite{Polchinski:2003bq}. 

\begin{tcolorbox}
{\bf Completeness Conjecture} \;\cite{Polchinski:2003bq}
{\it 
\newline
\newline
A theory with a gauge symmetry, coupled to gravity, must have states of all possible charges (consistent with Dirac quantization) under the gauge symmetry. 
}
\end{tcolorbox}

It is important to note that the Completeness Conjecture says nothing about what the mass of the charged states needs to be. They could all be extremely heavy with a mass around the Planck mass. 

The argument in terms of generalized global symmetries only requires a single charged state, but if we extend it to discrete generalized global symmetries then we require all possible charges in order to break also any discrete symmetries \cite{Gaiotto:2014kfa}.  This is also related to the point that the Einstein-Maxwell action (\ref{ccem}) actually has charged objects in the form of charged black holes.  Therefore, these states actually break the generalized global symmetry, but they would have very large charges. The relation between charged particles, Wilson lines, and black holes was also noted in \cite{Banks:2010zn}. In particular, it is possible to think of a Wilson line as the world-line of a probe particle, as in (\ref{cpwl}). The set of Wilson lines spans all the possible charges, which means we may consider $e^{n\int A_{\mu} dx^{\mu}}$  for all $n \in \mathbb{Z}$, and so is associated to a full set of charged probe particles. In field theory we may send the mass of this probe particle to infinity effectively decoupling it from the theory. But with gravity we cannot do that as it will eventually form a black hole, and so remain a physical state in the theory. 

The relations between gauge and global symmetries, charged particles and charged black holes, will thread much of this review and form crucial elements of the discussion around the Swampland. 

\subsection{The Weak Gravity Conjecture and charged black holes}
\label{sec:wgccbhs}

The Weak Gravity Conjecture, as stated in (\ref{ewgc}) and (\ref{mwgc}), was proposed in \cite{ArkaniHamed:2006dz} where also black hole based arguments were made for it. These arguments were aimed at emulating the global symmetries arguments for gauge symmetries. To understand the relation we must first consider how a black hole charged under a gauge symmetry differs from one charged under a global symmetry. 

It is simplest to restrict to four dimensions at this point. Let us then first state the Weak Gravity Conjecture in four dimensions.

\begin{tcolorbox}
{\bf Weak Gravity Conjecture (4-dimensions)} \;\cite{ArkaniHamed:2006dz}
{\it 
\newline
\newline
Consider a theory, coupled to gravity, with a $U(1)$ gauge symmetry with gauge coupling $g$
\be
\label{4dema}
S = \int d^4X \sqrt{-g} \left[ \frac{M_p^{2}}{2} R - \frac{1}{4g^2} F^2 + ... \right] \;.
\ee
\begin{itemize}
\item (Electric WGC) There exists a particle in the theory with mass $m$ and charge $q$ satisfying the inequality
\be
\label{ewgc}
m \leq \sqrt{2} g q M_p \;.
\ee
\item (Magnetic WGC) The cutoff scale $\Lambda$ of the effective theory is bounded from above approximately by the gauge coupling
\be
\label{mwgc}
\Lambda \lesssim g M_p \;.
\ee
\end{itemize}
}
\end{tcolorbox}

There are a few subtleties in this definition of the conjecture that are worth explicitly mentioning. The first is that we used the word {\it particle} in the electric Weak Gravity Conjecture. We take this to mean a state with a mass below the Planck mass $m < M_p$. It is sometimes considered to allow the state satisfying the Weak Gravity Conjecture to be much heavier than the Planck scale, in which case it should better be regarded as an extended state, say a charged black hole. This is a weaker version of the conjecture, and is in some sense trivial since, as we will see, extremal charged black holes themselves satisfy the inequality (\ref{ewgc}). Another subtlety is that the action (\ref{4dema}) by itself does not fix the gauge coupling $g$ since an arbitrary rescaling of $g$ can be absorbed into a definition of the gauge field in $F$. The normalization of the $U(1)$ gauge coupling is implicit in the action (\ref{4dema}) in the sense that the gauge field normalization is chosen to have a canonical coupling to matter current (for example appearing in a charged covariant derivative as $\left|\partial_{\mu} + i q A_{\mu} \right|^2$ with $q \in \mathbb{Z}$). Finally, the meaning of the cutoff scale $\Lambda$ in (\ref{mwgc}) is not made clear. However, we have seen already in the first encounter in section \ref{sec:fewgc}, and it will be argued for more generally in section \ref{sec:emergence}, that it is naturally associated to the mass scale of an infinite tower of states.

In this section we would like to present some arguments for this conjecture based on black hole physics. We will see that this is not simple. It is difficult to find a microscopic inconsistency directly within the low-energy effective theory associated to a violation of the Weak Gravity Conjecture. However, as discussed in the introduction, it is important to keep in mind that we should regard the arguments presented in this section as signposts towards how we might expect the ultraviolet theory to behave such that certain seemingly strange, though perhaps not clearly microscopically inconsistent, behaviour is avoided in the infrared. 

First, let us introduce some properties of black holes in the presence of a $U(1)$ gauge symmetry. The charged black hole solution to the action (\ref{4dema}) takes the same form as (\ref{schbh}) but with
\be
f\left(r\right) = 1 - \frac{2 M_{\mathrm{ADM}}}{r} + \frac{2g^2 Q^2}{r^2} \;,
\ee
where $Q$ is the quantized black hole charge. Since $f(r)$ is now quadratic, there are two horizons. Charged black holes satisfy an extremality bound where the two horizons coincide
\be
\label{extb}
M_{\mathrm{ADM}} \geq \sqrt{2} g Q M_p \;.
\ee
A black hole which saturates (\ref{extb}) is called extremal. Approaching extremality the outer horizon shrinks towards the inner horizon, and violating the extremality bound is therefore associated to a naked singularity. 

In the case of a global $U(1)$ symmetry we could theoretically create an infinite number of black holes with different charges and the same finite mass. For a gauge $U(1)$ symmetry the extremality bound implies that there are only a finite number of black holes below a given mass. More precisely, below a given mass $\Lambda$, there are $N_{\mathrm{BH}}$ possible black holes, with
\be
\label{nbh}
N_{\mathrm{BH}} \sim \frac{\Lambda}{g M_p} \;.
\ee
Eventually, increasing the charge any further necessarily implies an increase in the mass. Therefore, the entropy based argument against global symmetries, where one associates an infinite uncertainty to a black hole with a finite mass no longer works. Actually, it fails in an even more fundamental way since it is directly possible to measure the charge of the black hole by measuring the flux of the gauge field, and so it seems there is no uncertainty at all about the charge. 

These differences from the global symmetry case appear to become negligible however in the limit $g\rightarrow 0$. The number of black holes below a given scale diverges in this limit $N_{\mathrm{BH}} \rightarrow \infty$, and it becomes impossible to measure the black hole charge since the black hole sources no flux in that limit. This is something which will come up multiple times in the arguments for Swampland constraints: sending the coupling of a gauge symmetry to zero turns it into a global symmetry. This makes sense also from the perspective of the microscopic physics, the limit $g \rightarrow 0$ can be thought of as giving an infinite kinetic term to the gauge field which stops it becoming a propagating local field. However, the gauge symmetry remains as an exact selection rule on interactions, and so is behaving very much like a global symmetry. The magnetic Weak Gravity Conjecture can therefore be interpreted as capturing how such a continuous flow towards a forbidden global symmetry limit is continuously obstructed in quantum gravity.\footnote{In \cite{Grimm:2018ohb} it was argued that a similar interpretation arises also for the Swampland Distance Conjecture within string theory. See section \ref{sec:cymod}.}

If we accept the black hole arguments against global symmetries, we therefore can accept a black hole argument against the limit $g \rightarrow 0$ of a gauge symmetry. However, making this quantitative is not simple, and indeed we will see some of the issues with the global symmetry arguments arise again. We can consider first the problem that the Bekenstein-Hawking entropy may be violated due to a large uncertainty in the black hole charge. It was briefly argued in \cite{Banks:2006mm} that this argument still holds in the sense that for small gauge coupling $g$ it becomes increasingly difficult to measure the black hole charge. However, it is not clear how to quantify this. A related issue is that to measure any charge precisely requires an infinite amount of time, since we require the sphere measuring the flux to be at infinity. This has been quantified for example in \cite{Bousso:2017xyo}. Attempts at making the charge uncertainty argument for the magnetic Weak Gravity Conjecture quantitative and rigorous were made in \cite{Saraswat:2016eaz,Schadow}, but with limited success.   

The other argument in the global symmetry case was based on the presence of an infinite number of stable remnants. For the gauge symmetry case the number of possible remnants, with a mass below a scale $\Lambda$, is given by (\ref{nbh}). It is not clear how to find a rigorous problem due to these remnants, one approach is to consider a violation of entropy bounds. Indeed, in \cite{Banks:2006mm} it was argued that by putting such remnants in a box it is possible to violate the Covariant Entropy Bound \cite{Fischler:1998st,Bousso:1999xy}. It is informative to examine this issue in detail which we do in the following section.

\subsubsection{Covariant Entropy Bound and the Species Scale}
\label{sec:cebss}

Extracting a problem with a black hole remnant from a microscopic perspective usually amounts to treating them as fundamental particles in their own right. One can then consider, for example, running them in loops \cite{Susskind:1995da}. We will utilize an assumption that this is a valid approximation, so that whatever the internal structure of the remnants it does not modify their behaviour significantly relative to a fundamental particle. This necessarily implies that we should consider only remnants with a mass below $M_p$. In that case, we can consider a theory with $N_s$ different species of particles where, utilizing (\ref{nbh}) with $\Lambda =M_p$, we have
\be
N_s=\frac{1}{g}\;.
\ee 
While this relates a theory with $N_s$ species to black hole remnants, it is worth noting that the discussion in this subsection holds generally and independently of the origin of the species. 

The Covariant Entropy Bound (CEB) \cite{Fischler:1998st,Bousso:1999xy}, sometimes referred to as the Bousso Bound, is a way to precisely formulate the general principle of holography \cite{tHooft:1993dmi} (see \cite{Bousso:2002ju} for a review). It can be applied in an arbitrary spacetime, but we will only consider it for now in flat space. In this case it amounts to restricting the entropy $S$ inside a box or sphere of radius $R$ to be bounded by a quarter of its area $A\left(R\right)$,
\be
\label{fsceb}
S \leq \frac{A\left(R\right)}{4} \;.
\ee
The covariant entropy bound is closely related to the Bekenstein bound \cite{Bekenstein:1972tm,Bekenstein:1973ur,Bekenstein:1974ax} which states that given some matter with total mass-energy $E$, inside a sphere of radius $R$, the entropy of that matter $S_{\mathrm{matter}}$ is bounded by
\be
\label{bekb}
S_{\mathrm{matter}} \leq 2 \pi E R \;.
\ee
Note that a crucial property of the Bekenstein bound is that it does not involve $M_p$ and so the connection with gravity is unclear. Indeed, it can be applied directly in the quantum field theory limit $M_p \rightarrow \infty$. 

We would now like to consider how the entropy bounds behave in the case of $N_s$ species of particles. In particular, we would like to consider the Species Problem \cite{PhysRevD.34.373,Bekenstein:1993dz,Jacobson:1994iw,Bekenstein:2000sw,Veneziano:2001ah,Bousso:2002ju} which amounts to the possibility of violating the entropy bounds if $N_s$ is sufficiently large. Consider first the case of the Bekenstein bound in the gravity decoupling limit (\ref{bekb}). We can then consider $N_s$ species of particles which are non-interacting, and place them in a box. Their entropy should grow at least as $S_{\mathrm{matter}} \sim \log N_s$ and so for sufficiently large $N_s$ would violate the bound. However, it was shown in \cite{Casini:2008cr}, building on earlier ideas in \cite{Marolf:2003sq,Marolf:2004et}, that the bound is not violated even in the limit $N_s \rightarrow \infty$. This is explained by proposing that the entropy $S_{\mathrm{matter}}$ should be defined carefully as the relative entropy between the matter system and the vacuum state. So, at least when gravity is decoupled, having a large number of species may not lead to an inconsistency in the theory, at least not one related to entropy bounds.

If we now couple gravity in then the Bekenstein bound (\ref{bekb}) can be related to the covariant entropy bound (\ref{fsceb}) by requiring that the matter inside the sphere be gravitationally stable, so does not collapse to a black hole $2E \leq R$. To see how the species problem manifests in this case we consider $N_s$ species of relativistic weakly interacting fields and follow the analysis in \cite{Page:1982fj,Cohen:1998zx,Bousso:2002ju,Banks:2005bm,Ooguri:2018wrx}. We can associate to them a temperature $T$, such that the energy scales as $E \sim N_s R^3 T^4$, while the entropy behaves as $S \sim N_s R^3 T^3$. Imposing gravitational stability on the energy limits the temperature and therefore leads to a maximum entropy $S \sim N_s^{\frac14} R^{\frac32}$. The covariant entropy bound (\ref{fsceb}) then implies 
\be
\label{nsbceb}
N_s \lesssim R^2 \;.
\ee
Given a fixed size sphere with radius $R$, it appears that we can always find a sufficiently large $N_s$ which violates the bound. 

Some proposed resolutions to this problem are outlined in \cite{Bousso:2002ju}. There is a rather clear resolution however once we recall the species scale (\ref{spscb}). Indeed, for the value of $N_s$ which saturates the bound (\ref{nsbceb}) we find that the species scale is 
\be
\Lambda_s \sim \frac{1}{\sqrt{N_s}} \sim \frac{1}{R} \;.
\label{spceb}
\ee
This is the Hubble scale of the sphere, which also sets the minimum energy excitation in the sphere. Therefore, the species scale censors in this way increasing $N_s$ any further than $R^2$, beyond which the species cutoff would be lower than even the zero-point energies of the fields.

Returning to the question of black hole remnants, it is therefore not clear that the interpretation of having a large number of remnants or species necessarily leads to a violation of entropy bounds. Rather, a natural interpretation is that a large number of remnants would be associated with a lowered cutoff of any effective theory. Indeed, the magnetic Weak Gravity Conjecture proposes such a lowered cutoff (\ref{mwgc}). We will see in section \ref{sec:emergence} that indeed this lowered cutoff is closely related to (though not equal to) the species scale, thereby tying it to the entropy bounds analysis in this section. 

\subsubsection{Black hole discharge}
\label{sec:bhdisch}

We have seen so far that black hole arguments seem to point towards the magnetic Weak Gravity Conjecture (\ref{mwgc}). The electric Weak Gravity Conjecture (\ref{ewgc}) also has natural ties to black hole physics \cite{ArkaniHamed:2006dz}. Indeed, it can be formulated as the requirement that all black holes should be able to discharge themselves. 

\begin{figure}[t]
\centering
 \includegraphics[width=0.8\textwidth]{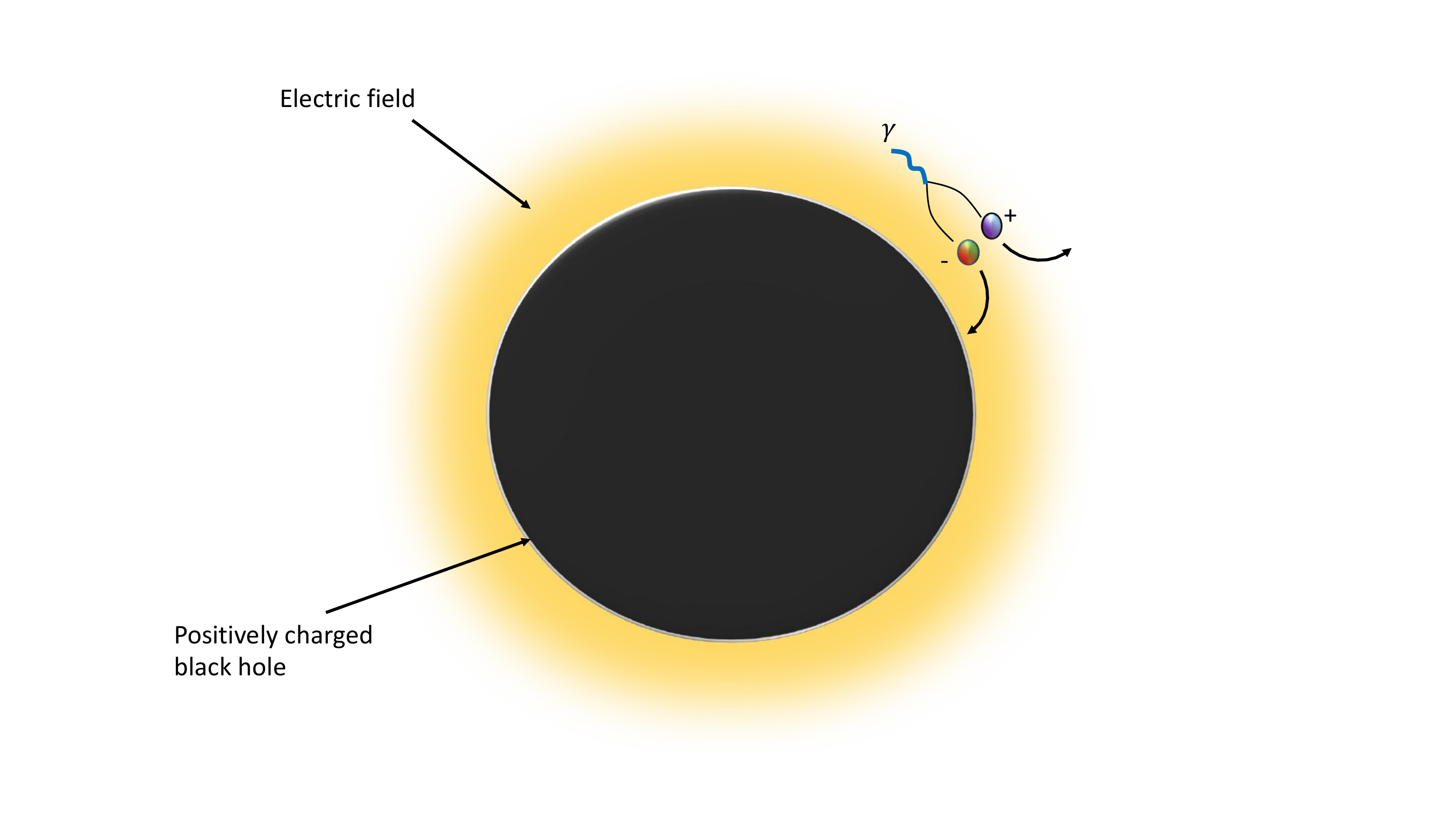}
\caption{Figure illustrating the discharge process of black holes. A pair of positively charged particle and its negatively charged anti-particle are created in the electric field outside the black hole. The anti-particle falls into the black hole, decreasing its charge, while the particle escapes to infinity.}
\label{fig:dbh}
\end{figure}

Recall that in the case of a global $U(1)$ symmetry there was nothing to discriminate between positive and negative charges on the black hole horizon, and therefore there cannot be a discharge process. In the case of a gauge $U(1)$ symmetry, the charge of the black hole induces a field around its horizon and this allows for a discharge process analogous to Hawking radiation whereby charged particles are emitted \cite{gibbons1975} (see \cite{Banks:2006mm} for a summary). There are two discharge processes and which one dominates depends on the mass of the black hole $M$, the charge of the black hole $Q$, the mass of the particle which is being emitted $m$ and its charge. The gauge coupling $g$ is normalized here such that the charge of the emitted particle is one $q=1$. Recall that we may associated a Hawking temperature to the black hole
\be
T_H = \frac{R_+ - R_- }{4\pi R_+^2}  \;,
\ee 
where $R_+$ and $R_-$ are the outer and inner horizons of the black hole.
If this temperature is much higher than the mass of the particle $T_H \gg m$, then the particle is thermally produced on the horizon. The electric field of the black hole induces a chemical potential for this process favouring the production of particles with opposite charge to the black hole. This becomes significant when $g^2 Q \sim M m$ \cite{gibbons1975,Banks:2006mm}. In such regimes therefore the black hole efficiently discharges itself thermally, it is a small or hot charged black hole. The other regime is that of a large or cold black hole $T_H \ll m$. In that case the dominant discharge comes from Schwinger pair production \cite{PhysRev.82.664} in the constant electric field of the black hole. This is particularly important for black holes which are extremal or near extremal. This discharge process becomes efficient when $m^2 R_+^2 \sim g Q$ \cite{gibbons1975,Banks:2006mm}. Both of these discharge processes are schematically illustrated in figure \ref{fig:dbh}.

A requirement for the discharge of black holes is that the theory must contain a charged particle that can be emitted. There is a constraint on the mass-to-charge ratio of the particle. To see this we first derive the general and simple statement that a charged object can only decay into a set of lighter objects if that set contains at least one particle with a higher charge-to-mass ratio than the original one. Let the initial object have a mass $M$, and charge $Q$. It decays into a set of particles with masses $m_i$ and charges $q_i$. By energy and charge conservation we have $M \geq \sum_i m_i$ and $Q = \sum_i q_i$. We therefore find
\be
\label{emcwgcdis}
\frac{M}{Q} \geq \frac{1}{Q} \sum_i m_i = \frac{1}{Q} \sum_i \left(\frac{m_i}{q_i}\right) q_i \geq \frac{1}{Q}\left.\left(\frac{m}{q}\right)\right|_{\mathrm{min}} \sum_i q_i = \left.\left(\frac{m}{q}\right)\right|_{\mathrm{min}}  \;.
\ee
For a black hole to decay we therefore require a particle to exist with a larger (or equal to) charge-to-mass ratio than the black hole. The black holes with the largest charge-to-mass ratio are those which saturate the extremality bound (\ref{extb}). Their decay therefore poses the strongest constraint on the theory, and this leads directly to the electric Weak Gravity Conjecture (\ref{ewgc}).  

The crucial question for the electric Weak Gravity Conjecture is therefore whether charged black holes must be able to decay. One motivation for this could be that we require the absence of charged remnants. If charged black holes can simply discharge themselves, then we should not expect any stable remnants. However, the remnant-based arguments seem to suggest that it may be sufficient to lower the cutoff scale of the theory, for example as captured by the magnetic Weak Gravity Conjecture (\ref{mwgc}). Indeed, we have mentioned that the scale of the magnetic Weak Gravity Conjecture $\Lambda \sim g M_p$ can actually often be associated with a tower of charged states. So far from wanting to forbid charged stable light states at weak coupling $g \ll 1$, we expect their presence. 

It is therefore, at this point, somewhat of an open question as to whether the discharge of black holes is something that should be expected to be a Swampland constraint. Proving that completely stable charged black holes carry an intrinsic inconsistency would therefore amount to the proof of the electric Weak Gravity Conjecture.

\subsubsection{Extremal black hole instabilities}
\label{sec:extbhins}

Charged black holes may have a self-built instability in the sense that they may not require a charged particle in order for them to decay. Rather, it is possible that charged black holes simply decay to smaller charged black holes. Such a decay would require that the charge-to-mass ratio of extremal black holes is not exactly one, but is slightly larger. Charged extremal solutions to Einstein-Maxwell theory, composed of the first two terms only in (\ref{4dema}), saturate the inequality of the electric Weak Gravity Conjecture (\ref{ewgc}) precisely. However, such a theory is not valid at the Planck scale, and so one expects additional massive structure such that when one integrates it out the low-energy effective theory would receive corrections in the form of higher derivative operators like $F^4$ and $R^2$. The extremal black hole solutions in such theories no longer need to saturate the inequality (\ref{ewgc}) precisely. It could be that they rather violate it, or it could be that they satisfy a strict inequality. Which one of these occurs depends on the structure of the higher derivative terms. 

\begin{figure}[t]
\centering
 \includegraphics[width=0.8\textwidth]{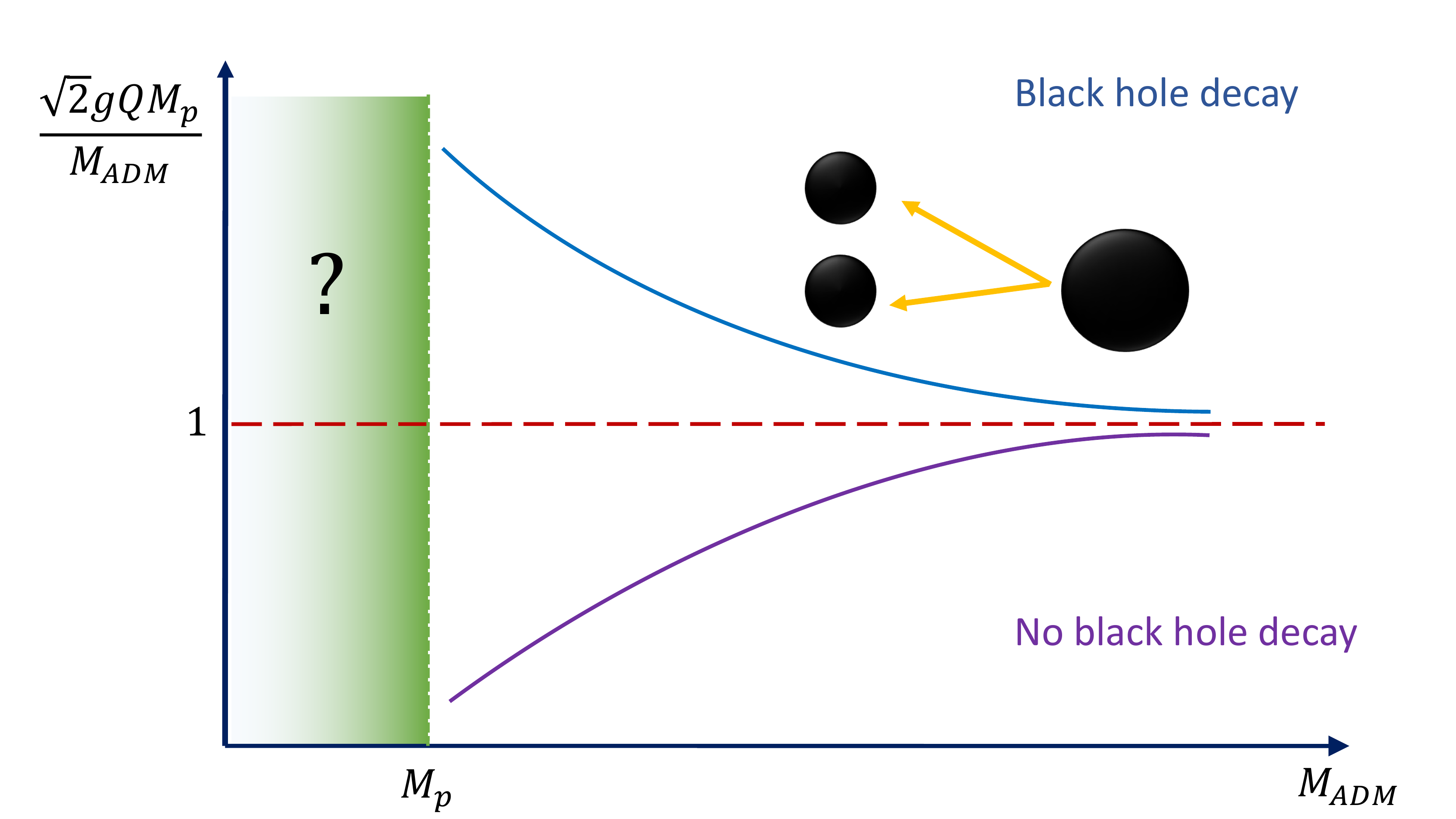}
\caption{Figure illustrating two possible curves for the maximal charge-to-mass ratio of black holes once higher derivative terms are included. In general, it is expected that the deviation from unity should increase for smaller black holes. Whether the charge-to-mass ratio increases or decreases depends on the particular structure of the higher derivative terms. If it increases then black holes can decay to smaller black holes, while such decays are forbidden in the other case. For black holes with mass approaching the Planck mass $M_p$ the semi-classical description breaks down and so the charge-to-mass ratio for objects near or below the Planck mass is not accessible through this methodology.}
\label{fig:ctmpl}
\end{figure}

If the charge-to-mass ratio of black holes satisfies a strict inequality version of (\ref{ewgc}), it is natural to expect that it becomes larger for lighter black holes. This is because larger black holes are extended solutions with larger radii and so are expected to be less sensitive to higher derivative terms. This way there would be an energetically allowed cascade of instability for charged black holes which can be followed down to masses above the Planck mass. Such a situation is illustrated in figure \ref{fig:ctmpl}. It is not clear if this would really solve any problems that may underly the Weak Gravity Conjecture, largely due to the fact that those problems are themselves not sharply defined. In some sense, proving that the higher derivative terms increase the charge-to-mass ratio of black holes would amount to a proof of a certain formulation of the Weak Gravity Conjecture, where the state satisfying the inequality is itself a black hole. However, it cannot be used in the regime where the state is a particle, and so would not prove the precise formulation presented in section \ref{sec:wgccbhs}.

The idea that higher derivative terms are structured in such a way that the charge-to-mass ratio of black holes is actually larger than one, was proposed already in the original paper \cite{ArkaniHamed:2006dz}. It was argued that this structure may be forced by the absence of superluminal propagation. Another approach proposed was that their structure may be constrained by analyticity of scattering amplitudes. This utilized the assumption that amplitudes should remain analytic even in the deep ultraviolet, which then leads to strong positivity constraints on higher derivative operators \cite{Adams:2006sv}. This approach was further developed in \cite{Cheung:2014ega,Cheung:2016wjt,Hamada:2018dde}, and a general proof was claimed in \cite{Bellazzini:2019xts}.\footnote{See also \cite{Chen:2019qvr} for related work.} In \cite{Cheung:2018cwt,Cheung:2019cwi} a slightly different approach based on general expectations that the entropy of a system should increase with light degrees of freedom was utilized to constrain the structure of higher derivative terms. The results all point towards the charge-to-mass ratio being raised above one due to higher derivative terms. 

Within string theory it is possible to calculate some of the higher derivative terms directly. Within the context of black holes this was done in \cite{Kats:2006xp,Giveon:2009da}. Again, their structure was found to support the idea that the charge-to-mass ratio is raised above one. 

\subsubsection{Arguments from bound states}
\label{sec:bstates}

In \cite{ArkaniHamed:2006dz} it was proposed that one could consider constructing stable remnant-type states not by considering black hole decay but rather directly by thinking about the forces acting on particles. This approach is more directly related to the name of the conjecture, specifically the simple argument shows that requiring the absence of stable gravitationally bound states implies that there should exist a particle on which gravity acts as the weakest force. 

Consider Einstein-Maxwell theory with charged particles, and consider the particle with the largest charge-to-mass ratio in the theory (which need not for now be larger than one). The interaction between two such particles, at a separation distance $r$ larger than their Compton wavelength, is that of two long-range Coulomb forces. An attractive force due to gravity $F_{\mathrm{Gravity}}$, and a repulsive one due to the electromagnetic field $F_{\mathrm{EM}}$, with magnitudes 
\be
\label{graemfroc}
F_{\mathrm{Gravity}} = \frac{m^2}{8 \pi M_p^2 r^2} \;, \;\; F_{\mathrm{EM}} = \frac{\left(g q \right)^2}{ 4 \pi r^2} \;.
\ee
The electric Weak Gravity Conjecture (\ref{ewgc}) can therefore be stated as $F_{\mathrm{EM}} \geq F_{\mathrm{Gravity}}$ for the particle with the largest charge-to-mass ratio. 

Let us consider what happens if the electric Weak Gravity Conjecture is violated, so if for the particle with the largest charge-to-mass ratio we still have $F_{\mathrm{EM}} < F_{\mathrm{Gravity}}$. Then two such (stationary) particles would attract rather than repel, and so form a bound state. The energy of this bound state would be strictly smaller than $2m$ because of the gravitational potential, but its charge would be exactly $2 q$. Therefore, the bound state would have a larger charge-to-mass ratio than the particle with the largest charge-to-mass ratio in the theory. By charge and energy conservation, as discussed in section \ref{sec:bhdisch}, the bound state cannot discharge by emitting particles and therefore would be completely stable. We can then consider adding more and more such particles. Since they are mutually attracting, there will always be such a stable bound state with an arbitrarily large charge. 

The stable bound states are in some sense analogous to the black hole remnants discussed at the start of this section, at least in that they are gravitationally bound states stable by their charge. However, they seem different in that they may be weakly coupled states, while remnants are naturally regarded as strongly coupled. As was the case for black hole remnants, it is not clear what microscopically goes wrong if such stable gravitationally bound states exist. Some arguments towards problems with such states were presented in \cite{Cottrell:2016bty}. These were based on the result that the binding energy for $N$ non-relativistic charged particles interacting under gravitational and electromagnetic Coulomb forces behaves as $N^3 \left(m^2 - 2 g^2q^2\right)^5$ \cite{1992PhR163J}. If the Weak Gravity Conjecture is violated this can be made arbitrarily negative for sufficiently large $N$. However, it was also acknowledged that a system with large $N$ will undergo gravitational collapse and form a black hole with a horizon. It therefore remains unclear whether there is a microscopic inconsistency, indeed it appears that the black hole formation leads back to the question of remnants and whether they are problematic or not.


\subsubsection{Arguments from AdS/CFT}

Holography in quantum gravity is best understood in anti de Sitter space through the AdS/CFT correspondence \cite{Maldacena:1997re}. It is therefore natural to consider if AdS/CFT can yield insights into the Swampland and in particular the Weak Gravity Conjecture. This is potentially a very promising approach, and has already yielded very interesting results, though it does face some intrinsic difficulties. It is restricted to AdS space, and the nature of the duality is strong-weak which means that weakly-coupled conformal field theories are dual to a non-Einstein regime of gravity. The Weak Gravity Conjecture is defined in the Einstein regime, and so probing it relies on understanding the strong coupling behaviour of the CFT. 

For early work on the Weak Gravity Conjecture and AdS/CFT see \cite{Hellerman:2009bu,Rattazzi:2010gj}. In \cite{Nakayama:2015hga} it was argued that the Weak Gravity Conjecture may be violated in regimes of CFTs which are not dual to Einstein gravity. It was also studied in \cite{Giombi:2017mxl} with a focus on understanding the implications of supersymmetry breaking. Particularly powerful results can be obtained in the context of AdS$_3$/CFT$_2$ duality by utilising modular invariance on the CFT side \cite{Benjamin:2016fhe,Montero:2016tif,Heidenreich:2016aqi,Bae:2018qym}. These results provided some of the most rigorous and general arguments for the Weak Gravity Conjecture in this context. In particular, they presented evidence for an interpretation of the magnetic Weak Gravity Conjecture scale (\ref{mwgc}) as being related to an infinite tower of states. In \cite{Montero:2017mdq} a relation between small gauge couplings and black hole thermodynamics was developed which lead to a bound on small gauge couplings, though this bound was much weaker than the magnetic Weak Gravity Conjecture scaling only logarithmically with the gauge coupling. In \cite{Urbano:2018kax} the Weak Gravity Conjecture was linked to thermalization properties of the boundary Conformal Field Theory. 

The implementation of information theoretic concepts, in particular entanglement entropy, within AdS/CFT has proven to be a productive line of research. These techniques have been applied in the context of the Weak Gravity Conjecture in \cite{Harlow:2015lma,Cottrell:2017ayj,Harlow:2018tng,Harlow:2018jwu,Montero:2018fns}. In particular, it was argued in \cite{Montero:2018fns} that exact stability of extremal black branes is in contradiction with information theoretic results, thereby implying that charged black holes should decay and so requiring the particle of the electric Weak Gravity Conjecture.

\begin{figure}[t]
\centering
 \includegraphics[width=0.9\textwidth]{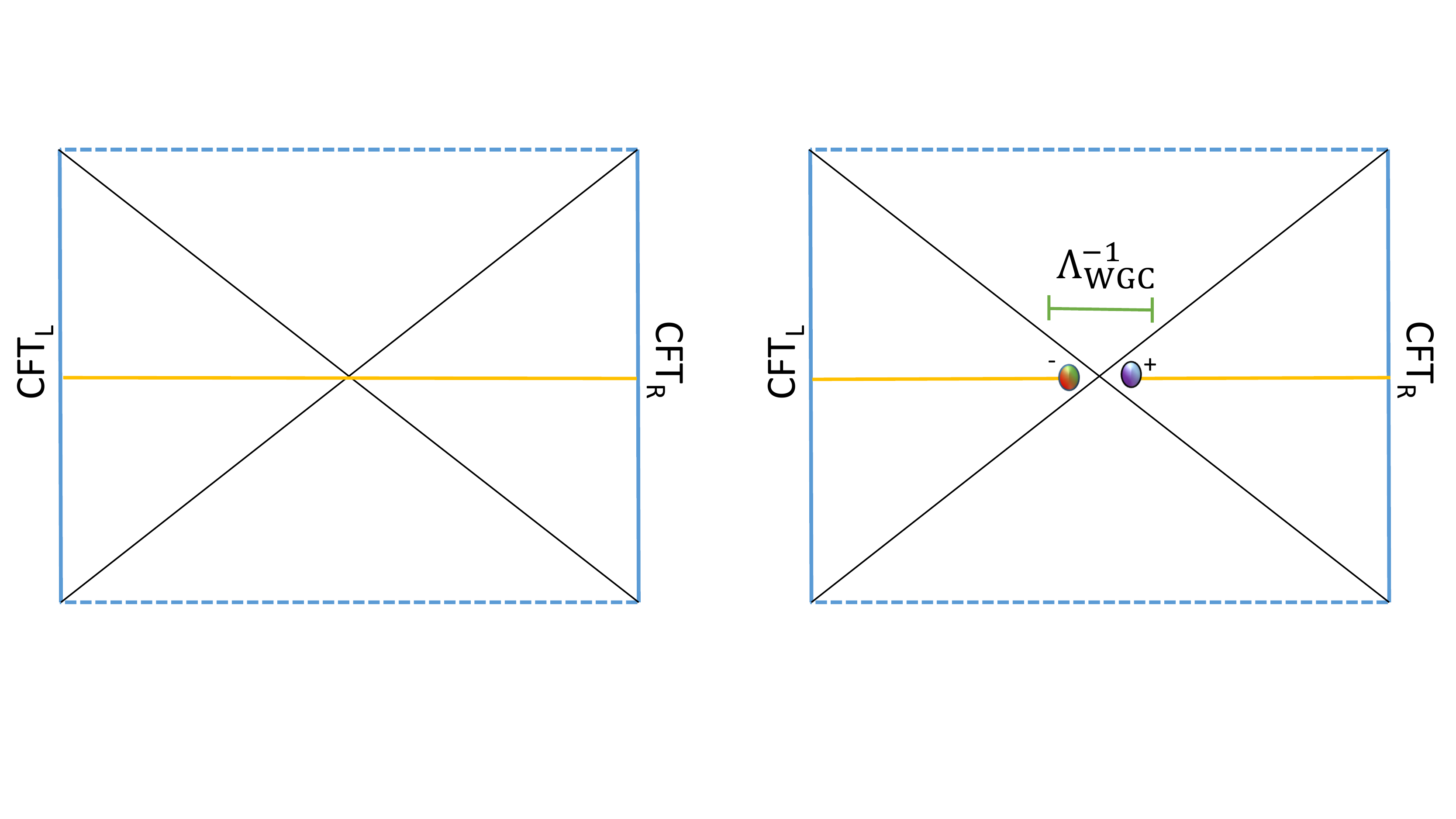}
\caption{Figure showing an AdS Schwarzschild geometry. There are left and right boundaries on which live two conformal field theories CFT$_L$ and CFT$_R$. It was argued in \cite{Maldacena:2001kr} that any state can be described in the direct product of the CFTs. The left figure shows a Wilson line connecting the two CFTs spoiling the factorization. The right figure shows a proposed resolution which restores factorisation by postulating that the Wilson line must end on charged particles at sufficiently short length scales set by the magnetic Weak Gravity Conjecture scale \cite{Harlow:2015lma}.}
\label{fig:adssch}
\end{figure}

In \cite{Harlow:2015lma,Harlow:2018tng,Harlow:2018jwu} the Weak Gravity Conjecture was related to a certain factorization problem in AdS/CFT. If we consider an AdS Schwarzschild geometry it has two boundaries on which live two dual Conformal Field Theories. It was argued in \cite{Maldacena:2001kr} that any state in the bulk can be described by a state in the direct product of the two boundary field theories. The problem studied in \cite{Harlow:2015lma} is that one may consider a Wilson line stretching between the two boundaries. A boundary description of this appears difficult to formulate in terms of states that are factorised between the two boundary theories, so the Wilson lines breaks the factorization. It was then proposed that this is resolved because the Wilson line actually terminates on massive charged states that would be present in the ultraviolet theory. This way, in the full theory, there are no Wilson lines stretching between the boundaries, and any state can be described purely within the product of the boundary theories. These massive charged states were then identified with the particles of the Weak Gravity Conjecture. This is illustrated in figure \ref{fig:adssch}.

While this argument suggests that the ultraviolet theory should contain the relevant charged states, it does not say anything about the relation between their mass and charge, as required for the Weak Gravity Conjecture. However, the fact that one requires the Wilson line to terminate on such states at sufficiently small length scales suggests that at sufficiently high energies we may actually consider the gauge field to be composed of the charged states. This picture of a composite gauge field was developed further in \cite{Harlow:2015lma} and was shown to lead naturally to a mass and charge relation consistent with the Weak Gravity Conjecture.  We postpone a more detailed discussion of this picture to section \ref{sec:emergence} where it will be posed within a larger framework of emergence.

\subsubsection{Further general arguments for the Weak Gravity Conjecture}
\label{sec:further}

In this section we aim to highlight further arguments for the Weak Gravity Conjecture. The electric and magnetic prefixes of the conjecture suggest that the two are related in some way, and so may support each other. This was argued to be the case in \cite{ArkaniHamed:2006dz} as follows. Starting from the electric Weak Gravity Conjecture (\ref{ewgc}) we can apply it to the magnetic dual field as a constraint on the mass of a magnetic monopole
\be
\label{mag1}
m_{\mathrm{mag}} \leq \frac{M_p}{g} \;.
\ee
Monopoles lead to a linearly divergent field and so the monopole field must be cutoff at some ultraviolet scale $\Lambda$. The monopole mass is then at least the energy stored in the field and so satisfies 
\be
\label{mag2}
m_{\mathrm{mag}} \sim \frac{\Lambda}{g^2} \;.
\ee
Equating the two expressions (\ref{mag1}) and (\ref{mag2}) gives the magnetic Weak Gravity Conjecture (\ref{mwgc}).

Another argument presented in \cite{ArkaniHamed:2006dz} is that the monopole field cutoff $\Lambda$ sets the monopole radius $R_{\mathrm{mag}}\sim \Lambda^{-1}$. One may then require that this monopole should not be a black hole, in that case we have $m_{\mathrm{mag}} \lesssim M_p^2 R_{\mathrm{mag}}$. Using (\ref{mag2}) then leads again to the magnetic WGC (\ref{mwgc}). 

It is worth noting that there are a few subtleties with these arguments. First, the monopole field profile is taken for the monopole solution neglecting gravity, and so cannot be trusted in the regime where the monopole is forming a black hole. Nonetheless, as order of magnitude estimates the analysis seems sufficient. It was also noted in \cite{Saraswat:2016eaz} that the monopole solution may have internal structure such that the naive identification of the monopole radius with the inverse cutoff of the effective theory can be incorrect. An explicit example of a Higgsed combination of $U(1)$ symmetries was presented. 

There have been studies regarding whether 1-loop corrections from charged particles could lead to an an inconsistency with black hole entropy \cite{Shiu:2017toy,Cottrell:2016bty,Fisher:2017dbc}. The idea being that these 1-loop effects would be very different if the particle satisfies or violates the Weak Gravity Conjecture. While the papers found a possible problem due in the entropy for large charge black holes, it was argued in \cite{Shiu:2017toy,Cottrell:2016bty} that this occurs in a regime where the computation breaks down. In \cite{Cottrell:2016bty} an argument was also presented for the magnetic Weak Gravity Conjecture (\ref{mwgc}) related to a Fermionic spectral density divergence.

In \cite{Hod:2017uqc} a relation between the Weak Gravity Conjecture and a relaxation bound on perturbed black holes was proposed. 

In \cite{Crisford:2017zpi,Crisford:2017gsb} a connection between the Weak Gravity Conjecture and the Cosmic Censorship Conjecture \cite{PhysRevLett.14.57} was proposed. The Cosmic Censorship Conjecture states that gravitational collapse should lead to a singularity which is always shielded by an horizon. It was shown that in Einstein-Maxwell theory in anti de Sitter space a violation of this conjecture could occur unless a charged particle satisfying the Weak Gravity Conjecture is added to the theory. This direction was extended to dilatonic black holes in \cite{Yu:2018eqq}.

\subsection{The Species Scale and black hole discharge}

In section \ref{sec:fespec} we introduced the idea of the species scale. In four dimensions it takes the form
\be
\label{spec4d}
\Lambda_s \sim \frac{M_p}{\sqrt{N}} \;,
\ee
with $N$ the number of states below $\Lambda_s$. There are a number of general arguments in support of $\Lambda_s$ being the strong-coupling scale of gravity, and we will discuss these further in section \ref{sec:emergence}. There is also an argument which follows if we require a black hole to fully discharge a discrete gauge symmetry \cite{Dvali:2007hz}, which fits in with this section. One considers a theory with $N$ species of bosonic fields, and an associated $\mathbb{Z}^N_2$ discrete gauge symmetry, which can act for example as a sign flip of each bosonic field. The fields are taken to all have equal mass $m=\Lambda_s$. The maximal charge under $\mathbb{Z}^N_2$ can be reached by forming a black hole from $N$ such fields, any further fields will not increase the charge. The black hole can thermally emit the fields when its temperature is of order their mass $T_{H} \sim \Lambda_s$, at which point its mass is $M_{\mathrm{BH}} \sim \frac{M_p^2}{\Lambda_s}$. However, the maximum number of particles, with mass $\Lambda_s$, that can be emitted by the black hole is 
\be
N_{\mathrm{max}} \sim \frac{M_{\mathrm{BH}}}{\Lambda} \sim \frac{M_p^2}{\Lambda_s^2} \;.
\ee
In order to fully discharge all the particles should be emitted back, which implies $N_{\mathrm{max}} \sim N$, leading to (\ref{spec4d}).

\subsection{Refinements of the Weak Gravity Conjecture}

There are a number of proposed Swampland criteria which are strongly influenced by the Weak Gravity Conjecture. They are typically proposals for constraints which are motivated by similar arguments in nature to those of the Weak Gravity Conjecture, so for example the requirement of discharge of extremal black holes. However, they are nearly always stronger than the Weak Gravity Conjecture, and so usually rely on additional assumptions. In this section we review these proposals. 

\subsubsection{Multiple $U(1)$ symmetries}
\label{sec:multiu1}

In the case when there are multiple $U(1)$ symmetries, the obvious generalization is that the Weak Gravity Conjecture should apply to each one in turn. However, it was argued in \cite{Cheung:2014vva} that a stronger requirement may be the relevant condition. The point is that if one considers black holes charged under multiple $U(1)$ symmetries, then the condition for their discharge is stronger than the condition necessary for black holes discharge for individual $U(1)$s. In the case of $N$ $U(1)$ symmetries, with gauge couplings $g_a$ where $a$ runs over the $U(1)$s, we can define a charge vector for particles
\be
{\bf q} \equiv \left( g_1 q_1, g_2 q_2, ... , g_N q_N \right) \;.
\ee
We also require multiple particles and so consider ${\bf q}_i$ with $i$ ranging over the set of particles, with masses $m_i$, that are required to satisfy the Weak Gravity Conjecture. We can define a charge-to-mass ratio vector then as
\be
\label{ctmrz}
{\bf z}_i \equiv \frac{{\bf q}_i M_p}{m_i} \;.
\ee
Now consider a black hole with charge vector ${\bf Q}$, mass $M$, and whose charge-to-mass ratio is denoted ${\bf Z}$. Then charge and energy conservation of its decay to particles, with $n_i$ of each type $i$, requires 
\be
{\bf Q} = \sum_i n_i {\bf q}_i \;,\;\; M \geq \sum_i n_i m_i \;.
\ee
This is in general a stronger condition than the requirement for a black hole charged under only a single $U(1)$ to decay. 

\begin{figure}[t]
\centering
 \includegraphics[width=0.9\textwidth]{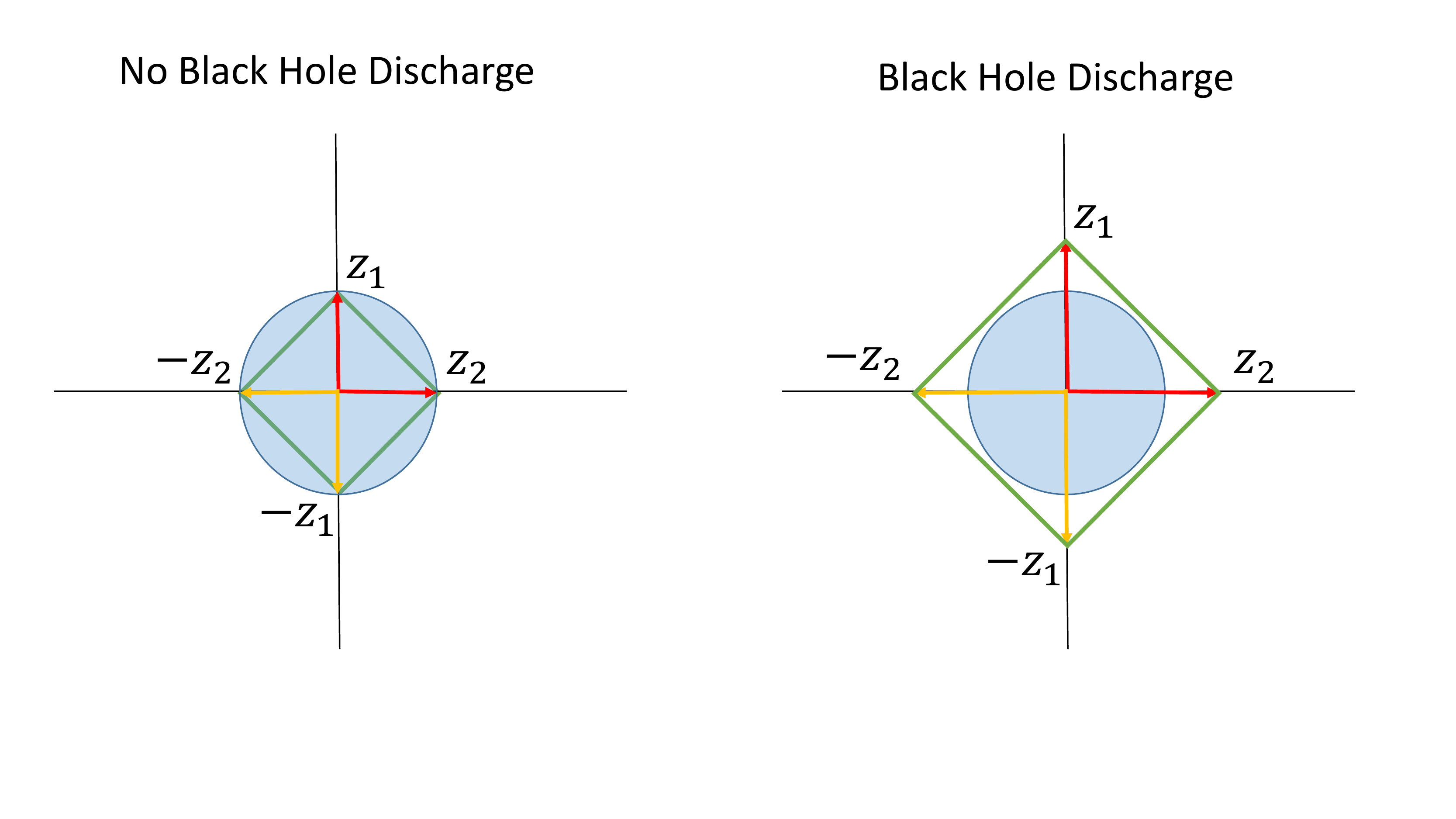}
\caption{Figure showing the charge-to-mass plane for the case of two $U(1)$ symmetries. The particles present in the theory are denoted by vectors ${\bf z}_1$ and ${\bf z}_2$ which define a convex hull shown by straight lines joining the vectors. In order for extremal black holes of arbitrary charge to discharge, the convex hull must include the unit disc. This, in particular, implies that $\left|{\bf z}_i\right| > 1$.}
\label{fig:CH}
\end{figure}

To see this it is useful to plot the charge-to-mass ratio vectors ${\bf z}_i$, this is done for the case of just two $U(1)$ symmetries in figure \ref{fig:CH}. Then the condition for an extremal black hole with an arbitrary charge to decay is that the convex hull, constructed by straight lines between the ${\bf z}_i$ vectors, should contain the unit circle. This means that the vectors ${\bf z}_i$ should actually be slightly larger than one in magnitude. In general, for $N$ $U(1)$ symmetries, their magnitude should be enhanced by a $\sqrt{N}$ factor. 

The proposal of \cite{Cheung:2014vva} is therefore the following generalization of the Weak Gravity Conjecture to multiple $U(1)$ symmetries.
\begin{tcolorbox}
{\bf Weak Gravity Conjecture (multiple $U(1)$s)} \;\cite{Cheung:2014vva}
{\it 
\newline
\newline
A theory with multiple $U(1)$s must have a spectrum of particles with charge-to-mass ratio vectors ${\bf z}_i$, defined as in (\ref{ctmrz}), whose convex hull includes the unit ball.
}
\end{tcolorbox}

\subsubsection{The Strong Weak Gravity Conjecture}
\label{sec:strongwgc}

In \cite{ArkaniHamed:2006dz} the Weak Gravity Conjecture was proposed in terms of three variants which differ by which particle in the theory must satisfy the electric Weak Gravity Conjecture (\ref{ewgc}). The proposals we
\begin{itemize}
\item ({\bf Smallest Charge WGC}) The WGC particle is the one with the smallest charge.
\item ({\bf Strong WGC}) The WGC particle is the one with the smallest mass.
\item The WGC particle is the one with the largest charge-to-mass ratio.
\end{itemize}
The last possibility is weaker than the Strong WGC, and is essentially the vanilla version of the Weak Gravity Conjecture. The first possibility was ruled out with a counter example from the heterotic string in \cite{ArkaniHamed:2006dz}, but this counter example was later found to be incorrect \cite{Heidenreich:2015nta}. In \cite{Heidenreich:2015nta,Heidenreich:2016aqi} a stronger, more detailed version of the first possibility was proposed, and we will discuss this in section \ref{sec:wgctower}.  In this section we will mostly focus on the second possibility, which is often termed the Strong Weak Gravity Conjecture.

The Strong WGC would naturally justify the use of the word {\it particle} in the definition of the electric WGC (\ref{ewgc}), since for example black holes are unlikely to be the lightest charged objects in the theory. However, the arguments presented for the Weak Gravity Conjecture so far always utilized only the charge-to-mass ratio, and this does not constrain in any way the absolute value of the mass by itself. The primary evidence for the Strong Weak Gravity Conjecture therefore comes from String Theory, and will be discussed in section \ref{sec:stringcomp}. Note that whether the Strong WGC holds or not is rather crucial for cosmological applications of the Weak Gravity Conjecture for axions, as will be discussed in section \ref{sec:wgcpform}. 

The Strong WGC has an interesting property which is that a theory satisfying it can, at least naively, be Higgsed down to an effective theory which violates it. This process was first considered in the context of the axion version of the WGC, which we will discuss in section \ref{sec:wgcpform}, while the gauge field version was studied in \cite{Saraswat:2016eaz}.\footnote{Indeed, some early proposed realisations of axion alignment, as discussed in section \ref{sec:wgcpform}, within string theory can be understood as Higgsing of linear combinations of 0-forms \cite{Hebecker:2015rya,Hebecker:2018fln}, while here we consider the 1-form version.} The idea is to consider a theory with two $U(1)$ symmetries, denoted $A$ and $B$, with equal gauge coupling $g$, and particles satisfying the electric WGC with charges $\left(0,1\right)$ and $\left(1,0\right)$ under the two symmetries. The WGC for multiple $U(1)$ implies that the mass of the particles $m$ is bounded by $m \leq g M_p$. One now considers an additional scalar Higgs field $H$ of charge $\left(N,1\right)$ which obtains an expectation value $v$. Then one linear combination of the gauge fields $V_H$ becomes heavy while another $V_L$ stays massless, with
\be
V_H = A + \frac{1}{N} B \;,\;\; V_L = B - \frac{1}{N} A\;.
\ee
The important point is that the normalized gauge coupling for the massless $U(1)$ is now 
\be
g_L = \frac{g}{N} \;.
\ee

\begin{figure}[t]
\centering
 \includegraphics[width=0.9\textwidth]{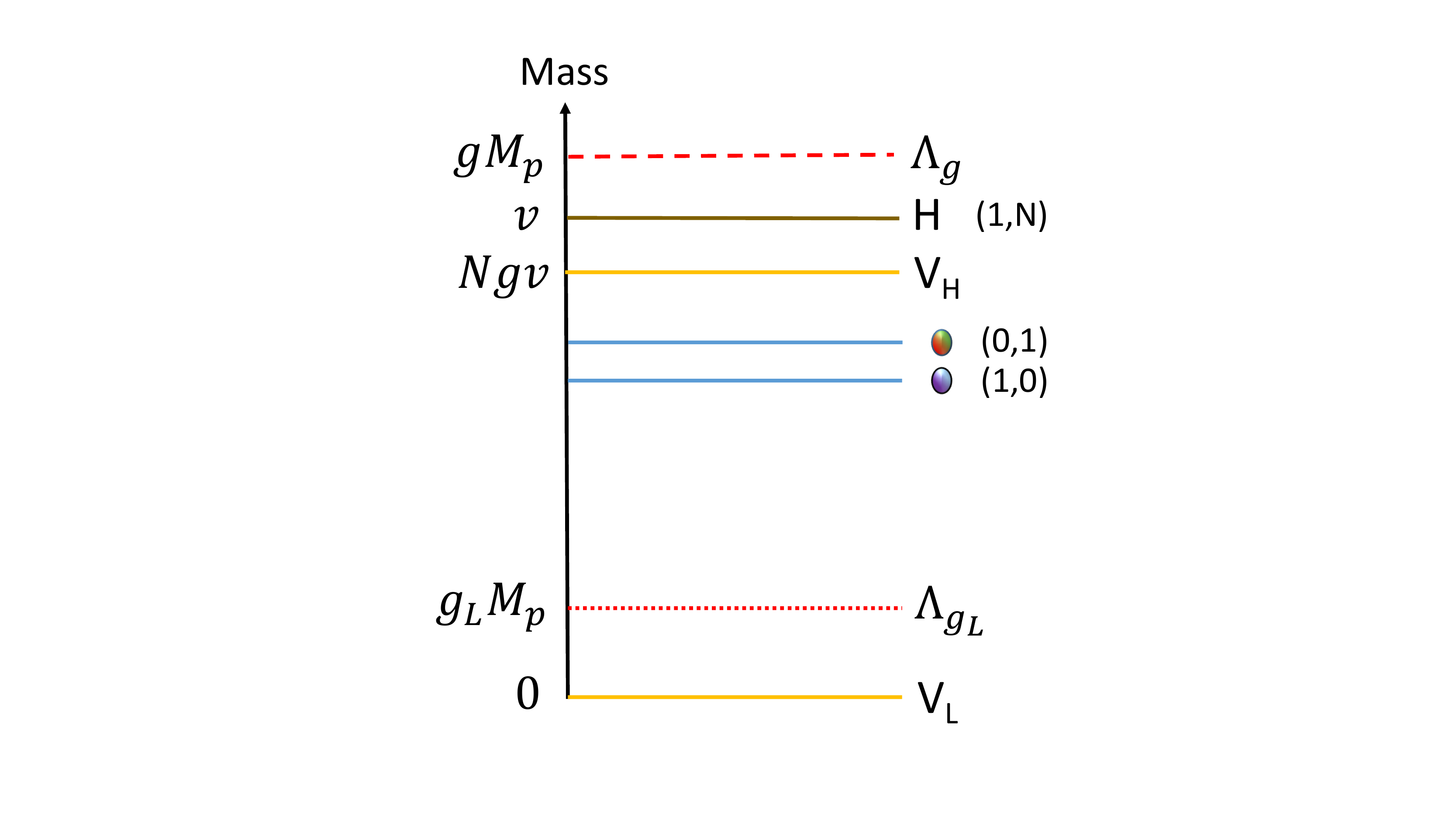}
\caption{Figure showing the spectrum of scales and fields in the Higgsing scenario proposed in \cite{Saraswat:2016eaz}. The Higgs field $H$ is given a large mass, a charge $\left(1,N\right)$ with $N \gg 1$, and a vacuum expectation value $v$. The resulting theory has a massive gauge field $V_H$ and a massless one $V_L$. The gauge coupling of the massless field $g_L$, is very small such that both the Strong and magnetic Weak Gravity Conjectures are violated, even though in the un-Higgsed phase the theory satisfies them.}
\label{fig:HSW}
\end{figure}

Let us consider the charged particles under $V_L$. The particle which had original charges $\left(1,0\right)$ now has a charge $q^{(1,0)}_L = 1$, while the particle with original charge $\left(0,1\right)$ has charge $q^{(0,1)}_L = N$. The latter particle, with original charges $\left(0,1\right)$, is very heavy but satisfies the electric WGC. The former particle, with charges $\left(1,0\right)$, can be taken lighter than the latter, but violates the electric WGC badly $m_{(1,0)} \gg g_L M_p$, for sufficiently large $N$. In fact, as discussed in \cite{Saraswat:2016eaz}, the magnetic WGC is also violated in this theory since there is no physical cutoff scale at $\Lambda \sim g_L M_p$.\footnote{Note however that in \cite{Furuuchi:2017upe} this last point was challenged by arguing that the monopoles do not obey the appropriate charge quantization conditions.} The spectrum of fields and scales in this scenario is shown in figure \ref{fig:HSW}

These arguments suggest that if the Strong and magnetic WGC holds then there should be some obstruction to realizing such a Higgsing scenario in quantum gravity. We will discuss this in detail in string theory settings in section \ref{sec:stringcomp}. From a more general perspective, it was suggested in \cite{Heidenreich:2017sim} that it would be strange to have a spectrum where a single field, the Higgs field, has parametrically larger charge, of order $N$, than all the other charged fields in the theory. The Higgs would also have to be parametrically super-extremal, since if its mass was of order its charge, then it would have a mass above the cutoff of the original ultraviolet theory.  

It is worth noting that the scenario of Higgsing by a large charge would also lead to a large discrete symmetry. It is possible that a version of the Weak Gravity Conjecture can be associated to this discrete symmetry. A general study of the relation between the Weak Gravity Conjecture to such remnant discrete symmetries was made in \cite{Craig:2018yvw}.

\subsubsection{The Weak Gravity Conjecture for dyons with $\theta$-angles}
\label{sec:wgcta}

So far we have primarily focused on electrically charged particles. We considered magnetically charged monopoles in section \ref{sec:further} but as massive solitonic solutions. A theory can also contain objects which carry both electric and magnetic charges, these are called dyons. In general, we may write a theory where the set of fundamental particles are dyons as long as the set of dyonic charges are mutually local. Mutual locality means that if we consider two particles with electric and magnetic charges $\left(q,p\right)$ and $\left(q',p'\right)$, then they are mutually local if
\be
\label{ml}
q p' - p q' = 0 \;.
\ee
Objects which are not mutually local must be described as a particle and a soliton, like an electric particle and a monopole soliton. 

We should consider the possibility that the Weak Gravity Conjecture particle could have both electric and magnetic charges, so that it could be a dyon. This possibility also becomes particularly interesting in the presence of a parity violating $\theta$-angle due to the angle modifying the electric charge of the dyon \cite{WITTEN1979283}. In \cite{Palti:2017elp} a general form of the Weak Gravity Conjecture was proposed for the case where the particle is a dyon and when there can be a non-trivial $\theta$-angle. The formulation was also for the case of multiple $U(1)$ symmetries. Consider the general action
\be
S = \int d^4x \sqrt{-g} \left[ \frac{R}{2} + {\cal I}_{IJ} F_{\mu\nu}^{I} F^{J,\mu\nu} +  {\cal R}_{IJ} F_{\mu\nu}^{I} \left(\star F\right)^{J,\mu\nu} \right] \;.     \label{N2actionEM}
\ee
Here the index $I$ runs over all the $U(1)$ gauge fields, with associated electric field strengths $F^I$. The $\star$ denotes the Hodge star $\left(\star F\right)^{\mu\nu}=\frac12\epsilon^{\mu\nu\rho\sigma}F_{\rho\sigma}$ with $\epsilon$ being the Levi-Civita symbol. The ${\cal I}_{IJ}$ is the gauge kinetic matrix. It encodes the gauge couplings of the various $U(1)$s as well as any kinetic mixing between them. The ${\cal R}_{IJ}$ is the CP-violating $\theta$-angle matrix. Note that if ${\cal R}_{IJ} \neq 0$ the magnetic field strength $G$ is not simply $\star F$ but rather
\be
G_I = {\cal R}_{IJ} F^J  -  {\cal I}_{IJ} \star F^J \;.
\ee 
It is also useful to introduce the matrix ${\cal M}$ 
\be
{\cal M} \equiv \left( \begin{array}{cc} {\cal I} + {\cal R} {\cal I}^{-1} {\cal R}& - {\cal R} {\cal I}^{-1} \\ -{\cal I}^{-1} {\cal R} & {\cal I}^{-1} \end{array} \right) \;. \label{Mdef}
\ee

The electric Weak Gravity Conjecture bounds the mass of a particle by its charge (\ref{ewgc}). We can then consider the bound on the mass when the particle is a dyon in the theory (\ref{N2actionEM}). To define the charge of the particle we introduce the vector
\be
\label{Qdef}
{\cal Q} \equiv \left( \begin{array}{c} p^I \\ q_I \end{array} \right) \;.
\ee
Here the $p^I$ and $q_I$ are integers which we will refer to as the magnetic and electric, respectively, charges of the particle. However, this is a slight abuse of notation. More precisely, the charges should be defined by measuring the flux through a sphere at infinity, as
\be
\frac{1}{4\pi} \int_{S_{\infty}} F^I =  P^I \;,\;\; \frac{1}{4\pi} \int_{S_{\infty}} \star F^I = Q^I \;.
\ee
These are then related to the quantized charges through
 \be
\left( \begin{array}{c} P^I \\ Q^I \end{array} \right)  =  \left( \begin{array}{c} p^I \\ \left( {\cal I}^{-1} \cdot {\cal R} \cdot p \right)^I - \left({\cal I}^{-1}\cdot q \right)^I \end{array} \right) \;.
\ee
We now introduce the notation 
\be
{\cal Q}^2 \equiv -\frac12 {\cal Q}^T {\cal M} {\cal Q} \;. \label{qdefq}
\ee
Using this, the Weak Gravity Conjecture bound on the mass $m$ of the particle was proposed in \cite{Palti:2017elp}.
\begin{tcolorbox}
{\bf Weak Gravity Conjecture (for general action (\ref{N2actionEM}))} \;\cite{Palti:2017elp}
{\it 
\newline
\newline
A theory with multiple $U(1)$s with a gauge kinetic matrix ${\cal I}_{IJ}$ and $\theta$-angle matrix ${\cal R}_{IJ}$, as in (\ref{N2actionEM}), should have a particle with mass $m$ satisfying the bound
\be
\label{wgcdy}
 {\cal Q}^2 M_p^2 \geq m^2 \;,
\ee 
with ${\cal Q}^2$ defined as (\ref{qdefq}).
}
\end{tcolorbox}

A first check on (\ref{wgcdy}) is that in the case of an electrically charged particle under a single $U(1)$ we have ${\cal Q}^2 = 2 g^2 q^2$, which reduces to (\ref{ewgc}). Another check is that the inequality (\ref{wgcdy}) matches the statement that gravity should be the weaker of the forces acting on the particle, as in section \ref{sec:bstates}. Here ${\cal Q}^2$ is the strength of repulsion between two such particles (see \cite{Ritz:2001jk,Andrianopoli:2006ub} for useful texts), and $m^2$ is the attraction. The condition of black hole discharge, as in section \ref{sec:bstates}, can also be naturally applied. Dyonic black hole solutions to the action (\ref{N2actionEM}) satisfy an extremality bound $M^2_{\mathrm{ADM}} \geq {\cal Q}^2 M_p^2$ (see \cite{DallAgata:2011zkh} for a review). However, which particles are needed for the discharge of black holes in the general action (\ref{N2actionEM}) is more complicated to determine. We postpone a discussion of this to section \ref{sec:wgcsca} where a further generalization of the action (\ref{N2actionEM}) will be discussed.

\subsubsection{The Weak Gravity Conjecture with Scalar Fields}
\label{sec:wgcsca}

The analysis of the Weak Gravity Conjecture so far has involved conditions only on gravity and gauge fields. It is interesting to consider how it should be modified by the presence of scalar fields. For the case of scalar fields with no potential, this was studied in \cite{Palti:2017elp}. One considers an extension of the action (\ref{N2actionEM}) to include scalar fields
\be
S = \int d^4x \sqrt{-g} \left[ \frac{R}{2} - g_{ij}\left(t\right)\partial_{\mu} t^i \partial^{\mu} t^{j} + {\cal I}_{IJ}\left(t\right)F_{\mu\nu}^{I} F^{J,\mu\nu} +  {\cal R}_{IJ}\left(t\right) F_{\mu\nu}^{I} \left(\star F\right)^{J,\mu\nu} \right]\;.  \label{genaction}
\ee
The $t^i$ denote scalar (or pseudo-scalar) fields, whose expectation values parameterize a moduli space with metric $g_{ij}$. All of the kinetic terms for the fields, $g_{ij}$, $ {\cal I}_{IJ}$ and $ {\cal R}_{IJ}$ are allowed to be functions of the scalar fields $t^i$. Importantly, also the mass of the Weak Gravity Conjecture particle $m$ is allowed to be a function of the scalars. It was then proposed that this leads to a new contribution to the Weak Gravity Conjecture bound.

\begin{tcolorbox}
{\bf Weak Gravity Conjecture with Scalar Fields (for general action (\ref{genaction}))} \;\cite{Palti:2017elp}
{\it 
\newline
\newline
A theory with scalar fields $t^i$, which have no potential, with general action as in (\ref{genaction}), should have a particle with mass $m\left(t\right)$ satisfying the bound
\be
\label{wgcsc}
{\cal Q}^2 M_p^2 \geq m^2 + g^{ij} \left( \partial_{t^i} m \right) \left( \partial_{t^j} m \right) M_p^2\;,
\ee 
with ${\cal Q}^2$ defined as (\ref{qdefq}).
}
\end{tcolorbox}

The most straightforward motivation for (\ref{wgcsc}) comes from the bound states argument as in section \ref{sec:bstates}. Recall, that this involved demanding that the repulsive force should be stronger than the attractive one on the particle with the largest charge-to-mass ratio in the theory. In the presence of massless scalar fields the particle may couple also to the scalars and this would lead to an additional attractive force. The strength of this additional force is then accounted for by the last term in (\ref{wgcsc}). So that the inequality is now directly stating that two Weak Gravity Conjecture particles should be self-repulsive rather than self-attractive. 

\begin{figure}[t]
\centering
 \includegraphics[width=0.8\textwidth]{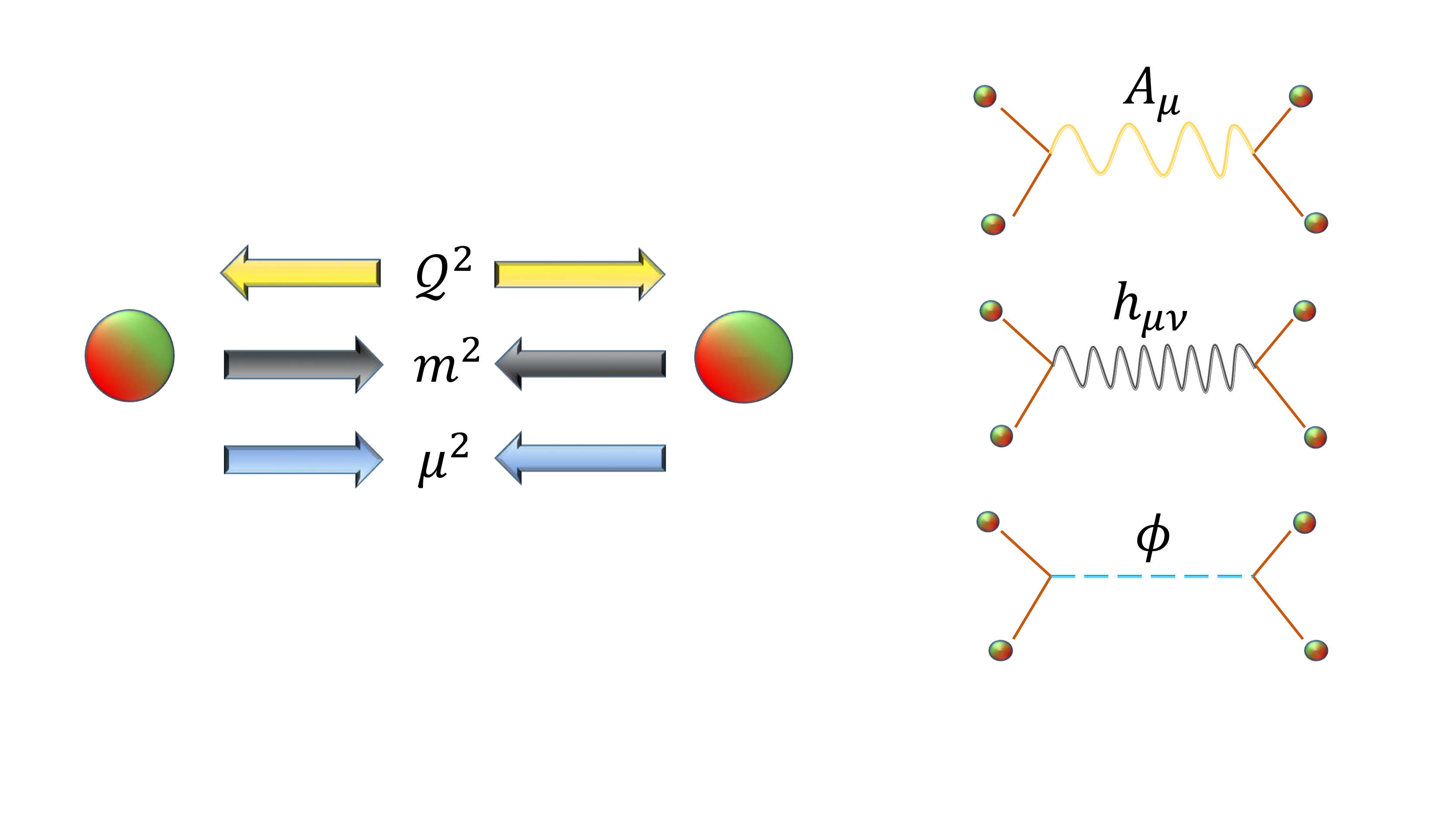}
\caption{Figure showing the long-range forces acting on the Weak Gravity Conjecture particle and the associated Feynman exchange diagrams. The repulsive electromagnetic force is mediated by the gauge field $A_{\mu}$ and acts with strength ${\cal Q}^2$, the attractive gravitational force is mediated by the graviton $h_{\mu\nu}$ and acts with strength $m^2$, and the attractive scalar force is mediated by a massless scalar $\phi$ with strength $\mu^2$. The Weak Gravity Conjecture (\ref{wgcsc}) is then the statement that the repulsive force should be at least as strong as the sum of the attractive forces.}
\label{fig:WGCSCF}
\end{figure}

We will refer to the scalar mediated force a number of times in this review, and so it is useful to discuss its derivation. Consider a canonically normalized scalar field $\phi$ which couples to a scalar particle $h$, with mass $m_0$, through the interaction
\be
\label{lagmmu}
{\cal L} \supset \left( 2 m_0 \mu \phi + m_0^2 \right)\left| h \right|^2 \;.
\ee
Here $\mu$ is a constant defining the interaction strength, and we have taken, with full generality, $\phi$ to have vanishing vacuum expectation value $\left< \phi\right>=0$. We can now consider the `general' mass term $m$ for $h$ as the function in the effective Lagrangian which multiplies $\left|h\right|^2$, so as given in (\ref{lagmmu}). Then we have
\be
\partial_{\phi} \left( m^2 \right) = 2 m \partial_{\phi} m = 2 m_0 \mu\;.
\ee
Evaluating this in the vacuum we see that the interaction strength $\mu$ is just the derivative of the mass term for $h$ evaluated in the vacuum
\be
\label{muvac}
\mu = \left< \partial_{\phi} m \right> \;.
\ee
It is simple to check that the same expression (\ref{muvac}) is true for a Fermionic particle $h$. More non-trivial is that (\ref{muvac}) also gives the relevant coupling for our purposes if $\phi$ was a pseudo-scalar (see \cite{Lust:2017wrl} for a discussion on this). The 3-point coupling $\phi \left|h\right|^2$ gives rise to a long-range Coulomb attractive self-force for $h$ mediated as an exchange of $\phi$. This force takes the form (see, for example \cite{TongQFT})
\be
\label{scaforc}
F_{\mathrm{Scalar}} = \frac{\mu^2}{4 \pi r^2} \;.
\ee 
Comparing the scalar mediated force (\ref{scaforc}) with the gravitational and electromagnetic forces (\ref{graemfroc}), we see that (\ref{wgcsc}) indeed captures their relative magnitude. This is illustrated in figure \ref{fig:WGCSCF}.

\subsubsection*{Black hole discharge with scalar fields}

The Weak Gravity Conjecture was motivated by the discharge of black holes. These black holes were Reissner-Nordstrom type. In the case where there are also scalar fields involved the black hole solutions can be modified, including their charge-to-mass ratio and extremality bound. This is because black holes may develop so-called secondary scalar hair, which means the solution, including the horizon, is sensitive to the values of scalar fields indirectly through their appearance in the gauge couplings of gauge fields. This is a very well studied effect, often termed the Attractor Mechanism, see \cite{DallAgata:2011zkh} for a review. The mechanism is typically studied for ${\cal N}=2$ supersymmetric solutions, for general reasons that will be discussed below, but actually holds also in the absence of supersymmetry. Indeed, it is possible to show that extremal black hole solutions to the action (\ref{genaction}) satisfy the following relation \cite{Ceresole:2007wx,DallAgata:2011zkh}
\be
{\cal Q}^2 = M_{\mathrm{ADM}}^2 + g^{ij} \left(\partial_i M_{\mathrm{ADM}}\right) \left(\partial_j M_{\mathrm{ADM}}\right) \;. \label{gWGCmadm}
\ee
Note that since the scalar fields may have a spatially varying profile, we should specify that (\ref{gWGCmadm}) holds at infinity. Indeed, on the extremal black hole horizon the scalar fields are fixed to their attractor values such that they solve $\partial_j M_{\mathrm{ADM}}=0$ \cite{Ceresole:2007wx}.
The equation (\ref{gWGCmadm}) is analogous to (\ref{wgcsc}), suggesting a connection to discharge of black holes. However, obtaining the precise requirement on particles in order for black holes to discharge in generality is not simple. In particular, the dependence of the black hole mass $M_{\mathrm{ADM}}$ on the scalar fields may in general be different to the dependence of the particle mass $m$, in which case it is difficult to establish a direct relation between the black hole statement (\ref{gWGCmadm}) and the particle one (\ref{wgcsc}). Nonetheless, the analogy between (\ref{gWGCmadm}) and (\ref{wgcsc}) is certainly strong motivation for the validity of (\ref{wgcsc}), especially once we note that the form of (\ref{wgcsc}) is almost uniquely fixed by requiring invariance under scalar fields reparameterisation and electric-magnetic duality.

\begin{figure}[t]
\centering
 \includegraphics[width=0.9\textwidth]{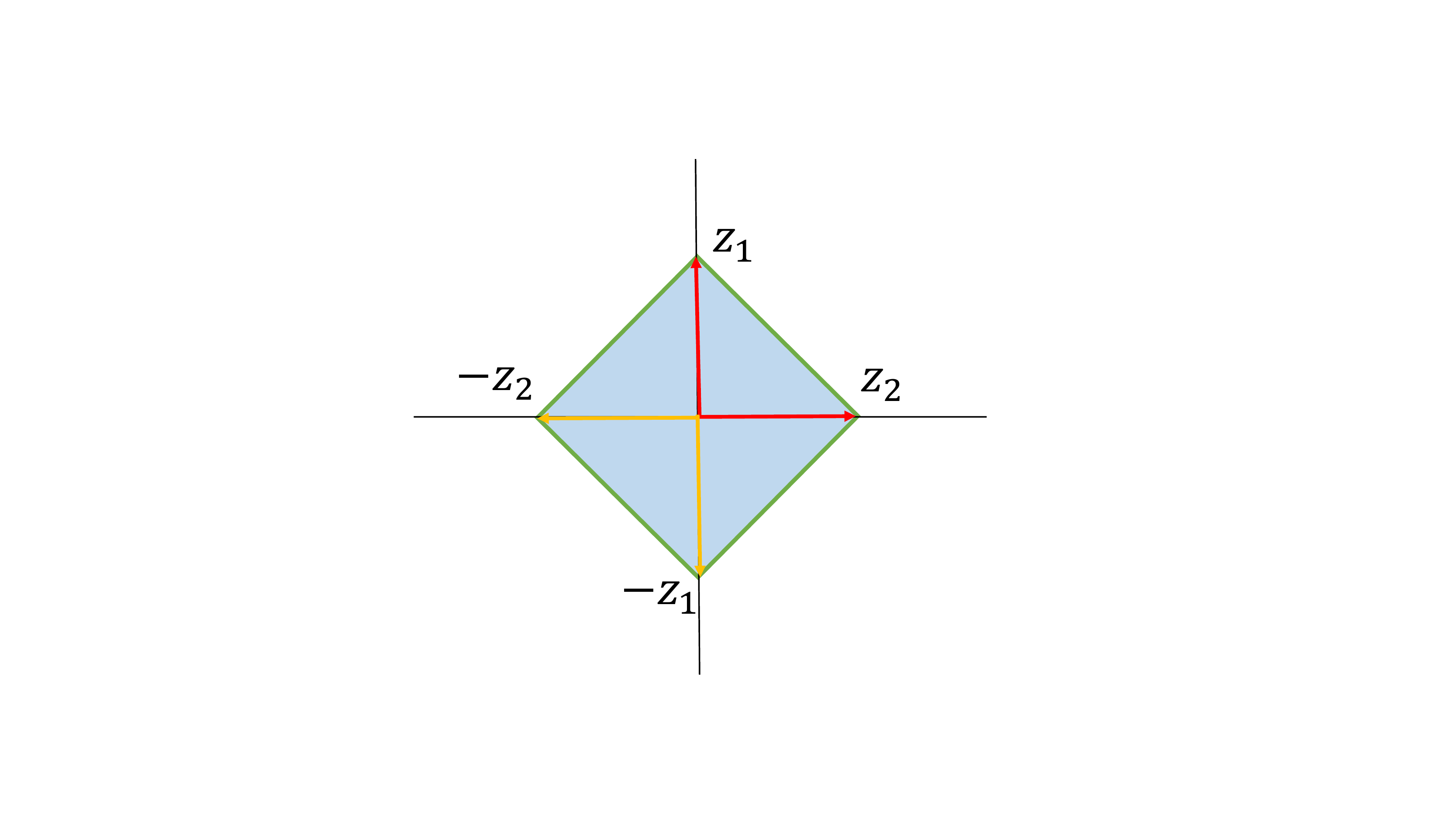}
\caption{Figure showing the charge-to-mass plane for dilaton black holes. The convex-hull shape, relevant for discharge, is modified compared to Reissner-Nordstrom black holes depicted in figure \ref{fig:CH}. This means that particles with charge-to-mass ratios, depicted by $z_1$ and $z_2$, which are the same as those of an extremal black holes are sufficient to allow for discharge.}
\label{fig:CHdbh}
\end{figure}

For simple black holes it is possible to study the condition for discharge quite explicitly. A simple black hole solution is the so-called dilaton black hole \cite{Horowitz:1991cd}. We will actually consider a slight variant where the action will have two symmetries $U(1)_0$ and $U(1)_1$ (see, for example \cite{Freedman:2012zz}). It is a restriction of the general action (\ref{genaction}) to two decoupled $U(1)$s, with related gauge couplings, and a single (canonically normalized) scalar field $\phi$. Specifically, we take
\be
{\cal I}_{00} \equiv -\frac{1}{4g_0^2} = - e^{-2\phi} \;,\;\; {\cal I}_{11} \equiv -\frac{1}{4g_1^2} = -e^{2\phi} \;,\;\; {\cal R}_{00}={\cal R}_{11}=0 \;.
\ee
We consider a black hole with purely electric charges $q_0$ and $q_1$. Note that the gauge couplings are inversely related. This means that we could actually map the electrically charged black hole solutions with two $U(1)$s to dyonic black hole solutions with just one $U(1)$. The analysis in this section can be viewed in terms of either of these pictures. With this choice of charges, the black holes have the explicit expressions
\be
\label{qdlbhe}
{\cal Q}^2 = 2 \left( q_0^2 g_0^2 + q_1^2 g_1^2\right) \;,\;\; M_{\mathrm{ADM}} = q_0 g_0 + q_1 g_1 \;,\;\; \partial_{\phi} M_{\mathrm{ADM}} = q_0 g_0 - q_1 g_1\;.
\ee
An interesting property of (\ref{qdlbhe}) is that the black hole mass is a linear function of the charges. If we plot the charge-to-mass ratio for such extremal black holes it no longer takes the form of a unit circle, as was the case for Reissner-Nordstrom black holes considered in section \ref{sec:multiu1}, but rather is a diamond. The convex hull condition for the discharge of the black hole then takes a different form. This is illustrated in figure \ref{fig:CHdbh}. 

At infinity, the scalar field $\phi$ is a free parameter and the relation (\ref{gWGCmadm}) indeed holds for all values of $\phi$. As we vary $\phi$ the charge-to-mass ratio of the extremal black hole varies but is bounded
\be
\label{ctmrange}
2\geq \frac{{\cal Q}^2}{M_{\mathrm{ADM}}^2} \geq 1 \;.
\ee
On the black hole horizon the scalar is fixed such that 
\be
\label{attval}
\left. \frac{g_1}{g_0} \right|_{\mathrm{Horizon}} = \frac{q_0}{q_1} \;.
\ee
A special type of black holes are ones where the scalar values are chosen at infinity to be equal to those on the horizon, these are called double extremal black holes \cite{Kallosh:1996tf}. Restricting to such black holes then fixes 
\be
{\cal Q}^2 = q_0 q_1 \;,\;\; M_{\mathrm{ADM}} = \sqrt{q_0 q_1} \;,\;\; \partial_{\phi} M_{\mathrm{ADM}} = 0\;.
\ee
Double extremal black holes are such that their mass is completely fixed in terms of the quantized charges. They also minimize the charge-to-mass ratio such that the lower inequality in (\ref{ctmrange}) is saturated. 

It is interesting to consider the case where $q_1=0$. Such black holes are not actually well understood because the horizon value of the gauge coupling cannot be finite (\ref{attval}). We may nonetheless proceed, probing the black hole from infinity. This is the setup studied in \cite{Heidenreich:2015nta}. In such cases we maximize the charge-to-mass ratio (\ref{ctmrange}) so that 
\be
\label{d4d1}
{\cal Q}^2 = 2M_{\mathrm{ADM}}^2\;.
\ee
It is important to note that there is a factor of 2 difference from the Reissner-Nordstrom case, as used in (\ref{wgcdy}). More generally, if we consider a dilatonic coupling like $e^{-\delta \phi} F^2$, and generalize to $d$-dimensions, then one finds \cite{Horowitz:1991cd,Heidenreich:2015nta}
\be
\label{gendd}
{\cal Q}^2 = 2\left(  \frac{d-3}{d-2} + \frac{\delta^2}{2} \right)M_{\mathrm{ADM}}^2 \;.
\ee
The case (\ref{d4d1}) is for $d=4$ and $\delta=1$, while the Reissner-Nordstrom case is for $\delta=0$. It was therefore proposed in \cite{Heidenreich:2015nta} that the Weak Gravity Conjecture particle should satisfy the bound (\ref{gendd}) on its charge-to-mass ratio.

\begin{tcolorbox}
{\bf Weak Gravity Conjecture in $d$-dimensions with dilaton} \;\cite{Heidenreich:2015nta}
{\it 
\newline
\newline
A $d$-dimensional theory with a $U(1)$ whose gauge coupling is given by a dilaton $g^2=e^{\delta \phi}$, should have a particle with mass $m$ and charge $q$ satisfying the bound
\be
\label{wgcscdd}
g^2 q^2 M_p^2 \geq \left(  \frac{d-3}{d-2} + \frac{\delta^2}{2} \right) m^2 \;.
\ee 
}
\end{tcolorbox}

Because the gauge coupling depends on $\phi$ exponentially, it is natural to expect that the mass of the Weak Gravity Conjecture particle should also depend on $\phi$ exponentially, with an exponent at least as large as the gauge coupling. Otherwise for sufficiently large $\phi$ one could make the charge smaller than any finite mass $m$. In that case, it is simple to check that (\ref{wgcsc}) is indeed satisfied, at least for weak coupling $g \ll 1$. However, as mentioned earlier, proving that (\ref{wgcsc}) is necessary for black hole discharge at strong coupling $g \sim 1$, or for non-dilatonic black holes, is much more difficult. This was noted already in \cite{Palti:2017elp}, and more recently elaborated upon in \cite{Lee:2018spm}.

\subsubsection*{Arguments from dimensional reduction}

Another motivating argument for (\ref{wgcsc}) comes from classical dimensional reduction \cite{Lust:2017wrl}. Consider, for example, a 5-dimensional theory 
\be
S_{5D} = \int_{M_4 \times S^1} d^5 X \sqrt{-G} \left[ \frac12 R^{(5)}  - \frac{1}{4g_5^2} F_{MN} F^{MN} - D_M H \left(D^M H \right)^* - M_H^2 H H^* \right] \;. \label{5daction}
\ee
The 5-dimensional gauge field is $A^M$, with gauge coupling $g_5$, There is a charged scalar $H$, with charge $q$ and mass $M_H$, which acts as the Weak Gravity Conjecture particle and so satisfies (\ref{ewgcd}) 
\be
g_5 q \geq \sqrt{\frac23}  M_H \;. \label{5dWGC}
\ee
If we reduce this theory on a circle, then we get an effective 4-dimensional action for the zero modes 
\be
S^0_{4D} = \int_{M_4} d^4 x \sqrt{-g} \left[ \frac12 R + {\cal I}_{IJ}  F_{\mu\nu}^{I}  F^{J,\mu\nu}  - \frac12 \left(\partial \varphi \right)^2 - \frac{1}{2 r^2g_5^2} \left(\partial a \right)^2 - \left| D h \right|^2 - m^2 h^* h\right] \;. \label{4dreducedaction}
\ee
The 4-dimensional fields are the dilaton $\varphi$, an axion $a$, a complex scalar $h$, the graviphoton $A_0^{\mu}$, and the zero mode of the gauge field $A_1^{\mu}$. The dilaton measures the circumference $r$ of the extra dimension
\be
r = e^{-\sqrt{\frac{2}{3}}\varphi} \;.
\ee 
The covariant derivative of the scalar field is given by
\be
D_{\mu} h = \left(\partial_{\mu} - i q A^1_{\mu} \right) h \;,
\ee
where $q$ is its charge, and the mass is given by
\be
m^2 =  \frac{M_H^2}{r} +\frac{q^2 a^2}{r^3} \;.
\ee
The gauge kinetic matrix reads
\be
{\cal I}_{IJ} = -\frac{r}{4 g_5^2} \left( \begin{array}{cc} \frac{r^2g_5^2}{2} + a^2 & -a \\ -a & 1 \end{array} \right) \;,\;\; \left({\cal I}^{-1}\right)^{IJ} = -\frac{8}{r^3} \left( \begin{array}{cc}  1 & a \\ a & \frac{r^2g_5^2}{2} + a^2 \end{array} \right)\;. 
\ee
The 4-dimensional theory has scalar fields which couple to the Weak Gravity Conjecture particle and so we should therefore consider the bound (\ref{wgcsc}) which reads
\be
\frac{2M_H^2 r}{3\left(a^2 q^2 + M_H^2 r^2 \right)}\left(3 g_5^2 q^2 - 2 M_H^2  \right) \geq 0 \;. \label{4dbo}
\ee
This precisely reproduces the 5-dimensional bound (\ref{5dWGC}). We find that the proposal for the Weak Gravity Conjecture with scalar fields is matched onto the Weak Gravity Conjecture without scalar fields in one dimension higher. This connection forms a motivation for the proposal (\ref{wgcsc}).

\subsubsection*{Relation to ${\cal N}=2$ supersymmetry}

The analysis so far as been classical in nature. Once we include quantum corrections it is clear that the conjecture (\ref{wgcsc}) can only apply in rather special circumstances. This is because the conjecture applies in situations where the scalar fields have no potential. However, since the fields must couple to at least the Weak Gravity Conjecture particle, at loop level they will generically obtain a potential. The exception to this generic rule is the case where there is an extended ${\cal N}=2$ supersymmetry (or more generally 8 supercharges). Such setups have scalar fields with no potential which therefore form a true moduli space. It is therefore expected that any statements involving scalar fields with no potential are only exact in the case of extended supersymmetry, and in other scenarios should be thought of as a classical leading order result that would need to be modified in some way once the scalar fields obtain a potential. 

The case of ${\cal N}=2$ also has natural candidates which satisfy, and saturate, the conjecture (\ref{wgcsc}). These are BPS states. In the case of extended supersymmetry one can define a central charge which is a function that depends on the charge of a state $Z\left(q\right)$, and on the scalar fields in the vector multiplets (see section \ref{sec:stringcomp} for an introduction). Then charged states in the theory, with charge $q$, have a lower bound on their mass
\be
\label{bpsboundsim}
M \geq \left|Z\left(q\right)\right| \;.
\ee
We will discuss the properties of BPS states and the Swampland in more detail in section \ref{sec:stringcomp}. For now we note that they saturate the bound (\ref{wgcsc}), but importantly, non-BPS states may still satisfy (\ref{wgcsc}). Therefore, the Weak Gravity Conjecture particle may not be a BPS state. From the perspective of forces and bound states this is apparent. Note, however, that black holes can themselves be BPS states.\footnote{In ${\cal N}=2$ supergravity BPS black holes are automatically extremal, but extremal black holes may or may not be BPS.} And it is a general result (see section \ref{sec:stringcomp}) that BPS states can only decay to other BPS states. So from the perspective of BPS black holes discharge the appropriate particle must be BPS. On the other hand, there are BPS black holes which simply cannot discharge. More generally, any BPS states charged under multiple $U(1)$s with some co-prime charges, can only decay to other BPS states over special loci in moduli space (so-called curves of marginal stability). Away from such loci, there are many BPS states, including black holes, which are absolutely stable. It is therefore not clear how black hole discharge should be interpreted in the case of ${\cal N}=2$ supersymmetry. See, for example, \cite{ArkaniHamed:2006dz,Palti:2017elp} for a discussion of such issues.

\subsubsection*{Massive generalizations}

In the case without extended supersymmetry one expects all scalars to gain some potential (see section \ref{sec:modsp}). Then it is not clear how (\ref{wgcsc}) should be generalized. One proposal was made in \cite{Lust:2017wrl}. The idea is to apply the bound states argument for a Yukawa type force for the scalar. Since such a force becomes trivial at distances much larger than the mass of the scalar, any bound states due to a scalar force would only be (quantum mechanically) meta-stable. It is unclear whether these should be forbidden. In any case, it is simple to implement the absence of meta-stable bound states. We can write (\ref{wgcsc}) for a single massless scalar field $\phi$, with coupling $\mu = \partial_{\phi} m$, as
\be
\label{wgctes}
{\cal Q}^2 M_p^2 \geq m^2 + \mu^2 M_p^2
\ee
Giving the scalar a mass $m_{\phi}$, but taking it to be much smaller than that of the Weak Gravity Conjecture particle $m_{\phi} \ll m$, and placing the two particles at the minimal distance where a classical force analysis is a decent approximation, set by the $m^{-1}$, the Yukawa force amounts to a replacement in (\ref{wgctes}) of
\be
\mu^2 \rightarrow \mu^2 - {\cal O} \left(\frac{\mu^2 m_{\phi}}{m} \right) \;.
\ee
This therefore, assuming the reasoning of the argument, sets the magnitude of the correction to the expression (\ref{wgctes}).

\subsubsection*{Relation to UV/IR mixing and naturalness}

The additional term in the Weak Gravity Conjecture in the presence of scalar fields (\ref{wgcsc}) naturally raises the interesting limit where scalar force term cancels the gauge force term. The physics of this limit was studied in \cite{Lust:2017wrl}. If we consider a single electrically charged particle of charge one, and a single massless real scalar $\phi$ with kinetic terms $\frac12 \left(\partial \phi \right)^2$, we may write (\ref{wgcsc}) as\footnote{Note that here we correct an unimportant factor of 2 missing in \cite{Lust:2017wrl}.}
\be
2\left( g^2 - \mu^2 \right) M_p^2 \equiv \beta^2 M_p^2 \geq m^2 \;,
\label{simineq}
\ee
with $\mu$ defined as in (\ref{muvac}). The limit of interest is the $\beta \rightarrow 0$ limit while keeping $g$, and therefore $\mu$, finite. This limit, if it exists in the parameter space of the theory, is interesting because it implies a small mass $m$, but keeps the ultraviolet cutoff scale due to the magnetic Weak Gravity Conjecture $g M_p$ parametrically higher. It therefore suggests a form of ultraviolet-infrared (UV/IR) mixing where quantum gravity physics requires a particle much lighter than the cutoff scale of the effective theory. 

The limit becomes even more interesting once we note that if $g^2\sim \mu^2 < \beta$ then a hierarchy $g \sim \mu \gg \beta$ is technically natural \cite{Lust:2017wrl}. So loop corrections from an arbitrary high mass sector do not necessarily destabilize the hierarchy of scales between the mass of the particle and the cutoff scale of the theory. Such a situation would be at odds with our ideas on naturalness. In this sense, the Weak Gravity Conjecture may shed light on the hierarchy problem by suggesting that our ideas about technical naturalness may be misleading because we assume a generic unconstrained ultraviolet completion, while it may be that the ultraviolet is constrained such that it does not disturb the hierarchy. Interestingly, the setup also has a smoking-gun experimental signature, specifically that if a particle is unnaturally light than we should also find that its coupling to scalar and gauge fields is almost precisely identical. 

In order to realize such scenarios in a realistic setting there are two main difficulties to overcome. Firstly, we would need to understand the physics of the scalar $\phi$. The inequality (\ref{simineq}) is only true in the case when $\phi$ is precisely massless. It can be checked that, since $\phi$ need only couple to the Weak Gravity Conjecture particle, its mass can naturally be much lower. This suggests that an appropriately modified, to account for a small mass, inequality should still hold. Another question is whether there could be an obstruction to the limit $\beta \rightarrow 0$ while $g$ finite. In ${\cal N}=2$ supergravity we know of explicit examples in string theory where this limit is possible, specifically the conifold locus, but perhaps in the absence of supersymmetry this region of parameter space is obstructed. 

Both of these issues may play a role in understanding how the Weak Gravity Conjecture (\ref{wgcsc}) is compatible with our notion of naturalness. But then they would raise further questions, such as why is the naturalness of the mass of a charged field dependent on the mass of some neutral scalar. And why would there be an obstruction to a particle having almost equal coupling to gauge and scalar fields. Perhaps it is the Weak Gravity Conjecture which must be violated in such settings, but then the question arises as to the consistency of the stable gravitationally bound states. In any case, it seems that the issues raised in \cite{Lust:2017wrl}, regarding the limit $\beta \rightarrow 0$, require some interesting new physical insight. 

\subsubsection{Scalar Weak Gravity Conjecture}
\label{sec:scalwgc}

The Weak Gravity Conjecture with scalar fields (\ref{wgcsc}) is partially motivated by the contribution of a scalar force (\ref{scaforc}). The original conjecture \cite{ArkaniHamed:2006dz} was stated as the idea that gravity should be the weakest force acting on at least one particle in the theory. It is natural to contemplate the idea that this should remain the case even when there are scalar forces acting on the particle. This leads to a Scalar Weak Gravity Conjecture proposed in \cite{Palti:2017elp}.

\begin{tcolorbox}
{\bf Scalar Weak Gravity Conjecture } \;\cite{Palti:2017elp}
{\it 
\newline
\newline
A theory with massless scalar fields $t^i$, whose kinetic terms are as in (\ref{genaction}), should have a state with mass $m$ satisfying the bound
\be
\label{swgc}
g^{ij} \left( \partial_{t^i} m \right) \left( \partial_{t^j} m \right) M_p^2 > m^2 \;.
\ee 
}
\end{tcolorbox}

Note that (\ref{swgc}) is stated with a strict inequality. The presence of massless scalar fields again naturally ties to ${\cal N}=2$ supersymmetry, and therefore we may look there for an initial motivation. Indeed, it was shown in \cite{Palti:2017elp} that (\ref{swgc}) is satisfied by BPS states. More precisely, there is at least one BPS states for every scalar field in a vector multiplet which satisfies (\ref{swgc}). We will show this in section \ref{sec:n2sugra}.

In \cite{Palti:2017elp} two slightly different versions of (\ref{swgc}) were proposed. One was that it should hold for the Weak Gravity Conjecture particle, so the one which is also charged under a $U(1)$ symmetry. This is actually a milder statement than (\ref{swgc}), and is better motivated. For example, by the BPS argument above. Also, while as a self-force a scalar always mediates an attractive interaction, between different states a scalar interaction can be repulsive. This lead to some, though not very strong, arguments for this milder version of (\ref{swgc}) in terms of bound states. The stronger version of (\ref{swgc}) is as stated above, that it should hold for any scalar field. This has no arguments from black hole discharge or absence of stable bound states, and so is on weaker footing in this sense. It is also quite likely that if that is to be the case, then the conjecture should be refined slightly such that the state that the scalar field couples to need not in general be a particle, but possibly an extended object. With the mass derivative replaced by the appropriate coupling strength relating to the tension. This is because scalar fields do not have a canonical dimension for the object that they couple to. We will discuss this form the perspective of string theory in more detail in section \ref{sec:stringcomp}.

Another motivation for (\ref{swgc}) is that it relates nicely to the Swampland Distance Conjecture. We will discuss this in more detail in section \ref{sec:distswgc}. This relation, however, holds most straightforwardly at large distances in field space and so does not support (\ref{swgc}) everywhere in field space. And, again, it naturally implies that the state the scalar should couple to may be extended in nature.  

In \cite{Lust:2017wrl} the behaviour of the Scalar Weak Gravity Conjecture under dimensional reduction was studied, and it was also suggested that the limit where the left hand side of (\ref{swgc}) tends to zero should be associated to a lowered cutoff scale. In \cite{Lee:2018spm} it was shown that in six dimensions the tower of states associated to the distance conjecture does not satisfy the Scalar Weak Gravity Conjecture, although there may be other states which do. This suggests that away from four dimensions the conjecture may need to be modified with some appropriate factor dependent on the number of dimensions. Or simply that it should only hold up to order-one factors. In \cite{Gonzalo:2019gjp} a version of the Scalar Weak Gravity Conjecture was proposed which involved the second derivative of the mass.\footnote{See also \cite{Palti:2017elp} for a similar suggestion.} The conjecture was also applied to the scalar itself, which lead to constraints on its potential.

\subsubsection{Tower versions of the Weak Gravity Conjecture}
\label{sec:wgctower}

Already in the first encounter with the Weak Gravity Conjecture in section \ref{sec:fewgc} we saw that the magnetic Weak Gravity Conjecture scale was associated with the mass scale of an infinite tower of states, in that case they were Kaluza-Klein and Winding towers. In fact, each state in those towers satisfied the electric Weak Gravity Conjecture. There are proposals that this may be general, so there should be (formally infinite) towers of states, with increasing charges and masses, which satisfy the Weak Gravity Conjecture \cite{Heidenreich:2015nta,Klaewer:2016kiy,Montero:2016tif,Heidenreich:2016aqi,Andriolo:2018lvp,Grimm:2018ohb}. The proposals differ slightly by some properties of the tower, and we will only discuss the cases of \cite{Heidenreich:2015nta,Montero:2015ofa,Heidenreich:2016aqi} and \cite{Andriolo:2018lvp} in this section. The proposal of \cite{Grimm:2018ohb} is better studied in the context of string theory in section \ref{sec:stringcomp}. The towers will play a central role in section \ref{sec:emergence}. 

\begin{figure}[t]
\centering
 \includegraphics[width=0.9\textwidth]{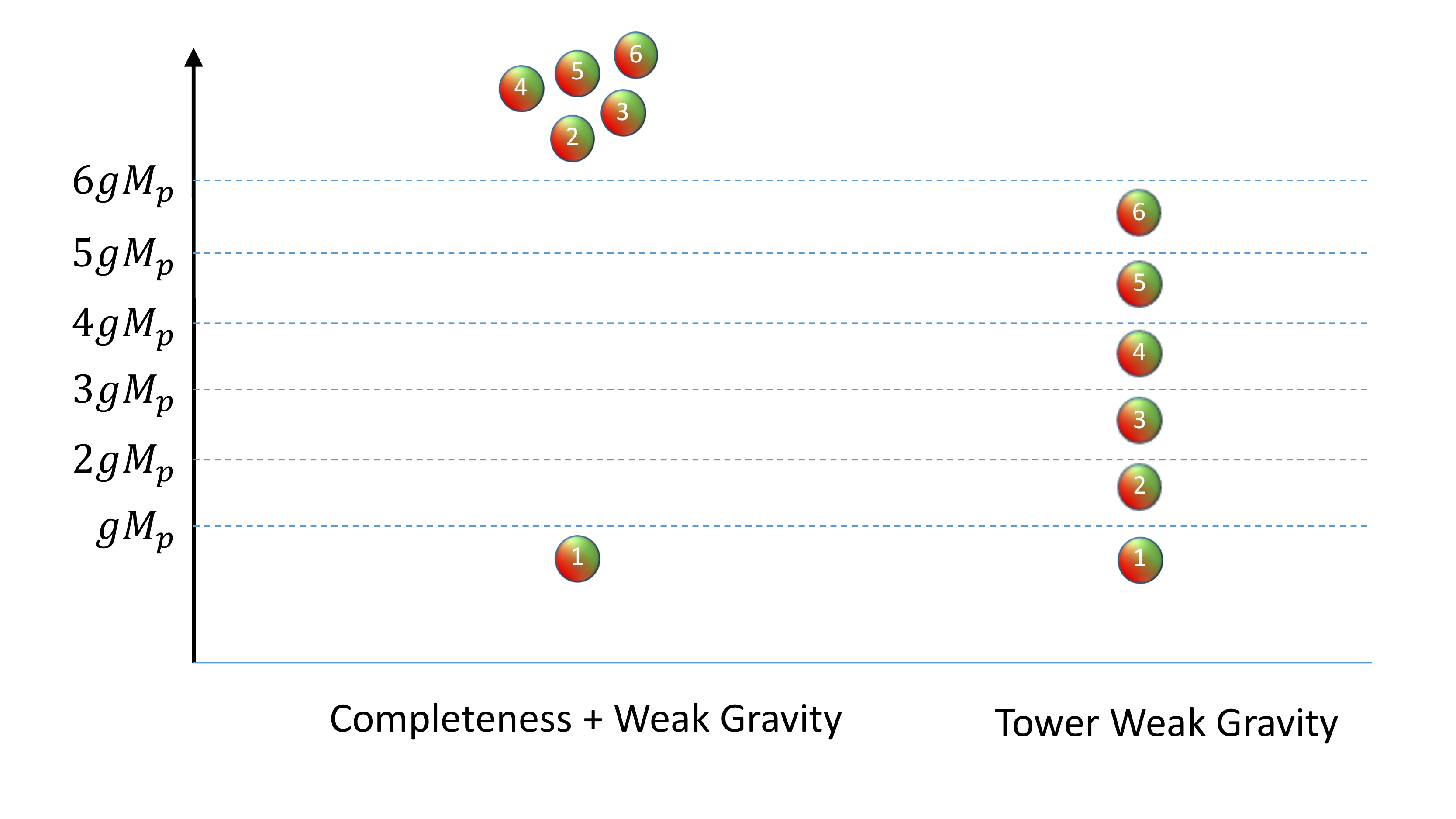}
\caption{Figure showing the increasing mass scales associated to $U(1)$ charges as dictated by the Weak Gravity Conjecture (\ref{ewgc}). The spectrum of charged states on the left respects the Completeness Conjecture and the Weak Gravity Conjecture. The spectrum on the right respects a tower version of the Weak Gravity Conjecture.}
\label{fig:two}
\end{figure}

The existence of towers of charged states is similar to the Completeness Conjecture discussed in section \ref{sec:complconj}. The conjecture would imply an infinite number of states filling the whole lattice of $U(1)$ charges. However, this places no constraints on the mass of the states. The tower conjectures all require that the mass of the states is restricted by their charge. This is illustrated in figure \ref{fig:two}.  

The first proposal of a tower version of the Weak Gravity Conjecture was made in \cite{Heidenreich:2015nta}. This was termed the Lattice Weak Gravity Conjecture. It was later adjusted to a Sub-Lattice Weak Gravity Conjecture \cite{Montero:2016tif,Heidenreich:2016aqi}\footnote{The refinement to a sub-lattice due to the presence of a discrete symmetry was noted first in the context of instantons \cite{Montero:2015ofa,Brown:2015lia}.}
\begin{tcolorbox}
{\bf Sub-Lattice Weak Gravity Conjecture } \;\cite{Heidenreich:2015nta} \;\; (refined to sub-Lattice in \cite{Montero:2016tif,Heidenreich:2016aqi})
{\it 
\newline
\newline
In a gauge theory with (possibly multiple) $U(1)$s, coupled to gravity, any spot on a charge sub-Lattice contains a super-extremal particle.
}
\end{tcolorbox}

Here super-extremal means particles whose charge is larger than their mass, which in our notation means in general (\ref{wgcdy}), or for simpler setups (\ref{ewgc}). The full lattice of charges are all the charges consistent with Dirac quantization. A sub-Lattice of that may have some spaces missing, while still retaining the lattice structure, which means adding the charges of any two populated sites gives a charge which is itself populated by a state. 

\begin{figure}[t]
\centering
 \includegraphics[width=0.9\textwidth]{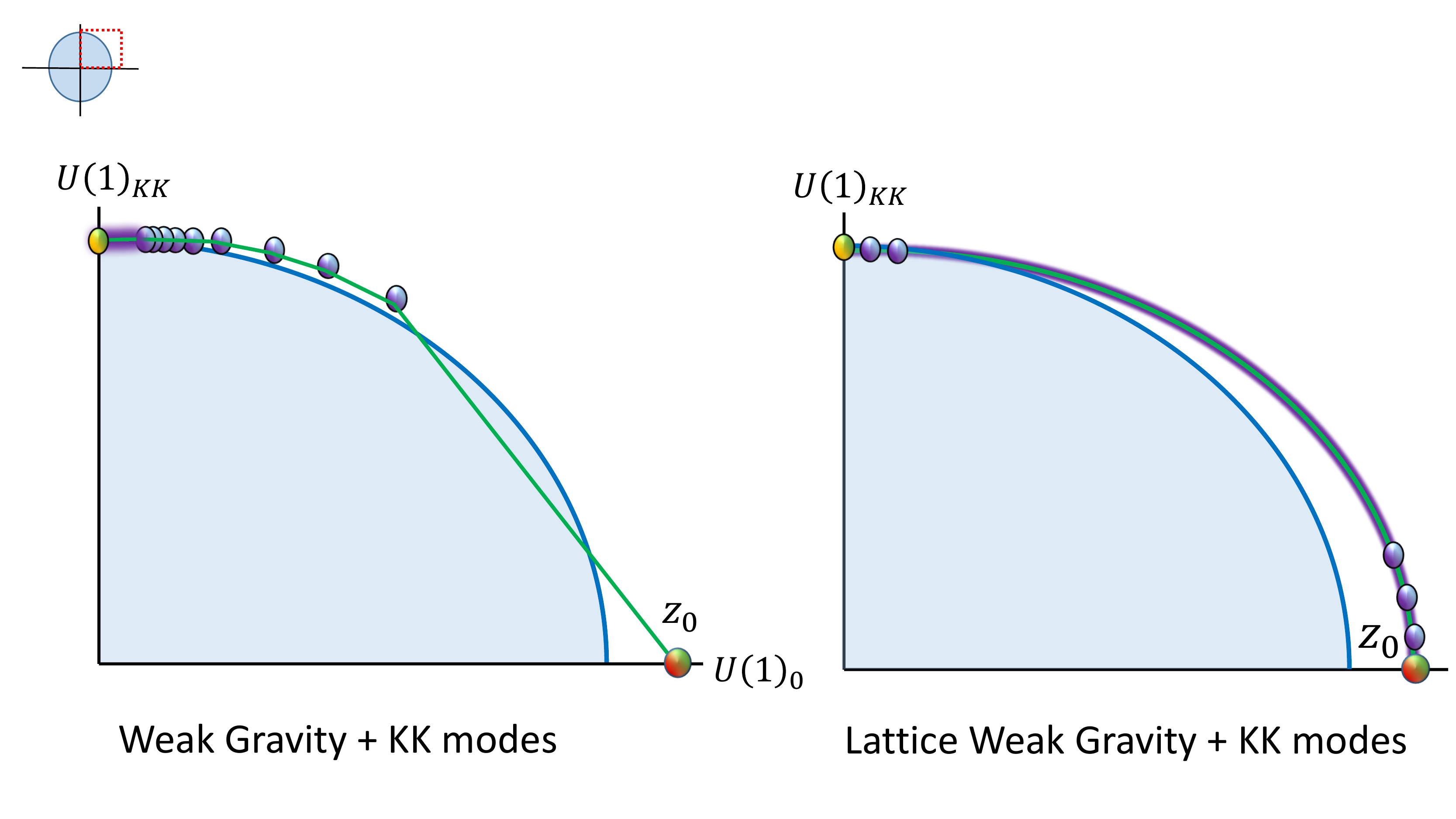}
\caption{Figure showing the top-right quadrant of the charge-to-mass plane for the zero mode of a higher dimensional $U(1)$, denoted $U(1)_0$, along with the Kaluza-Klein $U(1)_{KK}$. The charged black holes fill the unit circle. On the left we show the spectrum of charged states composed of a single higher dimensional state leading to a zero mode with charge-to-mass ratio $z_0$, along with all its KK modes. The convex hull, shown by a green line connecting the states, fails to contain the unit circle for sufficiently small extra-dimensional circle radius $R$. On the right we depict the spectrum when there is a full infinite tower of super-extremal higher dimensional states with increasing charges, as in the sub-Lattice Weak Gravity Conjecture. In this case, the convex hull contains the unit circle for any value of $R$.}
\label{fig:ellwgcl}
\end{figure}

We will present evidence for, and against, the conjecture from string theory in section \ref{sec:stringcomp}.\footnote{Note that in \cite{Lee:2019tst} a counter-example to the conjecture was proposed in four dimensions.} A general argument for the conjecture was presented in \cite{Heidenreich:2015nta,Heidenreich:2016aqi} based on dimensional reduction. The idea is to consider how the Weak Gravity Conjecture for multiple $U(1)$s, discussed in section \ref{sec:multiu1}, behaves under dimensional reduction. In particular, we would like to consider the particle spectrum required for the discharge of black holes in the lower dimensional theory which are charged under a $U(1)$ arising from the higher dimensional theory as well as the Kaluza-Klein $U(1)$ coming from the graviton component along a circle reduction. Using appropriate definitions of charges and masses, as in \cite{Heidenreich:2015nta,Heidenreich:2016aqi}, we can plot the spectrum of charge-to-mass ratios for such extremal black holes as the unit circle, similar to figure \ref{fig:CH}. The zero mode of the particle satisfying the Weak Gravity Conjecture in the higher dimensional theory will lead to a super-extremal state in the theory, with mass $m_0$, and charge-to-mass ratio $z_0 \geq 1$. There will then also be all the Kaluza-Klein modes of that particle, whose charge-to-mass ratio we denote $z_n$, with $n$ being the KK number. The pure KK modes, charged only under the KK $U(1)$, are exactly extremal and so lie on the unit circle.\footnote{More precisely, as seen in sections \ref{sec:fewgcst} and \ref{sec:wgcsca}, they saturate the bound (\ref{wgcsc}), or its simpler version (\ref{wgcscdd}).} The spectrum of charged states therefore sit on an ellipse as depicted in figure \ref{fig:ellwgcl}. The points on the ellipse become infinitely dense near the pure KK charged limit, and are most sparse near the zero mode. The point is now to consider the convex hull formed by these charged states. If we consider the case $z_0=1$, so the zero mode sits also on the unit circle. Then it is clear that the convex hull can never contain the unit circle. More generally, it was shown in \cite{Heidenreich:2015nta} that the requirement for the convex hull to contain the unit circle reads
\be
\label{m0rbound}
\left( m_0 R \right)^2 \geq \frac{1}{4z_0^2\left(z_0^2-1\right)} \;,
\ee
where $R$ is the radius of the circle. The point is then that for sufficiently small $R$, the bound (\ref{m0rbound}) can be violated. This can be avoided by positing that there should also have been a tower of charged states in the original higher dimensional theory. This means that the spectrum of charged states would take the form $\left(n,m\right)$ with $n,m \in \mathbb{Z}$, rather than $\left(q_0,m\right)$ where $q_0$ is the integer associated to the charge of the higher dimensional particle. Such a spectrum of charged states would form an infinitely dense set of points all along the ellipse, which can be seen by the symmetry between the two $U(1)$s, and therefore contain the convex hull for any value of $R$. This is illustrated in figure \ref{fig:ellwgcl}.
It is important to note however, that we cannot expect to trust the analysis for arbitrarily small radius $R$, which is the limit required to argue for an infinite tower of states. Indeed, the electric Weak Gravity Conjecture bound on $m_0$ is already expected to be near the magnetic Weak Gravity Conjecture cutoff scale of the theory, and so one expects an approximate constraint $m_0 R \gtrsim 1$. On the other hand, we may demand that even marginal black hole decay $z_0 \rightarrow 1$ should be stable under dimensional reduction. 

In \cite{Andriolo:2018lvp} another argument was presented for a necessary tower of particles satisfying the Weak Gravity Conjecture. The idea is to consider properties of higher derivative terms in the action, similar to the approach discussed in section \ref{sec:extbhins}. Then under the key assumption that those terms should arise primarily from integrating out charged states, it was argued that the dominant contribution should come from states which satisfy the Weak Gravity Conjecture.\footnote{It is interesting to note that the assumption of integrating out the tower of states being the dominant contribution to an operator matches nicely the emergence proposal discussed in section \ref{sec:emergence} \cite{danielcomment}.} This lead to a {\bf Tower Weak Gravity Conjecture} which posits the tower of super-extremal particles, but that they need not necessarily form a lattice of charges, differing in this respect from the Lattice Weak Gravity Conjecture. 

\subsubsection{The Weak Gravity Conjecture for $p$-forms}

The $U(1)$ gauge symmetries we have been considering are associated to a vector field $A_{\mu}$ transforming as $\delta A_{\mu} = \partial_{\mu} \lambda$ for some scalar gauge parameter $\lambda$. More generally, we may consider anti-symmetric tensors of rank $p$, $C^{(p)}_{[\mu_1\mu_2...\mu_p]}$, which transform as $\delta C^{(p)}_{[\mu_1\mu_2...\mu_p]} = \partial_{[\mu_1}\lambda^{(p-1)}_{\mu_2...\mu_p]}$. Here, as elsewhere in the review, the square brackets means anti-symmetrization. In this sense gauge fields are 1-form symmetries. A $p$-form field couples naturally to a $p-1$ dimensional object through the canonical pairing induced by integrating a $p$-form over the world-volume of the object
\be
\int_{{\cal W}_p} C^{(p)}_{[\mu_1\mu_2...\mu_p]} dx^{[\mu_1\mu_2...\mu_p]}  \;.
\ee 
Recall that the world-volume has one more dimension, due to time, than the spatial dimension of the object. In four spacetime dimensions, the relevant objects and form fields are shown in table \ref{tab:pforms}. 
\begin{table}[h]
\center
\begin{tabular}{|c|c|c|c|c|}
\hline
{\bf Object} &{\bf  p }&{\bf  Field Notation }& {\bf Field Name }& {\bf Dual}\\
\hline
Instanton & 0 &$C^{(0)} \equiv a$ & Axion & 2-form \\
\hline
Particle & 1 &$C^{(1)}_{\mu} \equiv A_{\mu}$ & Gauge field & Gauge field \\
\hline
String & 2 &$C^{(2)}_{[\mu\nu]} \equiv B_{\mu\nu}$ & 2-form & Axion\\
\hline
Domain wall & 3 &$C^{(3)}_{[\mu\nu\rho]} \equiv C_{\mu\nu\rho}$, & 3-form (non-dynamical) & ``Potential" \\
\hline
\end{tabular}
\caption{Table showing the different dimensional objects that can exist in four spacetime dimensions, the form fields they couple to, and the dual fields.}
\label{tab:pforms}
\end{table}

Each field has a magnetic dual field, which offers a dual description of the same physics. The type of the dual field can be obtained by acting with the Hodge star on, or in tensor form contracting with the Levi-Civita tensor $\epsilon_{\mu\nu\rho\sigma}$, the field strengths. More precisely, it is better defined as the variation of the action with respect to the original field (which for any dynamical field always contains the Hodge star dual from the kinetic terms). A particularly interesting case is for $p=3$, since the field strength is a 4-form which gives a scalar dual field strength. This shows that a 3-form is non-dynamical, since a scalar cannot be the field strength of a field, but it is still a useful field to consider as a way to write potentials. We will discuss this in more detail in sections \ref{sec:stringcomp} and \ref{sec:emergence}.

In \cite{ArkaniHamed:2006dz} the Weak Gravity Conjecture was proposed to hold for general $p$-forms in a general number of dimensions $d$. We define the analogue of the gauge coupling $g_p$ for general $p$-form field through 
\be
\label{pfkt}
{\cal L} \supset \frac{1}{2g_p^2} \left| F^{(p+1)} \right|^2 \;,
\ee
with $F^{(p+1)}_{\mu_1...\mu_{p+1}}\equiv \partial_{[\mu_1}C^{(p)}_{\mu_2...\mu_{p+1}]}$, and $\left| F^{(p+1)} \right|^2=\frac{1}{\left(p+1\right)!} F_{\mu_1...\mu_{p+1}}F^{\mu_1...\mu_{p+1}}$. Then in \cite{Heidenreich:2015nta} a precise formulation of the Weak Gravity Conjecture in terms of the charge and mass of black $p$-branes in $d$ dimensions was worked out.
\begin{tcolorbox}
{\bf Weak Gravity Conjecture for $p$-forms in $d$-dimensions} \;\cite{ArkaniHamed:2006dz,Heidenreich:2015nta}
{\it 
\newline
\newline
A $d$-dimensional theory with a $p$-form field with kinetic terms (\ref{pfkt}), should have a $p-1$-dimensional object with (quantized) charge $q_p$ and tension $T_p$ satisfying 
\be
\label{wgcpd}
\frac{p\left(d-p-2\right)}{d-2} T_p^2 \leq q_p^2 g_p^2 \left(M_p^d\right)^{d-2}\;.
\ee 
}
\end{tcolorbox}

The case for axions $p=0$ in (\ref{wgcpd}) is special, and is also the most widely studied, and we discuss it in the next section. 

\subsubsection{The Weak Gravity Conjecture for axions}
\label{sec:wgcpform}

The axion-instanton case $p=0$ is not covered by (\ref{wgcpd}), but can be argued for by dimensional reduction, as well as other methods, see for example \cite{ArkaniHamed:2006dz,Rudelius:2015xta,Brown:2015iha,Heidenreich:2015wga,Vittmann}. The notation is usually to denote the coupling for axions in terms of an axion decay constant $f$, the tension of the instanton as the action $S$, and the instanton number as $q$, so in the effective action one has the terms
\be
\label{actaxin}
{\cal L} \supset - f^2 \left(\partial a \right)^2 + \Lambda^4 \sum_q e^{-q S}\left( 1 - \cos{q a} \right) \;,
\ee
where $\Lambda$ is some appropriate mass scale. The axion periodicity is then $a \sim a + 2 \pi$. 
\begin{tcolorbox}
{\bf Weak Gravity Conjecture for axions} \;\cite{ArkaniHamed:2006dz}
{\it 
\newline
\newline
An axion with decay constant $f$ must couple to instantons with action $S$, such that  
\be
\label{wgcax}
f S \leq M_p \;.
\ee 
}
\end{tcolorbox}

\begin{figure}[t]
\centering
 \includegraphics[width=0.9\textwidth]{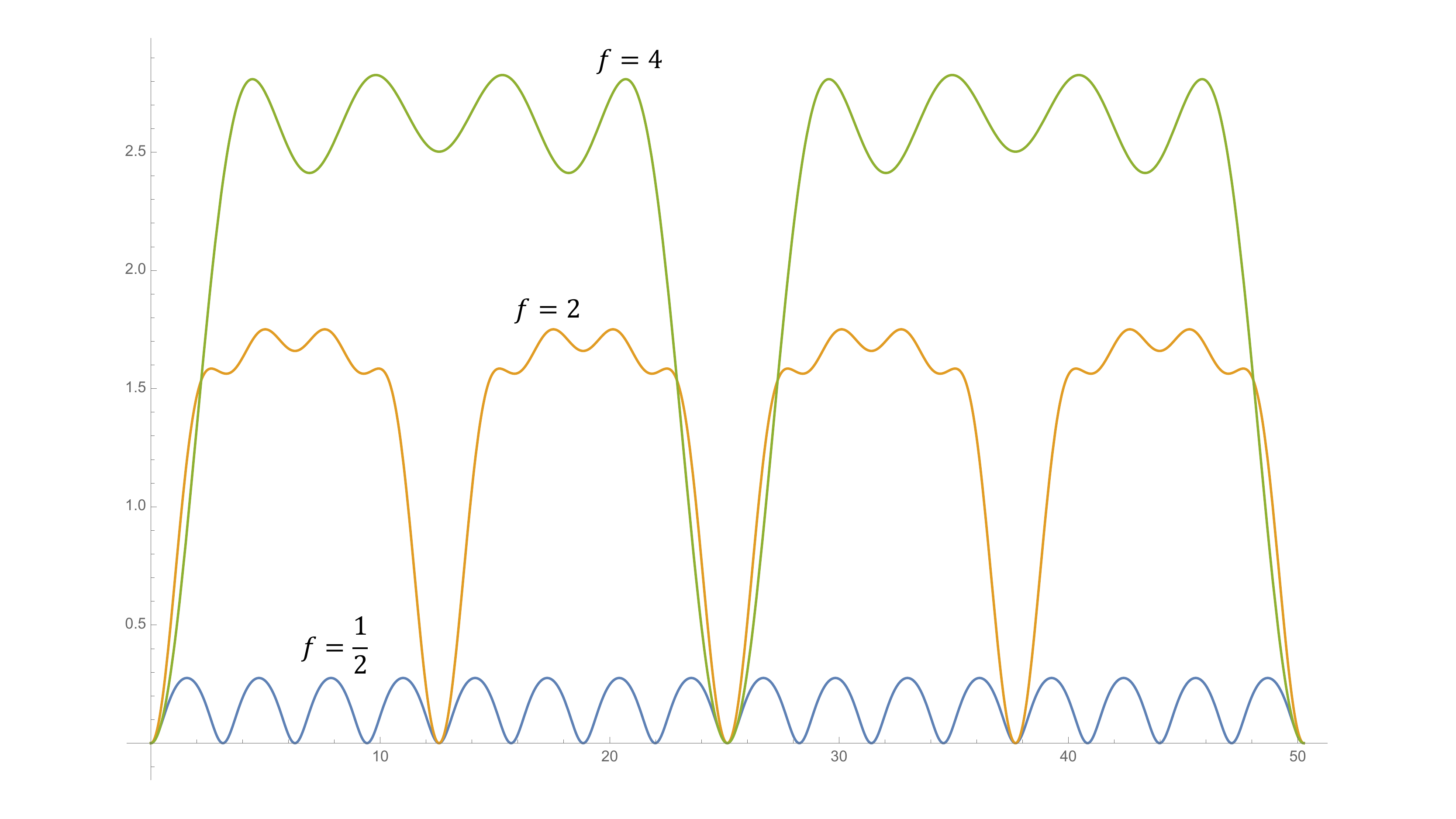}
\caption{Figure showing the potential for axions with different decay constants $f$. The first 4 instantons in the infinite expansion (\ref{actaxin}) are included, and the action $S$ is taken to saturate the Weak Gravity Conjecture bound (\ref{wgcax}). The overall periodicity of the axion is set by $2 \pi f$, but the maximum size of monotonic regions remains approximately constant and ${\cal O}\left(1\right)$ for any $f$.}
\label{fig:axplot}
\end{figure}

The constraint (\ref{wgcax}) has an interesting direct consequence. It imposes constraints on the instanton action $S$ and therefore on the potential of the axion. More precisely, it implies a bound on the size of monotonic regions in the axion potential, so regions in axion field space where the potential behaves monotonically in the axion expectation value. Since the action $S$ controlling the instanton expansion is bounded by $f^{-1}$, if we try to increase $f \gg 1$ then the instanton expansion in (\ref{actaxin}) becomes increasingly badly behaved in the sense that considering only the leading instanton is a bad approximation for the total axion potential. Including also higher instantons implies that while the overall periodicity of the axion potential is set by $2 \pi f$, and can be made large with $f \gg 1$, the magnitude of monotonic regions in the potential remains approximately constant. This is shown in figure \ref{fig:axplot}.

The obstruction of large monotonic regions in the potential due to the axionic Weak Gravity Conjecture (\ref{wgcax}) has lead to significant interest in it as a constraint on model of large-field inflation. We will discuss this in section \ref{sec:sdcinf}. For now we just note that one could consider the possibility of suppressing the higher instantons by more than just the exponential factor, so taking the scale $\Lambda$ in (\ref{actaxin}) to depend on $q$ in such a way that it is suppressed for higher instantons \cite{delaFuente:2014aca}. Though this would have to be model-dependent and also is unlikely to allow for parametrically large monotonic regions. 

There are also other consequences associated to the bound (\ref{wgcax}). In particular, axion potentials with large decay constants must have a magnitude which is of order $\Lambda^4$. This means that if we wish to describe the axion field in an effective theory, the energy density of its potential would have to be near the physics which sets the scale $\Lambda$. This typically leads to control problems in explicit examples. Indeed, in string theory, the region in parameter space with $f \gg 1$ is typically plagued with control issues. This lead to a proposal that it is not possible to reach $f \gg 1$ for string theory axions \cite{Banks:2003sx}.\footnote{Historically, this evidence for the axionic Weak Gravity Conjecture preceded, and was very much an influence on, the formulation of the gauge field Weak Gravity Conjecture.} There have been many studies related to this idea \cite{Svrcek:2006yi,Bachlechner:2014hsa,Long:2014dta,Long:2014fba,Bachlechner:2014gfa,Rudelius:2015xta,Montero:2015ofa,Bachlechner:2015qja,Shiu:2015uva,Ruehle:2015afa,Hebecker:2015rya,Brown:2015lia,Retolaza:2015sta,Peloso:2015dsa,Junghans:2015hba,Heidenreich:2015wga,Palti:2015xra,Bachlechner:2015cgq, Heidenreich:2015nta,Kooner:2015rza,Kappl:2015esy,Furuuchi:2015jfj,Choi:2015aem,Ibanez:2015fcv,Hebecker:2015zss,Conlon:2016aea,Retolaza:2016bpn,Heidenreich:2016jrl,Blumenhagen:2016bfp,Garcia-Valdecasas:2016voz,Heidenreich:2016aqi,Hebecker:2016dsw,Hebecker:2017wsu,Hebecker:2017uix,Montero:2017yja,Bachlechner:2017zpb,Bachlechner:2017hsj,Blumenhagen:2018hsh,Agrawal:2018mkd,Shiu:2018wzf,Hebecker:2018ofv,Hertog:2018kbz,Hebecker:2018yxs}, some of which we will discuss in more detail in section \ref{sec:stringcomp}.

In the case of the gauge field Weak Gravity Conjecture, there existed black hole solutions charged under the symmetry. In the case of the axionic Weak Gravity Conjecture we can also consider extended semi-classical solutions that are charged (under the magnetic dual two-form, see table \ref{tab:pforms}). The original study of these was in \cite{1988NuPhB.306..890G}. Some early work on such solutions in string theory is \cite{ArkaniHamed:2007js}, and see \cite{Hertog:2018kbz} for recent work. There is a detailed recent review \cite{Hebecker:2018ofv} on these objects and their relation to the Weak Gravity Conjecture, so here we can be brief. These are solutions to the Euclidean action with metric of the form
\be
ds^2 = \left(1+\frac{C}{r^4} \right)^{-1} dr^2 + r^2 d\Omega_3^2 \;.
\ee
Here $C$ is some constant which is calculable given a particular action. There are 3 classes of solutions which can be classified according to the sign of $C$.
\begin{itemize}
\item $C<0$\;: These are smooth wormhole solutions sometimes termed `gravitational instantons'. The total wormhole carries no charge under the dual two-form, but each end does. These solutions exist for just gravity and an axion, and also in the case of an additional dilaton $\phi$.
\item $C>0$\;: These are singular solutions, sometimes termed `cored gravitational instantons'. They carry charge under the dual two-form. The solutions only exist if there is a dilaton field $\phi$.   
\item $C=0$\;: These are smooth solutions, sometimes termed 'extremal gravitational instantons'. They carry charge under the dual two-form. The solutions only exist if there is a dilaton field $\phi$, and this field diverges at their core.
\end{itemize}

These solutions are related to instantons in a similar way that charged black holes are related to charged particles. Here, by instantons we do not mean the well-known gauge-theory instanton solutions that are present for non-Abelian gauge groups but some gravitational objects. They are best understood in the context of a quantum theory of gravity, and indeed have clear and well-studied meaning in string theory where they can be thought of as branes wrapping cycles in extra dimensions, see section \ref{sec:stringcomp}. Such an instanton may have charge of order one, while the extended solutions described above must have very large charges. This distinction between the microscopic instantons and the extended solutions is important since it is the former which have smallest action and should therefore be the leading contribution to the axion potential. Nonetheless, their relation has lead to studies of the contribution of the gravitational instanton solutions to axion potentials \cite{Montero:2015ofa,Hebecker:2016dsw}.

Also, analogously to how the Weak Gravity Conjecture for gauge fields can be determined from the charge-to-mass relation for charged black holes, the axionic Weak Gravity Conjecture may be related to the extended gravitational instantons solutions \cite{Heidenreich:2015nta}. Indeed, the charged solutions required the presence of a dilaton, and therefore one should consider a general dilaton-modified version of (\ref{wgcscdd}), which reads \cite{Heidenreich:2015nta}
\be
\label{wgcpddm}
\left[\frac{\alpha^2}{2}+\frac{p\left(d-p-2\right)}{d-2} \right]T_p^2 \leq q_p^2 g_p^2 \left(M_p^d\right)^{d-2}\;,
\ee 
where $\alpha$ is the appropriate dilatonic coupling. This now allows for the case $p=0$, which was then proposed as the relevant Weak Gravity Conjecture for axions with a dilaton. 

\subsubsection*{Multiple axions}

The gauge field Weak Gravity Conjecture for multiple gauge fields was studied in sections \ref{sec:multiu1} and \ref{sec:strongwgc}. The same considerations apply naturally to the axionic Weak Gravity Conjecture (\ref{wgcax}), and were first studied in \cite{Rudelius:2015xta,Brown:2015iha,Bachlechner:2015qja,Hebecker:2015rya,Brown:2015lia}. We may consider $N$ axions $a_i$ with $i=1,...,N$. These will in general have kinetic mixing, and so we can define the canonically normalized orthogonal axions as $\phi_i$. Following this diagonalization, many $\phi_i$ may appear in each instanton, and so we consider the general potential
\be
\label{multiaxpot}
V \sim \Lambda^4 \sum_a {\cal A}_a e^{-S_a} \cos \left( \sum_{j} \frac{\phi_j}{f_{aj}}\right) \;.
\ee
Here the index $a$ runs over the instantons contributing to the action, the ${\cal A}_a$ and $f_{aj}$ are defined as in (\ref{multiaxpot}), for canonical $\phi^i$, and we have considered only the leading instantons in the instanton sum. 
The generalization of the charge-to-mass ratio vectors for gauge fields (\ref{ctmrz}) are then \cite{Rudelius:2015xta}
\be
{\bf z}_a \equiv \sum_j \frac{M_p}{f_{aj} S_a} {\bf e}_j\;,
\ee
where the ${\bf e}_i$ are an orthonormal set of basis vectors. Using the ${\bf z}_a$ the generalization of the Weak Gravity Conjecture is then formulated that the convex hull should include the unit ball.

\begin{figure}[t]
\centering
 \includegraphics[width=0.9\textwidth]{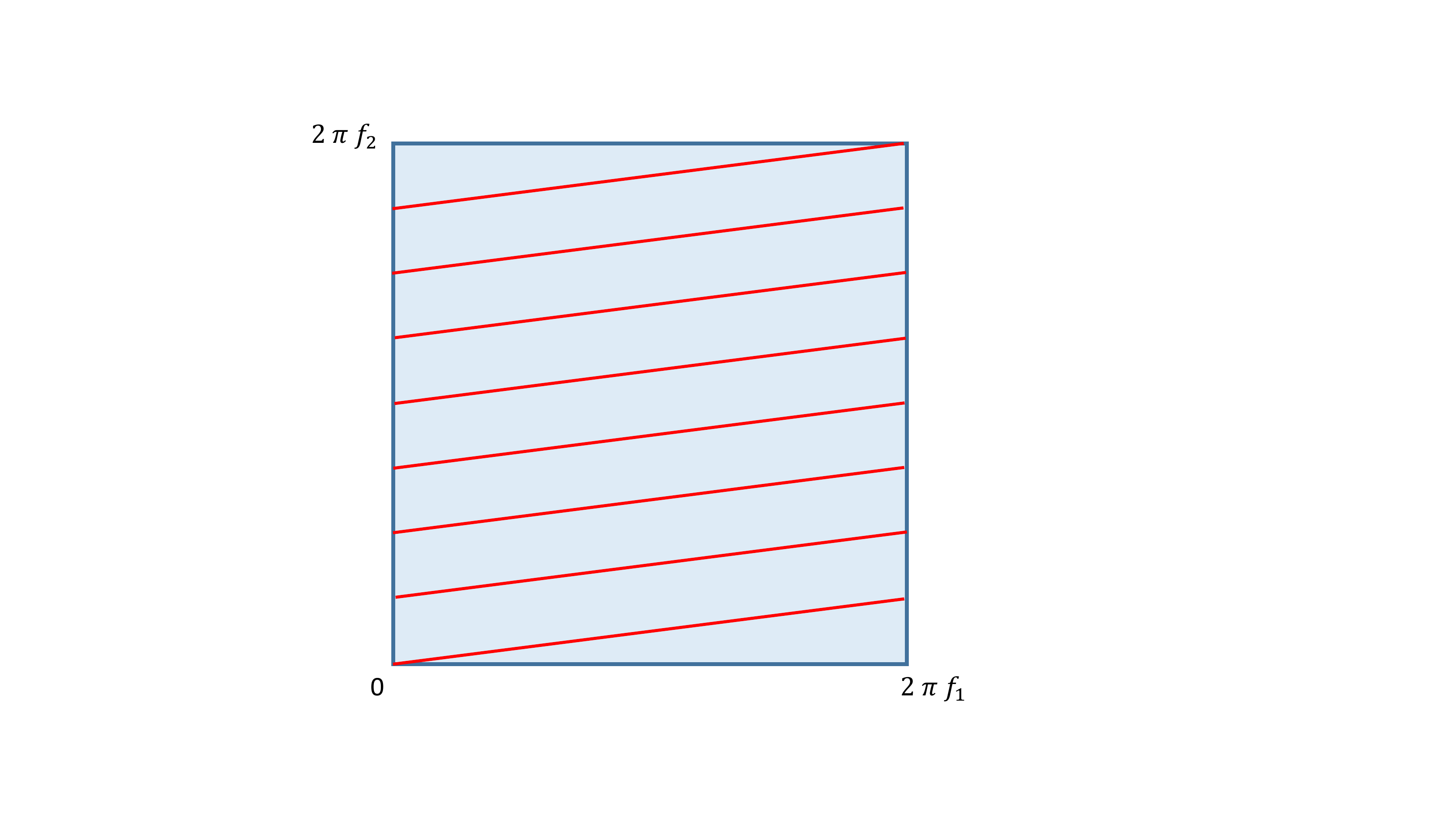}
\caption{Figure showing a two-dimensional axion field space with periodicities $2 \pi f_1$ and $2 \pi f_2$. The path in red is a linear combination of the axions which is closely aligned to one of them such that it is only periodic after multiple excursions of the periodicity $2 \pi f_1$.}
\label{fig:genaxal}
\end{figure}

\begin{figure}[t]
\centering
 \includegraphics[width=0.9\textwidth]{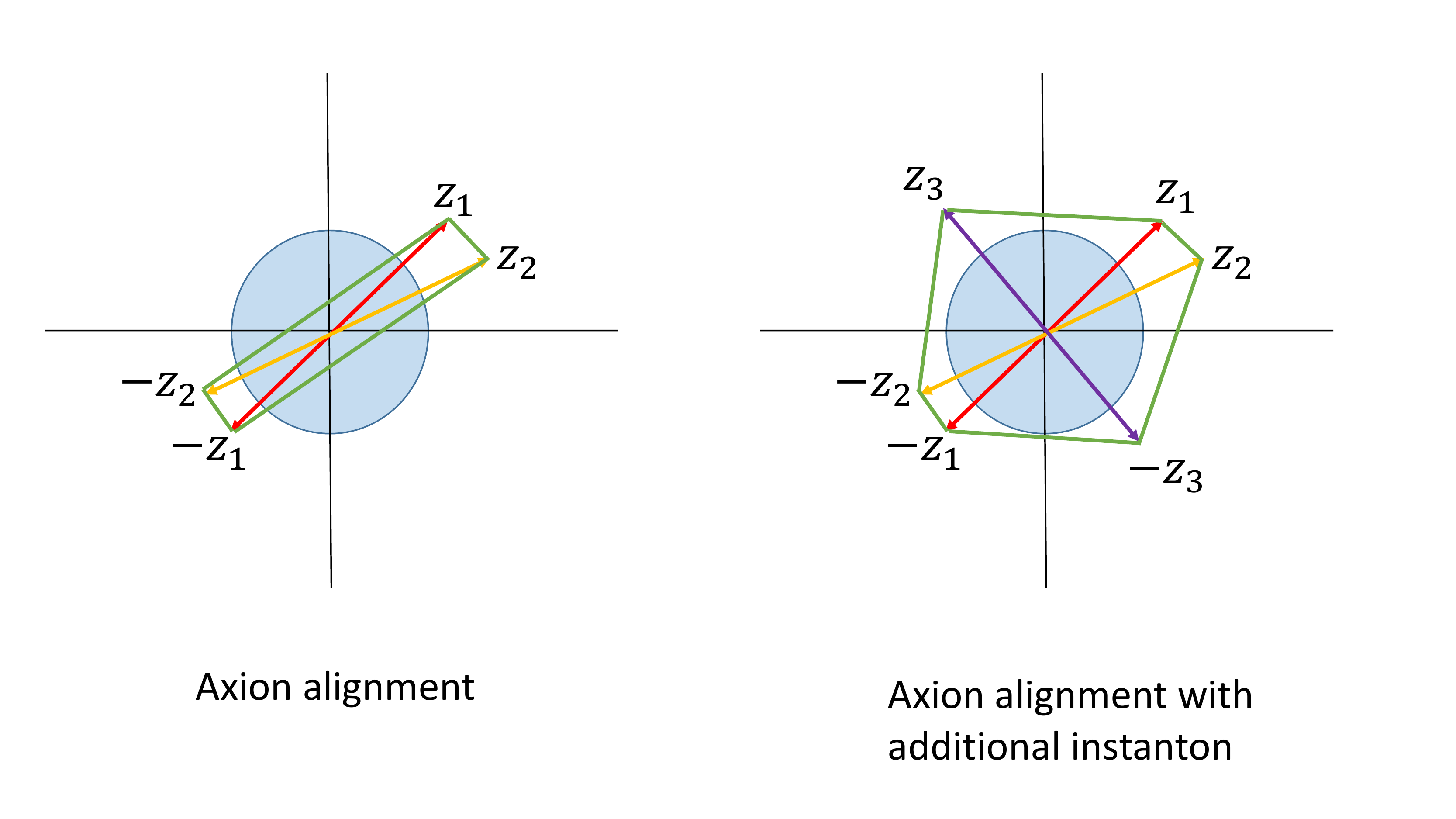}
\caption{Figure showing the convex hull for two aligned axions. The left figure shows that axion alignment with two instantons has a convex hull which does not contain the unit disc. The right figure shows how such a model can be made to consistent with the convex hull condition by adding another instanton.}
\label{fig:axalch}
\end{figure}

The multiple axion Weak Gravity Conjecture has interesting implications for the so-called axion alignment. The general idea behind axion alignment is that if we consider two or more axions, the axion field space is two dimensional and we can then identify an arbitrarily long one-dimensional path in it. The simplest such path is just a straight line at an angle to one of the axion directions, as illustrated in figure \ref{fig:genaxal}. 
Such long field paths are of interest for inflation models, and the first construction of an axion alignment model was motivated by this \cite{Kim:2004rp}. It is simplest to consider axion alignment for two axions, in which case the general potential for the canonically normalised orthogonal fields $\phi_i$ is a restriction of (\ref{multiaxpot})
\be
V \sim \Lambda^4 {\cal A}_1 e^{-S_1} \cos \left( \frac{\phi_1}{f_{11}} + \frac{\phi_2}{f_{12}} \right) + \Lambda^4 {\cal A}_2 e^{-S_2} \cos \left( \frac{\phi_1}{f_{21}} + \frac{\phi_2}{f_{22}} \right) \;.
\ee
The idea is to restrict $f_{aj} < 1$, but still induce a very large periodicity for a combination of $\phi_1$ and $\phi_2$. Note that if $\frac{f11}{f12}=\frac{f21}{f22}$ then the same linear combination of the $\phi_i$ appears in both instantons and therefore the orthogonal combination will be a completely flat direction. Then choosing $\frac{f11}{f12}\sim\frac{f21}{f22}$ leads to a direction which is not perfectly flat, but which nonetheless has an very long oscillation period \cite{Kim:2004rp}. Interestingly, the multi-axion Weak Gravity Conjecture forbids such a scenario, since if the two axion combinations in the instantons are almost aligned, the unit disc will not be contained in their associated convex hull \cite{Rudelius:2015xta,Brown:2015iha}. However, one can modify the theory by postulating a third instanton which allows the convex hull to contain the unit disc, but which has a very large action relative to the other instantons $S_3 \gg S_1 \sim S_2$. This way the Weak Gravity Conjecture is satisfied, but the axion potential is hardly modified \cite{Rudelius:2015xta,Brown:2015iha}. These possibilities are illustrated in figure \ref{fig:axalch}. 

\begin{figure}[t]
\centering
 \includegraphics[width=0.9\textwidth]{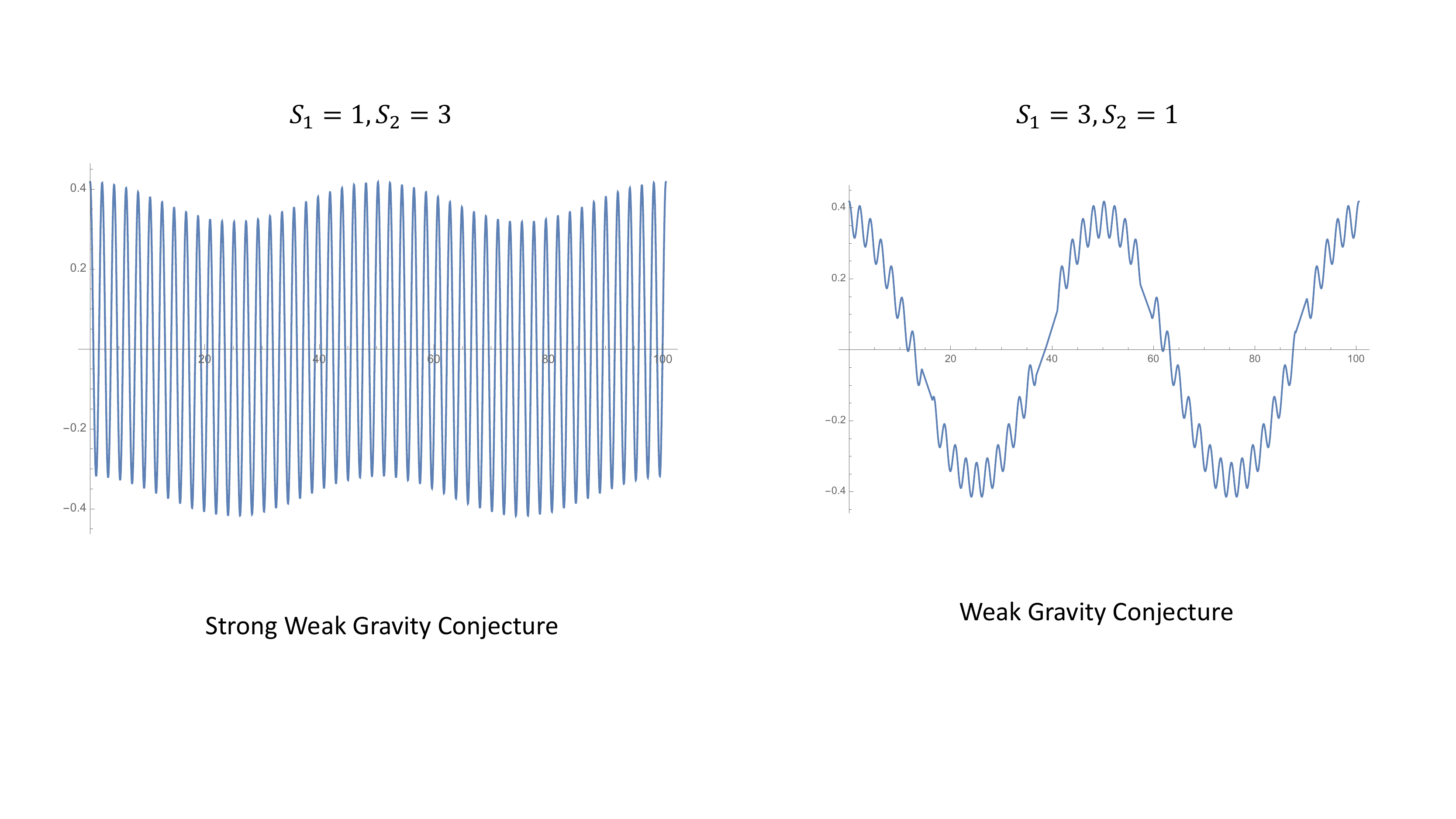}
\caption{Figure showing the two instanton potential (\ref{2axact}). The first axion satisfies the axion Weak Gravity Conjecture $S_1 f_1 \leq 1$, while the second does not $S_2 f_2 > 1$. The periodicities are taken to be $f_1=\frac13$ and $f_2=8$. The case on the left shows the form of the potential if the Strong Weak Gravity Conjecture holds and the dominant instanton is the one satisfying the Weak Gravity Conjecture. The case on the right shows the potential when the Strong form of the Weak Gravity Conjecture is violated.}
\label{fig:axtp}
\end{figure}

What this example shows is that how strongly the Weak Gravity Conjecture for axions restricts the axion potential is dependent on the relative magnitudes of the instantons. In fact, the crucial point is whether the an axion version of the Strong Weak Gravity Conjecture (see section \ref{sec:strongwgc}) is satisfied \cite{Rudelius:2015xta,Montero:2015ofa,Brown:2015iha,Heidenreich:2015wga,Hebecker:2015rya,Brown:2015lia}. One can ignore the ultraviolet origin and consider an effective low energy theory for the lightest axion mode $\phi$ which may appear in a number of different instantons
\be
\label{2axact}
V \sim  \Lambda^4 e^{-S_1} \cos \frac{\phi}{f_1} +  \Lambda^4 e^{-S_2} \cos \frac{\phi}{f_2} \;.
\ee
The axion Weak Gravity Conjecture (\ref{wgcax}) states that one instanton at least must satisfy $f S \leq M_p$. But we can have $f_1 S_1 \leq M_p$ and $f_2 S_2 \gg M_p$. Then we are free to take $f_1 < 1$ and $f_2 \gg 1$ while still controlling both instanton expansions. The form of the axion potential will then depend on whether $S_2 > S_1$ or $S_1 > S_2$. In the former case the dominant contribution to the potential has small period of oscillation, while in the latter case the dominant contribution has a long period. This is illustrated in figure \ref{fig:axtp}. The difference between the two possibilities is precisely the content of the Strong Weak Gravity Conjecture which says that the Weak Gravity Conjecture should be respected by the lightest field, or in terms of axions, the instanton with the smallest action. This would then determine $S_2 > S_1$, leading to the highly oscillatory potential. Indeed, this question is precisely analogous to the Higgsing scenario of \cite{Saraswat:2016eaz} presented in section \ref{sec:strongwgc}, and again is related to whether the Strong Weak Gravity Conjecture applies to the low energy effective theory. Since for the case of gauge fields general arguments for the Strong version of the Weak Gravity Conjecture were difficult to establish, the same should be true for the axionic version, and so tests of it in string theory play a crucial role. We will discuss these in detail in section \ref{sec:stringcomp}. 

Another interesting application of the multi-axion Weak Gravity Conjecture is to scenarios with many axions $N \gg 1$. Such theories have also been utilized to propose long field ranges in the context of inflation \cite{Dimopoulos:2005ac}. The idea is that if we consider $N$ axions with equal decay constants $f$, then the diagonal combination will have an effective decay constant of $\sqrt{N} f$ \cite{Dimopoulos:2005ac}. Recall, as discussed in section \ref{sec:multiu1}, that the convex hull condition implies that the charge vector magnitudes need to be enhanced by a factor of $\sqrt{N}$. For axions this means that the decay constants for each axion need to decrease by a factor of $\sqrt{N}$, so $f \sim \frac{1}{\sqrt{N}}$, which then precisely cancels the enhancement of the total decay constant \cite{Rudelius:2015xta,Montero:2015ofa,Brown:2015iha,Heidenreich:2015wga,Brown:2015lia}. However, in \cite{Bachlechner:2015qja} it was argued that a similar way out to alignment, with sub-leading instantons, can be implemented here and so similarly the question comes down to the Strong version of the Weak Gravity Conjecture. 

\subsubsection*{Instantons and Scalar Fields}

In the case of gauge fields the Weak Gravity Conjecture was argued to be modified due to scalar fields (\ref{wgcsc}). It is natural to propose that a similar modification should hold for the axion Weak Gravity Conjecture. One way to try and determine such a form is using string theory to determine the relation satisfied by instantons in the case of ${\cal N}=2$ supersymmetry, so the equivalent statement to BPS particles for the gauge Weak Gravity Conjecture. And then propose that it should hold as an inequality more generally. This was studied in \cite{Vittmann} where a general proposal was made for arbitrary axion kinetic terms. Here we just quote the result for a single canonically normalised scalar field $\phi$, axion with decay constant $f$, and instanton with action $S$
\be
f^2 S^2 + f^2 \left( \partial_{\phi} S \right)^2 M_p^2 \leq M_p^2 \;.
\ee   

\subsubsection*{The magnetic axion Weak Gravity Conjecture}

The axionic Weak Gravity Conjecture discussion so far has been in analogy with the electric Weak Gravity Conjecture (\ref{ewgc}). The magnetic Weak Gravity Conjecture (\ref{mwgc}) should also have an analogous statement for axions. This was studied in \cite{Hebecker:2017wsu,Hebecker:2017uix,Reece:2018zvv}. In \cite{Hebecker:2017uix} it was proposed that for an axion with decay constant $f$, there is a bound on the cutoff of the effective theory $\Lambda_f$ which takes the form
\be 
\Lambda_f \lesssim \sqrt{f M_p} \;.
\label{magf}
\ee
The scale $\Lambda_f$ was argued to be associated to light strings, which in \cite{Hebecker:2017uix} were needed in order for a black hole with two-form charge to discharge itself and avoid a remnant. 
There are some subtleties with this proposal, discussed in \cite{Hebecker:2017uix,Reece:2018zvv}, in particular the axion in question here should be an axion such that its associated string should have a singular core. This is opposed to say Nambu-Goldstone bosons which have associated string solutions which are smooth at their core.

\subsection{The Swampland Distance Conjecture}
\label{sec:sdc}

We first encountered the Swampland Distance Conjecture in section \ref{sec:fcsdc} in the context of compactification of string theory on a circle. In this section we will discuss it in more detail, and present some general arguments for it. The Swampland Distance Conjecture has been less studied from a general perspective relative to the Weak Gravity Conjecture, and most of the work has been on testing it in string theory, which will be discussed in section \ref{sec:stringcomp}. A further microscopic argument for the Swampland Distance Conjecture, along with the Weak Gravity Conjecture, will be presented in section \ref{sec:emergence}. Let us, for convenience, present the definition of the conjecture, first presented in \ref{sec:fcsdc}, also here. 

\begin{tcolorbox}
{\bf Swampland Distance Conjecture } \;\cite{Ooguri:2006in}
{\it 
\begin{itemize}
\item Consider a theory, coupled to gravity, with a moduli space $\cM$ which is parametrized by the expectation values of some field $\phi^i$ which have no potential. Starting from any point $P \in \cM$ there exists another point $Q \in \cM$ such that the geodesic distance between $P$ and $Q$, denoted $d\left(P,Q\right)$, is infinite. 
\item There exists an infinite tower of states, with an associated mass scale $M$, such that 
\be
\label{sdc}
M\left(Q\right) \sim M\left(P\right) e^{-\alpha d\left(P,Q\right) } \;,
\ee
where $\alpha$ is some positive constant.
\end{itemize}
}
\end{tcolorbox}

It is worth discussing the moduli space ${\cal M}$ in a bit more detail. We consider an action
\be
S = \int d^d x \sqrt{-g} \left[\frac{R}{2} - g_{ij}\left( \phi^i \right) \partial \phi^i \partial \phi^j + ... \right] \;.
\ee
The real scalar fields $\phi^i$ do not have a potential, and the index $i$ is kept arbitrary for now. The $\phi^i$ are coordinates on a moduli space ${\cal M}$, their kinetic terms define a metric on that space $g_{ij}$.\footnote{Care to note the difference between the moduli space metric $g_{ij}$ and the space-time metric which appears in the overall $\sqrt{-g}$ factor. This is an unfortunate clash of notation, and the two quantities are completely unrelated.} This metric has Euclidean signature since the kinetic terms for physical scalar fields should have fixed sign. The index range of $i$ defines the real dimension of the moduli space ${\cal M}$. The special case of canonically normalised fields $g_{ij}\left(\phi\right) = \delta_{ij}$, corresponds to a flat moduli space. The moduli space may still have a non-trivial topology though, for example, some of the fields may be compact $\phi \sim \phi + 2 \pi$.

A point $P$ in the moduli space corresponds to some specification of the expectation values for the scalar fields. The geodesic distance between two points $P$ and $Q$, which plays a role in the conjecture $d\left(P,Q\right)$, is then defined as usual
\be
d\left(P,Q\right) \equiv \int_{\gamma}  \left(g_{ij}\frac{\partial \phi^i}{\partial s} \frac{\partial \phi^j}{\partial s}\right)^{\frac12} ds \;.
\ee
Here $\gamma$ is the shortest geodesic connecting the points $P$ and $Q$, and $ds$ is the line element along that geodesic. 

The first part of the conjecture states that for a point $P \in \cM$ there exists some other point $Q \in \cM$ which is an infinite distance away. The simplest example of such a field space is just the real line $\cM = \mathbb{R}$, and this was the case discussed in section \ref{sec:fcsdc}. Then for any value of $\phi$, the points $\phi = \pm \infty$ are an infinite distance away. A simple example of a moduli space which violates the conjecture is a periodic scalar $\phi \sim \phi + 2\pi$, which defines a circle $\cM = S^1$, since the maximum distance between two points is just $2 \pi$. This does not mean that periodic scalars are forbidden in quantum gravity, only that they must be part of a larger moduli space. For the case of periodic pseudo-scalars, axions, this occurs because the axion decay constant $f$ is itself a scalar field.

In the case of a Kaluza-Klein reduction on a circle studied in section \ref{sec:fcsdc} we found that the mass of the tower behaves as 
\be
\label{kksdcm}
M\left(Q\right) = M\left(P\right)e^{-\alpha d\left(P,Q\right)} \;. 
\ee
Crucially, this held as an exact statement for any distance $d\left(P,Q\right)$. However, we cannot expect this to be a general property of moduli spaces in quantum gravity. If we take two points which are very close to each other $d\left(P,Q\right) \rightarrow 0$ then it may well be that actually $M\left(Q\right) > M\left(P\right)$. Indeed, if we consider any periodic direction in moduli space then it must be that $M$ must equally be a periodic function on $\cM$, $M\left( \phi \right) = M\left( \phi + 2\pi \right)$. Then for any point $P$ there must be a $Q$ with $d\left(P,Q\right)>0$ and $M\left(Q\right) \geq M\left(P\right)$. 

The relevant picture behind the Swampland Distance Conjecture is therefore that moduli spaces in string theory take the schematic form illustrated in figure \ref{fig:sdcg}. There is a bulk of moduli space and there are asymptotic regions near loci which are at infinite distance from the bulk. Of course, the actual geometry of the moduli space is much more complicated than that depicted in \ref{fig:sdcg}, as will be discussed in detail in section \ref{sec:stringcomp}, but schematically the picture is sufficient. The conjecture therefore is a statement about universal behaviour in the asymptotic regions (coloured red in figure \ref{fig:sdcg}). 

We may consider then how large does  $d\left(P,Q\right)$ need to be in order that the asymptotic exponential behaviour gives a good approximation for any starting point $P$. This is also tied to a bound on the magnitude of $\alpha$, since if $\alpha \ll 1$ we can have $d\left(P,Q\right) \gg 1$ without much change in the tower mass according to (\ref{sdc}). In other words, a sharp statement would have to be one which constrains $\alpha$ and $d\left(P,Q\right)$ separately. Although no explicit general statements about these more precise questions were made in \cite{Ooguri:2006in}, it is natural to interpret the conjecture as saying that $\alpha \sim \cO\left(1\right)$ for sufficiently large distances $d\left(P,Q\right) \gg 1$.  

To make these things more explicit and sharply defined, in \cite{Baume:2016psm,Klaewer:2016kiy} a Refined Swampland Distance Conjecture was proposed. The conjecture has two parts, and the first matches on to the discussion so far, while the second refers to fields with a potential.\footnote{The second part was implicit in the analysis of \cite{Baume:2016psm,Klaewer:2016kiy}, since it applied to fields with potentials, but was not stated explicitly.}  Each part may hold separately but they are also intrinsically related as we will discuss below.
\begin{tcolorbox}
{\bf Refined Swampland Distance Conjecture } \;\cite{Baume:2016psm,Klaewer:2016kiy}
{\it 
\begin{itemize}
\item Consider a theory, coupled to gravity, with a moduli space which is parametrized by the expectation values of some fields which have no potential. Let the geodesic distance between any point $P \in \cM$ and another point $Q \in \cM$ be denoted $d\left(P,Q\right)$. There exists an infinite tower of states with mass scale $M$ such that 
\be
\label{rsdc}
M\left(Q\right) < M\left(P\right) e^{-\alpha \frac{d \left( P,Q \right)}{M_p}} \;,
\ee 
if $d \left(P,Q\right) \gtrsim M_p$. 
\item The first statement holds even for fields with a potential, not just for moduli, where the moduli space is replaced with the field space in the effective theory. 
\end{itemize}
}
\end{tcolorbox}

Since the conjecture involves the Planck mass in an important way we have reinstated it. The first part of the Refined Swampland Distance Conjecture is closely related to the original formulation (\ref{sdc}), and some looser version of it is in some sense implicit in \cite{Ooguri:2006in}. It has though a sharp inequality sign in (\ref{rsdc}). And the approximate sign was meant in a tight way, so that the exponential behaviour sets in at 1-2 $M_p$, and there is no way to delay it, say to $d \left( P,Q \right) \sim 10 M_p$. It also has some stronger implications for the behaviour before the exponential decrease appears since it must not be such that the mass increases too much in that regime. This places an implicit lower bound on $\alpha$, since a very mild behaviour at $d \left( P,Q \right) \sim M_p$ would not be sufficient to counter any possible increase in the mass at small $d \left( P,Q \right)$. Note that it also weakens in one way the original statement (\ref{sdc}) because it allows for the possibility that $M$ may decrease faster than exponentially in $d \left( P,Q \right)$.

The second part of the conjecture is substantially different to the original formulation in \cite{Ooguri:2006in} since it is also a statement about the type of potentials which may appear in quantum gravity. The two parts are closely linked because in attempting to define a field distance conjecture for fields with a potential, it is not clear that an asymptotic infinite distance point exists within the effective theory. It may have an energy density associated to it which is above the cutoff scale of the theory. More precisely, the cutoff scale $\Lambda$ of an effective theory limits any energy densities and in particular it restricts the value of the fields $\phi^i$ to be such that the potential is bounded $V\left( \phi \right)^{\frac14} < \Lambda$. Therefore, much like the two parts in the original conjecture, making the precise finite distance first statement is prerequisite for the second statement. 

Finally, there is a general comment that is useful to make regarding both versions of the distance conjecture. The tower of states is usually taken to be particle-like in nature. However, examples from string theory suggest that we should also allow them to be light extended objects. So for example, tower of light string states. One may still treat these as particle-like below the length scale of the objects, but it is not clear that this is the correct thing to do. So we should keep in mind that by a tower of light states, we may also allow for light extended objects.  

Note that, for convenience, we will often refer to both the Swampland Distance Conjecture and its refined version as the distance conjecture. 

\subsubsection{Relation to the Weak Gravity Conjecture} 
\label{sec:distswgc}

There are a number of relations between the distance conjectures and the Weak Gravity Conjecture. In \cite{Klaewer:2016kiy} it was noted that since a periodic scalar cannot satisfy the exponential behaviour of the distance conjecture the Refined Swampland Distance Conjecture would require that periodic scalars cannot have period distances larger than $M_p$. This matches nicely to the constraint imposed by the axionic Weak Gravity Conjecture on the axion decay constant (\ref{wgcax}).

 It was also argued in \cite{Baume:2016psm} that we may interpret (\ref{wgcax}) in a different way as a constraint on the scalar field which controls the axion decay constant. In particular, within a supersymmetric setting the scalar superpartner of the axion, the saxion, controls both the instanton action and the axion decay constant. Schematically, the action takes the form
\be
S = \int d^4x \sqrt{-g}\left[ \frac{R}{2} - f\left(u\right)^2 \left(\partial u \right)^2 - f\left(u\right)^2 \left(\partial a \right)^2 + \Lambda^4 {\cal A} e^{-u} \cos \left( a \right) + ...  \right] \;.
\ee
The constraint (\ref{wgcax}) then translates into one on the kinetic term of the saxion
\be
f\left(u\right) u \leq 1 \;.
\ee
If $f$ was a constant, then this would be a direct sharp constraint on the geodesic distance along the $u$ direction in field space $d\left(P,Q\right) \leq 1$. We can extend the distance further only if $f\left( u\right)$ falls off at least as fast as $\frac{1}{u}$. This therefore implies that the proper distance grows beyond $M_p$ only at best as $d \sim \log u$. This logarithmic behaviour is tied to the exponential behaviour of the distance conjectures, as $u$ typically controls the mass scale of towers of states.

A relation between the distance conjecture and the Scalar Weak Gravity Conjecture (\ref{swgc}) was pointed out in \cite{Palti:2017elp}. Let us consider the Scalar Weak Gravity Conjecture applied to a single canonically normalised scalar field $\phi$. It states that the mass of the field should satisfy
\be
\label{swgccan}
\left|\partial_{\phi} m\right| > m \;.
\ee
We may now consider varying $\phi$ and ask that (\ref{swgccan}) should be maintained. It is manifest that for sufficiently large $\phi$ any power-law behaviour $m \sim \phi^p$ would be violated. Instead we require $m \sim e^{-\alpha \phi}$ with $\alpha > 1$. We therefore recover the exponential behaviour of the mass in the distance conjectures (\ref{sdc}) and (\ref{rsdc}). In other words, a particle must have exponentially decreasing mass if gravity is to remain the weakest force acting on it even for large scalar expectation values. 

As discussed in section \ref{sec:scalwgc}, the Scalar Weak Gravity Conjecture does not require that the same particle satisfy it throughout all of field space. We may then consider a simple way to avoid the exponential behaviour of the mass by introducing to particles whose masses are oscillatory in $\phi$, such that there is always one which satisfies the conjecture. For example,
\be
m_1 = \left|\cos  \phi \right| \;,\;\; m_2 = \left|\sin  \phi \right| \;. 
\ee
However, the period of $\phi$ is $2\pi$ and if we tried to parametrically increase it we would need to introduce more fields. It can be checked that the size of the region with $\left|\partial m\right| > m$ for $m = \left|\cos \frac{\phi}{N} \right|$ stays constant for large $N$. Therefore, we need $N$ particles which are interchanging their roles in order to satisfy the Scalar Weak Gravity Conjecture over a field distance in $\phi$ of $2 \pi N$. For an infinite distance we would need an infinite tower of states. This was shown in \cite{Palti:2017elp}, where it was also pointed out that this is a way to think of axions in string theory, leading to a relation between axions and infinite towers of states which we will discuss in more detail in section \ref{sec:stringcomp}. 

So far we have discussed relations between electric versions of the Weak Gravity Conjecture and the distance conjectures. However, in many ways it is the magnetic Weak Gravity Conjecture (\ref{mwgc}) which is closest in nature to the distance conjectures. As discussed in section \ref{sec:wgctower}, it is natural to associate the magnetic Weak Gravity Conjecture scale $g M_p$ with the mass scale of an infinite tower of states. We may then identify this tower with the tower of the distance conjectures. This relation between the towers was first discussed in \cite{Heidenreich:2016aqi,Klaewer:2016kiy}, and we will study it in more detail in section \ref{sec:emergence}. It suggests that gauge couplings should be functions of scalar fields, which is certainly the case in string theory, such that the relation between the value of the coupling $g$ and the canonically normalised field $\phi$, is exponential 
\be
\label{gep}
g \sim e^{-\phi} \;. 
\ee

The Weak Gravity Conjecture also provides an insight into a possible further refinement of the Refined Swampland Distance Conjecture \cite{Baume:2016psm}. Recall the type of potentials one obtains for small instanton action, as depicted in figure \ref{fig:axplot}, or for axion alignment, as depicted in figure \ref{fig:axtp} (we refer here to the potential respecting the Strong Weak Gravity Conjecture). We note that their structure is such that the overall periodicity of the potential can be made parametrically large, while the size of monotonic regions in the potential remains constant and around the Planck scale. It is not clear that such scenarios can be realized in string theory, but if they can then they would suggest that the Refined Swampland Distance Conjecture should hold while restricting the field distance to lie within monotonic regions of the potential. This possible refinement was noted already in \cite{Baume:2016psm}. 

\subsubsection{Relation to spatial field variations}
\label{sec:spatfv}

Consider a situation where a scalar field varies its expectation value spatially. We let the spatial variation be restricted to a region of size $R$, and consider a variation $\Delta \phi$, from one side of the region to the other, for a canonically normalised scalar field. If we keep $R$ fixed, and try to increase $\Delta \phi$ we eventually reach an obstruction which is that energy in the spatial gradient of the scalar field, coming from the kinetic terms, will have a Schwarzschild radius larger than $R$ leading to a collapse of the system into a black hole. This leads to an interesting way to probe a gravitational censorship of large field variations. The relation between the censorship of large spatial field variations and the Swampland Distance Conjecture was studied in \cite{Klaewer:2016kiy}, while its relation to the axionic Weak Gravity Conjecture was studied in \cite{Dolan:2017vmn}, and more generally to censorship of large field variations in \cite{Draper:2019zbb}.  

A more precise version of the above estimate on a bound by gravitational collapse was made in \cite{Nicolis:2008wh}. It was shown that if gravity is to remain in the Newtonian regime, then a free scalar field cannot undergo a super-Planckian variation. A free scalar field is one which is not feeling an effective potential. If we consider a general potential, then a scalar field can have a super-Planckian spatial field variation within the Newtonian regime, but the maximum variation $\Delta \phi$ is obtained for a logarithmic spatial profile $\phi \sim \log r$, where $r$ is the radial coordinate. A similar logarithmic bound was shown also for a general class of strongly curved backgrounds in \cite{Klaewer:2016kiy}. An interesting implication of these results is that for super-Planckian spatial variations the energy density in the field due to the kinetic terms $\rho$, has to vary at least exponentially $\rho \sim e^{\alpha \Delta \phi}$, for some $\alpha$. This can be seen by noting that if $\phi \sim \log r$, then $\partial_r \phi \sim \frac{1}{r} \sim e^{-\phi}$. 

One class of solutions which can induce a potential for a scalar field are localized solutions charged under a $U(1)$ gauge symmetry. In such cases, if the gauge coupling of the $U(1)$ depends on a scalar field $\phi$ then the scalar field will develop a spatial gradient which can support a trans-Planckian variation. In \cite{Klaewer:2016kiy} such solutions were studied and it was argued that far away from the source, the gauge coupling $g\left(\phi\right)$ is of similar magnitude to the energy density in the scalar field kinetic terms. Then the exponential increase in the energy density discussed above implies that if the solutions is to be self-consistent then the cutoff scale of the theory set by the magnetic Weak Gravity Conjecture scale $\Lambda \sim g M_p$ must also increase exponentially and stay above the energy density. This then leads to the relation (\ref{gep}) thereby motivating the distance conjecture from the Weak Gravity Conjecture. 
 
\subsubsection{Implications for inflation and primordial gravitational waves} 
\label{sec:sdcinf}

The distance conjecture (\ref{sdc}) and (\ref{rsdc}), and its close relation the axionic Weak Gravity Conjecture (\ref{wgcax}), have important implications for models of inflation. Indeed, much of the work on the Swampland is motivated precisely by the resulting constraints on inflation, as a way to formalize and quantify the general sense of difficulty of constructing models of large field inflation in string theory. Models of large field inflation are those where the inflaton $\phi$ undergoes a super-Planckian excursion in field space. See \cite{Baumann:2009ds} for a review of inflation, and \cite{Baumann:2014nda} for a review on its potential realizations in string theory. See also \cite{Blumenhagen:2018hsh}. for a review related to the distance conjecture, and \cite{Blumenhagen:2017cxt,RompineveSorbello:2017lqb,Cicoli:2017axo,Cicoli:2018tcq,Dias:2018pgj,Landete:2018kqf,Lehners:2018vgi,Dias:2018ngv,Matsui:2018bsy,Kinney:2018nny,Schimmrigk:2018gch,Chiang:2018lqx,Scalisi:2018eaz} for (an incomplete list of) an analysis of specific inflation models with respect to the distance conjecture.

Let us first recall some basic elements of single-field slow-roll inflation. The inflaton is canonically normalised and denoted as $\phi$, with potential $V\left(\phi\right)$.\footnote{Note that in this section $\phi$ has kinetic terms $\frac12 \left(\partial \phi\right)^2$, rather than $ \left(\partial \phi\right)^2$ as in other parts. } The potential slow-roll parameters are
\be
\epsilon_V = \frac{M_p^2}{2} \left( \frac{\partial_{\phi}V}{V} \right)^2 \;,\;\; \eta_V =  M_p^2 \frac{\partial^2_{\phi}V}{V}   \;.
\ee
These are related to the Spectral tilt $n_s$ and the tensor-to-scalar ratio $r$ as
\be
n_s-1 \simeq 1 + 2 \eta_V - 6 \epsilon_V \;,\;\; r \simeq 16 \epsilon_V \;.
\ee
The parameter $r$ measures the ratio of the power of tensor perturbations during inflation to the power in scalar perturbations. The current best constraints on the parameters are  \cite{Akrami:2018odb}
\be
n_s = 0.9649\pm 0.0042 \;,\;\; r < 0.064 \;.
\label{spetrbou}
\ee
The tensor-to-scalar ratio is a particularly interesting parameter since it relates the magnitude of the inflaton field excursion to the energy scale of inflation. Specifically, for 60 e-foldings of inflation, the correct magnitude of the scalar perturbations requires
\be
E_{\mathrm{inf}} = V^{\frac14} \simeq 8 \times 10^{-3} \left(\frac{r}{0.1} \right)^{\frac14} M_p \;,
\label{enr}
\ee
while the Lyth bound states \cite{Lyth:1996im}
\be
\frac{\Delta \phi}{M_p} \gtrsim 0.25 \left(\frac{r}{0.01} \right)^{\frac12} \;.
\label{Lythb}
\ee
The two relations (\ref{enr}) and (\ref{Lythb}) imply that large $r$ is in (exponential) tension with the distance conjecture. This is because large inflaton excursions lead to a tower of states becoming exponentially light, but on the other hand we also need a high energy scale of inflation which should nonetheless be lower than the mass scale of the tower of states. 

\begin{figure}[t]
\centering
 \includegraphics[width=0.9\textwidth]{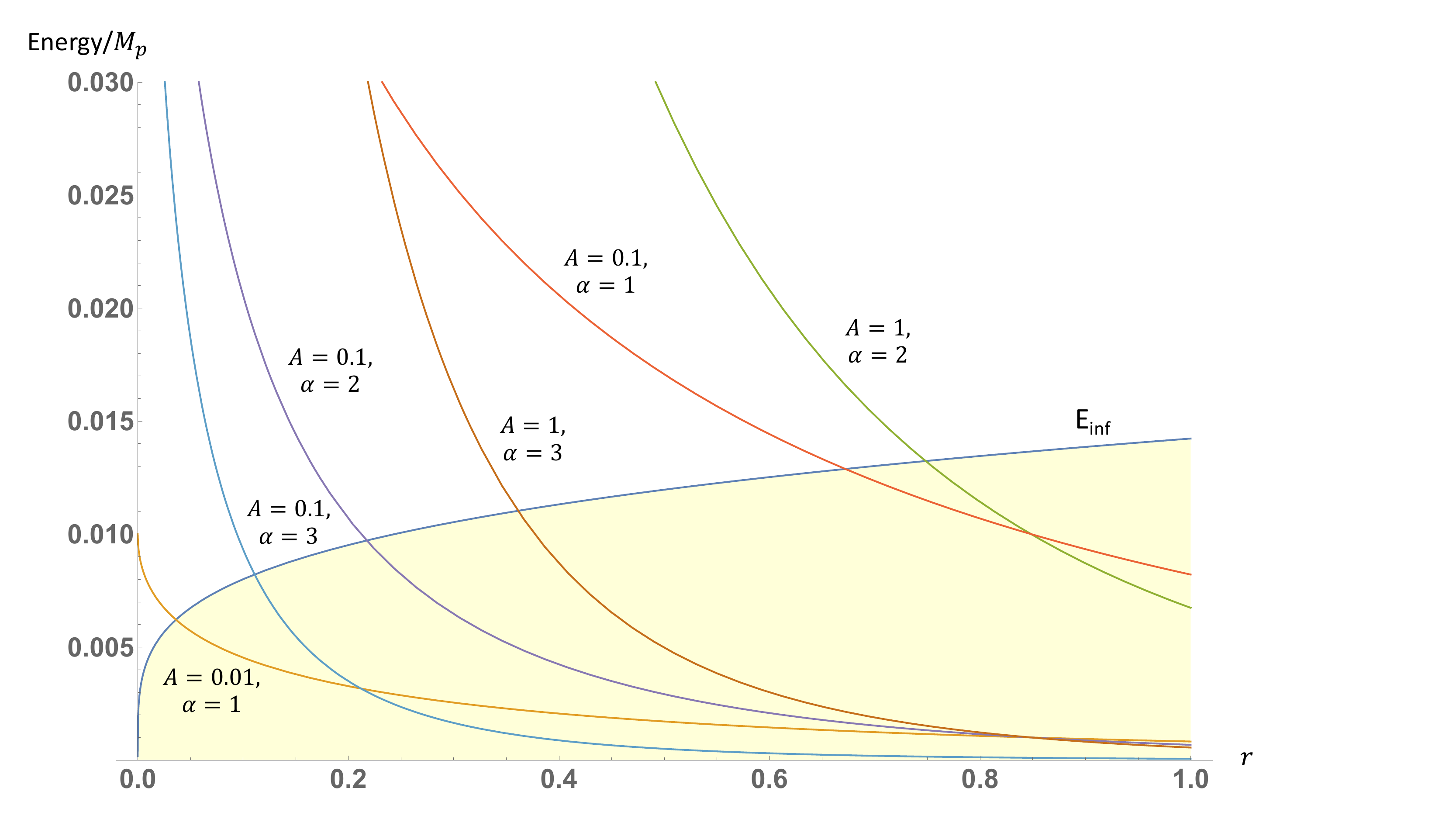}
\caption{Figure showing the energy scale of inflation $E_{\mathrm{inf}}$ and the mass scale of the tower of states of the distance conjecture, parameterized as $A e^{-\alpha \frac{\Delta \phi}{M_p}} M_p$ for different values of $A$ and $\alpha$, as a function of the tensor-to-scalar ratio $r$. The value of $r$ is bounded by the crossing point of $E_{\mathrm{inf}}$ and the tower mass scales.}
\label{fig:bndr}
\end{figure}

To give a quantitative impression of this tension, let us define the mass scale of the tower of states of the distance conjecture to be
\be
\Lambda_{dc} = A e^{-\alpha \frac{\Delta \phi}{M_p}} M_p \;.
\label{quandc}
\ee
Then we can obtain bounds on the parameters $A$ and $\alpha$ in (\ref{quandc}) for a given magnitude of $r$ by requiring $\Lambda_{dc} > E_{\mathrm{inf}}$.\footnote{Slightly weaker bounds can be obtained by requiring the cutoff to be larger than only the Hubble scale during inflation $\Lambda_{dc} > H_{\mathrm{inf}}$.} These are shown in figure \ref{fig:bndr}. 
While the exponential nature of the mass scales of the tower of states is important, in practice, much of the tension in building controlled models in string theory which leads to significant $r$ is actually due to the parameter $A$, which corresponds to the starting mass scale for the tower of states. Overall, while a sharp bound on $r$ from the distance conjecture would require sharp bounds on $A$ and $\alpha$, it is clear that order one values for these can yield bounds that are comparable to observational bounds.\footnote{Of course, such bounds may be evaded in models where the magnitude of $r$ is not bounded by the field excursion, see for example \cite{BenDayan:2009kv,Wolfson:2016vyx}.} 

The distance conjecture also places important bounds on specific models of inflation. For example, power-law inflation models $V\left(\phi\right) \sim \phi^p$, for some $p$, have $E_{\mathrm{inf}}\sim 7 \times 10^{-3} p^{\frac14} M_p$. While say for $p=1,2,3,4$ we have $\Delta \phi \sim 11M_p, 15M_p, 19M_p, 22M_p$. Then for $A=1$ we have that $\Lambda_{dc} > E_{\mathrm{inf}}$ restricts $\alpha < 0.4,0.3,0.24,0.2$. These models give $r=0.07,0.13,0.20,0.27$, which shows that the bounds from the Lyth bound (\ref{Lythb}), as shown in figure \ref{fig:bndr}, are actually often far too weak and so underestimate the strength of the distance conjecture. 

A related model is natural inflation \cite{Freese:1990rb}. This is based on a potential as in (\ref{actaxin}), but requires $f \sim 10 M_p$ in order to be compatible with spectral tilt bounds (\ref{spetrbou}) \cite{Baumann:2014nda}. Such a large axion decay constant goes against the Weak Gravity Conjecture for axions (\ref{wgcax}), and more generally violates the distance conjecture as discussed in section \ref{sec:distswgc}.  

\begin{figure}[t]
\centering
 \includegraphics[width=0.9\textwidth]{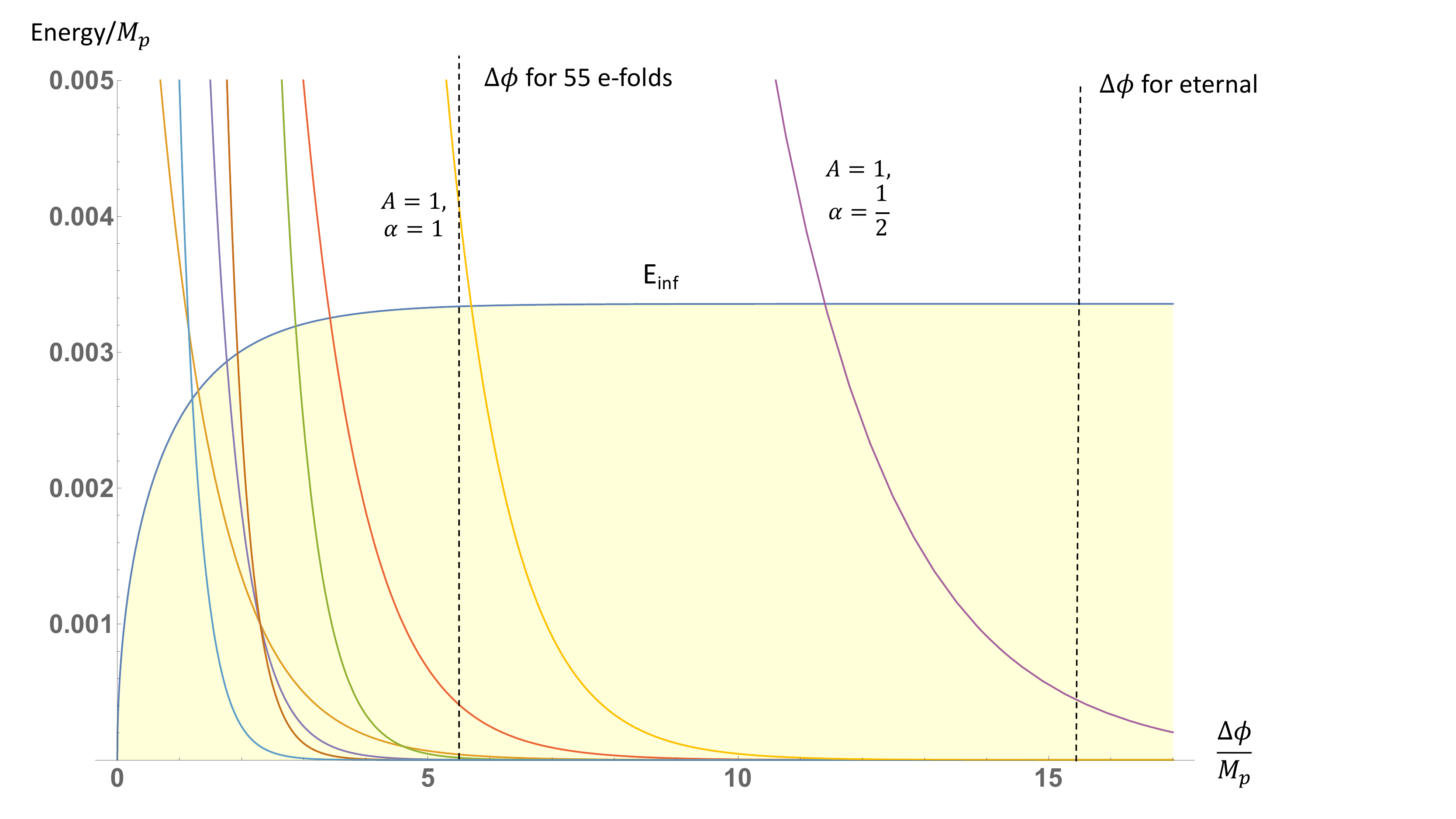}
\caption{Figure showing the energy scale of Starobinsky inflation $E_{\mathrm{inf}}$ and the mass scale of the tower of states of the distance conjecture, parameterized as $A e^{-\alpha \frac{\Delta \phi}{M_p}} M_p$, as a function of the inflaton excursion $\Delta \phi$. The values of $A$ and $\alpha$ for most of the curves are the same as in figure \ref{fig:bndr} (by colour), while two new curves are added. The values of $\Delta \phi$ required to achieve 55 e-foldings and to achieve eternal inflation are shown.}
\label{fig:starbounds}
\end{figure}

Even when $r$ may be very small the constrains can be important. A canonical example is Starobinsky inflation \cite{STAROBINSKY198099}, which is one of the best fitting models to current data. While this model has $r \sim 0.003$, it is still a large field inflation models, with 55 e-foldings requiring $\Delta \phi > 5 M_p$, and an eternal inflation scenario requiring $\Delta \phi > 15 M_p$. The mass scale of inflation is $E_{\mathrm{inf}}  \sim 3\times10^{-3} M_p$, and requiring $\Lambda_{dc} > E_{\mathrm{inf}}$ places strong bounds on $A$ and $\alpha$ as shown in figure \ref{fig:starbounds}.

Starobinsky inflation is part of a more general set of inflation models termed $\alpha$-attractors \cite{Kallosh:2013yoa,Galante:2014ifa,Roest:2015qya}. In \cite{Scalisi:2018eaz} an analysis of the cosmological implications of the distance conjecture with respect to these models was performed.

Finally, it is important to keep in mind that inflation is a dynamical process, while the distance conjecture is a statement about distances in field space. It is therefore important to study the implications of this difference in the context of bounds on inflation. This has been performed in \cite{Landete:2018kqf,Schimmrigk:2018gch} in the context of the Swampland, and it was argued that the onset of the exponential behaviour of the Refined distance conjecture (\ref{rsdc}) may be delayed by this difference by an order one factor. Such differences should be accounted for in a precise quantitative analysis of bounds on inflation. 
 
\section{Swampland Tests in String Theory}
\label{sec:stringcomp}

In section \ref{sec:map} we discussed Swampland constraints from a general quantum gravity perspective. In this section we will review how the ideas of the Swampland manifest in string theory. In section \ref{sec:first} we introduced some basics of string theory which were sufficient for a first encounter with the Swampland upon considering a simple compactification on a circle. It will not be possible to review this section at the same level of technical detail, simply because it is much more advanced material and the literature is much larger and also less sharply understood. However, in keeping with the introductory theme of this article, we will try to at least present the key concepts and also some simple examples which are worked through in more detail. Some introductory textbooks which are relevant to the content of this section are \cite{Blumenhagen:2013fgp,Polchinski:1998rq,Polchinski:1998rr}, and more advanced reviews are \cite{Blumenhagen:2006ci,Grana:2005jc,Douglas:2006es}.

\begin{figure}[t]
\centering
 \includegraphics[width=0.9\textwidth]{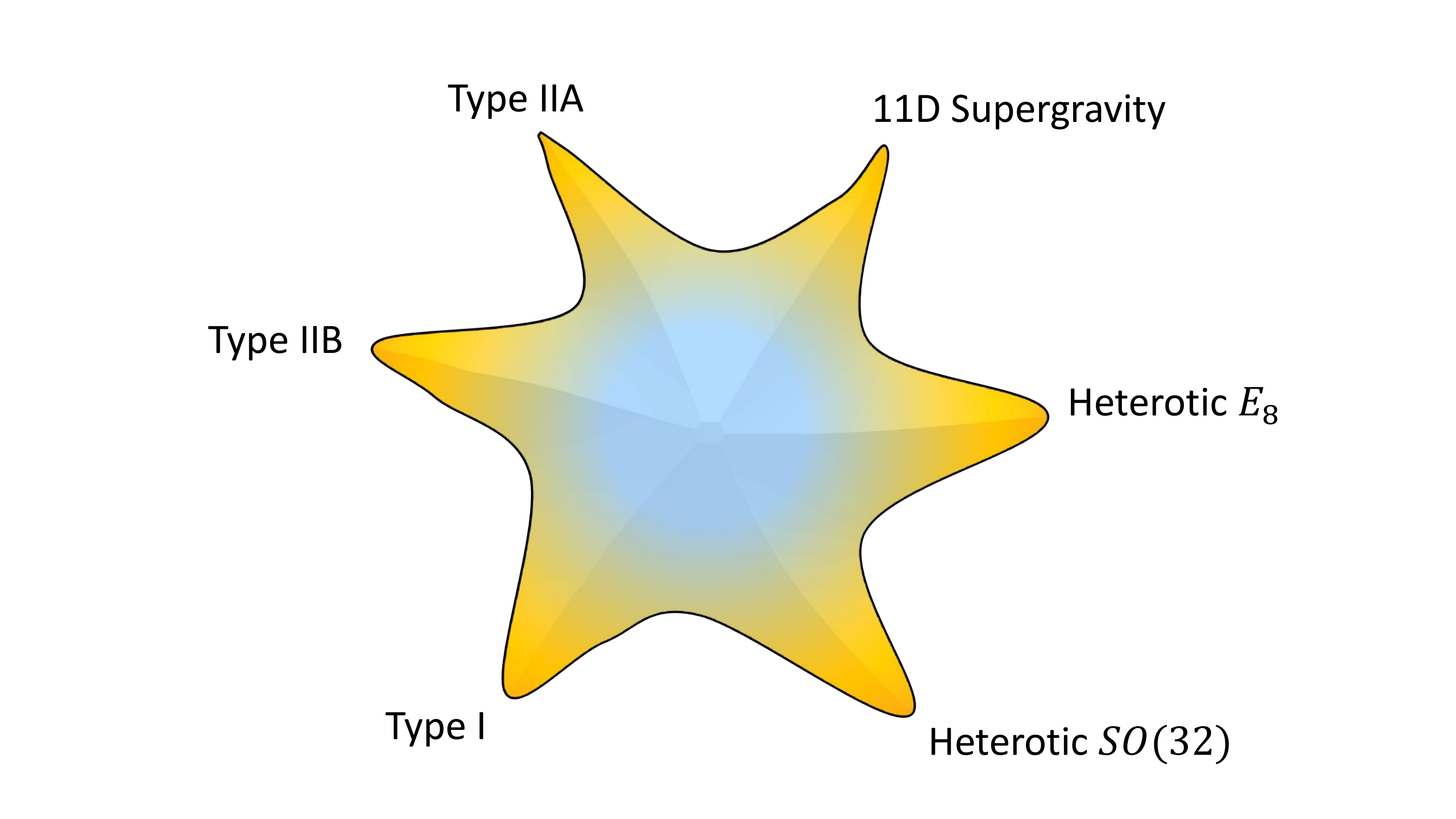}
\caption{Figure showing the parameter space of M-theory. The different well-understood limits correspond to the 5 string theories and 11-dimensional supergravity.}
\label{fig:mtheory}
\end{figure}

In section \ref{sec:first} we introduced the bosonic string. We noted the presence of a tachyon mode but did not pay it too much attention. The tachyon, as well as other non-perturbative instabilities, can be removed if we consider a supersymmetric string, a superstring. There are five types of superstring theories: Type I, Type IIA, Type IIB, Heterotic $SO(32)$ and Heterotic $E_8$. However, they are all linked by dualities and so are capturing different parameterizations of the same theory, often termed M-theory. The different parameterizations are good in different regimes of the theory, and so we may consider a single theory with a parameter space, and in different corners of the parameter space the superstring theories offer good descriptions. This is illustrated in figure \ref{fig:mtheory}.  

In (\ref{steff}) we presented an effective action for the massless spectrum of the bosonic string. The superstring effective actions are 10-dimensional and contain a different spectrum of massless fields. All the string theories contain a (classically) massless graviton and dilaton. We present the massless spectrum for anti-symmetric forms in table \ref{tab:stringthe}, where we also include 11-dimensional supergravity, which is proposed to be the low-energy limit of M-theory. In the case of the bosonic string, the anti-symmetric Kalb-Ramond field $B_{[\mu\nu]}$ naturally integrates over the string world-sheet. The superstrings have different degree anti-symmetric tensors and they similarly naturally couple to different dimensional objects. The $C^{(p)}$ field in table \ref{tab:stringthe} are termed Ramond-Ramond fields, and the objects they couple to are termed D-branes \cite{Polchinski:1995mt}. A $Dp$-brane is a D-brane which is $p+1$ dimensional, and these are also shown in table \ref{tab:stringthe}. 
\begin{table}[h]
\centering
\begin{tabular}{|c|c|c|}
\hline
Massless Field &  Electric Charged Object & Magnetic Dual \\
\hline
\multicolumn{3}{|c|}{{\bf Type IIA String Theory}} \\
\hline
$B^{(2)}_{\mu\nu}$ & F1-String & NS5-brane\\
\hline
$C^{(1)}_{\mu}$ & D0-Brane & D6-brane\\
\hline
$C^{(3)}_{\mu\nu\rho}$ & D2-Brane & D4-brane\\
\hline
$C^{(9)}_{\mu\nu...\rho}$ & D8-Brane & - \\
\hline
\multicolumn{3}{|c|}{{\bf Type IIB String Theory}} \\
\hline
$B^{(2)}_{\mu\nu}$ & F1-String & NS5-brane\\
\hline
$C^{(0)}$ & D(-1)-Brane & D7-brane\\
\hline
$C^{(2)}_{\mu\nu}$ & D1-Brane & D5-brane\\
\hline
$C^{(4)}_{\mu\nu\rho\sigma}$ & D3-Brane & D3-Brane \\
\hline
\multicolumn{3}{|c|}{{\bf Type I String Theory}} \\
\hline
$C^{(2)}_{\mu}$ & D1-Brane & D5-brane\\
\hline
$A^i_{\mu}$, $SO(32)$ & \multicolumn{2}{|c|}{F1 String}   \\
\hline
\multicolumn{3}{|c|}{{\bf Heterotic String $SO(32)$}} \\
\hline
$B^{(2)}_{\mu\nu}$ & F1-String & NS5-brane\\
\hline
$A^i_{\mu}$, $SO(32)$ & \multicolumn{2}{|c|}{F1 String}   \\
\hline
\multicolumn{3}{|c|}{{\bf Heterotic String $E_8$}} \\
\hline
$B^{(2)}_{\mu\nu}$ & F1-String & NS5-brane\\
\hline
$A^i_{\mu}$, $E_8 \times E_8$ & \multicolumn{2}{|c|}{F1 String}   \\
\hline
\end{tabular}
\caption{Table showing the bosonic massless spectrum of different superstring theories. Anti-symmetric forms couple to associated extended objects, which are labelled along with their magnetic duals. Some superstring theories contain non-Abelian gauge groups.} 
\label{tab:stringthe}
\end{table}

The type II string theories also contain other localized objects called orientifold planes, which we denote by $Op$. They have the same dimension as the associated $Dp$ branes, and have opposite sign charge under the associated field. More precisely their charges are related as $Q\left(Op\right) = -2^{p-4} Q\left(Dp\right)$, see for example \cite{Blumenhagen:2006ci}.\footnote{Type I string theory can be thought of as type IIB string theory with an $O9$-plane. To cancel the charge one needs to add $32$ $D9$ branes, which lead to the $SO(32)$ gauge group.} Orientifold planes are non dynamical in the weak-coupling limit of the string theories, they correspond to fixed loci for a quotient of the string theory of the form 
\be
\cO = \cO_{F} \;\sigma_O \;,
\label{oplaneac}
\ee 
where $\cO_F$ is an internal involution action on the string worldsheet fields and $\sigma_O$ is an external involution on the geometry.

In section \ref{sec:first} we discussed string theory on a circle. This means we considered a background which is topologically $\mathbb{R}^{1,d} \times S^1$. We then considered the effective theory for the massless modes in $\mathbb{R}^{1,d}$. Constructing this effective theory is termed a compactification, in this case on a circle. More generally, a compactification refers to a considering the background to be a non-compact space, times a compact space. And the effective lower dimensional theory is constructed at an energy scale below the Kaluza-Klein scale of the compact dimensions. There are some exceptions to these rules, which we will discuss later. 

In the case of the bosonic string on a circle we saw that the radius of the circle was a scalar field $\phi$ (\ref{Rphi}). This scalar had no potential classically. In the absence of supersymmetry, it is expected that quantum corrections will lead to a potential for the field. However, if we consider compactifications which preserve supersymmetry in the lower-dimensional theory, then the resulting scalar fields which parameterize the extra-dimensional geometry can remain massless, and are termed moduli. The expectation values of the exactly massless scalar fields in the theory define a field space called a moduli space.  In general, only compactifications which preserve at least 8 supercharges are expected to have true moduli spaces in the sense that there are scalar fields which have no potential even at the non-perturbative level. Compactifications which preserve 4 supercharges, the minimal supersymmetry in four dimensions, may have moduli spaces at the perturbative level, though it is natural to expect that there is no moduli space (other than a point) at the non-perturbative level (see section \ref{sec:modsp}). If we consider compactifications to four dimensions, we therefore have a qualitative difference between vacua which preserve ${\cal N} \geq 2$ supersymmetry and ${\cal N}=1$ supersymmetry. In particular, the best understood ${\cal N}=1$ vacua are those which exhibit a potential for the (would-be) moduli such that they are fixed in a controlled regime. This is called moduli stabilization, and is simplest and most actively studied for type IIA and IIB string theories. 

Moduli stabilization will play a central role in much of the Swampland discussions and so we will spend some time introducing a simple case of moduli stabilization in type IIA string theory in section \ref{sec:simpleiia}. More complicated, but much more widely studied, moduli stabilization scenarios are realized in type IIB string theory, and will be the topic of section \ref{sec:typeiibor}. We will then utilize these introductory sections to review tests of the Swampland constraints discussed so far, namely the Weak Gravity Conjecture and the Swampland Distance Conjecture, in string theory.

However, the Weak Gravity Conjecture in string theory is most naturally introduced in the context of the Heterotic string, to which we now turn.

\subsection{The Weak Gravity Conjecture and Heterotic Strings}
\label{sec:wgchet}

In \cite{ArkaniHamed:2006dz} the first string theory example test of the Weak Gravity Conjecture was for the Heterotic string on a six-torus $T^6$. We refer to \cite{Blumenhagen:2013fgp} for an account of such a setup from the worldsheet perspective, as in section \ref{sec:first}, and will be relatively brief here. The six-torus is just a direct product of circles and so some of the discussion in section \ref{sec:fecompga} for the reduction of the bosonic string on a circle will apply. Since we have 6 circles we obtain in four dimensions 12 gauge fields from the metric and Kalb-Ramond field. The Heterotic string also has a ten-dimensional gauge group which is generically broken to its Cartan sub-algebra by Wilson lines on the circles, yielding a further 16 $U(1)$ gauge fields. Overall there are therefore 28 $U(1)$ gauge fields in four dimensions, and they are within an ${\cal N}=4$ supergravity theory preserving 16 supercharges (half the maximal supersymmetry). 

We will focus here on the 16 $U(1)$ which come from the breaking of the $SO(32)$ gauge group in the Heterotic string. An analysis of this with respect to the Weak Gravity Conjecture was performed in \cite{Heidenreich:2015nta}. The first massive excitation is the $SO(32)$ spinor which has a mass
\be
m^2 = \frac{4}{\alpha'} \;.
\ee
Upon reduction on $T^6$ it leads to states with charges populating the 16-dimensional charge vectors
\be
{\bf q} = \left( \pm \frac12, ...., \pm \frac12 \right)\;,
\ee
where the number of minus signs is even. In the Einstein frame, and in Planck units, the gauge coupling for any of the $U(1)$s is 
\be
g^2 = \frac{2}{\alpha'} \;.
\ee
Note that $\alpha'$ depends on the dilaton expectation value through a relation analogous to (\ref{mpmsdire}), which then is the field-dependence of the gauge coupling. Because there is a dilatonic coupling, the appropriate formulation of the Weak Gravity Conjecture bound is \cite{Heidenreich:2015nta}
\be
\label{hetwgcbsp32}
m^2 \leq g^2 \left|{\bf q}\right|^2 = \frac{8}{\alpha'} \;,
\ee
which is indeed satisfied. Note that the Weak Gravity Conjecture bound will actually be modified due to the Wilson lines, as for example the axion $a$ in (\ref{4dreducedaction}). The bound (\ref{hetwgcbsp32}) can be thought of as the limit of small Wilson line expectation values, but it would be interesting to test it for arbitrary Wilson lines. The analysis can be generalized for the full 28 $U(1)$s, so including the $U(1)$s from the $T^6$, as in \cite{ArkaniHamed:2006dz}. 

\begin{figure}[t]
\centering
 \includegraphics[width=0.9\textwidth]{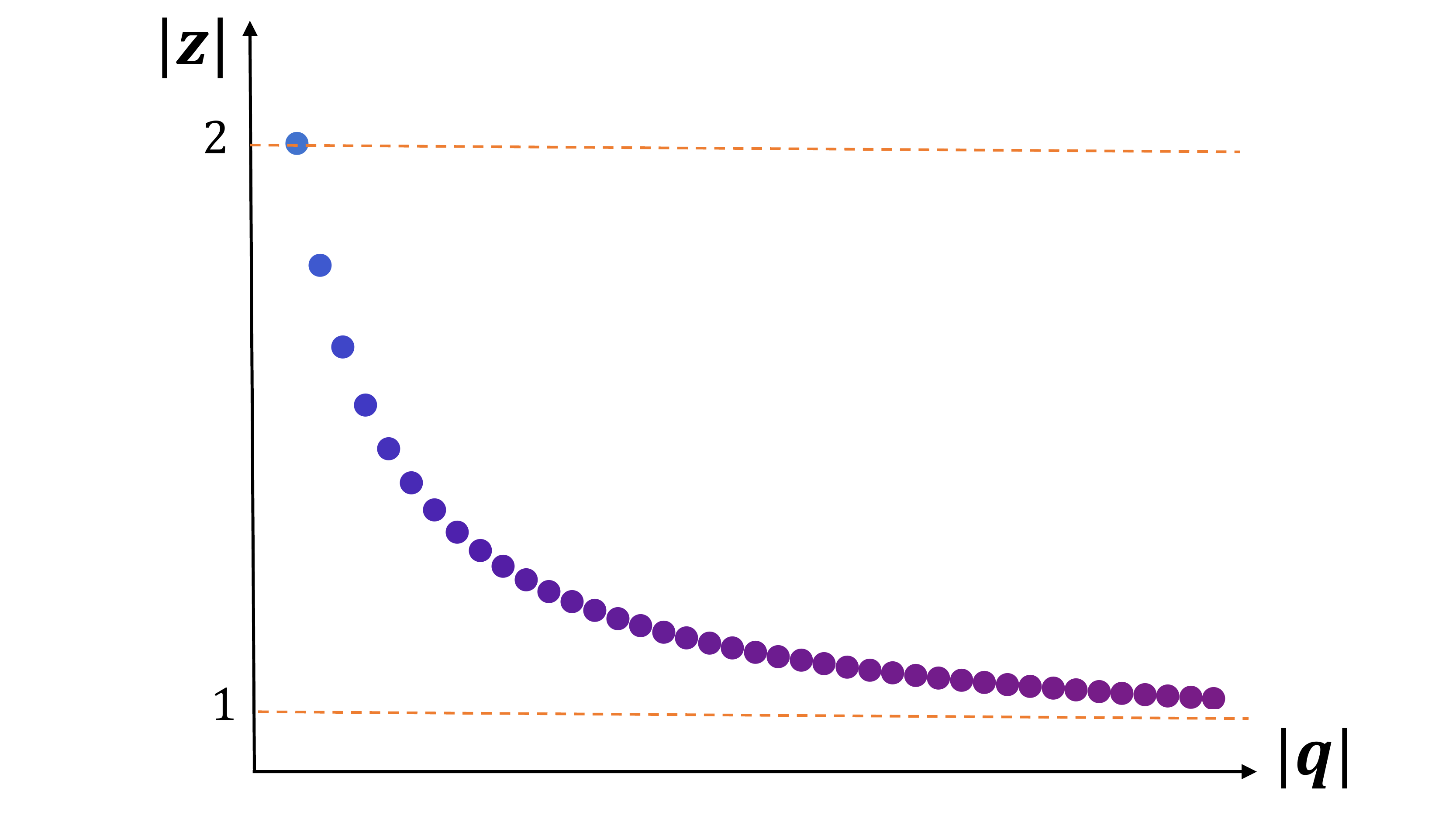}
\caption{Figure illustrating the charge-to-mass ratio for the oscillator modes of the Heterotic string. The spectrum approaches the Weak Gravity Conjecture bound from above.}
\label{fig:cmrhet}
\end{figure}

We can also consider the higher oscillator modes which have a mass
\be
m^2 = \frac{2}{\alpha'} \left( \left|{\bf q}\right|^2 - 2\right) \;,
\ee
where now the possible charges are
\be
{\bf q} = \left(q_1 + \frac{c}{2},...,q_{16}  + \frac{c}{2} \right) \;
\ee
where $q_i \in \mathbb{Z}$ and $c=0,1$. The charges are further constrained to satisfy $ \left|{\bf q}\right|^2 \in 2 \mathbb{N}$. Then the charge-to-mass ratio ${\bf z}$ behaves as
\be
\left|{\bf z}\right|^2 = \frac{\left|{\bf q}\right|^2}{\left|{\bf q}\right|^2-2} \;.
\ee
This approaches the Weak Gravity Conjecture bound $\left|{\bf z}\right|^2 \geq 1$ for large charges. The spectrum is illustrated in figure \ref{fig:cmrhet}. 

We note that the spectrum of charged states satisfying the Weak Gravity Conjecture for the Heterotic string on $T^6$ forms a lattice, and so satisfies the sub-Lattice Weak Gravity Conjecture (see section \ref{sec:wgctower}). This also implies that it satisfies the Strong Weak Gravity Conjecture (see section \ref{sec:strongwgc}). Note that while for $T^6$ the states actually form the full lattice of charges, in the case of orbifold compactifications some charged states can be projected out leading to only a sub-Lattice \cite{Heidenreich:2016aqi,Brown:2015lia}. It was further argued in \cite{ArkaniHamed:2006dz} that the existence of the spectrum of charged states satisfying the Weak Gravity Conjecture is implied by modular invariance of the world-sheet theory for the Heterotic string on an arbitrary manifold.\footnote{see \cite{Heidenreich:2016aqi,Montero:2016tif} for follow-up studies discussing subtleties with this.} It was also argued in \cite{Heidenreich:2016aqi} that this extends to any $U(1)$ coming from the NS-NS sector of any perturbative string theory.  

More advanced tests of the Weak Gravity Conjecture for Heterotic strings $U(1)$s were performed in \cite{Ibanez:2017vfl,Aldazabal:2018nsj}. In \cite{Ibanez:2017vfl} studies of the Higgsing scenario of \cite{Saraswat:2016eaz}, discussed in section \ref{sec:strongwgc} were made. It was argued that there is an obstruction to parametrically large violations of the Strong Weak Gravity Conjecture. In \cite{Aldazabal:2018nsj} studies of Heterotic scenarios with multiple $U(1)$s were made, and in particular it was conjectured that in a certain class of compactification whenever there is a Fayet-Iliopoulos parameter for a $U(1)$ there exists a field which can adjust its expectation value to restore supersymmetry. This was tied to the convex hull condition, as discussed in section \ref{sec:multiu1}. In \cite{Lee:2018urn,Lee:2018spm,Lee:2019tst} tests of the Weak Gravity Conjecture in rich and general Heterotic settings in both six and four dimensions were performed by utilising Heterotic/F-theory duality. 

In this section we mostly focused on the Heterotic string on $T^6$, which leads to a highly supersymmetric theory in four dimensions. In general, the less supersymmetry preserved by the compactification of a string theory, the richer the possible effective four-dimensional theories, and so the stronger the tests of the Swampland. Indeed, the most challenging tests for the various versions of the Weak Gravity Conjecture are in a minimally supersymmetric setting, so ${\cal N}=1$ or ${\cal N}=0$ in four dimensions (though see sections \ref{sec:wgciios} and \ref{sec:dcwith8} for highly non-trivial tests of the magnetic Weak Gravity Conjecture for extended supersymmetry). In such settings, there is a potential induced for the scalar fields in the theory, and so to test the Swampland constraints one must understand this potential and the resulting vacuum structure of the theory. This is particularly important for axion versions of the Weak Gravity Conjecture since they are most interesting when the axions have a potential. We therefore go on now to discuss such minimally supersymmetric settings, and the resulting potentials that can be generated in string theory.

\subsection{Bosonic sector of ${\cal N}=1$ Supergravity}

Compactifications of string theory to four dimensions which preserve ${\cal N}=1$ supersymmetry in the vacuum can be described in terms of an effective theory which is an ${\cal N}=1$ supergravity. Here we simply summarize some useful properties of such theories, and refer to \cite{Wess:1992cp} for a thorough introduction. The field in such theories fall into supersymmetric multiplets. Two multiplets which are of particular relevance are chiral multiplets, whose bosonic content consists of two real scalar fields, and Vector multiplets whose bosonic content is a gauge field. We will mostly be concerned with the scalar fields sector and this can be described using a K{\"a}hler potential $K$, a superpotential $W$ and D-terms. We will almost always study backgrounds with vanishing D-terms, and so will not discuss them here. The bosonic fields of chiral multiplets can be paired into a complex scalar fields $T^i$, where the $i$ index runs over the different chiral multiplets in the theory. The K{\"a}hler potential is then a real function of the $T^i$, while the superpotential is holomorphic in them. The effective theory for the bosonic fields is then given by
\be
S_{{\cal N}=1} = \int d^4 x \sqrt{-g}\left[ \frac{R}{2} - g_{i\bar{j}} \partial T^i \partial \overline{T}^{\bar{j}} + V\left(T^i \right) + ...\right] \;,
\ee
where the field space metric $g_{ij}$ and the potential $V$ are given in terms of the K{\"a}hler potential and superpotential as
\be
g_{i\bar{j}} = \partial_{i} \partial_{\bar{j}} K \;, \;\; 
V = e^{K} \left[ g^{i\bar{j}} D_i W \overline{D}_{\bar{j}} \overline{W}  - 3 \left|W\right|^2 \right] \;,
\label{gVfromKW}
\ee
and we have introduced the covariant derivative
\be
D_i = \partial_i + \left(\partial_i  K\right) \;.
\ee

The gauge fields kinetic terms are given as in (\ref{N2actionEM}), where $\cI_{IJ}$ and $\cR_{IJ}$ are the imaginary and real parts of the gauge kinetic function 
\be
f_{IJ}\left(T^i\right) \equiv \cR_{IJ} + i \cI_{IJ}  \;.
\ee
The gauge kinetic function is a holomorphic function of the scalar fields in the chiral multiplets. 

Both the superpotential and gauge kinetic function enjoy strong non-renormalisation properties. The superpotential receives no quantum corrections in perturbation theory, though can receive non-perturbative corrections. The gauge kinetic function receives only 1-loop corrections in perturbation theory. 

\subsection{Simple type IIA compactifications}
\label{sec:simpleiia}

In this section we will study some simple compactifications of type IIA string theory to four dimensions. The aim is to introduce some basic concepts of compactifications so that when considering more complicated compactifications we will be able to refer back to such concepts. While these are simple settings, we will see that a substantial amount regarding Swampland constraints can be exhibited using these examples. 

The section will utilize the technology of differential forms, and we refer to \cite{Nakahara2003geometry} for an introduction to this.

\subsubsection{Type IIA on $T^6$}
\label{sec:iiaont6}

In this section we consider a compactification of type IIA to four dimensions on a six-dimensional torus $\cM_6=T^6$. We parameterize the six directions of $T^6$ in terms of coordinates $x^i$, $i=1,..,6$, and define the associated 1-forms 
\be
\eta^i = dx^i \;.
\ee
We impose that the torus should be invariant under a discrete $\mathbb{Z}_2 \times \mathbb{Z}_2$ symmetry which acts as
\bea
& &\left( \eta^1,\eta^2,\eta^3,\eta^4,\eta^5,\eta^6 \right) \rightarrow \left( -\eta^1,-\eta^2,\eta^3,-\eta^4,-\eta^5,\eta^6 \right) \;, \nn \\
& &\left( \eta^1,\eta^2,\eta^3,\eta^4,\eta^5,\eta^6 \right) \rightarrow \left( \eta^1,-\eta^2,-\eta^3,\eta^4,-\eta^5,-\eta^6 \right) \;.
\label{z2z2s}
\eea
This restricts the possible deformations of the $T^6$ to ones which respect the symmetry (\ref{z2z2s}). In particular, it means that we have restricted the six-torus to a product of two tori $T^6 \rightarrow  T^2 \times T^2 \times T^2$. To see this one can note that there are no invariant one-forms under (\ref{z2z2s}), and three invariant two-forms
\be
\omega_1 = -\eta^{14} \;, \;\; \omega_2 = -\eta^{25} \;, \;\; \omega_3 = -\eta^{36} \;.
\label{omdef}
\ee
We have introduced the notation $\eta^{i...k} = \eta^{i}\wedge ... \wedge \eta^{k}$. These two-forms are volume forms for the volume of each two-torus. Each two-torus then has two deformation parameters associated to it, its overall volume and the ratio of the radius of its two constituent circles. Denoting by $R_i$ the radius of each circle in the $T^6$ we can explicitly write the possible deformations of it as
\bea
t^1 &=& \frac{R_1 R_4}{\alpha'} \;,\;\; t^2= \frac{R_2 R_5}{\alpha'} \;,\;\; t^3= \frac{R_3 R_6}{\alpha'} \;, \nn \\
\tau_1 &=& \frac{R_4}{R_1} \;,\;\; \tau_2 = \frac{R_5}{R_2} \;,\;\; \tau_3 = \frac{R_6}{R_3} \;.
\label{defrs}
\eea
The geometric deformations of the torus are scalar fields in the four-dimensional effective theory, which are analogous to the circle radius $\phi$ in (\ref{Rphi}). This compactification, and in particular the related one studied in section \ref{sec:iias3s3}, has been extensively studied, see for example \cite{Camara:2005dc,Aldazabal:2007sn,DallAgata:2009wsi}. 

Note that the symmetry (\ref{z2z2s}) picks out a sub-sector of all the possible geometric deformations. It is possible to impose this strictly by considering an orbifolding of $T^6$ by the symmetry. This will lead to some subtleties, which we discuss later, so for now we just consider restricting to the sector respecting (\ref{z2z2s}). Such a restriction is said to be a consistent truncation of the full set of geometric deformations, and will be something we often utilize. The point is that dynamics within the sector respecting (\ref{z2z2s}) will not induce dynamics in the sector violating (\ref{z2z2s}), since the former preserve the symmetry while the latter break it. More generally, consistent truncations can be employed when the compactification manifold has a group, or coset, structure. See, for example, \cite{Camara:2005dc,Aldazabal:2007sn,Grana:2006kf,Cassani:2009ck} for discussions on this point. We will henceforth denote the compactification as
\be
\cM_6 = \frac{T^6}{\mathbb{Z}_2 \times \mathbb{Z}_2} \;,
\ee 
and in practice almost everything we write holds for either the consistent truncation only without orbifolding, or for actually orbifolding the torus.

The compactification preserves supersymmetry, and so it is useful to write the geometry in a way which manifests the supersymmetry. To do this we consider the possible forms on the space that are invariant under (\ref{z2z2s}). In (\ref{omdef}) we constructed the possible two-forms, the possible three-forms are then
\bea
\begin{array}{cc}
\alpha_0 = \eta^{123} \;,& \beta^0 = \eta^{456} \;,\\
\alpha_1 = \eta^{156} \;,& \beta^1 = \eta^{423} \;,\\
\alpha_2 = \eta^{426} \;,& \beta^2 = \eta^{153} \;,\\
\alpha_3 = \eta^{453} \;,& \beta^3 = \eta^{126} \;.
\end{array}
\eea
Note that the forms are not all independent but come in pairs related by Hodge duality. More generally, such bases can be normalised such that
\be
\int_{\cM_6} \alpha_I \wedge \beta^J = \delta_I^J \;.
\ee
Similarly, we can construct four-forms 
\be
\tilde{\omega}^1 = \eta^{2536} \;, \;\; \tilde{\omega}^2 = \eta^{1436} \;, \;\; \tilde{\omega}^3 = \eta^{1425} \;.
\ee
And these form an orthonormal basis with the two-forms such that
\be
\int_{\cM_6} \omega_i \wedge \tilde{\omega}^j = \delta^i_j \;.
\ee
 
A compactification manifold which preserves supersymmetry has some universal structures on it which come from the requirement that it should support nowhere vanishing spinors in order to construct the associated supersymmetry generators, see for example the review \cite{Grana:2005jc}. It is said to have a reduced structure group, and in this case it has $SU(3)$-structure. This means that it always supports a real two form $J$, the K{\"a}hler form, and a complex three-form $\Omega$. These forms can be used to formulate the geometry of the manifold in a manifestly supersymmetric way. The geometric deformations of the manifold are then associated to expansions of $J$ and $\Omega$ in the basis of forms on the manifold. In this case, the appropriate expansions are
\bea
\label{jott6}
J &=& t^1 \omega_1 + t^2 \omega_2 + t^3 \omega_3 \;,  \\ \nn
\Omega &=& \left[\alpha_0 + \left(\frac{u_1}{s} \right) \alpha_1 + \left(\frac{u_2}{s} \right) \alpha_2+ \left(\frac{u_3}{s} \right) \alpha_3 \right] 
+ i\left[-\tau_1 \tau_2 \tau_3 \beta^0 + \tau_1 \beta^1+ \tau_2 \beta^2+ \tau_3 \beta^3  \right] \;.
\eea 
Here we have introduced the combinations
 \be
\tau_1 \tau_2 \tau_3 = e^{6 \phi} \frac{\left(u_1 u_2 u_3\right)^2}{\left(t^1 t^2 t^3 \right)^3} \;,\;\; \tau_1 = e^{2 \phi}\frac{u_2 u_3}{t^1 t^2 t^3} \;,\;\; \tau_2 = e^{2 \phi}\frac{u_1 u_3}{t^1 t^2 t^3} \;,\;\; \tau_3 = e^{2 \phi}\frac{u_1 u_2}{t^1 t^2 t^3} \;,
\ee
and
\be
e^{\phi} = \frac{\left(t^1 t^2 t^3\right)^{\frac12}}{\left(s u_1 u_2 u_3 \right)^{\frac14}} \;.
\ee
In (\ref{jott6}) there are 3 parameters in the expansion of $J$, and 3 parameters in the expansion of $\Omega$. These capture exactly the same possible geometric deformations of the manifold as in (\ref{defrs}), but are just packaged in a different way more appropriate for supersymmetry. Note that we have also introduced a real scalar $s$, but one can check that out of $u_i$ and $s$ only 3 combinations appear in (\ref{jott6}). The notation is that the deformations of $J$, the $t_i$, are called K{\"a}hler moduli, while the deformations of $\Omega$, the $u_i$, are called complex-structure moduli. 

So far we have preserved ${\cal N}=2$ supersymmetry, but we would like to break it further to ${\cal N}=1$. This is done through the introduction of an orientifold plane, in this case an $O6$ plane. The geometric involution, $\sigma_O$ in (\ref{oplaneac}), is taken to act as
\be
\sigma_O \left(\eta^{1,2,3}\right) = \eta^{1,2,3} \;, \;\; \sigma_O \left(\eta^{4,5,6}\right) = -\eta^{4,5,6} \;.
\label{sigoac}
\ee
The $O6$ plane is then space-time filling and wraps the fixed locus of $\sigma_O$. The forms transform under the orientifold action as
\bea
\sigma_O \left(\omega_i\right) &=& - \omega_i \;,\;\; \sigma_O \left(\tilde{\omega}^i \right)= \tilde{\omega}^i \;, \nn \\
\sigma_O \left(\alpha_I\right) &=& \alpha_I \;,\;\; \sigma_O \left(\beta^I\right) = -\beta^I \;.
\eea
Note that since the $O6$-plane in a negative charge source for the Ramond-Ramond forms, we should introduce appropriate positive sources to cancel the net charge. In this case, we can place 4 $D6$ branes on top of each $O6$ plane.\footnote{Note that the $D6$ branes will have open-string moduli associated to them, which we will not discuss at this point.}

The resulting effective ${\cal N}=1$ theory contains massless bosonic scalar fields which arise from the geometric deformations of the metric, or equivalently $J$ and $\Omega$, from the 10-dimensional dilaton field denoted here $\phi$, and from the higher dimensional anti-symmetric forms, as in table \ref{tab:stringthe}. The general expression for this effective action for orientifold compactifications was determined in \cite{Grimm:2004ua}. First we define the (scalar parts of the) superfields as
\bea
J_c \equiv B^{(2)} + i J = i T^i \omega_i \;, \nn \\
\Omega_c \equiv C^{(3)} + i \mathrm{Re\;} \left(s \Omega \right) = i S \alpha_0 - i U^i \alpha_i \;.
\label{def10JO}
\eea
We expand these in components
\be
T^i = t^i + i v^i  \;, \;\; U^i = u^i + i \nu^i \;,\;\; S = s + i \sigma \;.
\label{compsupiia} 
\ee
The real parts of the superfields match onto (\ref{jott6}). The field $s$ is denoted the four-dimensional dilaton, to distinguish it from the ten-dimensional dilaton $\phi$.  The $\nu^i$ are seen to come from the ten-dimensional Kalb-Ramond form, while the $\nu^i$ and $\sigma$ come from the ten-dimensional Ramond-Ramond three-form $C^{(3)}$.  There are no other scalar fields because of the combination of involutions (\ref{z2z2s}) and (\ref{sigoac}). For example, the internal orientifold action in (\ref{oplaneac}) acts as $\cO_{F}\left(C^{(1)}\right) = -C^{(1)}$, which means that it is projected out with legs in the external directions that are filled by the O-plane, and there are no odd 1-forms to expand it in. Similarly, only the odd three-forms appear in the expansion of $C^{(3)} $.

To specify the effective action for the scalar fields, composed of only kinetic terms, we then need to specify the K{\"a}hler potential as in (\ref{gVfromKW}), which in general reads \cite{Grimm:2004ua} 
\be
K = - \mathrm{log\;} \left(\frac43 \int_{\cM_6} J \wedge J \wedge J \right) - 2\mathrm{log\;} \left( i \int_{\cM_6} \left(s\Omega\right) \wedge \left(s\Omega^*\right) \right) \;.
\label{gen10ea}
\ee
So for this specific case takes the form
\be
\label{kpott6}
K = - \log s - \sum_{i=1}^3 \log u_i - \sum_{i=1}^3 \log t_i \;. 
\ee
This gives the effective action which is analogous to the simple one in (\ref{ddim}), but now comes from a superstring compactification  to four dimensions. The scalar fields have no potential, so the superpotential vanishes $W=0$. More precisely, it only vanishes perturbatively, and there are non-perturbative contributions which will be discussed in section \ref{sec:wgciist}. It is possible to induce a perturbative superpotential by turning on fluxes, to which we now turn.

\subsubsection{10-dimensional solution with fluxes}
\label{sec:10diiaflux}

In section \ref{sec:iiaont6} we considered type IIA string theory on $T^6$. We derived an effective four-dimensional theory which arises below the Kaluza-Klein scale of the $T^6$. This four-dimensional theory then has associated four-dimensional equations of motion. The idea of dimensional reduction is that these equations of motion are such that solutions to them are also, a simple subset of, solutions to the ten-dimensional equations of motion. This, in particular, implies that any vacuum of the four-dimensional effective theory can be uplifted to a ten-dimensional vacuum. For the simple case in section \ref{sec:iiaont6} this can be readily seen, the different expectation values of the four-dimensional moduli fields correspond to different shapes for the torus, but any shaped torus solves the ten-dimensional Einstein equations because it is a Ricci-flat manifold. 

As we consider increasingly complicated compactifications of string theory we will see that this ideal connection between the four-dimensional and ten-dimensional equations will become increasingly difficult to establish. It is therefore crucial to keep in mind that it is always the ten-dimensional physics which is the more fundamental and correct, at least as long as the Kaluza-Klein scale is below the String scale. The four-dimensional effective theory is meant to capture simple solutions to the ten-dimensional equations of motion but it may not always do so correctly. 

A setup where this becomes particularly interesting is the case when fluxes are turned on. This means giving expectation values to the field strengths of the various massless anti-symmetric fields. In such cases the ten-dimensional equations of motion become more complicated, and so we must take extra care in thinking about how the four-dimensional effective theory is capturing these equations. The aim of this section, and of section \ref{sec:iias3s3}, is to provide an analysis of a simple case where we can match the four-dimensional and ten-dimensional analyses. This can then act as a reference point for the more complicated setups discussed later. 

Let us first consider the fields in the ten-dimensional supergravity theory. From table \ref{tab:stringthe} we see that we have a dilaton $\phi$ and also the Kalb-Ramond form which we denote $B_2$. In the case where this form develops an expectation value which leads to a non-vanishing field strength we denote this field strength $H^f_3$, and $B_2$ remains the component which is not contributing to $H^f_3$. So the full ten-dimensional field strength is denoted by $\hat{H}_3$ and takes the form
\be
\hat{H}_3 = H_3^f + d B_2 \;.
\ee
We can do the same for the Ramond-Ramond $C_1$ and $C_3$ fields, with flux backgrounds $F_2^f$ and $F^f_4$. The three-form $C_3$ is special because it can have a non-vanishing field strength along the external non-compact directions which does not break the Lorentz symmetry, which we denote by $f$. Finally, we may consider the Hodge dual of the ten-dimensional field strength of $C_9$, which is a scalar quantity we denote by $\hat{F}_0$. There is no propagating field associated to this field-strength, and so it is pure flux only $F_0^f$, and is known as the mass parameter of massive type IIA supergravity \cite{Romans:1985tz}.  

The case we will consider is type IIA string theory compactified to four-dimensional anti-de Sitter space. For now, the internal manifold will be kept general and we only demand that the vacuum preserves ${\cal N}=1$ supersymmetry, which means that the internal manifold $\cM_6$ should have a K{\"a}hler two-form $J$ and three-form $\Omega$, as in section \ref{sec:iiaont6}. The most general such solution to the ten-dimensional equations was found in \cite{Behrndt:2004km,Lust:2004ig}. In terms of the fields discussed above, it takes the form
\bea
\hat{F}_2 &=& \frac12 \frac19 \sqrt{\frac{108}{5}} \hat{F}_0 J \;, \nn \\
\hat{F}_4 &=& f \mathrm{d Vol}_4 + \frac12 \frac{3 \hat{F}_0}{5} J \wedge J \;, \nn \\
\hat{H}_3 &=& \frac12 \frac{4 \hat{F}_0}{5} e^{2 \phi} s \mathrm{Im\;} \Omega \;.
\label{10dsusy}
\eea
While the Bianchi Identity gives
\be
f = \frac12 \sqrt{\frac{108}{5}} \hat{F}_0 e^{\phi} \;.
\label{10bi}
\ee
Here the hats denote that these are the 10-dimensional forms, which are given by
\bea
\hat{F}_0 &=& F^f_0 \;, \nn \\
\hat{F}_2 &=& F^f_2 + F_0^f B_2 \;, \nn \\
\hat{F}_4 &=& f \mathrm{d Vol}_4 + F^f_4 + dC_3 + B_2 \wedge F_2^f + \frac12 F_0^f B_2 \wedge B_2 \;, \nn \\
\hat{F}_6 &=& F^f_6 - \left( H^f_3 + d B_2 \right) \wedge C_3 + B_2 \wedge F_4^f + \frac16 F_0^f B_2\wedge B_2 \wedge B_2\;, \nn \\
\hat{H}_3 &=& H^f_3 + d B_2 \;.
\eea
Here $\mathrm{d Vol}_4$ denotes the four-dimensional volume form, and $\hat{F}_6$ does not appear in (\ref{10dsusy}) because it is dual to $f$, but will be utilized later.
More precisely, (\ref{10dsusy}) are the supersymmetry equations, and solving them and also the Bianchi Identity gives a solution to the Einstein equation and the equations of motion for the fields. Note that the expression for the Bianchi identity (\ref{10bi}) already utilized the solution of the supersymmetry equations (\ref{10dsusy}). Without doing this it instead can be written as
\be
d \hat{F}_2 = \hat{F}_0 \hat{H}_3  \implies d F_2^f = F_0^f H_3^f \;.
\label{binoloc}
\ee
Upon adding localized sources the only thing which gets modified is the Bianchi identify which receives a contribution proportional to the sum of the charges \cite{Acharya:2006ne,Koerber:2008rx}
\be
d \hat{F}_2 =  \hat{F}_0 \hat{H}_3+ \sum \delta_{\mathrm{loc}} \;.
\label{bi10dloc}
\ee
Therefore, as long as the charges cancel locally the 10-dimensional equations remain the same. We can implement this by placing the D6 branes on top of the O6 planes.

\subsubsection{Type IIA on $T^6$ with flux}
\label{sec:iiat6flux} 

Let us now consider the presence of fluxes from the four-dimensional effective theory perspective. We will keep the manifold the same but now allow for non-vanishing fluxes. That means we can give a non-vanishing expectation value to the field strengths in the internal manifold, which we choose to be
\bea
F_6^f &=& e_0 \epsilon_6 \;, \nn \\
F_4^f &=& e \left( \tilde{\omega}_1 + \tilde{\omega}_2 + \tilde{\omega}_3 \right) \;, \nn \\
F_2^f &=& q \left( \omega_1 + \omega_2 + \omega_3 \right)\;, \nn \\
F_0^f &-& - m \;, \nn \\
H_3^f &=& h_0 \beta^0 + h \left(\beta^1 + \beta^2 + \beta^3 \right) \;.
\label{fluxch}
\eea
where $\epsilon_6 = \eta^{123456}$ is the volume form on the manifold. The fluxes are quantized, so the parameters $e_0$, $e$, $q$, $m$, $h_0$ and $h$ are integers. We have chosen the fluxes to be compatible with the involutions (\ref{z2z2s}) and (\ref{sigoac}). 

Note that in (\ref{fluxch}) we have turned on $F_6^f$. This is dual to the external Freund-Rubin flux $f$ in the ten-dimensional solution of section \ref{sec:10diiaflux}, and they are related through the equation of motion for the external component of $\hat{C}_3$
\be
\label{ftoF6}
f = - \frac{1}{t_1 t_2 t_3 e^{-\phi}} \int \hat{F}_6 \;.
\ee

The fluxes will induce a potential in the effective four-dimensional action. This can be understood physically as the response of the extra-dimensional geometry to the energy density associated to the fluxes. Supersymmetry constrains this potential to be formulated as a superpotential in the ${\cal N}=1$ framework, and it takes the general form \cite{Louis:2002ny,Grimm:2004ua}
\be
\label{fluxwnodj}
W = \int_\cM \left[ F^f_6 + J_c \wedge F^f_4 + \frac12 J_c \wedge J_c \wedge F^f_2 +  \frac16 J_c \wedge J_c \wedge J_c F^f_0 + \Omega_c \wedge H^f_3 \right] \;.
\ee
For the choice of fluxes (\ref{fluxch}) the superpotential potential takes the form
\be
W = e_0  + i e \left( T^1 + T^2 + T^3\right) - q \left( T^1 T^2 + T^1 T^3 + T^2 T^3\right)  + i m T^1 T^2 T^3 + i h_0 S - i h \left( U^1 + U^2 + U^3 \right) \;.
\label{wt6fl}
\ee

We can now look for a vacuum to the action with the K{\"a}hler potential (\ref{kpott6}) and superpotential (\ref{wt6fl}). We look for a supersymmetric vacuum, which solved the F-terms $D_i W=0$. To simplify the solution we consider a vacuum where $T^1=T^2=T^3$ and $U^1=U^2=U^3$, which gives \cite{Camara:2005dc}
\bea
t &=& \frac{1}{|m|}\sqrt{\frac{-5\left(e m + q^2 \right)}{3}}\;, \;\; v= -\frac{q}{m}\;, \nn \\
u  &=&  \frac{2}{3\left(-hm\right)|m|}\sqrt{\frac{-5\left(e m + q^2 \right)^3}{3}}\;, \;\; \nu= -\frac{1}{3h} \left(e_0 + \frac{3 e q}{m} +\frac{2 q^3}{m^2}- h_0 \sigma \right) \;, \nn \\
s &=& \frac{2}{3\left(h_0m\right)|m|}\sqrt{\frac{-5\left(e m + q^2 \right)^3}{3}}\;.
\label{solt6flux}
\eea
We observe that all the moduli fields $s,t,u$ have been fixed by the fluxes. We require $h_0 m > 0$, $h m <0$ and $e m + q^2 < 0$ in order for them to be at physical values. Out of the 7 axions, the three $v^i$ have all been fixed, while because only one linear combination of the $\nu^i$ and $\sigma$ appears in the perturbative action, there remain 3 unfixed axions. We note that taking $\left|e m + q^2\right| \gg 1$ allows us to go to large volume and weak string coupling which makes the supergravity analysis trustable. 

We still need to consider the four-dimensional version of the Bianchi identity (\ref{bi10dloc}), which are known as the tadpole constraints. They are an integrated version of the local Bianchi identity, so are less refined and only capture the overall net charge conservation for the Ramond-Ramond fields. In this case, since $d\hat{F}_2=0$, they read
\bea
N^0_{D6} &+& \frac12 h_0 m = 16 \;, \nn \\
N^i_{D6} &+& \frac12 h m = 0 \;.
\eea
There is a constraint for each 3-cycle homology class. The contribution of 16 comes from the negative O6-plane charge, while $N_{D6}$ refers to the number of D6 branes which are wrapping the associated 3-cycle. Note that since $h m < 0$, it is possible to satisfy the tadpole constraints by adding D6 branes rather than anti-D6 branes, and this means that the vacuum remains supersymmetric. 

We have therefore encountered our first moduli stabilization scenario. It is actually a simple version of the more general analysis in \cite{DeWolfe:2005uu}. We see that for different choices of flux integers we have different vacua. In fact, since the fluxes $e$ and $q$ are unconstrained, we have formally an infinite number of vacua in the theory. 

Let us consider now the relation between the four-dimensional vacuum and the ten-dimensional one. We saw that it is possible to satisfy the four-dimensional tadpole constraints by adding localized charged sources. However, the ten-dimensional Bianchi Identity (\ref{bi10dloc}) is not satisfied because it is a local equation and the flux on the left hand side is diffuse throughout the cycle, while the right hand side has localized sources. It is usually stated that in such a situation the Bianchi Identity is satisfied in a smeared approximation, which means one considers smearing the localized sources over the cycle so that they can cancel the flux, see for example \cite{Grana:2006kf}. 

It would be very useful to understand the effect of the smearing approximation on the solution. In general this is difficult to understand, but in this case we can gain some insight by looking at a fully exact solution, to which we now turn.

\subsubsection{Type IIA on $S^3 \times S^3$ with flux}
\label{sec:iias3s3}

In order to find a compactification which solves the 10-dimensional Bianchi identity, and equations of motion, exactly we need to deform our geometry from a torus to a twisted torus. We can keep everything the same as in section \ref{sec:iiaont6}, but we impose that the basis forms $\eta^i$ are no longer closed but satisfy
\be
d \eta^i = - \half \omega^i_{ij} \eta^j \wedge \eta^k \;.
\ee
We consider the case where the $\omega^i_{ij}$ are constants, which means that the manifold is homogenous. The manifold can be viewed as a Lie group $G$ where the $\omega^i_{ij}$ are the structure constants. Imposing the nilpotency of the exterior derivative implies
\be
\omega^i_{[jk}\omega^l_{m]n} = 0 \;.
\ee
This can also be understood as the Jacobi identity for the structure constants. The discrete symmetry (\ref{z2z2s}) implies that we should fix 
\bea
& &\omega^1_{56}=\omega^2_{64}=\omega^3_{45}=a \;, \nn \\
& &\omega^1_{23}=\omega^4_{53}=\omega^4_{26}=\omega^5_{34}=\omega^2_{31}=\omega^5_{61}=\omega^6_{42}=\omega^6_{15}=\omega^3_{12}=b \;.
\eea
With this choice note that we can go to a rescaled basis $\hat{\eta}$
\be
\eta^{1,2,3} = \frac{1}{b} \hat{\eta}^{1,2,3} \;,\;\; \eta^{4,5,6} = \frac{1}{\sqrt{ab}} \hat{\eta}^{4,5,6} \;,
\ee
in which all the non-vanishing $\hat{\omega}^{i}_{jk}=1$. To see what this choice implies for the manifold we can calculate the Killing form
\be
K_{jl} = \hat{\omega}^{i}_{jk}\hat{\omega}^{k}_{li} = -2\; \mathbb{I} \;.
\ee
This is the Killing form of $SO(4) \simeq SO(3) \times SO(3)$. Therefore, we see that this manifold is nothing but a (squished) product of two three-spheres $S^3 \times S^3$. This can be seen explicitly through a change of basis 
\bea
\xi^1 &=& b \eta^1 + \sqrt{ab}\eta^4 \;, \;\; \hat{\xi}^1 = b \eta^1 - \sqrt{ab}\eta^4 \nn \;, \\
\xi^2 &=& b \eta^2 + \sqrt{ab}\eta^5 \;, \;\; \hat{\xi}^2 = b \eta^2 - \sqrt{ab}\eta^5 \nn \;, \\
\xi^3 &=& b \eta^3 + \sqrt{ab}\eta^6 \;, \;\; \hat{\xi}^3 = b \eta^3 - \sqrt{ab}\eta^6 \;. 
\label{xibas}
\eea
In which case the structure constants are given by
\be
d \xi^i = - \half \epsilon_{ijk} \xi^j \wedge \xi^k \;,\;\; d \hat{\xi}^i = - \half \epsilon_{ijk} \hat{\xi}^j \wedge \hat{\xi}^k \;.
\ee
So the $\xi$ and $\hat{\xi}$ are the basis forms on the two $S^3$s. Note that the isometry group is also $SU(2) \times SU(2)$.
We further would like to impose an orientifold projection. The geometric action $\sigma_O$ of the projection is taken as
\be
\sigma_O\left(\eta^{1,2,3} \right) = \eta^{1,2,3} \;, \;\;\sigma_O\left(\eta^{4,5,6} \right) = -\eta^{4,5,6} \;,
\ee
We therefore see that the manifold is 
\be
{\cal M} = \frac{S^3 \times S^3}{\mathbb{Z}_2} \;.
\ee
The orientifold action interchanges the two $S^3$, which can be seen from (\ref{xibas}).

The basis forms now satisfy the differential properties
\bea
d \omega_1 &=& a \beta^0 + b \left( \beta^1 - \beta^2 - \beta^3 \right) \;, \nn \\
d \omega_2 &=& a \beta^0 + b \left( -\beta^1 + \beta^2 - \beta^3 \right) \;, \nn \\
d \omega_3 &=& a \beta^0 + b \left( -\beta^1 - \beta^2 + \beta^3 \right) \; \nn \\
d \tilde{\omega}^i &=& 0 \;.
\label{diff1}
\eea
Note that there is no linear combination of the $\omega_i$ which is closed, this means that the manifold contains no four-cycles. For the odd forms we have
\bea
d \alpha_0 &=& a \left( \tilde{\omega}^1 +  \tilde{\omega}^2 +  \tilde{\omega}^3 \right) \;, \nn \\
d \alpha_1 &=& b \left( \tilde{\omega}^1 -  \tilde{\omega}^2 -  \tilde{\omega}^3 \right) \;, \nn \\
d \alpha_2 &=& b \left( -\tilde{\omega}^1 +  \tilde{\omega}^2 -  \tilde{\omega}^3 \right) \;, \nn \\
d \alpha_3 &=& b \left( -\tilde{\omega}^1 -  \tilde{\omega}^2 +  \tilde{\omega}^3 \right) \;, \nn \\
d \beta^I &=& 0 \;.
\label{diff2}
\eea
There is one orientifold odd and one orientifold even harmonic three-form, which correspond to the two $S^3$s.

Since all we have done is modify the differential, but not the algebraic, properties of the basis forms, the effective theory is only modified through the superpotential. Indeed, the general superpotential for a manifold with curvature that still has $SU(3)$ structure was derived in \cite{Grimm:2004ua,Derendinger:2004jn,Villadoro:2005cu,House:2005yc,Grana:2005ny,Benmachiche:2006df} and reads
\be
W = \int_{\cM_6} \left[ F^f_6 + J_c \wedge F^f_4 + \frac12 J_c \wedge J_c \wedge F^f_2 +  \frac16 J_c \wedge J_c \wedge J_c F^f_0 + \Omega_c \wedge \left( H^f_3 + d J_c \right) \right] \;.
\label{wt6dj}
\ee
Note the last term which is missing in (\ref{fluxwnodj}). 

The primary aim of deforming the geometry is to solve the 10-dimensional Bianchi identity locally. That means that we cancel localized sources by themselves, by placing D6 branes on any O6 planes, in this case (\ref{bi10dloc}) reduces to (\ref{binoloc}) and gives the constraint
\be
3 a q = - m h_0 \;,\;\; b q = m h \;.
\ee 
We can solve this by restricting 
\be
h_0 = - 3 a \;,\;\; h = b \;,\;\; q = m \;.
\label{flt}
\ee
The resulting effective superpotential then reads
\bea
W &=& \left(e_0 + 3 e \right)  -m \left( T^2 T^3 + T^1 T^3 + T^1 T^2 \right) + i m T^1 T^2 T^3 \nn \\
& & + \left(T^1+i \right)\left[-a S + b \left( U^1 - U^2 - U^3 \right) + i e\right] \nn \\
& & + \left(T^2+i \right)\left[-a S + b \left( -U^1 + U^2 - U^3 \right)+ i e\right] \nn \\
& & + \left(T^3+i \right)\left[-a S + b \left( -U^1 - U^2 + U^3 \right)+ i e\right] \;.
\label{wa}
\eea
The vacuum solution now takes the form \cite{Camara:2005dc}
\bea
t &=& \frac{\sqrt{3}5^{\frac16}}{2^{\frac53} m^{\frac13}} \left(e_0 + 3 e + 2 m \right)^{\frac13} \;,\;\;\;\; v= -1 - \frac{1}{2^{\frac53} 5^{\frac13}m^{\frac13}} \left(e_0 + 3 e + 2 m \right)^{\frac13}  \nn \\
u &=& \frac{3\sqrt{3}m^{\frac13}}{2^{\frac73}5^{\frac16} b} \left(e_0 + 3 e + 2 m \right)^{\frac23} \;, \;\; \nu = \frac{1}{10 b} \left( 10 e - 10 a \sigma + 10 m + 2^{\frac23} 5^{\frac13}m^{\frac13} \left(e_0 + 3 e + 2 m \right)^{\frac23} \right) \nn \\
s &=& \frac{\sqrt{3}m^{\frac13}}{2^{\frac73}5^{\frac16} a} \left(e_0 + 3 e + 2 m \right)^{\frac23} \;.
\label{susysol}
\eea
One can then check that this four-dimensional solution now solves exactly all the ten-dimensional equations in section \ref{sec:10diiaflux}. An important difference from the solution in section \ref{sec:iiat6flux} is that the number of perturbatively massless axions is now 1 not 3.

\begin{figure}[t]
\centering
 \includegraphics[width=0.9\textwidth]{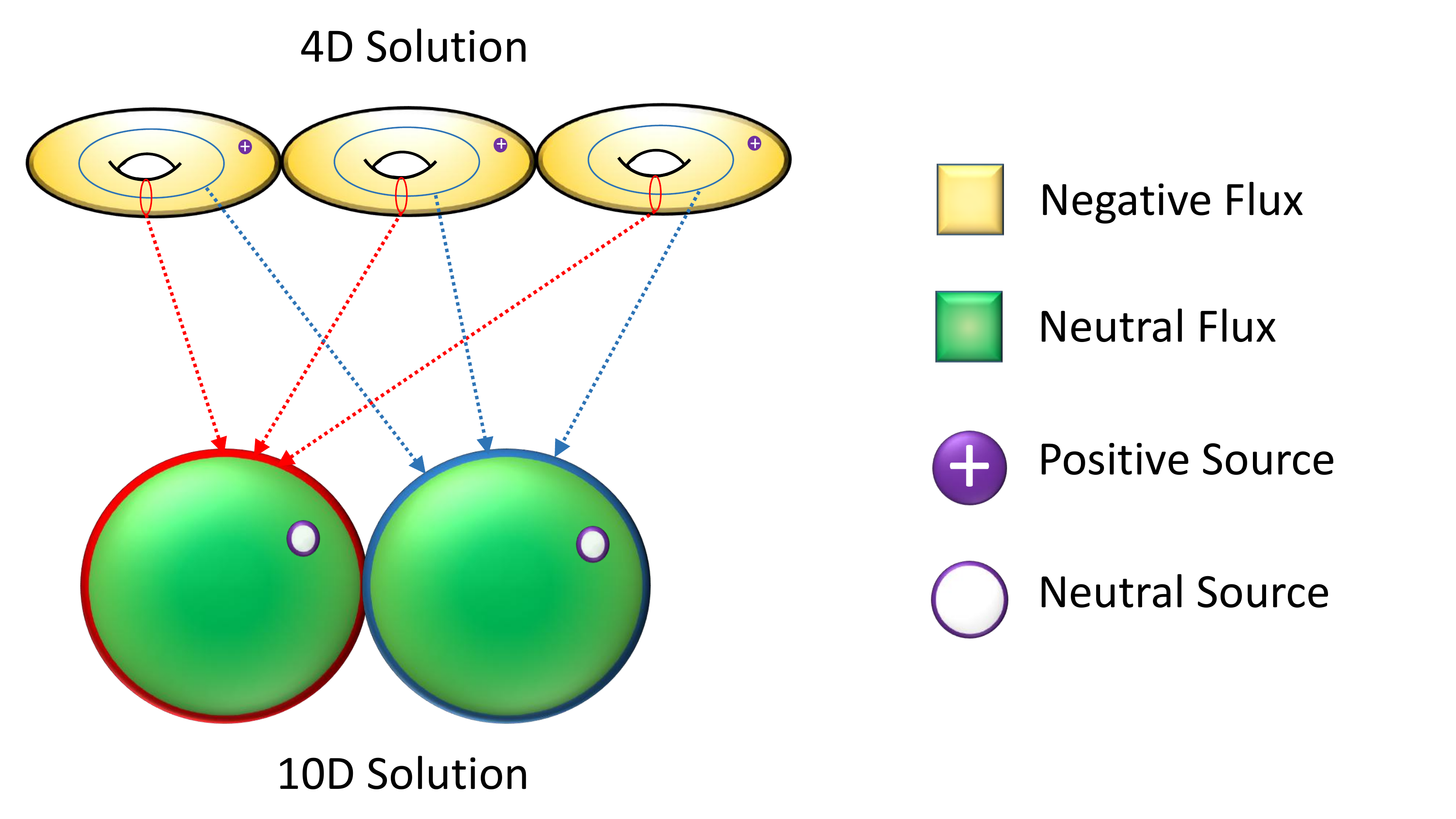}
\caption{Figure illustrating a 4-dimensional vacuum for $T^2 \times T^2 \times T^2$ with flux and charged localized sources cancelling tadpoles, and a 10-dimensional solution on $S^3 \times S^3$ where the charges cancel locally in the Bianchi identity. The solutions are topologically different, but can be mapped to each other with each $S^3$ composed of circles on the $T^2$s that are then cut and twisted together. This geometric map corresponds to having the same fields in the effective theory but adding a deformation to the superpotential. In particular, this deformation changes the number of perturbatively massless axions from three to one, corresponding to the topology change.}
\label{fig:tt6s3}
\end{figure}

We have therefore obtain two related solutions, in section \ref{sec:iiat6flux} a solution on $T^6$ obtain in an effective four-dimensional theory (\ref{solt6flux}), which does not uplift to an exact ten-dimensional solution. While in this section we obtained a solution in an effective four-dimensional theory (\ref{susysol}), which does uplift to an exact ten-dimensional one. The two are related in the four-dimensional effective theory by a superpotential deformation, the last term in (\ref{wt6dj}), and by different four-dimensional tadpole constraints.\footnote{These are the type of changes associated to fluxes, and indeed the relation in the type IIB dual is such that the $S^3\times S^3$ is dual to $T^2 \times T^2 \times T^2$ with H-flux turned on \cite{Kachru:2002sk,Camara:2005dc}.} In the ten-dimensional geometry the two solutions are related by a map from $T^2 \times T^2 \times T^2 \rightarrow S^3 \times S^3$ which changes the topology. This is illustrated in figure \ref{fig:tt6s3}. 

It is useful to compare some coarse features which differ between the 4-dimensional solution in section \ref{sec:iiat6flux} and the ten-dimensional solution in this section. First the number of perturbatively massless axions has changed. Axions are topologically classified and so the fact that they changed is a four-dimensional probe of the ten-dimensional topology difference. Second, the four-dimensional solution had the property that a parametric separation of scales between the anti de-Sitter radius and the Kaluza-Klein radius can be achieved \cite{DeWolfe:2005uu}. This means it could be treated as effectively four-dimensional. On the other hand, the ten-dimensional solution has no such parametric separation. It is therefore not truly four-dimensional but rather can be thought of as an effective theory for a consistent truncation of modes. 


It is worth noting that the $S^3 \times S^3$ solution is special in that it is a compactification to four dimensions where all the moduli are stabilized and there is an exact ten-dimensional uplift. There are a few other such compactifications in type IIA. The first example was studied in \cite{House:2005yc}, which was the coset $\frac{SU(3)}{U(1)\times U(1)}$. Other type IIA cosets were studied in \cite{Caviezel:2008ik,Koerber:2008rx,Cassani:2009ck}, and in \cite{Grana:2006kf} more general group manifolds were studied. In \cite{Micu:2006ey} a similar procedure was performed for M-theory compactifications on the cosets $\frac{SO(5)}{SO(3)}$ and $\frac{SU(3)\times U(1)}{U(1)\times U(1)}$.

\subsection{Type IIB Compactification on Calabi-Yau orientifolds}
\label{sec:typeiibor}

In section \ref{sec:simpleiia} we considered compactifications of type IIA string theory on simple manifolds. A much larger class, of Ricci flat manifolds are Calabi-Yau manifolds. See \cite{Greene:1996cy,Hubsch:1992nu} for an introduction. Moduli stabilization in compactifications on Calabi-Yau manifolds of type IIA in the presence of fluxes were studied in \cite{DeWolfe:2005uu}, and in a qualitatively different setup in \cite{Palti:2008mg}. The effective action for compactifications of type IIA on Calabi-Yau orientifolds was derived in \cite{Louis:2002ny,Grimm:2004ua}, and the general expressions (\ref{gen10ea}) and (\ref{fluxwnodj}) continue to hold. However, the number of moduli fields can vary significantly. More precisely, the number of K{\"a}hler moduli is counted by the number of two-cycles in the Calabi-Yau which corresponds to the Hodge number $h^{(1,1)}$, and the number of complex-structure moduli is counted by the number of three-cycles, or more precisely by the Hodge number $h^{(2,1)}$. Further, the  functional form of the $\Omega$ three-form can be very complicated. As in the simple $T^6$ case, Calabi-Yau manifold compactifications in type IIA do not solve the ten-dimensional equations of motion, and a similar issue with the local Bianchi identity arises \cite{Acharya:2006ne}. We therefore expect that the qualitative behaviour of such solutions is not completely different to the simple cases studied in section \ref{sec:simpleiia}. 

While the type IIA picture allows for simple scenarios of moduli stabilization, by far the most attention has been given to type IIB compactifications. This is because while they are more complicated, and arguably under less good control, they exhibit a rich spectrum of physically interesting possibilities. Compactifications of type IIB on Calabi-Yau manifolds will form the focus of this section. There exist a number of good reviews on such compactifications, see in particular \cite{Douglas:2006es}, and so we will be relatively brief. 

\subsubsection{The 10-dimensional solution with fluxes}
\label{sec:iib10d}

There is no currently known ten-dimensional solution analogous to the one in section \ref{sec:10diiaflux} for type IIB compactifications which exhibits complete moduli stabilization for some manifolds. Instead, there is a solution for which the K{\"a}hler moduli are classically flat directions, but the complex-structure moduli are fixed. This is the famous Giddings-Kachru-Polchinski (GKP) solution \cite{Giddings:2001yu}.\footnote{See \cite{Taylor:1999ii,Dasgupta:1999ss,Curio:2000sc} for some predecessors.} The solution involves background fluxes for the field strengths $F_5$, $F_3$ and $H_3$. The metric takes the forms of a warped product
\be
ds^2 = e^{2A} \eta_{\mu\nu} dx^{\mu} dx^{\nu} + e^{-2A}\tilde{g}_{mn} dy^m dy^n \;,
\ee
where the $x^{\mu}$ are external coordinates, while the $y^n$ are internal ones. The scalar function $A$ is called a warp factor and is a function of the internal coordinates. In general, the solution will involve dilaton gradients, and so is best understood in the context of F-theory \cite{Vafa:1996xn}, or more precisely in terms of the M-theory solution \cite{Becker:1996gj}. However, in the weak coupling type IIB limit, it was shown that the internal manifold is conformally Calabi-Yau, which means the metric only differs from a Calabi-Yau metric by the warp factor. This is an extremely powerful result because we know how to construct a huge range of Calabi-Yau manifolds. It sits in contrast to the ten-dimensional solution for type IIA in section \ref{sec:10diiaflux} for which Calabi-Yau manifolds are not a solution. 

A crucial part in the solution is played by the axio-dilaton 
\be
S = C_0 + i e^{-\phi} \;,
\ee
where $\phi$ is the ten-dimensional dilaton. Which combines with the three-form fluxes into 
\be
G_3 \equiv F_3 - S H_3 \;.
\ee
The ten-dimensional solution is such that it requires $G_3$ to satisfy an imaginary-self-dual (ISD) condition
\be
\label{isd}
\star G_3 = i G_3 \;. 
\ee
Here the Hodge star is associated to the internal manifold metric. The ISD condition (\ref{isd}) implies that if the flux background is fixed, then there is a constraint on the metric. This constraint is the ten-dimensional realization of moduli stabilization of the complex-structure moduli. 

\begin{figure}[t]
\centering
 \includegraphics[width=0.9\textwidth]{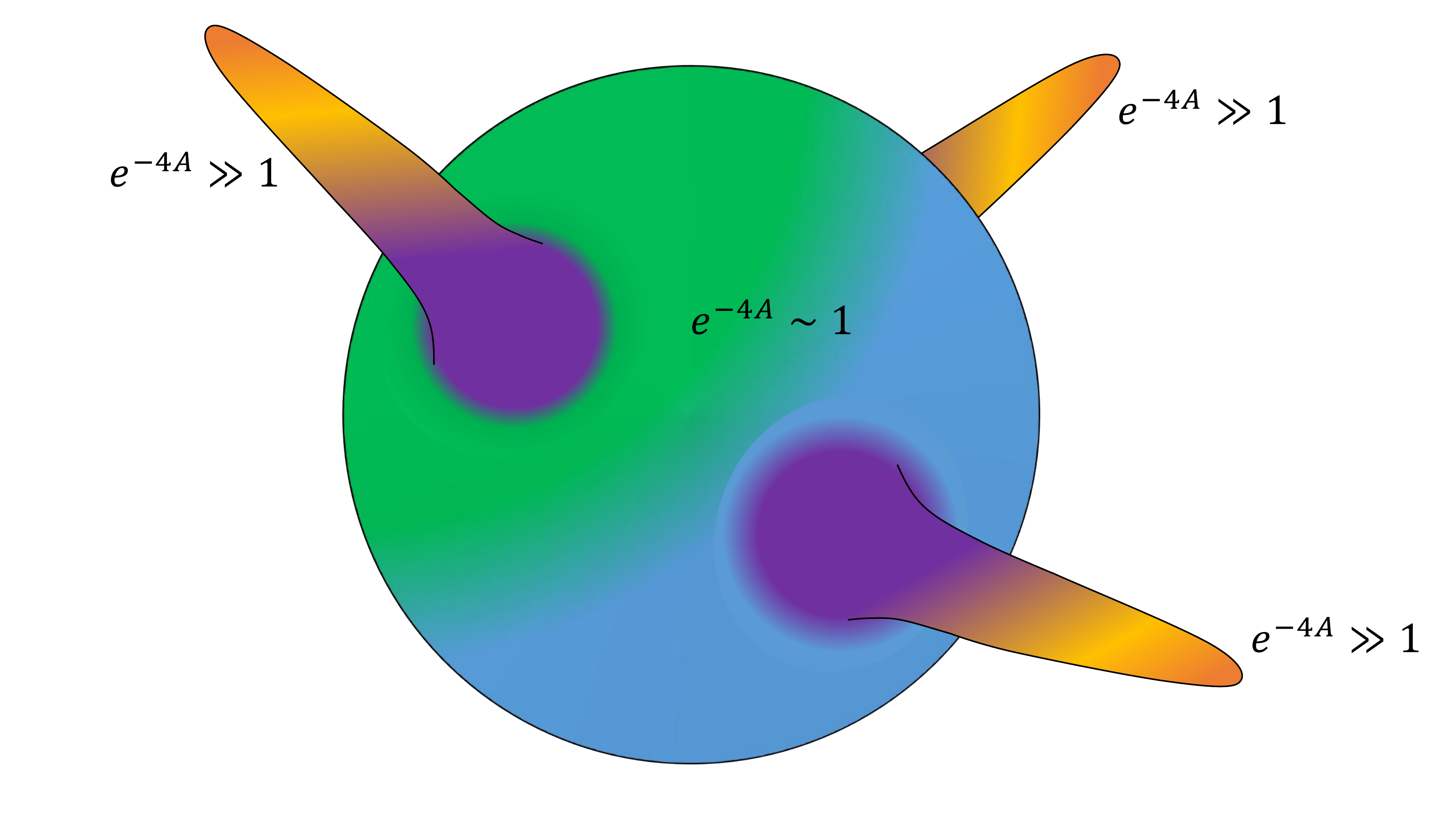}
\caption{Figure illustrating a warped Calabi-Yau with flux in type IIB string theory. If the fluxes are chosen such that a three-cycle becomes exponentially small then the Calabi-Yau develops a long warped throat. All energy scales are warped down in the bottom of the throat, including the string scale. Since there are many three-cycles in the Calabi-Yau, there may be many such throats of different lengths.}
\label{fig:wcyt}
\end{figure}

Since the Ricci-flat metric on a Calabi-Yau manifold is not known explicitly, the GKP solution similarly is not explicit. However, in certain limits in the Calabi-Yau complex-structure moduli space, specifically the conifold limit where a three-cycle shrinks to small size, there does exist a local solution for which the metric is explicitly known. This is the Kelbanov-Strassler (KS) solution \cite{Klebanov:2000hb}. More precisely, the KS solution is a non-compact solution of type IIB supergravity, which is then expected to flow into the compact Calabi-Yau GKP solution. Since the three-cycle size corresponds to a complex-structure modulus, it is determined by fluxes through the ISD condition (\ref{isd}). A particularly interesting feature of this is that the fluxes can fix the cycle size to be exponentially small in the flux.  Denoting the cycle size by $z$, the $F_3$ flux through it by $M$, and the $H_3$ flux through the symplectic dual three-cycle by $-K$, on finds \cite{Giddings:2001yu}
\be
z \sim e^{-\frac{2\pi K}{g_s M}} \;, 
\ee
where $g_s=e^{\phi}$ is the string coupling. The KS solution is then such that the warp factor is proportional to the three-cycle size, and this means that it is also exponential in the fluxes in the local region near the small three-cycle
\be
e^{-4A} \sim e^{\frac{8\pi K}{g_s M}} \;.
\ee
This means that the region in the Calabi-Yau near the small three-cycle becomes strongly warped and stretches out into a type of throat. Since the warp factor effects the four-dimensional metric, it implies that all energy scales are warped down along with it. This is a string theory version of the scenario \cite{Randall:1999ee}, and means that at the bottom of the KS throat scales such as the string scale and Kaluza-Klein scale become exponentially small, see for example \cite{Burgess:2006mn} for an analysis of this. This is illustrated in figure \ref{fig:wcyt}.  

\subsubsection{Four-dimensional effective theory}
\label{sec:4dnoscale}

We consider now the four-dimensional effective theory from flux compactifications in IIB which is valid in the bulk of the Calabi-Yau, so away from the strongly warped throats. We will therefore neglect the warping, and refer to \cite{Giddings:2005ff}, and follow-up work, for studies of its effects on the effective theory. The K{\"a}hler potential and superpotential take the general form \cite{Gukov:1999ya,Giddings:2001yu,Grimm:2004uq}
\bea
\label{iibkw}
K &=& -2\log \left[ \int_{\cM_6} J \wedge J \wedge J \right] - \log \left[ i \int_{\cM_6} \Omega \wedge \overline{\Omega} \right] - \log\left[ -i\left( S - \overline{S} \right) \right] \nn \\  
W &=& \int_{\cM_6} G_3 \wedge \Omega \;.
\eea
The K{\"a}hler moduli superfields $T_i$ take the form
\be
T_i = \tau_i + i \nu_i  \;,
\ee
where $\tau_i$ are volumes of four-cycles $\cC^4_i$, as measured by $\int_{\cC^4_i} J \wedge J$, while the $\nu_i$ arise from reduction of the Ramond-Ramond $C_4$ on the same four-cycles. They are counted by the Hodge number $h^{(2,2)}$. The complex-structure moduli fields are denoted $z_a$ and come from the expansion of the holomorphic three-form $\Omega$, they are counted by the Hodge number $h^{(2,1)}$. We refer to \cite{Grimm:2004uq,Jockers:2004yj,Grimm:2005fa,Conlon:2006gv} for more details on the type IIB effective four dimensional theory.  

The ten-dimensional solution discussed in section \ref{sec:iib10d} was such that the complex-structure of the Calabi-Yau is fixed through the imaginary-self-dual condition (\ref{isd}), while the K{\"a}hler moduli are classically unfixed amounting to an overall rescaling symmetry of the solution. This is manifest in the four-dimensional theory because the superpotential (\ref{iibkw}) only depends on the complex-structure moduli. Consider the supergravity scalar potential (\ref{gVfromKW}), because of the split of the K{\"a}hler potential between the different moduli (\ref{iibkw}), the sum over the supergravity fields splits into three contributions
\be
V = e^K \left[ \underbrace{g^{a\bar{b}} D_a W \overline{D_b W}}_{\mathrm{complex-structure\;moduli}} +  \underbrace{g^{S\bar{S}} D_S W \overline{D_S W}}_{\mathrm{dilaton}}  +  \underbrace{g^{i\bar{j}} D_i W \overline{D_j W}}_{\mathrm{Kahler\;moduli}} - 3 \left|W\right|^2 \right] \;.
\ee
The K{\"a}hler moduli do not appear in the superpotential, and further they satisfy an identity \cite{Grimm:2004uq,Douglas:2006es} $g^{ij}K_i K_j =3$, which means that the scalar potential takes the so-called no-scale form
\be
V = e^K \left[ g^{a\bar{b}} D_a W \overline{D_b W} + g^{S\bar{S}} D_S W \overline{D_S W} \right] \;.
\ee
The potential is positive definite, which means that the solution to four-dimensional flat space, as in section  \ref{sec:iib10d}, should satisfy 
\be
\label{susyiibcs}
D_aW = D_SW = 0 \;.
\ee
In the vacuum therefore, the K{\"a}hler moduli completely drop out of the potential and are classically flat directions. It is also possible to check that the equations (\ref{susyiibcs}) match onto the imaginary-self-dual condition (\ref{isd}). We therefore find that the four-dimensional effective theory matches at the classical level the ten-dimensional solution quite well away from the warped regions, not as precisely or as explicitly as the one studied in section \ref{sec:iias3s3}, but this is expected given its vastly more complicated nature.

Note that the equations (\ref{susyiibcs}) are the supersymmetry equations for the complex-structure moduli and dilaton. The supersymmetry equations for the K{\"a}hler moduli are $\left(D_i K\right) W=0$. Therefore we see that whether supersymmetry is broken or preserved depends on the value of the superpotential in the vacuum, as fixed by (\ref{susyiibcs})
\be
\label{w0}
W_0 = \left< W\right> \;.
\ee
If $W_0 = 0$ then the vacuum is supersymmetric, otherwise supersymmetry is broken. 

Since the K{\"a}hler moduli are flat directions at leading order, their potential is dominated by the sub-leading corrections. A clear analysis of these is presented in \cite{Conlon:2005ki}. The corrections come from quantum effects, controlled by the string coupling, and higher derivative terms in the supergravity action. The K{\"a}hler potential receives all types of corrections, but the superpotential only receives non-perturbative quantum corrections. Therefore, we can write 
\be
K = K_{\mathrm{tree}}  + K_{\mathrm{pert}} \;, \;\; W =  W_{\mathrm{tree}}  + W_{\mathrm{non-pert}} \;.
\ee
Now inserting this into the supergravity formula for the scalar potential, we find
\be
V = V_{\mathrm{tree}} + W_0^2 K_{\mathrm{pert}} + W_0 W_{\mathrm{non-pert}} + ... \;. 
\ee
We therefore observe that there are two possibilities 
\bea
\label{w0kklt}
W_0 &\ll& \frac{W_{\mathrm{non-pert}}}{K_{\mathrm{pert}}} \;, \\
W_0 &\gg& \frac{W_{\mathrm{non-pert}}}{K_{\mathrm{pert}}} \;.
\label{w0lvs}
\eea
The first case (\ref{w0lvs}) leads to K{\"a}hler moduli stabilisation within the Kachru-Kallosh-Linde-Trivedi (KKLT) scenario \cite{Kachru:2003aw}. The second case (\ref{w0lvs}) leads to K{\"a}hler moduli stabilisation within the Large Volume Scenario (LVS) \cite{Balasubramanian:2005zx}. 

In any moduli stabilisation scenario the non-perturbative superpotential for K{\"a}hler moduli plays a crucial role. The K{\"a}hler moduli superfields appear in the superpotential as
\be
W = W_G\left(z\right) + \sum_M A_M\left(z\right) e^{-2 \pi a^i_M T_i} \;.
\ee
Here $W_G\left(z\right)$ corresponds to the superpotential involving the complex-structure moduli (\ref{iibkw}). The sum over the index $M$ accounts for the various possible non-perturbative contributions to the superpotential. The exponential form in the K{\"a}hler moduli superfields is fixed due to the required unbroken discrete shift symmetry of the axionic component of the superfield $\nu_i$ coming form the Ramond-Ramond four-form. The $a^i_M$ are constant factors in the exponent whose form depends on the effect leading to the potential. There are two such possibilities, one is coming from instanton effects. These are D3 branes that are wrapping the four-cycles associated to the K{\"a}hler moduli \cite{Witten:1996bn}. In that case the $a^i_M$ are the wrapping numbers of the instanton on the various four-cycles in the Calabi-Yau. See \cite{Blumenhagen:2009qh} for a review of instanton effects in type II string theory. The other possible effect is gaugino condensation on D7 branes that are wrapping the associated four-cycles leading to a non-Abelian gauge theory in four dimensions. In that case the non-perturbative potential is the Affleck-Dine-Seiberg superpotential \cite{Affleck:1983mk}, and the $a$ associated to it would be the $N_c^{-1}$ where $N_c$ is the number of D7 branes wrapping the four-cycle (or equivalently the rank of the $SU(N_c)$ gauge group). The pre-factors $A_M\left(z\right)$ will in general depend on the complex-structure moduli. Calculating them is very difficult as they are related to certain one-loop calculations. 

Because the K{\"a}hler moduli are only fixed by quantum effects, it is often the case that they are treated within an effective theory where the complex-structure moduli have already been fixed by the potential (\ref{iibkw}) and are integrated out leaving an effective constant contribution to the superpotential (\ref{w0}). In that case one works with an effective K{\"a}hler and superpotential of the form
\bea
\label{kkltwef}
K _{\mathrm{eff}} &=& - 3\log \left(T + \overline{T}  \right) \;, \nn \\
W_{\mathrm{eff}} &=& W_0 + Ae^{-2\pi a T} \;.
\eea
The K{\"a}hler potential is just the restriction of (\ref{iibkw}) where we consider a single K{\"a}hler modulus for simplicity. In the KKLT scenario we then have a requirement that $W_0$ be sufficiently small such that $W_0 \sim Ae^{-2\pi a T}$ while still maintaining the geometric regime $\mathrm{Re\;}T \gg 1$. More precisely, in \cite{Denef:2004dm} it was argued that one requires
\be
\left( \frac{\log \left|W_0\right|^{-1}}{2\pi k} \right) \gg 1\;,
\ee
where $k$ is a constant parameter which increases with the number of K{\"a}hler moduli (at least as large as $h^{(1,1)}$). 
That means that we require an exponentially small $W_0$. Since the superpotential (\ref{iibkw}) contains quantized fluxes, such a small $W_0$ can only be generated by some very tuned cancelling between many numbers which are of order one. While it seems difficult to achieve such a tuning, the idea is that because there may be a very large number of flux choices, two fluxes $F$ and $H$ for each three-cycle class in the Calabi-Yau, there will exist some flux choice which will allow for this. In practice, this is rather difficult to show explicitly. In \cite{Denef:2004dm} an explicit flux choice was shown to lead to $W_0 \sim 10^{-2}$. However, it was also argued that statistically one expects sufficiently small $W_0$ to be possible given enough complex-structure moduli, although they are unlikely to be fixed near the large complex-structure limit, which means calculating this explicitly is non-trivial.\footnote{See \cite{Blumenhagen:2016bfp,Wolf:2017wmu,Blumenhagen:2019qcg} for alternative suggestions on how to construct exponentially small $W_0$ near the conifold locus.}

If we do not want to assume the existence of a sufficiently tuned $W_0$, then the possibility (\ref{w0lvs}) is more natural and one then is lead to the Large Volume Scenario \cite{Balasubramanian:2005zx}. If we consider working in the effective theory where the complex-structure moduli and dilaton are integrated out, then the Large Volume Scenario is based on an effective theory with
\bea
\label{lvswef}
K _{\mathrm{eff}} &=&  -2 \log \left[ \left(T_b + \overline{T}_b  \right)^{\frac32} - \left(T_s + \overline{T}_s  \right)^{\frac32} + \frac{\xi}{2g_s^{\frac32}} \right] \;, \nn \\
W_{\mathrm{eff}} &=& W_0 + A_b e^{-2\pi a_b T_b} + A_s e^{-2\pi a_s T_s} \;.
\eea
It is a theory with two K{\"a}hler moduli in superfields $T_b$ and $T_s$, and the volume of the Calabi-Yau takes a particular form such that the resulting K{\"a}hler potential is as in (\ref{lvswef}).
The last  term in the K{\"a}hler potential comes from (a supersymmetric completion of) the $R^4$ term in the ten-dimensional theory \cite{Becker:2002nn}. The constant $\xi$ is of order one and is fixed by the Euler number of the Calabi-Yau. After fixing the axionic direction $\nu_s$, the scalar potential for this effective theory takes the form \cite{Balasubramanian:2005zx}
\be
\label{lvssp}
V = \frac{\lambda \sqrt{\tau_s}e^{-4\pi a_s \tau_s}}{\cV} - \frac{\mu\tau_s e^{-2\pi a_s \tau_s}}{\cV^2} + \frac{\nu}{\cV^3} \;.
\ee
where $\cV \sim \tau_b^{\frac32} - \tau_s^{\frac32}$ is the Calabi-Yau volume, and $\lambda$, $\mu$ and $\nu$ are order one constants. It was then shown that this potential has a minimum which leads to a non-supersymmetric anti de-Sitter vacuum where the K{\"a}hler moduli are fixed as
\bea
\label{lvsm}
\tau_s \sim \left(\frac{4 \nu \lambda}{\mu^2} \right)^{\frac23} \;,\;\; \cV \sim \frac{\mu}{2\lambda}  \left(\frac{4 \nu \lambda}{\mu^2} \right)^{\frac13} e^{2\pi a_s  \left(\frac{4 \nu \lambda}{\mu^2} \right)^{\frac23}}\;.
\eea
The striking point being that the Calabi-Yau volume is exponentially large, leading to the name Large Volume Scenario. 
Note that there are a number of variants on this basic scenario, see for example \cite{Cicoli:2008va}. 

The Large Volume Scenario avoids the fine tuning requirement on $W_0$ of KKLT. This comes at the price of being less generic, in that it cannot be applied to every Calabi-Yau. A more difficult issue is that it relies fundamentally on the leading perturbative corrections to the scalar potential taking the form as in (\ref{lvssp}). In particular, in (\ref{lvssp}) there are no terms which behave as $\cV^{-1}$ or $\cV^{-2}$ without also the exponential factor in $\tau_s$. Since $\cV$ is exponentially large, these terms would dominate over the last term in (\ref{lvssp}) and may substantially modify the result. In the presence of ${\cal N}=2$ supersymmetry, it is known that such corrections are absent \cite{Becker:2002nn}, and this motivates the original proposal. In a general non-supersymmetric, or ${\cal N}=1$ setting, there is no local symmetry forbidding such terms. It therefore remains an open problem to show that these terms are absent in general. However, no such dangerous leading terms have been found in string theory so far, despite extensive attempts to understand them quantitatively, see for example \cite{Berg:2005ja,Berg:2007wt,Cicoli:2007xp,Cicoli:2008va,Conlon:2010ji,GarciaEtxebarria:2012zm,Pedro:2013qga,Grimm:2013gma,Berg:2014ama,Minasian:2015bxa,Grimm:2013bha,Grimm:2014xva,Grimm:2014efa,Grimm:2015mua,Grimm:2017okk,Grimm:2017pid,Haack:2018ufg}. It therefore may be that such leading terms are forbidden, or sufficiently suppressed, in string theory due to some as yet unknown reason, or simply that they have not been calculated due to the technical difficulty of the problem.

The Large Volume Scenario is also a good setting to exhibit an important property of type IIB compactifications relative to the type IIA scenarios which can be lifted to ten-dimensional solutions as studied in section \ref{sec:iias3s3}. It is possible to parametrically separate the anti de-Sitter scale from the Kaluza-Klein. The Kaluza-Klein radius behaves with the volume as $\cV^{\frac16}$, while the potential (\ref{lvssp}) goes as $\cV^{-3}$ and so leads to an anti de-Sitter radius of $\cV^{\frac32}$. 

This completes the very brief summary of moduli stabilisation in type IIB, although we will return to the issue of de Sitter vacua in section \ref{sec:testsdS}. There are a number of other possible moduli stabilisation scenarios, but we will not discuss them here and refer to the cited reviews. It is clear that the account was much less explicit than the type IIA case in section \ref{sec:simpleiia}. This is a manifestation of the general statement that moduli stabilisation in type IIB string theory is much more complicated than in type IIA. On the other hand, the type IIB setting offers a richer set of possible scenarios. It is worth noting however that it is also possible to realize in type IIA KKLT and LVS type scenarios \cite{Palti:2008mg}. The implications for studies of the Swampland are that the richer settings in type IIB allow for a wider range of tests of the Swampland. This will be particularly central to the discussion regarding de Sitter vacua. This comes at the cost of having a less clear connection to the underlying string theory, and so introduces the recurring question of whether these settings are really probing the nature of string theory or string-motivated effective quantum field theories. 

A significant step towards understanding better these vacua would be to uplift them to higher-dimensional solutions. No such uplifts are known, and attempts at doing so is a rather difficult field which we will not review in detail here. There are some interesting general results we should mention. Consider the supersymmetric anti-de Sitter KKLT vacuum. If we place a spacetime-filling D3 brane at some point in the internal extra dimensions it will feel a potential. This can be understood in a number of ways, see for example \cite{Baumann:2006th,Koerber:2007xk,Marchesano:2009rz}. The fact that D3 branes feel a potential can be used to deduce some general statements about the internal manifold, for example it cannot be Calabi-Yau. Rather, it should be a manifold with $SU(3)\times SU(3)$ structure group \cite{Koerber:2007xk}. Ideally, the solutions would actually be uplifted to an F-theory, or more precisely an M-theory setup, in the same sense that \cite{Becker:1996gj} is an uplift to M-theory of the solution of \cite{Giddings:2001yu}. One can use the existence of a D3 potential to constrain properties of such solutions. In the M-theory setup this would imply that spacetime-filling M2 branes would feel a potential in compactifications of M-theory to three dimensions. The consequences of this requirement were studied in \cite{Condeescu:2013hya}.

Finally, we note that moduli stabilisation in the Heterotic string may be possible on manifolds which are not Calabi-Yau. However, there is no analogue of the Calabi-Yau with fluxes compactification, and the group manifold constructions are difficult to uplift to ten-dimensional solutions. See \cite{Lukas:2015kca,Klaput:2012vv,Anderson:2011cza} for the cutting edge in Heterotic moduli stabilisation. 

\subsubsection{The Racetrack Scenario}
\label{sec:reacetrack}

The racetrack moduli stabilisation scheme \cite{KRASNIKOV198737,Dixon:1990ds,raceross,Kaplunovsky:1997cy,deCarlos:1992kox} is a variation on the KKLT setup (\ref{kkltwef}), but where one considers $W_0=0$, and two or more non-perturbative effects. It is much older than the KKLT proposal, and is also a somewhat universal mechanism, in that it can be used in any of the string theories (or M-theory). The simplest racetrack scenario, for a single modulus $S$, considers a superpotential with two non-perturbative effects
\be
W = A e^{-\frac{2\pi S}{N}} + B e^{-\frac{2\pi S}{M}} \;.
\label{racetrackW}
\ee
This superpotential can, for example, be induced by gaugino condensation of an $SU(N)\times SU(M)$ gauge theory.
It has a supersymmetric minimum at
\be
S = \frac{N M}{M-N} \log \left( -\frac{M B}{N A} \right) \;.
\ee
The idea is then to consider $N M \gg \left(M-N\right)$ so that the modulus is fixed in a perturbative regime where other corrections may be sufficiently controlled. 

The racetrack scenario is powerful due to its versatility, but is rather difficult to realize explicitly and in a controlled way. It is almost completely four-dimensional in nature and its ten-dimensional uplift is therefore even more difficult to establish than the KKLT setting. Note that it is possible also to consider a combination of the racetrack and KKLT scenarios, see for example \cite{Kallosh:2004yh}.

\subsection{The Weak Gravity Conjecture in type II string theory}
\label{sec:wgciist}

Most of the work on testing the various versions of the Weak Gravity Conjecture has been performed in the type II string theory setting. In fact, most of the work has been on understanding the axionic version of the Weak Gravity Conjecture (\ref{wgcax}). We will discuss this in section \ref{sec:wgcax}, but first we consider the $U(1)$ version. In type II string theories there are two types of $U(1)$ symmetries, those coming from closed-strings and those supported on D-branes, which we will refer to as open-string $U(1)$s. At strong coupling, like in M-theory or F-theory, this distinction becomes less clear but for perturbative string theories it is a useful split. 

\subsubsection{Closed-string $U(1)$s}
\label{sec:clstru1}

The closed-string $U(1)$s which come from the NS-NS sector, so the metric and Kalb-Ramond field, lead to similar physics to that studied in sections \ref{sec:fewgcst} and \ref{sec:wgchet}. The states charged under them are Kaluza-Klein and Winding modes, or in the context of the Heterotic string also oscillator modes. In type II string theory there are also $U(1)$s which come from the RR sector. This means they are associated to the anti-symmetric $C^{(p)}$-forms in table \ref{tab:stringthe}, dimensionally reduced using the appropriate internal forms to yield gauge fields in four dimensions. The charged particles under these $U(1)$s are non-perturbative states in string theory, specifically D-branes wrapping cycles. Note that this shows that the Weak Gravity Conjecture is really a statement about non-perturbative quantum gravity, it is violated in perturbative string theory. Studies of the Weak Gravity Conjecture in the context of closed-string $U(1)$s were made in \cite{Hebecker:2015zss,Heidenreich:2016jrl,Montero:2017yja,Grimm:2018ohb,Lee:2018urn,Lee:2018spm,Vittmann,Grimm:2018cpv,Lee:2019tst}. Note that some of these can also be considered studies of open-string $U(1)$s, as considered in section \ref{sec:wgciios}, because at strong coupling the distinction becomes rather blurred. In keeping with the theme of this review, we will study some simple explicit example cases as an illustration of general principles. 

Let us consider the simple case of $\frac{T^6}{\mathbb{Z}_2 \times \mathbb{Z}_2}$ as studied in section \ref{sec:iiaont6}. Like in that section, we will not consider necessarily orbifolding by the $\mathbb{Z}_2 \times \mathbb{Z}_2$ symmetry, but just utilize it as a consistent truncation of the theory to states respecting it. For now we will also not impose the orientifold projection. 

In the type IIA case we can consider the RR forms $C^{(1)}$ and $C^{(3)}$. The former gives rise to a $U(1)$ upon restricting the index to four dimensions, while the latter leads to multiple $U(1)$s from an expansion in the two-forms (\ref{omdef}), so we obtain four $U(1)$ gauge fields
\be
\label{iiaas}
C^{(1)} = A^0  \;,\;\; C^{(3)} = A^i  \wedge \omega_i \;. 
\ee
To obtain the gauge couplings of the $U(1)$s we can dimensionally reduce the ten-dimensional action, including the kinetic terms for the RR fields, this yields (see, for example \cite{Klaewer}) 
\bea
S^{\mathrm{4D}}_{\mathrm{IIA}} &=&  \int \sqrt{-g} \left[ \frac{R}{2} - \sum_{i=1}^3 \frac14 \left(\partial \log t^i \right)^2  - \sum_{i=1}^3 \frac{1}{4\left(t^i\right)^2} \left(\partial v^i\right)^2  \right. \nn \\ 
& & \left. - \frac{t^1 t^2 t^3}{4} \left|F^0\right|^2 - \frac{ t^2 t^3}{4 t^1} \left|F^1\right|^2 - \frac{ t^1 t^3}{4 t^2} \left|F^2\right|^2 - \frac{t^1 t^2}{4 t^3} \left|F^3\right|^2  \right] \;.
\label{actiia4dgauge}
\eea
Here we have taken the moduli fields as defined in (\ref{def10JO}), and only focused on the K{\"a}hler moduli and axions. The field strengths are defined as $F^0 = dA^0$ and $F^i = dA^i$, with Hodge duals $\tilde{F}^i$. We have set the axion expectation values to vanish for simplicity, $v^i=0$. 

Note that, as in previous cases, the gauge couplings are exponential in the canonically normalised fields. The charged states under $A^0$ are D6-branes wrapping the whole torus, while under $A^i$ they are D4 branes wrapping the cycles $\cC_i$ defined as $\int_{\cC_i} \omega_j = \delta_{ij}$. This can be seen by reducing the DBI and CS action components for Dp-branes (see \cite{Johnson:2003gi} for example)
\bea
S_{DBI} &=& -T_p \int_{\Sigma_{p+1}} d^{p+1}\xi \sqrt{-\mathrm{det} \left(G+ B + 2 \pi \alpha'\cF \right)} \;, \nn \\
S_{CS} &=& - T_p \int_{\Sigma_{p+1}} \left(\sum_k C^{(k)} \right) \mathrm{Exp} \left[ 2 \pi \alpha' \cF + B\right]\;, \label{dbranecs}
\eea
where $\Sigma_{p+1}$ is the world-volume of the D-brane, $G$ is the pullback of the metric to the brane, $B$ the pullback of the Kalb-Ramond form, and $\cF$ is any world-volume gauge flux. The results can be readily seen, the DBI term yields the volume of the cycle wrapped by the D-brane, and going to the Einstein frame yields a further volume factor $\left(t^1 t^2 t^3 \right)^{-\frac12}$, so in Planck units the mass is given by 
\bea
m^2_{D6} &=& t^1 t^2 t^3 \;, \;\; m^2_{D4_1} = \frac{t^2 t^3}{t^1}   \;, \;\; m^2_{D4_2} = \frac{t^1 t^3}{t^2} \;, \;\; m^2_{D4_3} = \frac{t^1 t^2}{t^3} \;, \nn \\
m^2_{D0} &=& \frac{1}{t^1 t^2 t^3} \;, \;\; m^2_{D2_1} = \frac{t^1}{t^2 t^3}   \;, \;\; m^2_{D2_2} = \frac{t^2}{t^1 t^3} \;, \;\; m^2_{D2_3} = \frac{t^3}{t^1 t^2} \;. 
\label{iiadmass}
\eea
We can therefore identify the electric and magnetic states and see that they indeed satisfy the Weak Gravity Conjecture.\footnote{Note that since we have dilatonic couplings we should utilize the appropriately modified Weak Gravity Conjecture as in section \ref{sec:wgcsca}.} In fact, the wrapped branes are actually BPS states, and so saturate the inequality. 

It is interesting to note that making the torus cycles large $t^i \rightarrow \infty$, goes to weak coupling. Then, perhaps surprisingly, the states which become light are D2 branes which are wrapping the large cycles. This shows that it is crucial to always go to the Einstein frame when evaluating the Weak Gravity Conjecture. Then we can understand that while the intuition that D2 branes wrapping large cycles should be massive is correct, also gravity becomes weaker for larger extra dimensions, and it does so at a faster rate than the D2 brane states become massive. So effectively, keeping the Planck mass constant, the wrapped D2 branes become light.

The expressions for the masses (\ref{iiadmass}) are true for single wrapped branes. But they actually set the mass scale for a tower of states which are composed of bound states of branes. We can focus on the electric D0 and D2 states. Bound states of multiple D0 branes are expected to exist in type IIA string theory, indeed they should form the Kaluza-Klein modes for the uplift to M-theory \cite{Witten:1995ex}. It is possible to also argue for them more precisely (see for example \cite{Johnson:2003gi}) by giving such a bound state momentum along an adjacent circle in the torus, and then T-dualising along that circle. The D0-D0 bound state then becomes a D1-F1 bound state, which are strings known to exist in type IIB string theory. The D2-D2 bound states can then be argued for by T-dualising the D0-D0 ones along two directions. D2-D0 bound states are particularly interesting. We can consider a single D2-brane and $n$ D0-branes. Now looking at the Chern-Simons action for D-branes (\ref{dbranecs}) we note that  in the presence of non-vanishing world-volume flux a D2 brane couples to $C^{(1)}$, which is the field sourced by D0 branes. This is a manifestation of the fact that a D2-D0 bound state is non-supersymmetric and undergoes a transition where the D0 dissolves into one unit of world-volume flux on the D2 brane. So a bound state with $n$ D0 branes corresponds to a D2 brane with $n$ units of worldvolume flux. We therefore see that indeed there are infinite towers of states starting at the Weak Gravity Conjecture scale, along the lines discussed in section \ref{sec:wgctower} and in \cite{Heidenreich:2015nta,Klaewer:2016kiy,Heidenreich:2016aqi,Andriolo:2018lvp,Grimm:2018ohb}.

The setting we considered is toroidal, which allows for a simple identification of the appropriate wrapped D-brane states. In more complicated settings, especially preserving ${\cal N} \leq 2$ supersymmetry, identifying a tower of states is very difficult in general. Already for ${\cal N}=2$ the spectrum of half-BPS states in the theory (supersymmetric wrapped branes) depends on the position in moduli space. However, we will utilize a useful and general technique in section \ref{sec:dcwith8} that will allow to identify such a tower even for complicated geometries \cite{Grimm:2018ohb}. The physics of it will actually play an important role throughout much of this review and so it is worth discussing in this simple setting first. In the action (\ref{actiia4dgauge}) the fields $v^i$, coming from reduction of the Kalb-Ramond form (\ref{omdef}), are classically flat directions. At the quantum level they are lifted by worldsheet instanton corrections to the K{\"a}hler potential, and so are actually periodic axions in the large volume regime $t^i \gg 1$. Now consider the D-brane actions (\ref{dbranecs}). We see that if we shift the axions by a period, the $B$-field shifts, but the gauge invariant combination which appears is $2 \pi \alpha' \cF + B$.  This means that the physics of a discrete $B$-field shift is the same as that of turning on worldvolume flux $\cF$, which is the same as inducing D0 charge on a D2 brane. Then there is a special tower of states, namely one D2 brane with $n$ dissolved D0 branes, whose charges are generated by axion shifts from a single state, the un-fluxed D2 brane. We can denote set of charges in this tower a {\it monodromy orbit} \cite{Grimm:2018ohb}.\footnote{Actually, as discussed in \cite{Grimm:2018ohb}, the argument holds even for multiple D2 branes, and states associated to non-Abelian bundles on them, because the axion shifts are always tensoring the bundle by a line-bundle which preserves the stability of the bundle at large volume.} The important point is that the monodromy structure generalizes to complicated geometries, essentially since axions are topological in nature, and so the tower associated to monodromy orbits will be accessible in general. Indeed, type IIA on an arbitrary Calabi-Yau manifold, rather than a torus, was studied in \cite{Corvilain:2018lgw} using these techniques. 

Note that there are also other types of towers of states associated with the Weak Gravity Conjecture scale, specifically Kaluza-Klein modes. For a homogenous torus $t^1=t^2=t^3=\frac{R^2}{\alpha'}$ their mass behaves as $m_{KK} \sim \frac{1}{\left(R M_s\right)^3}\frac{1}{\left(R M_s\right)}$ (where we utilized (\ref{MsMprel})). Comparing with (\ref{iiadmass}) we see that the Kaluza-Klein tower is actually lighter than the wrapped D-branes. This is common in the Swampland, typically there are multiple towers of states becoming light at weak couplings or large distances. However, the Kaluza-Klein modes are not charged under the $U(1)$ symmetries, and so are less relevant from the perspective of the Weak Gravity Conjecture.

Let us consider the simple deformations of the toroidal setting discussed in section \ref{sec:simpleiia}. First we note that imposing the orientifold projection (\ref{sigoac}) actually projects out the gauge fields. This is not a general property of type IIA orientifold compactifications, but due to the fact that there are no orientifold even two-forms in this particular toroidal setting. Let us therefore not impose the orientifold projection, and instead consider the map $T^2 \times T^2 \times T^2 \rightarrow S^3 \times S^3$, as in section \ref{sec:iias3s3}. In the $S^3 \times S^3$ theory, there are no even-dimensional cycles for the D-branes to wrap. There are still two-forms, to reduce the $C^{(3)}$ field to give gauge fields, and so it might appear that there is a puzzle from the perspective of the Weak Gravity Conjecture. However, without the orientifold projection there are also new axion fields coming from the expansion of $C^{(3)}$ in the three-forms
\be
C^{(3)} \supset \theta_0 \beta^0 + \theta_i \beta^i  \;.
\ee
From the differential relations (\ref{diff1}) we see that the 10-dimensional kinetic term for $C^{(3)}$ takes the form
\be
d_{10} \left( \theta_0 \beta^0 + \theta_i \beta^i  + A^i \wedge \omega_i \right) \;, 
\label{d10iiaa0}
\ee
where $d_{10}$ denotes the 10-dimensional exterior derivative. By splitting the exterior derivative into its four-dimensional and internal parts $d_{10}= \partial_4 + d$, we can write (\ref{d10iiaa0}) as
\bea
& &\left(\partial_4 \theta_0 + a A^1 + a A^2 + a A^3 \right) \beta^0 \; \nn \\
&+& \left(\partial_4 \theta_1 + b A^1 - b A^2 - b A^3 \right) \beta^1 \; \nn \\ 
&+& \left(\partial_4 \theta_2 - b A^1 + b A^2 - b A^3 \right) \beta^2 \; \nn \\
&+& \left(\partial_4 \theta_3 - b A^1 - b A^2 + b A^3 \right) \beta^3 \;.    
\eea
These kinetic terms are actually St\"uckelberg mass terms for the three gauge fields. Three of the $\theta$ axions are eaten to give a mass to the gauge fields, while the last remaining axion is associated with one linear combination of the two $S^3$s. We therefore see that the map $T^2 \times T^2 \times T^2 \rightarrow S^3 \times S^3$ removed the wrapped D-brane states but also at the same time broke the $U(1)$ gauge symmetries, retaining consistency with the Weak Gravity Conjecture. In fact, this St\"uckelberg breaking could have led to a remnant discrete symmetry, in which case one could have identified appropriate states charged under this discrete gauge symmetry. This is a simple example of a general relation in string theory between geometry, gauge fields and charged states, which has been studied in great detail, see \cite{Grimm:2010ez,Camara:2011jg,Grimm:2011tb,BerasaluceGonzalez:2011wy,BerasaluceGonzalez:2012zn,Berasaluce-Gonzalez:2013bba,Mayrhofer:2014haa,Mayrhofer:2014laa,Cvetic:2015moa,Kimura:2016crs} for example. It is worth noting that there is a mirror version of this story in type IIB associated with turning on H-flux. 

We will return to the Weak Gravity Conjecture for closed-string $U(1)$s in section \ref{sec:dcwith8} in the context of the Swampland Distance Conjecture with ${\cal N}=2$ supersymmetry. 

\subsubsection{Open-string $U(1)$s}
\label{sec:wgciios}

Recall that we denote the $U(1)$ gauge symmetries that are supported on D-branes as open-string $U(1)$s. This is because the objects charged under them are open strings ending on the D-branes. Consider a compactification of type IIA/B string theory on $\mathbb{R}^{1,(9-n)}\times T^{n}$, where the torus is taken to have all equal radii circles with radius $R$. Then a $U(1)$ gauge symmetry can be induced by a Dp-brane with $p \geq 9-n$ which is filling the non-compact spacetime. Consider two D-branes, labelled $A$ and $B$, then a fundamental open string stretching between them has charge $+1$ under the $U(1)_A$ and $-1$ under $U(1)_B$. See \cite{Blumenhagen:2013fgp,Johnson:2003gi,Marchesano:2007de,Maharana:2012tu} for a review on spectra of strings on branes. The first way we could try to violate the electric Weak Gravity Conjecture (\ref{ewgc}) is to take a single spacetime filling D-brane. In that case a fundamental string with both ends ending on the brane will only lead to a $U(1)$ neutral state. However, such a setting is actually not consistent because D-branes source R-R fields, and the brane therefore acts as a positive charge in the compact $T^n$ without any negative charges to act as sinks for the R-R fields. One way to solve this is to add an anti-brane, however this will lead to an unstable configuration where the branes will annihilate against each other into closed strings. A stable way to introduce negative charges is through orientifold planes. However, then there will be unoriented strings in the spectrum, which can be thought of as stretching from the D-brane to its image under the orientifold involution, with charge $2$ under the $U(1)$. 

\begin{figure}[t]
\centering
 \includegraphics[width=0.9\textwidth]{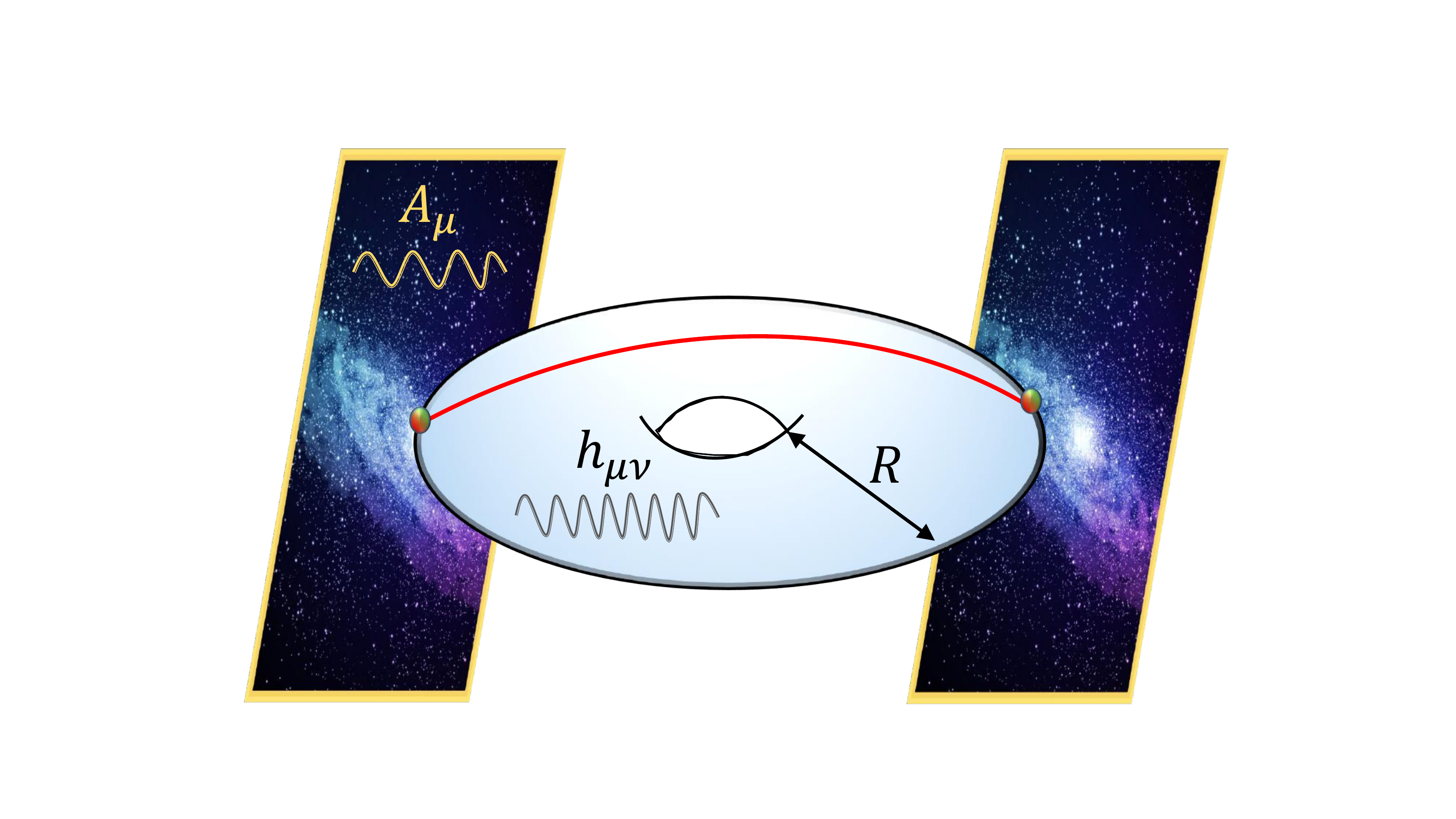}
\caption{Figure illustrating the Weak Gravity Conjecture for $U(1)$ gauge symmetries realized on spacetime filling branes. The mass of the charged state is bounded by the maximum separation distance between the branes. Increasing the radius of the extra dimensions allows this mass to increase as measured in string units. However, gravity propagates in the extra dimension and so becomes weaker as they increase in size. The mass of the particle in Planck units therefore actually decreases, and the Weak Gravity Conjecture is satisfied for any radius larger than the string scale.}
\label{fig:brsep}
\end{figure}

We can attempt to violate the Weak Gravity Conjecture by moving the D-brane away from the orientifold, the string would then stretch over some spatial distance making it massive, while its charge would remain constant. However, the distance of separation is limited by the radius of the compact extra dimensions $R$ and so the maximum mass for the string state is $m \sim \left( R M_s \right) M_s$. On the other hand, making the extra dimensions larger reduces the strength of gravity. It can be checked that this effect always counters the increased mass of the string as long as $n>2$ \cite{ArkaniHamed:2006dz}. We will return to the interesting $n \leq 2$ cases below, but first let us look at the four-dimensional spacetime example, so $n=6$. In that case the spacetime filling brane is a D3 brane and its gauge coupling is $g = \sqrt{g_s}$, where $g_s$ is the string coupling. The relation between the string scale and the Planck scale can be most easily deduced by examining the Ricci scalar pre-factor in direct dimensional reduction of the action (\ref{steff}) with $D=10$ 
\be
\label{MsMprel}
M_s^{2} g_s^{-2} \left(R M_s\right)^6 \sim M_p^2\;.
\ee
The stretched string mass is then
\be
m^2 \sim \left(R M_s\right)^2 M_s^2 \sim \frac{g_s^2 M_p^2 }{\left(R M_s\right)^4} \;.
\ee
The Weak Gravity Conjecture ratio is therefore
\be
\frac{m^2}{g^2 M_p^2} \sim \frac{g_s}{ \left(R M_s\right)^4} \;.
\ee
The controlled weak-coupling geometric regime is $g_s \ll 1$ and $RM_s \gg 1$, and therefore we see that the Weak Gravity Conjecture is satisfied. The general physics of this is illustrated in figure \ref{fig:brsep}. 

The case of $n=2$ is such that the Weak Gravity Conjecture bound is independent of $R$ at leading order, and therefore checks would need to rely on calculating the order one factors. This is particularly difficult because a co-dimension two object will induce a backreaction on the metric which does not die off but increases logarithmically, and this would have to be accounted for in such a calculation. The case $n=1$ is interesting since it is the only one where the dependence on $R$ is such that for large radius the Weak Gravity Conjecture is naively violated. The backreaction on the space is extremely strong though, and so one should consider the full solution. In \cite{ArkaniHamed:2006dz} a rough analysis of this was done by considering a D3 brane as sourcing an $AdS_5 \times X$ where $X$ is some internal space. This is matching onto the case of a stack of $N$ D3 branes leading to $AdS_5 \times S^5$. Then the Weak Gravity Conjecture implies that the AdS radius $L_{AdS}$ must not be larger than the radius of the internal space $R_X$. This is an interesting, since in section \ref{sec:iias3s3} we saw that in compactifications of string theory to $AdS_4$ which are upliftable to local 10-dimensional solution indeed there is no such separation of scales. The same is true for $AdS_5 \times S^5$ compactifications. We will return to this point in section \ref{sec:scsepcon}.

We saw that the electric Weak Gravity Conjecture (\ref{ewgc}) can be argued to hold quite generally for open-string $U(1)$s. More difficult to see is how the tower of states associated with the magnetic Weak Gravity Conjecture scale (\ref{mwgc}) manifests for open-string $U(1)$s. In the heterotic string case studied in section \ref{sec:wgchet}, the string oscillator modes had increasing charges with respect to the $U(1)$. This is not the case for fundamental strings in type II where the oscillator modes have the same charge under the D-brane $U(1)$s. Instead, the appropriate states should be non-perturbative in nature. This matches the duality picture since type I or II - Heterotic duality is a strong-weak coupling duality. One of the most powerful ways to treat non-perturbative type II string theory is through its uplift to M-theory (for type IIA) and F-theory (for type IIB). We refer to \cite{Denef:2008wq,Weigand:2010wm,Weigand:2018rez} for some reviews on this uplift. Indeed, under such an uplift the open-string $U(1)$s are seen to originate from the M-theory three-form $C^{(3)}$, with charged objects being wrapped M2-branes. This unifies them with the closed-string $U(1)$s studied in section \ref{sec:clstru1}, and a similar analysis of wrapped branes can be made.   

\begin{figure}[t]
\centering
 \includegraphics[width=0.9\textwidth]{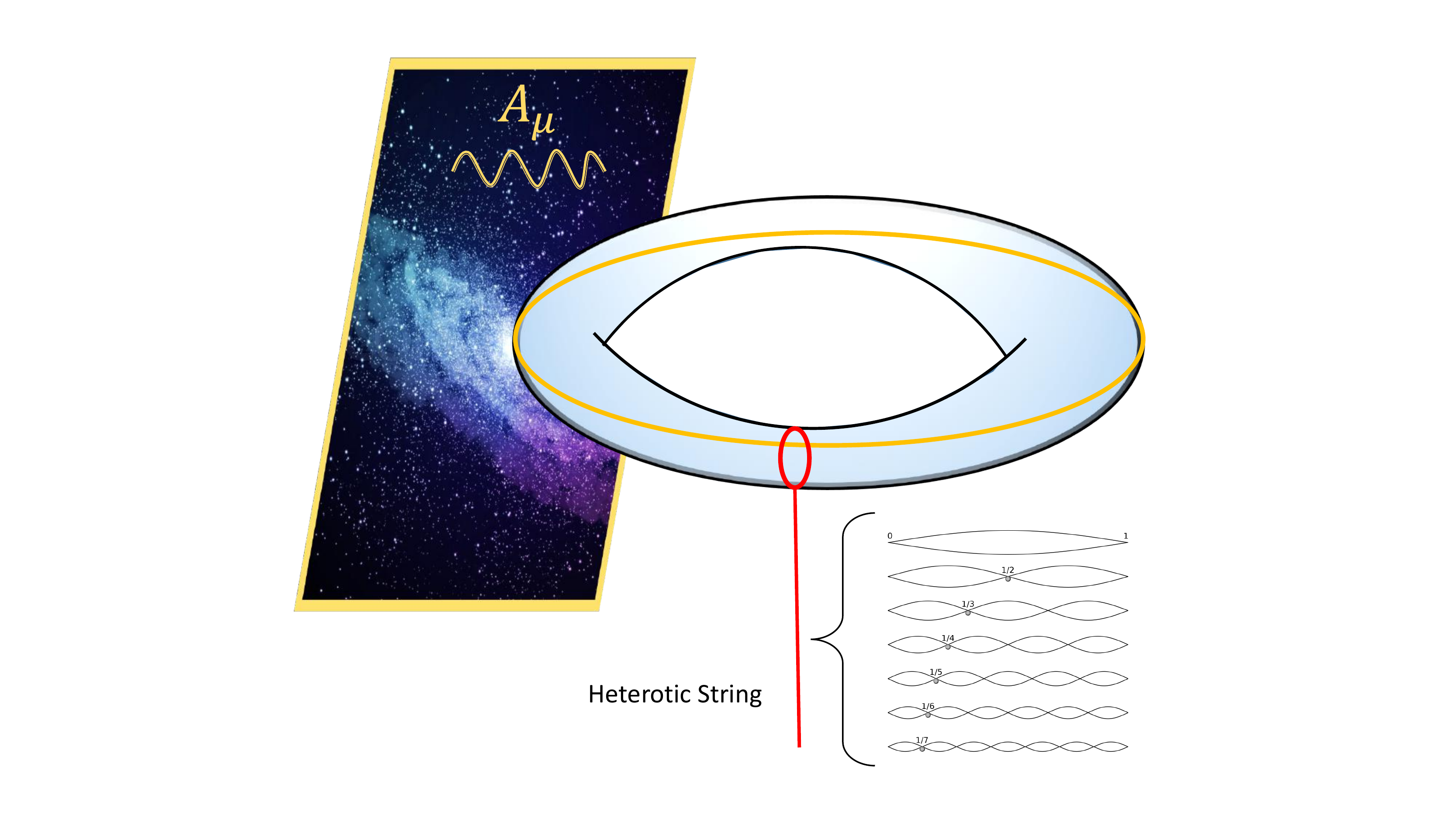}
\caption{Figure illustrating the tower of charged states for a $U(1)$ on a D7-brane in type IIB as studied in \cite{Lee:2018urn,Lee:2018spm,Lee:2019tst}. Going to weak gauge coupling implies making a cycle large, which for fixed extra-dimensional volume means a different cycle becomes small. A D3-brane wrapping this small cycle gives rise to a light string, which is actually the fundamental Heterotic string. The tower of charged states are then the oscillator modes of this string.}
\label{fig:ftheory}
\end{figure}

Such an analysis was performed in the context of the Weak Gravity Conjecture in \cite{Lee:2018urn,Lee:2018spm,Lee:2019tst}. It was shown that indeed at weak coupling $g \rightarrow 0$, and infinite tower of states becomes light with a characteristic mass scale of $m \sim g$. The core idea of the physics is as follows. Consider a D7-brane in type IIB string theory filling six-dimensional spacetime, and wrapping a 2-cycle in the extra dimensions. The D7 gauge coupling depends on the size of the cycle that it is wrapping, and going to weak coupling means making the cycle large. However, in considering the Weak Gravity Conjecture, we would like to work in Planck units which for a fixed string scale means keeping the volume of the extra dimensions constant. This means that some other 2-cycle in the extra dimensions has to become small at the same time. Now one can consider a D3 brane wrapping this small cycle, and this would then give rise to an effective string whose tension is controlled by the 2-cycle volume, and so is light. Uplifting this configuration to F-theory, the dilaton of type IIB is geometrized into an additional 2-cycle, and the wrapped D3 brane is uplifted to a wrapped M5-brane (in the M-theory dual). The light string is then nothing but the Heterotic string, and the tower of charged states due to the Heterotic oscillators is then mapped this way to type IIB. This scenario is illustrated in figure \ref{fig:ftheory}. In \cite{Lee:2019tst} the four-dimensional version of this analysis was performed. A notable point is that in the six-dimensional case the sub-Lattice Weak Gravity Conjecture, see section \ref{sec:wgctower}, was satisfied by the tower of states, but in four dimensions it was found that the tower of charged states did not form a lattice.

While most of the explicit tests of the Weak Gravity Conjecture in string theory have been performed with some supersymmetry preserved at the string or Kaluza-Klein scale, some settings with a high scale of supersymmetry breaking were studied in the context of type I string theory in \cite{Bonnefoy:2018mqb}. There it was argued that the Weak Gravity Conjecture may be violated at tree-level, but loop corrections take the form such that it is respected at the quantum level provided the string coupling is not too small.

\subsubsection{Closed-string Axions}
\label{sec:wgcax}

Most work on the Weak Gravity Conjecture in string theory has been performed in the context of the axionic version discussed in section \ref{sec:wgcpform}. This is mostly due to the fact that this is the relevant version of the Weak Gravity Conjecture in relation to inflation, specifically in the context of natural inflation \cite{Freese:1990rb}. Let us consider the simple setup of type IIA string theory on $T^6$ studied in section \ref{sec:iiaont6} as a starting point. The bosonic components of the superfields are given by (\ref{compsupiia}). The K{\"a}hler potential (\ref{kpott6}) does not involve the imaginary parts of the superfields, and so since the superpotential is vanishing, these fields enjoy a classical shift symmetry. This symmetry is broken at the quantum level by non-perturbative effects. These can be either instantons or gaugino condensation on stacks of spacetime-filling branes, we will focus on the former. The relevant instantons for $\sigma$ and the $\nu^i$ are D2 branes wrapping three-cycles in the $T^6$. While for the $v^i$ they are string worldsheet instantons. If ${\cal N} \geq 2$ supersymmetry is preserved, then the instantons will correct only the K{\"a}hler potential, which means the fields will remain as classical flat directions, but without a shift symmetry. In the case when only ${\cal N}=1$ supersymmetry is preserved, say by orbifolding and orientifolding, the instantons can contribute to the superpotential. The fields will then develop a periodic potential, and so are denoted axions. The instanton action is just the worldvolume action of the D2 brane, and the contribution to the superpotential is fixed by holomorphy to be of the type
\be
W \supset A_{S} e^{-2\pi S} + \sum_i A_{U_i} e^{-2 \pi U_i} \;.
\label{Wschemins}
\ee
It is important to note that not all instantons contribute to the superpotential, so some of the terms in (\ref{Wschemins}) could be missing. They are expected though to contribute corrections to the K{\"a}hler potential. Indeed, the worldsheet instantons associated to the K{\"a}hler axions $v^i$ do not contribute to the superpotential in the absence of fluxes. The question of which instantons contribute to the superpotential is a subtle one relating to the zero-mode structure, and is discussed in detail in \cite{Blumenhagen:2009qh}. The $A$ prefactors in (\ref{Wschemins}) are in general functions of the K{\"a}hler superfields $T^i$. 

Let us consider the axion $\sigma$, superpartner to the dilaton $s$. Comparing the K{\"a}hler potential (\ref{kpott6}) with the general expression (\ref{actaxin}), we can read off the axion decay constant
\be
2\pi f_{\sigma} = \frac{1}{2s} = \frac{g_s}{2 \left( \frac{R_1}{\sqrt{\alpha'}} \right) \left( \frac{R_2}{\sqrt{\alpha'}}  \right) \left( \frac{R_3}{\sqrt{\alpha'}}  \right)} \;.
\label{tt6fsig}
\ee
We can then consider the axion Weak Gravity Conjecture (\ref{wgcax}), which states
\be
2 \pi f_{\sigma} s = \frac{1}{2} \leq 1 \;,
\ee
and so is obviously satisfied. The same follows trivially for the other axion decay constants. Because the axionic Weak Gravity Conjecture is satisfied, the general discussion in section \ref{sec:wgcpform} regarding the break down of the instanton expansion if $f_{\sigma} > 1$ applies here.  However, in the explicit string theory realization we can say more. From (\ref{tt6fsig}), it is manifest that in the weakly-coupled $g_s \ll 1$ and geometric $R_i M_s \gg 1$ regime, we find $f_{\sigma} \ll 1$. Therefore, there is obstruction to reaching $f_{\sigma} > 1$ in a controlled regime already at the perturbative level, so without considering the non-perturbative instanton effects. 

\begin{figure}[t]
\centering
 \includegraphics[width=0.9\textwidth]{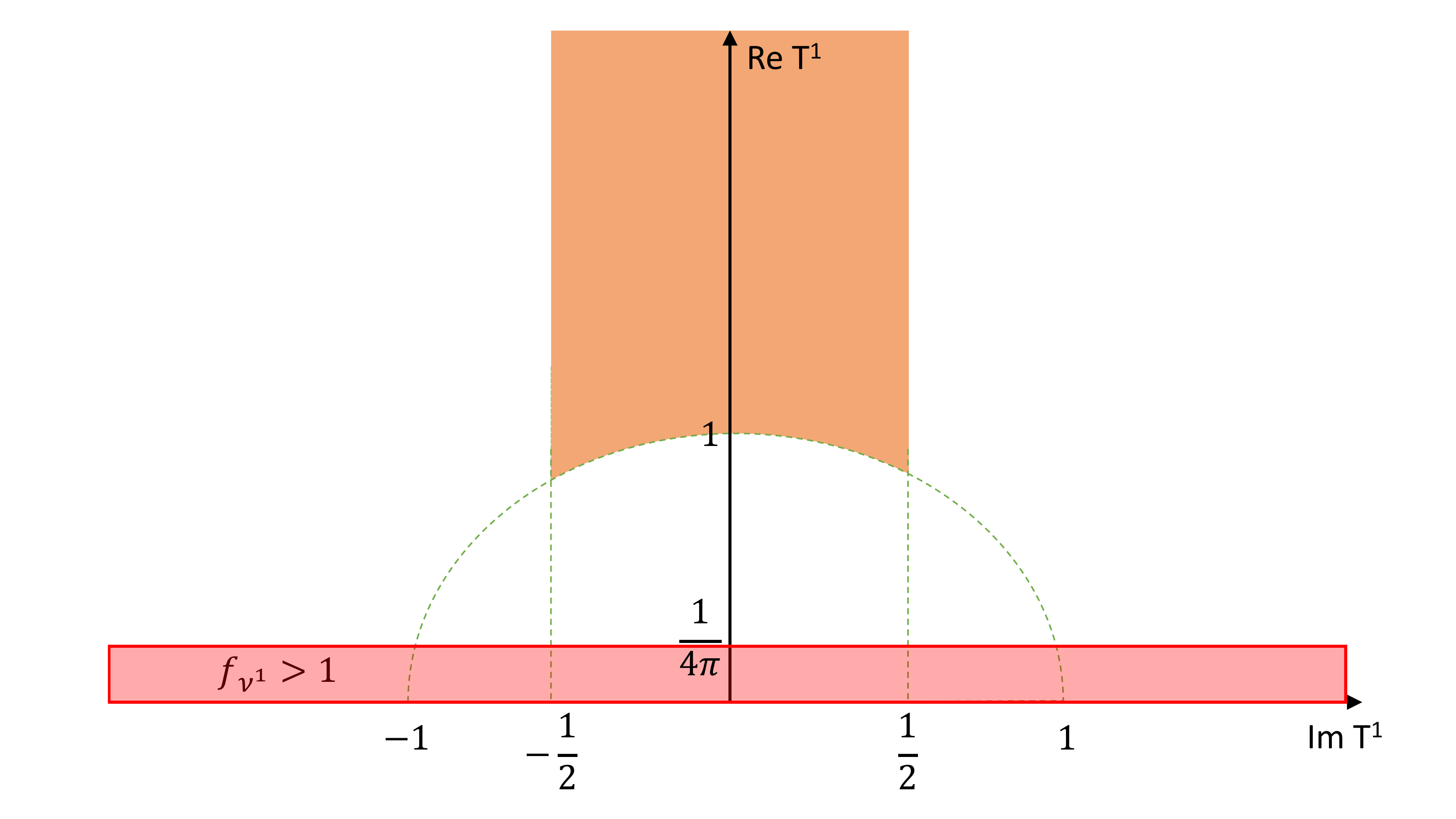}
\caption{Figure showing the fundamental domain for the field $T^1$ controlling the axion decay constant, under the $SL\left(2,\mathbb{Z}\right)$ duality group. The region $f_{v^1} >1$ corresponds to $\mathrm{Re\;} T^1 < \frac{1}{4 \pi}$, which is outside the fundamental domain. }
\label{fig:iifundom}
\end{figure}

Let us consider the axions $v^i$, they have decay constants of type
\be
2 \pi f_{v^1} = \frac{1}{2 t^1} = \frac{1}{2 \left( \frac{R_1}{\sqrt{\alpha'}}  \right) \left( \frac{R_4}{\sqrt{\alpha'}}  \right)} \;.
\ee
Again we have in the geometric regime $f_{v^1} \ll 1$. However, if we are in the pure $T^6$ setting we may actually consider going to the $R_i M_s \ll 1$ regime in string theory. In that case the instanton expansion breaks down. But the four-dimensional effective theory should be invariant under duality groups of the internal dimensions, and in particular T-duality along the $R_1$ and $R_4$ directions send $t^1 \rightarrow \frac{1}{t^1}$. Therefore, we should recover the same physics as the geometric regime. More generally, T-duality combines with axions shifts of the $v^i$ to form an $SL\left(2,\mathbb{Z}\right)$ duality group acting as
\be
T^1 \rightarrow \frac{aT^1 - i b}{i cT^1 + d} \;,\;\; \left\{a,b,c,d \right\} \	\in \mathbb{Z} \;,\;\; ad - bc =1 \;.
\ee
The physically inequivalent set of value of $T^1$ is given by the fundamental domain, as illustrated in figure \ref{fig:iifundom}. 
We see that $f_{v^1} > 1$ is actually outside of the fundamental domain, and so is simply not a physically relevant region of parameter space.\footnote{It is interesting to note that we actually have $f_{v^1} \leq \frac{1}{4\pi}$, which is an order of magnitude stronger than what is required by the Weak Gravity Conjecture. This stronger constraint can be understood as the requirement that the axion periodicity $P$ should be bounded by $P < 1$, while the constraint $f < 1$ would only imply $P < 2 \pi$. This nicely connects to the Refined Swampland Distance Conjecture (\ref{rsdc}).} The same is true for the other axion decay constants, assuming the purely toroidal setting. While the duality group is special to the toroidal setting, the restriction of $f < 1$ in the weak-coupling and geometric regimes of string theory is universal. This was first studied systematically in \cite{Banks:2003sx}. And then explored much more widely and in greater detail, see for example \cite{Svrcek:2006yi,Ben-Dayan:2014lca,Bachlechner:2014hsa,Long:2014dta,Long:2014fba,Bachlechner:2014gfa,Rudelius:2015xta,Montero:2015ofa,Bachlechner:2015qja,Shiu:2015uva,Ruehle:2015afa,Hebecker:2015rya,Brown:2015lia,Retolaza:2015sta,Peloso:2015dsa,Junghans:2015hba,Heidenreich:2015wga,Palti:2015xra,Bachlechner:2015cgq, Heidenreich:2015nta,Kooner:2015rza,Kappl:2015esy,Furuuchi:2015jfj,Choi:2015aem,Ibanez:2015fcv,Hebecker:2015zss,Conlon:2016aea,Retolaza:2016bpn,Heidenreich:2016jrl,Blumenhagen:2016bfp,Garcia-Valdecasas:2016voz,Heidenreich:2016aqi,Hebecker:2016dsw,Hebecker:2017wsu,Hebecker:2017uix,Montero:2017yja,Bachlechner:2017zpb,Bachlechner:2017hsj,Blumenhagen:2018hsh,Agrawal:2018mkd,Shiu:2018wzf,Hebecker:2018ofv,Hertog:2018kbz,Hebecker:2018yxs}. All of the results are consistent with the absence of a closed-string fundamental axion decay constant that is super-Planckian.\footnote{In \cite{Kooner:2015rza} it was argued that one may be able to obtain and open-string super-Planckian decay constant in warped throats in type IIB compactifications.} However, we now go on to discuss the more complicated possibility of inducing an effective large axion decay constant.  

\subsubsection*{Axion alignment from fluxes}

In section \ref{sec:wgcpform} we discussed the idea of inducing an effective large axion decay constant from aligning two axions that individually have sub-Planckian decay constants. Let us consider this in string theory. We will first consider the simplest and most explicit realizations in type IIA string theory, as originally studied in \cite{Palti:2015xra,Baume:2016psm} (see \cite{Hebecker:2018fln,Goswami:2018pvk} for more recent work). We start with the scenario of type IIA on $T^6$ with flux, as discussed in section \ref{sec:iiat6flux}. However, to simplify the setup we will consider projecting the three complex-structure superfields to a single one $U^i \rightarrow U$, and similarly for the K{\"a}hler moduli $T^i \rightarrow T$. This can be done through a further $\mathbb{Z}_3$ identification of the three tori, or more generally we can consider a Calabi-Yau compactification with a single complex-structure modulus and an arbitrary K{\"a}hler moduli sector since it will not influence our discussion \cite{Palti:2015xra}. We therefore have the four-dimensional effective theory
\bea
K &=& - \log s - 3 \log u + K\left(T + \overline{T}\right) \;,  \\  \nn
W &=&  W_T\left(T\right) + i h_0 S - i h_1 U \;.
\label{simaxaliiaa}
\eea
Here $W_T$ is the part of the superpotential which is independent of $S$ and the $U$.
We have two axions $\sigma=\mathrm{Im\;} S$ and $\nu=\mathrm{Im\;} U$, and one combination of them obtains a perturbative mass from the fluxes in the superpotential. The remaining combination remains classically massless, but quantum non-perturbative effects, as in (\ref{Wschemins}), will induce a periodic potential for it. The interesting point is that the fluxes $h_0$ and $h_1$ allow for control over how the light axion is embedded in the two-dimensional fundamental axion field space. They therefore allow for a string theory realization of the alignment scenario, discussed in section \ref{sec:wgcpform}, through fluxes.\footnote{Axion alignment through fluxes in string theory was first studied in \cite{Hebecker:2015rya} in the type IIB context.} The decay constant of the effective classically massless axion $\psi$ in each of the two instantons, denoted by $f_{\psi}^s$ and $f_{\psi}^u$, is \cite{Palti:2015xra}
\be
f_{\psi}^u = \left(f^2_{\nu} + \left(\frac{h_1}{h_0} f_{\sigma}\right)^2 \right)^{\frac12} \;,\;\; f_{\psi}^s =  \left(f^2_{\sigma} + \left(\frac{h_0}{h_1} f_{\nu}\right)^2 \right)^{\frac12}  \;.
\label{fsfuiia}
\ee
We therefore see that by taking, say $h_1 \gg h_0$ we can have $f_{\psi}^u \gg f_{\sigma}$, realizing a parametric enhancement of the effective axion decay constant relative to a fundamental one. 

\begin{figure}[t]
\centering
 \includegraphics[width=0.9\textwidth]{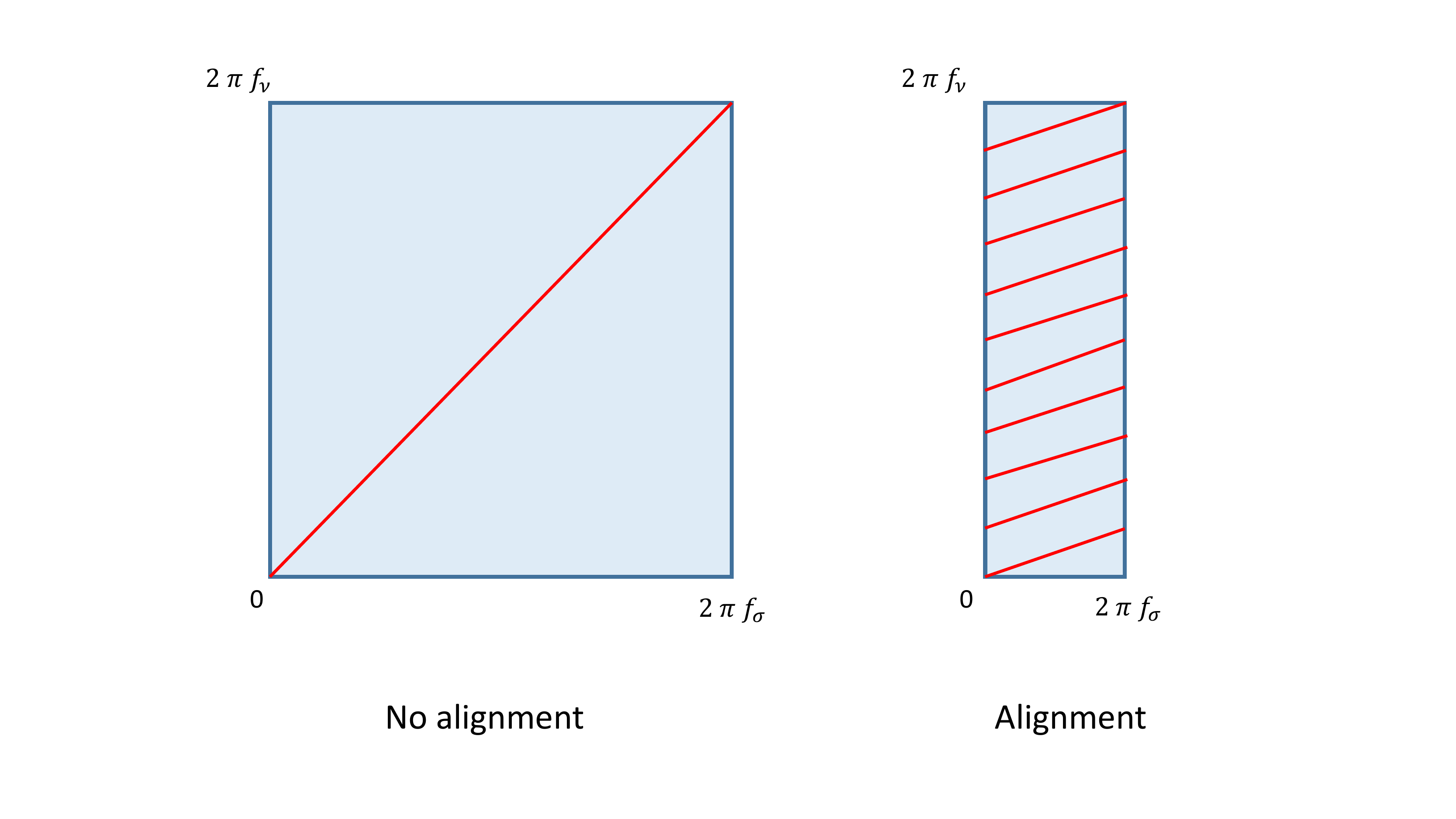}
\caption{Figure showing the two-dimensional axion field space and, in red, the axion combination which is perturbatively massless. Two cases are shown, the first with no alignment, and the second with alignment. As the light axion becomes aligned, parametrically increasing its periodicity relative to one of the fundamental decay constants, the same decay constant also becomes smaller. Overall the net effect leads to no increase in the periodicity of the light axion. This string theory setting should be compared with the naive field theory expectation as shown in figure \ref{fig:genaxal}.}
\label{fig:iiaaxal}
\end{figure}

Let us consider a supersymmetric vacuum of the theory (\ref{simaxaliiaa}). The F-term conditions $D_SW=D_UW=0$ imply
\be
\frac{f_{\nu}}{f_{\sigma}}  = \frac{\sqrt{3}s}{u} = \frac{h_1}{\sqrt{3}h_0} \;.
\label{ffixiiaf}
\ee
We can then check that the instanton with the enhanced decay constant, say for $h_1 \gg h_0$, is the dominant one since $s \gg u$. It therefore may appear that indeed we have obtained a parametrically enhanced axion decay constant, however this is not so. Indeed, substituting (\ref{ffixiiaf}) into (\ref{fsfuiia}) we find 
\be
f_{\psi}^u = 2 f_{\sigma}\;.
\ee
So while the decay constant was enhanced with respect to $f_{\sigma}$ it is actually not enhanced with respect to $f_{\nu}$. So what has happened is that the fluxes had two effects, one was aligning the light axionic direction close to one of the fundamental axions, but the other was fixing the moduli fields which control the decay constants, and the two effects cancelled precisely \cite{Palti:2015xra}. This is illustrated in figure \ref{fig:iiaaxal}. The dynamical adjustment of the moduli fields to censor an enhancement of the field ranges in the low-energy effective theory is an effect that we will encounter multiple times. 
In \cite{Palti:2015xra,Baume:2016psm} this effect of cancellation of enhancement was shown to hold more generally, for different Calabi-Yau and non-Calabi-Yau settings. However, there is no general proof, and in particular, already aligning a combination of more than two axions is more complicated to analyze. Nonetheless, the existence of this censorship mechanism is quite striking.

In \cite{Palti:2015xra} examples were presented where there was an enhancement of a periodicity in an instanton of the effective axion, but this instanton was sub-leading to the un-enhanced instanton. Such scenarios would manifest the strong version of the Weak Gravity Conjecture, leading to the situation shown in figure \ref{fig:axtp}. While such a setup would not lead to possible models of inflation, it is interesting to study nonetheless. Indeed, in \cite{Hebecker:2018fln} a very simple realization of such a scenario was proposed and termed axion {\it mis-alignment}. The idea is to construct the perturbatively massless axion combination to be almost aligned with the diagonal in the two-axion field space. This would still lead to a large overall periodicity, but each instanton would have an un-enhanced periodicity. Explicitly, in the action (\ref{simaxaliiaa}), it corresponds to taking $h_0$ and $h_1$ both large and co-prime. There is no cancellation obstruction in such a setting, and therefore an enhancement is only limited by the tadpole constraints and also K{\"a}hler moduli backreaction \cite{Hebecker:2018fln}. These constraints are quite strong and so it is debatable whether this could lead to a super-Planckian overall periodicity. However, there is yet another important aspect of axion alignment, or mis-alignment, to which we now turn. 

\subsubsection*{Axions and Discrete Torsion}

(Note that this section is not a review of published work, and is mostly new material.) The axions in string theory come from the expansion of the RR forms in terms of harmonic forms. These are topologically counted by the Hodge or Betti numbers of the manifold. It is therefore natural to expect that axion alignment should also have some topological formulation. In this subsection we aim to introduce such a formulation, mostly based on \cite{hebunpub}. Consider first the simple scenario of a type IIA axion arising from an harmonic three-form, and we ask how it appears in the instantons associated to Euclidean D2 branes wrapping three-cycles. We can check that instantons wrapping homologous cycles have the same action. Define the three-form $\alpha^{\psi}$ such that the axion arises from an expansion $C^{(3)}=\psi \alpha^{\psi}$. Then $\psi$ is an axion only if $\alpha^{\psi}$ is harmonic, so it is closed but not exact. Instantons which feature the axion in the action will be branes wrapping a cycle $\cC$ such that
\be
S_{\mathrm{inst}} \supset \int_{D2} C^{(3)} = \psi \int_{\cC} \alpha^{\psi} \;.
\ee 
If $\cC$ is a boundary, $\cC = \partial \cD$, then we have 
\be
\psi  \int_{\cC} \alpha^{\psi} = \psi \int_{\cD} d \alpha^{\psi} = 0 \;.
\ee
And so instantons wrapping cycles which are homologous, have the same axion action and periodicity. 

Now consider turning on H-flux, say $h_0$ and $h_1$ as in (\ref{simaxaliiaa}). This gives a mass to an axion. We can also understand this topologically as the statement that axions are actually counted by harmonic forms in twisted K-theory, which means harmonic with respect to the operator
\be
d_H = d - H \;.
\ee
From the expression for the superpotential (\ref{wt6dj}), it is manifest that perturbatively massless axions arise from forms that are closed under $d_H$. More interesting is that also forms which are exact under $d_H$ are not relevant. To see this, we can focus on the $H$ part of the operator and note that an exact form means that we can write $\alpha^{\psi} = p H$, where $p$ is a constant (is integer in terms of integer homology). Then 
\be
\psi  \int_{\cC} \alpha^{\psi} = p \psi \int_{\cC} H = 0 \;.
\ee
The reason that this vanishes is because there is a Freed-Witten anomaly for a D2 brane wrapping a cycle with H-flux through it \cite{Freed:1999vc}. We therefore see that indeed axions are determined by $d_H$ cohomology. We can also check that instantons which are cohomologous in $d_H$ homology lead to the same action. Let us define a cycle $\cC$ to be a boundary in $d_H$ homology if 
\be
\int_{\cC} \alpha^{\psi} = -\int_{\cM} H \wedge \alpha^{\psi} \;.
\ee
Now since we require $\alpha^{\psi}$ to be $d_H$ closed we have $H \wedge \alpha^{\psi} = 0$. In the case of normal homology, we can physically see that an instanton wrapping a homologically trivial cycle, so a boundary, is unstable to shrinking to a point which explains why it can not yield an action which includes the axion. There is a similar process in the case of $d_H$ homology, because a $D2$ brane wrapping a three-cycle whose Poincar\'e dual is a multiple of the $H$-flux suffers from an instability of decay where it annihilates against the $H$-flux into RR-flux (see \cite{Evslin:2006cj} for a review of twisted K-theory). A nice way to see this is utilising generalized calibrations, which give the action of a D-brane in terms of a calibration form $\Theta$ defined to satisfy the constraint (see, for example \cite{Martucci:2005ht})
\be
d_H \Theta = F \;.
\ee
Here both $\Theta$ and $F$ are a formal sum of different degree forms, where $F = \sum_i F^{(i)}$. Then the energy $M$ of a brane wrapping a cycle $\cC$ is\footnote{We drop here contributions from world-volume flux on the brane for simplicity.}
\be
M = \int_{\cC} \Theta \;.
\ee
Then if $\cC$ is exact in $d_H$ homology we can write
\be
M = \int_{\cM} \Theta \wedge d_H \chi =  \int_{\cM} d_H \Theta \wedge \chi = \int_{\cM} F \wedge \chi \;,
\ee
where $\chi$ is some formal sum over even-dimensional forms. 
This shows that the mass of such a brane is equal to an integral over RR-flux, thereby allowing for the brane-flux transition. 

The brane-flux transition is allowed if the brane can annihilate against the full H-flux, which means that $p$ above should be an integer. So more precisely, the axion-instanton system is classified by the integer $d_H$ homology. Now, in order to have some sort of alignment or mis-alignment mechanism we require that the axion appears in two different instantons with different periodicities. But since there is just one perturbatively massless axion, there is only one $d_H$ cohomology class, and only one dual $d_H$ homology class, and therefore one might deduce that all instantons should have the same action for this axion. While this is true in real homology, it is not quite true in integer homology. A single real cohomology class may nonetheless have multiple integer homology classes, the difference is accounted for by torsion homology. Therefore, we reach the conclusion that any type of axion alignment must be topologically classified by discrete torsion in $d_H$ homology. 

In the case of $T^6$, or Calabi-Yau, with flux indeed turning on $H$-flux can be understood as inducing torsion homology in $d_H$. This explains why axion alignment or misalignment can be realized in such settings. However, if we look at compactifications which are upliftable to 10-dimensional solutions then we need to satisfy the Bianchi identity (\ref{bi10dloc}) exactly locally, which means that the local sources should cancel. One then sees that the $H$-flux must become exact in such cases. If $H$ is exact then $d_H$ cohomology becomes equivalent to $d$ cohomology, and so axion alignment in such settings has to correspond to a compactification on a manifold which carries torsion homology. The example case in section \ref{sec:iias3s3} of $S^3 \times S^3$ carries no such torsion homology and so cannot realize axion alignment. It would be interesting to study example which do have torsion homology, for example a product of Lens spaces, to see how axion alignment manifests in such cases. More generally, an important lesson to learn is that for RR axions which are topologically protected, axion alignment similarly has a topological formulation and so compactifications which must change their topology when uplifted to a fully backreacted local 10-dimensional solution can also expect to have any conclusions regarding axion alignment modified by such backreaction.

\subsubsection*{Axion alignment in type IIB string theory}

Explicit studies of axion alignment scenarios in type IIB can be split according to which axions, RR or complex-structure, are being utilized. An alignment utilising  two RR axions was studied in \cite{Ben-Dayan:2014lca,Rudelius:2015xta}. While alignment models utilising the complex-structure moduli, which develop axion-like directions in certain limits, were studied in \cite{Hebecker:2015rya,Blumenhagen:2016bfp}. The models are rather complicated due to the underlying complexity of moduli stabilisation in IIB relative to IIA. However, some similar effects can be seen. For example, in \cite{Hebecker:2018fln} it was shown that aligning two axions manifests similar cancellations as in type IIA. By contrast, it was argued that mis-aligned scenarios, see above, are easier to implement in a type IIB context. Indeed, it was agued that such enhancement of the overall axion periodicity is likely to be realizable to some extent, though not in a parametrically controlled way. In \cite{Hebecker:2018yxs} mis-alignment was also argued to be present when considering axions which are down warped throats. Alignment of a large number of axions, $N$, in type IIB was studied in \cite{Bachlechner:2014gfa,Junghans:2015hba,Long:2016jvd}. Explicit studies of Calabi-Yau spaces revealed only moderate enhancement relative to the fundamental decay constants, around a factor of 2-3.  

All the alignment scenarios in type IIB are within a KKLT type moduli stabilisation framework, and so are not upliftable to 10-dimensional solutions. Partially due to this, they are significantly more involved and complicated than the simple type IIA cases discussed above.

\subsection{Bosonic sector of ${\cal N}=2$ Supergravity}
\label{sec:n2sugra}

Aspects of the Swampland related to scalar fields are most sharply tested in the context of extended ${\cal N}=2$ supersymmetry because such theories exhibit exact moduli spaces. We therefore first present a quick introduction to ${\cal N}=2$ supergravity. A sufficiently complete account can be found in \cite{Ceresole:1995ca,Andrianopoli:1996cm}. The multiplets consist of $n_V$ vector multiplets with bosonic content of a complex scalar field, denoted $t^i$ with $i=1,...,n_V$, and a gauge field, denoted $A^i$.\footnote{Note the clash of notation between the complex vector multiplet scalars $t^i$ and the real parts of the K{\"a}hler superfields in the type IIA setting of sections \ref{sec:iiaont6} and \ref{sec:wgciist}.} There are $n_H$ hypermultiplets, with bosonic content of four real scalar fields denoted $l^{\lambda}$, with $\lambda=1,...,4n_H$. There is a gravity multiplet which contains the bosonic fields of a graviton and a graviphoton $A^0$. We can combine all the gauge fields as $A^I$ with the index $I=\left\{0,i\right\}$.

The action takes the form 
\be
S_{{\cal N}=2} = \int d^4x \sqrt{-g} \left[ \frac{R}{2} - g_{ij}\partial_{\mu} t^i \partial^{\mu} \overline{t}^{j} - h_{\sigma \lambda} \partial_{\mu}l^{\sigma}\partial^{\mu} l^{\lambda} + {\cal I}_{IJ} {\cal F}_{\mu\nu}^{I} {\cal F}^{J,\mu\nu} +  {\cal R}_{IJ} {\cal F}_{\mu\nu}^{I} \left(\star {\cal F}\right)^{J,\mu\nu} \right] \;.     \label{N2action}
\ee
The gauge kinetic functions are the real and imaginary parts of a complex matrix, $\cR_{IJ} =\mathrm{Re\;} N_{IJ}$ and $\cI_{IJ} =\mathrm{Im\;} N_{IJ}$. The metrics $g_{ij}$ and $h_{\sigma\lambda}$ are on two separate manifolds, so the total moduli space splits into vector multiplets and hypermultiplets $\cM = \cM_V \times \cM_H$. The hypermultiplets manifold is a quaternionic K{\"a}hler manifold and the vector multiplets span a special K{\"a}hler manifold. We will mostly focus on the vector multiplet part.\footnote{In string theory, the hypermultiplet manifold is special quaternionic which means that it contains a special K{\"a}hler subspace. This is a manifestation of mirror symmetry, and means that much of our discussion on the vector multiplet space can be applied to the special K{\"a}hler submanifold in the hypermultiplet sector.} The geometric structure on the vector multiplet field space is determined through the periods $\left\{X^I,F_I\right\}$ which are holomorphic functions of the scalar fields $t^i$. The K{\"a}hler potential for the scalar field-space metric takes the form 
\be
K = -\log i\left( \overline{X}^I F_I - X^I \overline{F}_I \right) \;. \label{N2KahlerPot}
\ee
It is usually, not always but for our purposes we can assume so, possible to determine the periods in terms of a prepotential $F$ through $F_I=\partial_{X^I} F$. Then the general expression for the symplectic matrix takes the form
\be
N_{IJ} = \overline{F}_{IJ} + 2i \frac{\mathrm{Im} F_{IK} \mathrm{Im} F_{JL} X^K X^L }{\mathrm{Im} F_{MN} X^MX^N} \;,
\ee
where $F_{IJ} = \partial_I F_J$. We can carry over the definitions in section \ref{sec:wgcta}, up to the expression for ${\cal Q}^2$ (\ref{qdefq}).

It is convenient to introduce an index $\cI$ which ranges over both electric and magnetic components, so has a range $2 n_V +2$. Then we can define the period vector ${\bf \Pi}$, which in an appropriate local coordinate basis can be written as
\be
{\bf \Pi} = \left( \begin{array}{c} X^0 \\ X^i \\ F_j \\ F_0  \end{array} \right)\;,
\label{pervec}
\ee
with components $\Pi^{\cI}$. Note that in (\ref{pervec}) we take the electric index as increasing down the vector, while the magnetic index is increasing from the bottom moving up. We can also write the vector $\cQ$ as ${\bf q}$ with components $q^{\cI}$. This vector space has a natural symplectic form on it $\eta$ which can be used to construct symplectic inner products. In the coordinate basis where the period vector takes the form (\ref{pervec}) the symplectic matrix is
\be
\eta = \left( \begin{array}{ccccc} 0 & 0 &... & 0 & 1 \\ 0 & 0 & ... &1 & 0  \\ 0 & -1&... &  0 & 0  \\ -1 & 0 &... & 0 & 0  \end{array} \right) \;.
\ee
It is important to keep in mind that the expressions for $\eta$ and ${\bf \Pi}$ are a local choice of basis on the vector-multiplet moduli space. It is convenient to utilize this basis for calculations, but other basis choices are possible, and in general one has to patch such bases over different regions in moduli space. This can be captured nicely by thinking about the origin of the vector multiplets in type IIB string theory on a Calabi-Yau manifold. The Calabi-Yau supports a nowhere vanishing holomorphic three-form $\Omega$. We can then pick a basis of three-cycles $\Gamma^{\cI}$ and define the period vector as the integral of $\Omega$ in this basis
\be
\Pi^{\cI} = \int_{\Gamma^{\cI}} \Omega \;.
\ee
The symplectic inner product can then be defined in terms of the basis intersections
\be
\eta_{\cI \cJ} = \Gamma^{\cI} \cdot \Gamma^{\cJ} \;.
\ee
We can also write a general expression for the K{\"a}hler potential
\be
K = - \log \left[i\int_{CY} \Omega \wedge \overline{\Omega} \right] = - \log \left[ i {\bf \Pi}^T \cdot \eta \cdot \overline{{\bf \Pi}} \right] \;.
\label{genKexpPi}
\ee
While the basis of cycles may change over different points in moduli space, the form $\Omega$ is defined over the whole space.

An important property of extended supersymmetry is the existence of BPS states. In particular, given a charge vector ${\bf q}$, we can define the central charge
\be
Z\left( {\bf q} \right) \equiv e^{\frac{K}{2}} \left(  {\bf q} \cdot \eta \cdot {\bf \Pi} \right) \;.
\label{Zcc}
\ee
Then any charged state in the theory with mass $M$ satisfies the bound
\be
M\left( {\bf q} \right) \geq \left| Z\left( {\bf q} \right) \right| \;.
\ee
The bound is saturated by BPS states. An important identity involving the central charge is
\be
{\cQ}^2 = \left| Z\right|^2 + g^{ij} D_i Z \overline{D}_j \overline{Z} \;, \label{si}
\ee
where the covariant derivative acts as
\bea
D_i \psi^j &=& \partial_{z_i} \psi^j  + \Gamma^{j}_{ik} \psi^k  + \frac{p}{2}\left(\partial_{z_i} K \right) \psi^j \;,
\eea
on an object $\psi^j$ with K{\"a}hler weight $p$ ($Z$ has weight 1). Another useful identity is
\be
{\cal Q}_F^2 = \left|Z\right|^2 - g^{ij} D_i Z \overline{D}_j \overline{Z} \;. \label{si2}
\ee
Here ${\cal Q}_F^2$ is defined in the same way as in (\ref{qdefq}) but with $N_{IJ} \rightarrow F_{IJ}$. 

These identities imply certain properties of BPS states. As discussed in section \ref{sec:wgcsca}, the expression (\ref{si}) captures the forces acting on BPS states, and shows that BPS states feel no self-force. The expression (\ref{si}) implies that gravity acts as the weakest force on BPS states \cite{Palti:2017elp}.  The matrix $\cI_{IJ}$ is negative definite. The matrix $\mathrm{Im}\left(F\right)_{IJ}$ has $n_V$ strictly positive eigenvalues and one strictly negative eigenvalue. So there is a basis where $n_V$ BPS states have scalar forces acting strictly stronger than gravity, and one of them has gravity acting strictly stronger. The odd one out is due to the graviphoton which has no scalar superpartners. This shows that in ${\cal N}=2$ the Scalar Weak Gravity Conjecture (\ref{swgc}) is satisfied by BPS states.

It is possible to write a potential in ${\cal N}=2$ supergravity, but it is constrained by the structure of the gauge field sector. Specifically, a potential can only be induced by gauging isometries in the moduli space $\cM$. We can replace the derivatives in (\ref{N2action}) with gauge covariant derivatives
\bea
\nabla t^i &=& \partial t^i  + A^I k_I^i\left(t\right) \;,\nn \\
\nabla l^{\lambda} &=& \partial l^{\lambda}   + A^I k_I^{\lambda}\left( l\right) \;,
\eea
where the $k_I^i$ and $k_I^{\lambda}$ are Killing vectors on the moduli space. The induced potential is then a function of the Killing vectors $V\left( k_I^i,k_I^{\lambda} \right)$. See \cite{Andrianopoli:1996cm,Palti:2006yz} for a discussion (including so-called magnetic gauging). 

We will comment here only a particularly interesting relation to flux compactifications, see \cite{Louis:2002ny} for a general analysis of such relations. Consider type IIB string theory with RR fluxes, as studied in section \ref{sec:4dnoscale}. Then since the dilaton and K{\"a}hler moduli do not appear in the superpotential, the scalar potential takes the form
\be
V_{RR} = e^K \left( g^{i\bar{j}} D_i W \overline{D_j W} + \left| W \right|^2 \right) \;.
\ee 
But we see that the superpotential is related to the central charge 
\be
e^{\frac{K}{2}} W\left( {\bf q}\right) = Z \left( {\bf q}\right)  \;,
\ee
where the charge vector ${\bf q}$ corresponds to the RR fluxes. Then we see that using (\ref{si}) we can write
\be
V_{RR} = {\cal Q}^2 \;.
\label{VRRQ2}
\ee
So the potential which is induced is nothing but the charges contracted with the gauge coupling matrices. We will return to this relation in section \ref{sec:emergence}. 

Finally, note that sometimes we will utilize the shorthand for the symplectic inner product
\be
S\left({\bf A}, {\bf B} \right) \equiv {\bf A} \cdot \eta \cdot {\bf B} \;.
\label{Sipdef}
\ee

\subsection{The Distance Conjecture with 8 supercharges}
\label{sec:dcwith8}

Having set up the formalism of ${\cal N}=2$ supergravity, the theories with gravity with the minimal supersymmetry which exhibit exact moduli spaces, we can consider the distance conjectures, as discussed in section \ref{sec:sdc}. In section \ref{sec:fcsdc} we studied a particularly simple scenario of the distance conjecture where the field is associated with the size of a circle. The next step up is to consider the $T^6$ setting of section \ref{sec:iiaont6}. It is convenient to formulate it in the ${\cal N}=1$ language of that section. Let us consider the K{\"a}hler moduli field space, spanned by the complex scalar fields of the $T^i$. As discussed in detail in section \ref{sec:clstru1}, the real parts of the superfields control closed-string $U(1)$s gauge couplings. There are infinite towers of states, associated to branes wrapping even-dimensional cycles with a mass scale set by those gauge couplings. Writing the gauge couplings (\ref{actiia4dgauge}) in terms of the canonically normalised K{\"a}hler moduli fields we see that they are exponential in nature. Therefore, the distance conjecture maps directly to the tower versions of the Weak Gravity Conjecture in such settings, as discussed generally in section \ref{sec:distswgc}, and all the results of section \ref{sec:clstru1} apply equally to the distance conjecture. Note that going to large distances in K{\"a}hler moduli space, which using the notation of section \ref{sec:clstru1} means $t^i \rightarrow \infty$, also exponentially lowers the mass scale of Kaluza-Klein towers. This is typical behaviour of the distance conjecture in string theory, where many different towers may become light at infinite distance. The behaviour of the imaginary parts of the $T^i$, the axions $v^i$, also satisfies the distance conjecture. In this case we can map to the discussion in section \ref{sec:wgcax} regrading the axionic Weak Gravity Conjecture (\ref{wgcax}) where we argued, utilising duality, that the periodicity of the $v^i$ is sub-Planckian. This is an example of the general discussion of this relation in section \ref{sec:distswgc}.

The results above for the K{\"a}hler moduli apply also to the dilaton superfield $S$, with the associated branes being D0 (and dual D6) branes. Let us now consider the complex-structure moduli space spanned by the $U_i$ in section \ref{sec:iiaont6}. The axionic parts are again periodic and so respect the distance conjecture. The real parts $u_i$, control the sizes of three-cycles in the space. T-duality along an odd number of directions in the internal space will exchange them with the K{\"a}hler moduli. However, it will also take us to type IIB string theory. The NS sector, of fundamental strings, is the same in both theories, and so the towers of states of Kaluza-Klein and Winding modes will satisfy the distance conjecture. The towers of wrapped D-brane states change between IIA and IIB and so we must look for them directly in the type IIA setting. In this case, we see that branes wrapping three-cycles lead not to particle states in four dimensions but to instantons, strings and space-time filling states. The tension of these objects, say the strings from wrapped D4 branes, will depend on the $u_i$ moduli. The moduli measure the size of three-cycles and so, say in the limit $u_i \rightarrow 0$, the tension of the strings will decrease exponentially in the proper distance.\footnote{The opposite limit in complex-structure moduli space is related by duality.} The distance conjecture will then be satisfied by these light extended objects. There actually will be two types of towers of light states, first the excitations of the light strings, and second, towers of light extended objects themselves. We may consider either one to be related to the distance conjectures, as discussed in general in section \ref{sec:sdc}. Note that another way to think of the distance conjecture for complex-structure moduli is by considering spacetime filling D6 branes wrapping three-cycles. Then the distance conjecture is again related to the Weak Gravity Conjecture for the D6 gauge field which becomes weakly-coupled at large three-cycle volumes.																																																									

To generalize the result to more complicated setting. first let us consider the cases with more than 8 supercharges, in any dimension. There is a general result which is that the moduli space of a supergravity in any dimension which has more than 8 supercharges is a coset 
\be
\cM = \frac{G}{H} \;.
\ee
The explicit forms for the known supergravities of $G$ and $H$ can be found in \cite{Cecotti:2015wqa}. Due to this restricted coset structure, it is possible to utilize group theoretic arguments to show that the Swampland Distance Conjecture (\ref{sdc}) holds in such settings \cite{Cecotti:2015wqa}. More precisely, to show that the mass of BPS states decreases exponentially with proper distance, but not to show that those BPS states are actually in the theory. Exploiting the coset structure of moduli spaces was a key element of the examples presented in original proposal \cite{Ooguri:2006in}. In particular, it was argued that dualities in the moduli space relate any infinite distance point with the large volume limit. And therefore at least there should be states dual to the Kaluza-Klein states of the large volume limit. 

\subsubsection{Calabi-Yau moduli spaces}
\label{sec:cymod}

The case where there are exactly 8 supercharges preserved in the vacuum is special, because it still has an exact moduli space, but this moduli space is not a coset manifold. Indeed, known ${\cal N}=2$ moduli spaces span a vast spectrum, not least due to each Calabi-Yau manifold supporting such a moduli space. Nonetheless, we will see that the structures of the distance conjectures can still be shown to arise within such moduli spaces \cite{Grimm:2018ohb,Blumenhagen:2018nts,Lee:2018urn,Grimm:2018cpv,Corvilain:2018lgw,Lee:2019tst,Joshi:2019nzi}. In particular, in \cite{Grimm:2018ohb,Grimm:2018cpv} a formalism was developed which allows to study infinite distances in ${\cal N}=2$ moduli spaces with complete generality. We outline this below, before discussing more specific examples of infinite distances. 

We focus here on the vector-multiplet moduli space $\cM_V$. In string theory, we can consider type IIB string theory compactified on a Calabi-Yau to four dimensions and then the moduli space is the complex-structure one. Let us define a locus in $\cM_V$ through the constraint $z=0$, where $z$ is some local coordinate in the moduli space around this locus. We are interested in how the period vector ${\bf \Pi}$, see section \ref{sec:n2sugra}, transforms upon circling the point. In general, it may transform with an action of a matrix $T$ as
\be
{\bf \Pi}\left(z e^{2 \pi i}\right) = T \cdot {\bf \Pi}\left(z \right) \;.
\label{monTdef}
\ee
$T$ is called the Monodromy matrix, and if it is not just the unit matrix then there is a monodromy about the point $z=0$. The monodromy matrix is composed of constant rational numbers. The order of the monodromy is determined by the smallest power of $T$ which gives back the unit matrix. If no power does, then the monodromy is said to be of infinite order.  In the case when it is of infinite order it is possible to introduce the matrix\footnote{More precisely, there may be also a finite order part on top of the infinite order one which is not important for our purposes and we will not discuss here, see \cite{Grimm:2018ohb} for a discussion on this.} 
\be
N = \log T \;.
\ee
The matrix $N$ is nilpotent, which means $N^{n+1}=0$ for some integer $n$. For a moduli space of a Calabi-Yau threefold the maximum $n$ is 3. There is a theorem which gives the leading order behaviour of the period matrix ${\bf \Pi}$ near the region $z=0$. The Nilpotent Orbit Theorem states that \cite{Schmid}, 
\be
{\bf \Pi}\left(z,\xi\right) = \mathrm{Exp\;}\left[ \frac{1}{2\pi i} (\log z) N \right] {\bf A}\left(z,\xi \right) \;.
\label{nilorbgen}
\ee
Here, $\xi$ denote other coordinates on the moduli space. The crucial part of the theorem is that the vector ${\bf A}$ is holomorphic in $z$, so can be expanded as
\be
 {\bf A}\left(z,\xi \right) = {\bf a}_0\left(\xi\right) + \sum_{n=1}^{\infty} {\bf a}_n \left( \xi \right) z^n \;.
\ee
Therefore the leading behaviour, to an exponentially good approximation, of the $z$ dependence of period vector as $z \rightarrow 0$ is given by the Nilpotent orbit
\be
{\bf \Pi}_{\mathrm{nil}}\left(z,\xi\right) = \mathrm{Exp\;}\left[ \frac{1}{2\pi i} (\log z) N \right] {\bf a}_0\left(\xi \right) \;.
\label{nilorb}
\ee

We can change coordinates to 
\be
t = \frac{1}{2\pi i} \log z \;.
\ee
So the point $z=0$ corresponds to $t \rightarrow +i\infty$.
Now we define an integer $d$ as the maximum integer such that
\be
N^d {\bf a}_0 \neq 0 \;. 
\ee
Then by using (\ref{nilorb}) in the expression for the K{\"a}hler potential (\ref{genKexpPi}), we recover the leading behaviour of the moduli space metric as
\be
g_{t\bar{t}} = \frac{d}{4\left(\mathrm{Im\;} t\right)^2}  + \cO\left( \frac{1}{\left(\mathrm{Im\;} t\right)^3} \right)\;.
\ee
Now consider the proper distance $\Delta \phi$ between two points along this one-parameter approach where, for simplicity, we fix $\mathrm{Re\;} t=0$. So from $\mathrm{Im\;} t_i$ to $ \mathrm{Im\;} t_f$, we have
\be
\Delta \phi = \int_{\mathrm{Im\;} t_i}^{ \mathrm{Im\;} t_f} \frac{\sqrt{d}}{2} d \left( \log  \mathrm{Im\;} t \right)+ \mathrm{subleading} = \frac{\sqrt{d}}{2} \log \left(\frac{ \mathrm{Im\;} t_f}{ \mathrm{Im\;} t_i} \right) + \mathrm{subleading}  \;.
\ee
Where the subleading contributions, in particular, remain finite as $ \mathrm{Im\;} t_f \rightarrow \infty$. we therefore observe a logarithmic divergence of the proper distance approaching $z=0$, or equivalently $\mathrm{Im\;} t \rightarrow \infty$, as long as $d > 0$. In fact, there is a theorem which states that $d>0$ is not only sufficient but is also the necessary condition for $z=0$ to be at infinite distance \cite{wang1}. We therefore find the result that any locus which is an infinite distance in Calabi-Yau complex-structure moduli space must have an infinite order monodromy associated to it, and the proper distance approaching the locus diverges logarithmically in $\mathrm{Im\;}t$. 

We can also study the form of the central charge, and therefore the mass of BPS states, as $z \rightarrow 0$ or $\mathrm{Im\;}t \rightarrow \infty$. Utilising (\ref{nilorb}) in the expression for the central charge we find
\be
Z\left( {\bf q} \right) \sim \frac{\sum_{j=0}^{j=d} \left(\mathrm{Im\;} t \right)^{j} S\left({\bf q},N^j{\bf a}_0 \right)}{\left(\mathrm{Im\;} t \right)^{\frac{d}{2}}} + \cO\left( e^{2 \pi i t}\right)\;.
\label{asycech}
\ee
The behaviour of the mass of BPS states therefore depends on how their charge ${\bf q}$ contracts with $N^j {\bf a}_0$. We can split the possible charged states into those which become massless as $\mathrm{Im\;}t \rightarrow \infty$, and those which retain a finite mass. The set of massless states $\cS$ can be defined as \cite{Grimm:2018ohb}
\be
\cS \equiv \left\{ \mathbf{q} : \;S\left({\bf q},N^j {\bf a}_0  \right) = 0 \;,\; \mathrm{for\;all}\; j \geq \frac{d}{2} \right\} \;. \label{masslesscondI1} 
\ee
This set then splits further into two sets of states, denoted type G and type F defined as\footnote{In \cite{Grimm:2018ohb} states of type G and type F were denoted type I and type II respectively.}
\bea
  \cS_{\rm G} &\equiv& \left\{ \mathbf{q} \in \cS:\; S_j\left({\bf q},N^j {\bf a}_0  \right) \neq 0 \;\mathrm{for\;some}\; j < \frac{d}{2}\right\} \;, \label{masslesscondI2} \\
  \cS_{\rm F} &\equiv& \left\{ \mathbf{q} \in \cS:\; S\left({\bf q},N^j {\bf a}_0  \right) = 0 \;,\; \mathrm{for\;all}\; j \geq 0 \; \right\} \;. \label{masslesscondII}
\eea
We see that states with charges of type G become massless as a power-law in $\mathrm{Im}\; t$, while states of type F do so exponentially in $\mathrm{Im}\; t$. We also can see that only type F states can become massless at finite distance in moduli space. In fact there is a nice physical distinction between type G and type F states, the former carry charge under the graviphoton while the latter do not.\footnote{The graviphoton was, somewhat carelessly, stated to be $A^0$ earlier, but actually it is a combination of all the $A^I$ associated to the central charge.}

We therefore find that all BPS states of type G become massless exponentially fast in the proper distance when approaching any infinite distance locus in any Calabi-Yau moduli space
\be
\frac{M_{\mathrm{BPS}}\left(t_f\right)}{M_{\mathrm{BPS}}\left(t_i\right)} \sim \left(\frac{\mathrm{Im\;} t_i}{\mathrm{Im\;} t_f }\right)^{\alpha} \sim e^{-\frac{2 \alpha\Delta \phi}{\sqrt{d}}} \;,
\label{bpsmassa}
\ee
where $\alpha = \frac32, 1, \frac12$ depending on the state.
This is strong evidence for the distance conjectures (\ref{sdc}). However, there remains to show that there are an infinite number of type G states in the theory. This was argued for in the case of $d=3$ infinite distance loci in \cite{Grimm:2018ohb}, and more generally for other types, though not all, of loci in \cite{Grimm:2018cpv}. In this type IIB setting these are D3 branes which are wrapping supersymmetric three-cycles (special Lagrangian cycles) in the Calabi-Yau, which shrink to zero size at infinite distance in field space. These states are also charged under the $U(1)$ gauge symmetries in the vector multiplets. They therefore also play the role of the tower of states associated to the magnetic Weak Gravity Conjecture as discussed in section \ref{sec:wgctower}. Indeed, the gauge couplings of many of the $U(1)$s vanish at the infinite distance locus $\mathrm{Im\;}t \rightarrow \infty$. It is possible to similarly associate a type G or type F to gauge fields. We will consider only type G gauge fields here, for a more general analysis see \cite{Grimm:2018ohb}. For $d=3$ there are two such fields, with gauge couplings $g_{(0)}$ and $g_{(1)}$, while for $d<3$ there is only one type G gauge field, with gauge coupling $g_{(0)}$. The rate at which the gauge couplings vanish was calculated in \cite{Grimm:2018ohb} and is presented in table \ref{tab:gauvan}.
\begin{table}
\center
\begin{tabular}{| c | c | c |}
\hline
$d$  & $g_{(0)}$ & $g_{(1)}$  \\
\hline
 \rule[-.3cm]{0cm}{0.8cm} $1$ & $\frac{1}{\left(\mathrm{Im\;} t\right)^{\frac12}}$ &  - \\
\hline
 \rule[-.3cm]{0cm}{0.8cm} $2$ &$\frac{1}{\left(\mathrm{Im\;} t\right)}$  &  - \\
\hline
 \rule[-.3cm]{0cm}{0.8cm} $3$ & $\frac{1}{\left(\mathrm{Im\;} t\right)^{\frac32}}$ &  $\frac{1}{\left(\mathrm{Im\;} t\right)^{\frac12}}$  \\
\hline
\end{tabular}
\caption{Table showing the rate of the vanishing of gauge couplings approaching infinite distances $\mathrm{Im\;}t \rightarrow \infty$ in vector multiplets moduli space . There are two classes of gauge fields, with differing rates of vanishing depending on the type of infinite distance locus determined by the integer $d$. The behaviour of the gauge couplings matches the mass scale of the tower of BPS states for some $\alpha$ in (\ref{bpsmassa}), as predicted by the magnetic Weak Gravity Conjecture (\ref{mwgc}).}
\label{tab:gauvan}
\end{table}
We see that, as discussed in section \ref{sec:distswgc}, the gauge couplings behave exponentially in the proper field distance. Also the magnetic Weak Gravity Conjecture scale (\ref{mwgc}) indeed sets the mass scale of the tower of states, for an appropriate $\alpha$ in (\ref{bpsmassa}). Similarly, the relation between the gauge couplings and field distances in string theory implies that also the evidence in \cite{Lee:2018urn,Lee:2018spm,Lee:2019tst} for the Weak Gravity Conjecture, discussed in section \ref{sec:wgciios}, acts equally as evidence for the distance conjecture.

We have seen that the monodromy matrix $T$ plays a crucial role in the moduli space behaviour near infinite distances. It was argued in \cite{Grimm:2018ohb,Grimm:2018cpv} that it also plays an important role in the tower of states which becomes exponentially light. First note that $T$ is defined through its action on the period vector (\ref{monTdef}), but it also has a natural action on the charges of the BPS states. This can be seen by looking at the expression (\ref{Zcc}) for the central charge and how it transforms under monodromy
\be
\left|Z\left({\bf q}\right)\right| = e^{\frac{K}{2}}\left|{\bf q} \cdot \eta \cdot {\bf \Pi } \right| \xrightarrow{T}  e^{\frac{K}{2}}\left|{\bf q} \cdot \eta \cdot T \cdot {\bf \Pi } \right|  = e^{\frac{K}{2}}\left|\left(T^{-1} \cdot {\bf q} \right) \cdot \eta \cdot {\bf \Pi }\right| =  \left|Z\left(T^{-1} \cdot{\bf q}\right)\right|  \;. \label{bpsmonorb}
\ee
Since $T$ is of infinite order, it generates an infinite orbit within the charged BPS states. It was suggested in \cite{Grimm:2018ohb} that this infinite orbit is related to the infinite tower of states. Evidence for this includes showing that for $d=3$ that states in the monodromy orbit are indeed states present in the theory. We have already encountered a simple case of this in section {\ref{sec:clstru1}) with monodromy generated D2-D0 bound states. 

Another point emphasized in \cite{Grimm:2018ohb} is that infinite distance loci in moduli space are where an effective global symmetry exists. This is true for some gauge fields since their gauge coupling vanishes, but perhaps more intrinsically is true also for the axionic elements in the complex-structure moduli. The infinite tower of states which becomes massless at such loci can then be understood as a quantum gravity obstruction to inducing such a global symmetry. This yields an interpretation of the Swampland Distance Conjecture as quantifying the quantum gravity obstruction to global symmetries, matching a similar interpretation of the magnetic Weak Gravity Conjecture, as discussed in section \ref{sec:wgccbhs}. 

The discussion has so far focused on one-parameter approaches to infinite distances in moduli space. In \cite{Grimm:2018cpv} multi-parameter approaches were studied. Indeed, it is useful to see a picture of a quantum gravity moduli space, which we present in figure \ref{fig:modCY}. In general, there are multiple infinite distance loci, and they may also intersect. In \cite{Grimm:2018cpv} it was shown that there are certain rules for which type of infinite distance loci can intersect each other. 
\begin{figure}[t]
\centering
 \includegraphics[width=0.8\textwidth]{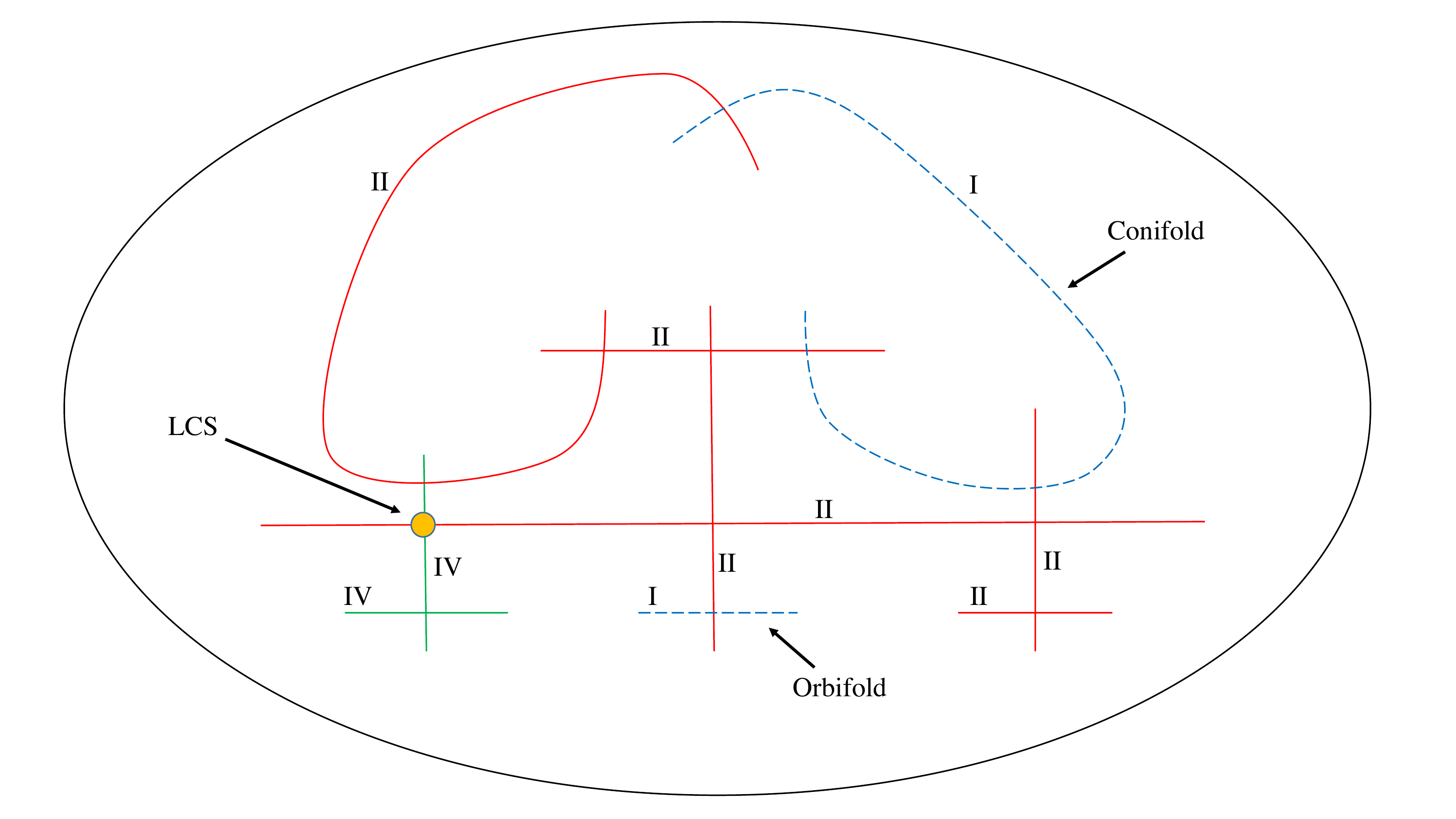}
\caption{Figure, taken from \cite{Grimm:2018cpv}, showing an example moduli space, of the two parameter Calabi-Yau $\mathbb{P}^{1,1,2,2,2}[8]$ as studied in \cite{Candelas:1993dm}. Each infinite distance locus is denoted by a solid line and assigned a type labelled by $\mathrm{II}$, $\mathrm{III}$, or $\mathrm{IV}$ (corresponding to $d=1,2,3$ respectively). We also show special finite distance loci with dashed lines, and these are associated to type I. Some well-known loci are labelled explicitly, the finite distance conifold and orbifold loci, and the infinite distance large complex-structure point.}
\label{fig:modCY}
\end{figure}
A particularly accessible region where multiple infinite distance loci intersect is the large volume limit of Calabi-Yau manifolds. This multiple parameter infinite distance region in field space was studied in type IIA string theory in \cite{Corvilain:2018lgw}.

So far we have discussed loci at infinite distance in moduli space. This is the most relevant setting for the Swampland Distance Conjecture (\ref{sdc}), but are not tests of the Refined Swampland Distance Conjecture (\ref{rsdc}). Specifically, it is important to understand precisely at what distance the exponential behaviour of the tower of states emerges. This analysis was performed for Calabi-Yau moduli spaces in \cite{Blumenhagen:2018nts}. Different possible geodesics were studied within the one-parameter moduli space of the quintic Calabi-Yau. It was found that geodesic lengths were such that the exponential behaviour in the tower of states appeared at distance which is at most $\Delta \phi \sim 1.4 M_p$. Also three two-parameter examples were studied, and it was found that the critical distances was $1.3 M_p$, $1.1 M_p$ and $1.1 M_p$.  Paths, not necessarily geodesic, even on a 101-dimensional moduli space was studied, as the mirror of the quintic. It was shown that each modulus contributed a reduced possible distance relative to the one and two-parameter examples, such that the total path distance before the exponential behaviour remained closely bound to $M_p$.  

\subsection{The Distance Conjecture with ${\cal N}=1$ supersymmetry}
\label{sec:dcaxmon}

Away from the ${\cal N}=2$ vacuum setting, scalar fields will have potentials. In such settings, the appropriate Swampland statement to test is the Refined Swampland Distance Conjecture (\ref{rsdc}). There are two qualitatively different types of scalar fields with potentials with respect to the distance conjecture. The first type are those for which the distance conjecture is realized through the field-space metric and the potential is not playing any significant role. For example, we may consider the K{\"a}hler moduli of $T^6$ as studied in section \ref{sec:iiaont6}, which are parameterizing the sizes of two-cycles. If we turn on flux, as in section \ref{sec:iiat6flux}, a potential is induced for the K{\"a}hler moduli. This does not change the fact that tower of wrapped branes become exponentially light and so the distance conjecture is still satisfied. 

Having said this, testing the conjecture even for such moduli is fundamentally different from the ${\cal N}=2$ extended supersymmetry setting because it is much more difficult to fully explore the moduli space as done in section \ref{sec:cymod}. Specifically, moving to strong coupling regimes is difficult to control. Nonetheless, some results are obtainable. In particular, in \cite{Gonzalo:2018guu} a study was performed over even strongly-coupled regions by utilising duality symmetries in the context of the Heterotic string. The results were consistent with the distance conjectures. 

The second type of scalar fields, which form the main focus of this section, are those whose potential plays a central role in their field space, in particular so-called monodromy axions. The point is to consider breaking the axion periodicity completely. So this would be an effective theory of the type
\be
\cL \supset - f^2 \left(\partial a \right)^2 + m^2 a^2 \;.
\label{effaxmonaty}
\ee
A simple example of this was seen in section \ref{sec:iiat6flux}, where the superpotential (\ref{wt6fl}) breaks the periodicity of many of the axions. The potential therefore decompactifies the axion field space, allowing for formally infinite distances $a \rightarrow \infty$. In fact, the axion periodicity is a gauge symmetry in string theory, and so cannot be broken in the full theory. However, like all gauge symmetries, it can be effectively broken within a low-energy effective theory. What this means is that there will always be a discrete symmetry which is the remnant of the axion periodicity, but this symmetry will act also on the flux parameters. For example, in the superpotential (\ref{wt6fl}) this manifests as the symmetry acting as
\be
\sigma \rightarrow \sigma + 1 \;,\;\; e_0 \rightarrow e_0 + h_0 \;.
\label{exaxperflux}
\ee
As well as other such symmetries, one for each axion whose periodicity is broken by fluxes. The general form of these symmetries can be found in \cite{DeWolfe:2005uu} for example.

Monodromy axions were understood in string theory since early work on flux compactifications, see for example \cite{Louis:2002ny}. However, they became particularly studied once it was realized that they are very interesting candidates for inflatons in large field inflation \cite{Silverstein:2008sg}.\footnote{Axion monodromy models are also interesting in non-inflationary phenomenological contexts, for example \cite{Graham:2015cka}.} The key reason is that their potential is protected by the underlying discrete gauge symmetry, and so it is possible to trust the form of the potential over large field distances. This lead to a significant effort to incorporate them into models of inflation in string theory. Such efforts can be decomposed into two types: those which utilize localized objects, branes, to break the axion periodicity, see for example \cite{McAllister:2008hb,Flauger:2009ab,Palti:2014kza,Escobar:2015ckf,Retolaza:2015sta}. And those which use fluxes to do so, see for example \cite{Marchesano:2014mla,Hebecker:2014eua,Arends:2014qca,Ibanez:2014kia,Blumenhagen:2014gta,Blumenhagen:2014nba,Franco:2014hsa,Ibanez:2014swa,Blumenhagen:2015qda,Buratti:2018xjt}. We will discuss the flux case first, and comment on the brane case later. 

As in previous sections, we first will discuss a very simple realization of axion monodromy. Specifically, within type IIA on $T^6$ with flux as in section \ref{sec:iiat6flux}. This was studied in \cite{Baume:2016psm}. We will, for simplicity as in section \ref{sec:wgcax}, identify the superfields so that there is one field in each sector, and also slightly rescale the fluxes to match the conventions of \cite{Baume:2016psm}. So the effective theory we consider is an ${\cal N}=1$ supergravity with 
\bea
K &=& - \log s - 3 \log u -3 \log t \;,  \\  \nn
W &=&  e_0 + i e T - q T^2 + \frac{i}{6} T^3 + i h_0 S - i h_1 U \;.
\label{simaxmoniia}
\eea
This theory has a supersymmetric anti-de Sitter minimum at
\be
h_0 s_0 = -\frac{h_1 u_0}{3} = \frac{m t_0^3}{15} = \frac{2}{9}\sqrt{\frac{10}{3}} p^{\frac32}\;,\;\; \rho_0 = \frac{2q}{3m^{\frac23}} \left(3 p + \frac{2q^2}{m^{\frac43}} \right) \;,\;\; v_0 = - \frac{2q}{m^{\frac23}} \;,
\ee
where we have introduced the axion combination 
\be
\rho \equiv e_0 - h_0 \sigma + h_1 \nu \;,
\label{rhodef}
\ee
and the flux combination
\be
p \equiv - \frac{e}{m^{\frac13}} - \frac{2q^2}{m^{\frac43}} \;.
\ee
The $0$ indices on the fields denote that this is their value in the minimum of the potential. 

The field variation that we are interested in is along the massive axionic combination $a$ defined as
\be
a = \rho - \rho_0 \;.
\ee
Formally, this is an infinite field space since there is no longer a periodic structure to the axions. The axion decay constant associated to the axion $a$, as in the action (\ref{effaxmonaty}), takes the form
\be
f = \frac12 \left( \left(h_0 s\right)^2 + \frac13 \left(h_1 u\right)^2 \right)^{-\frac12} \;.
\ee
As expected, $f$ has no explicit dependence on the axion $a$, which means that the proper field distance, denoted $\Delta \phi$, appears to grow linearly with $a$. This is in stark contrast to moduli fields whose field distance only grows logarithmically. However, in \cite{Baume:2016psm} it was pointed out that one must account for the backreaction of the axion potential on the axion decay constant $f$ when calculating the proper field distance. This can be done, to a first approximation, by minimizing the potential with respect to the fields that are not $a$, as a function of $a$. So solving
\be
\partial_T V = \partial_u V = \partial_s V = 0 \;,
\ee
for the moduli fields as a function of the axion expectation value $a$. This gives a solution
\be
-\frac{1}{3} h_1 u = h_0 s \simeq \left[ \left( 0.38 a \right)^4 + 0.05 p^3 a^2 + \left(h_0 s_0 \right)^4 \right]^{\frac14} \;.
\ee
The crucial point is that for large axion expectation value we find $f \sim \frac{1}{a}$, and so the proper distance grows only logarithmically in the axion expectation value. This is shown in figure \ref{fig:axback}. Further, since the decay constant, or equivalently the moduli fields, control the mass scale of an infinite tower of states, as discussed in section \ref{sec:dcwith8}, we indeed find that the distance conjecture is satisfied even for the monodromy axion. We also find that this system respects the refined distance conjecture (\ref{rsdc}), since the logarithmic growth of the proper distance with the axion expectation value occurs at sub-Planckian proper distance in field space. Remarkably, this is independent of the choice of fluxes. Indeed, in \cite{Baume:2016psm}, it was shown that this flux independence holds even for complicated Calabi-Yau K{\"a}hler potentials, and for compactifications on twisted-tori (as in the example of section \ref{sec:iias3s3}).  
\begin{figure}[t]
\centering
 \includegraphics[width=0.8\textwidth]{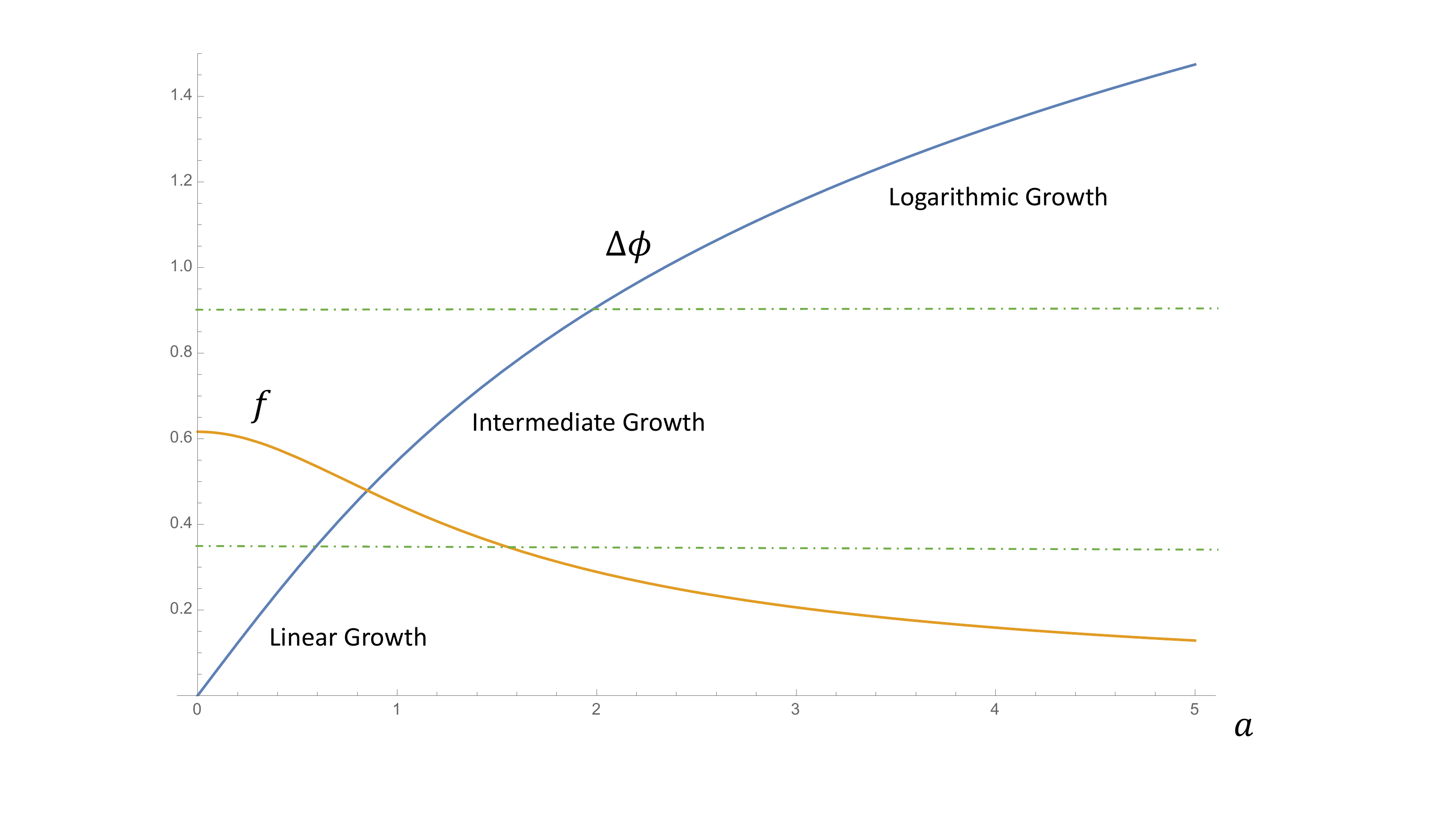}
\caption{Plot showing the axion decay constant $f$ and the proper distance $\Delta \phi$ as a function of the axion expectation value $a$ (all in Planck units), for an axion monodromy setup in type IIA string theory. The initial linear growth $\Delta \phi \sim a$ is modified due to backreaction of the potential energy at $\Delta \phi < 1$, and reaches logarithmic growth $\Delta \phi \sim 0.7 \log a$ by around $\Delta \phi \sim 2$. The plot is shown for flux choices $p=h_0=h_1=1$, but the behaviour of $\Delta \phi$ is independent of this choice.}
\label{fig:axback}
\end{figure}

It is striking that one of the simplest realization of axion monodromy indeed also respects the refined distance conjecture. However, as with other string theory examples of Swampland constraints, it is natural to question if more complicated axion monodromy models may violate the conjecture. One way that this could occur is by realizing an axion monodromy model with a sufficiently large hierarchy of masses between the axion mass $m_a$ and the masses of the moduli which control the decay constant, $m_f$. This may nullify the backreaction sufficiently to maintain a linear growth of the proper field distance with the axion expectation value.\footnote{Note that backreaction was studied much earlier in the context of flattening the effective axion potential in \cite{Dong:2010in}.} The analysis of the magnitude of backreaction effects in flux-based axion monodromy models has been performed in, for example \cite{Hebecker:2014kva,Blumenhagen:2015kja,Hebecker:2015tzo,Valenzuela:2016yny,Blumenhagen:2017cxt,Blumenhagen:2018hsh,Brown:2016nqt,Buratti:2018xjt}. The results depend on the specific models, with some cases possibly allowing for backreaction control \cite{Hebecker:2014kva}, however in the most explicit examples the analysis shows that it is not possible to induce a sufficiently large hierarchy of masses to stop the effect seen in the simple IIA models of backreaction effects becoming important already at sub-Planckian proper distances. Note that the interaction between backreaction, the distance conjecture, and mass hierarchies is general and applies also for non-axionic directions. See for example, \cite{Andriot:2018tmb} for recent studies along this direction. 

A qualitatively different approach to axion monodromy is by utilising localized branes to break the axion periodicity. This is the case for probably the best known model of axion monodromy which is embedded in a KKLT-type setup \cite{McAllister:2008hb}. In that scenario a pair of NS5/anti-NS5 branes are placed in a warped throat and lead to axion monodromy for the axion coming from the RR two-form $C^{(2)}$. Although qualitatively different, similar backreaction issues arise also in brane axion monodromy models. For early work on this see \cite{Conlon:2011qp}. Careful analysis of such brane monodromy models was performed in \cite{McAllister:2016vzi,Kim:2018vgz}. It was found that backreaction issues indeed do become important due to a build up of lower brane charge by the axion (the same effect discussed in section \ref{sec:clstru1} in the D2-D0 context). 

Recently, an axion monodromy model was proposed in \cite{Buratti:2018xjt} which claims to control backreaction sufficiently to allow for a parametrically super-Planckian proper distance by the axion. This comes at the price of requiring the axion to have a spatially varying expectation value. It would be interesting to see if indeed such spatial variations could help.

We note that it is possible to induce a large mass hierarchy between the massive moduli and multiple light fields, so that there is an effective light moduli space. See in particular \cite{Hebecker:2017lxm,Cicoli:2018tcq,Goswami:2018pvk}. However, as shown in \cite{Hebecker:2017lxm}, the field space is such that again the proper distance of geodesics grows only logarithmically at super-Planckian values. 

Finally, it is worth noting that within axion monodromy models there are certain interactions between swampland constraints. In particular, the relation between fluxes and axion periodicities, for example (\ref{exaxperflux}), means that it is possible for the system to develop an instability where a membrane is nucleated that reduces the flux by a unit and thereby decreases the potential \cite{Kaloper:2011jz}. The bound on how far up the axion potential it is possible to go before this decay occurs depends on the membrane tension, and this is bounded by the Weak Gravity Conjecture, which then relates it to the distance conjecture. This bound was analyzed in \cite{Kaloper:2011jz,Ibanez:2015fcv,Hebecker:2015zss,Brown:2016nqt}. Certain interesting constraints on such models were derived, in particular a bound on the allowed field distances was found in \cite{Hebecker:2015zss}
\be
\Delta \phi < \left(\frac{M_p}{m}\right)^{\frac23} \left( \frac{2 \pi f}{M_p}\right)^{\frac13} M_p\;.
\ee
Here $m$ is the mass of the monodromy axion, and $f$ its decay constant. This constraint does not by itself constrain the proper distance to be sub-Planckian, since it is possible to consider $m \ll M_p$.

\section{The Emergence Proposal}
\label{sec:emergence}

Ideally, we would like to derive the Swampland constraints from some underlying microscopic physics. In this section we present a proposal which goes some way towards such an objective. We denote the inter-related collection of ideas as the {\it Emergence Proposal}. The proposal is based on ideas and results in \cite{Harlow:2015lma,Heidenreich:2017sim,Grimm:2018ohb,Heidenreich:2018kpg,Ooguri:2018wrx}.\footnote{Quantitatively we will only match the results in \cite{Heidenreich:2017sim,Grimm:2018ohb,Heidenreich:2018kpg}. The idea in \cite{Harlow:2015lma} of the role of emergent gauge fields in the Swampland is important, but differs quantitatively and qualitatively from the picture we present. In \cite{Ooguri:2018wrx} the idea of emergent potentials was introduced, but we will develop it significantly here.} In order to present a coherent proposal, in this section we will make general statements that are closely related to, but not exactly, those made in \cite{Harlow:2015lma,Heidenreich:2017sim,Grimm:2018ohb,Heidenreich:2018kpg,Ooguri:2018wrx}. We will try to emphasize the differences from the specific results in the published papers when possible. We will also unify and extend the results, both in breadth and detail. With these caveats in mind, the emergence proposal can then be stated as follows. 
\begin{tcolorbox}
{\bf The Emergence Proposal } \;(based on ideas in \cite{Harlow:2015lma,Heidenreich:2017sim,Grimm:2018ohb,Heidenreich:2018kpg,Ooguri:2018wrx})
\newline
\newline
{\it 
The dynamics (kinetic terms) for all fields are emergent in the infrared by integrating out towers of states down from an ultraviolet scale $\Lambda_s$, which is below the Planck scale.
}
\end{tcolorbox}
Practically, what this means is that the renormalisation group boundary conditions at a certain scale $\Lambda_s$ are fixed for all fields such that their kinetic terms are vanishing. Assuming a weakly-coupled description in the infrared, this practically leads to the same results as taking the classical contribution to the kinetic terms in the ultraviolet to be sub-dominant to the 1-loop renormalisation group running down from the ultraviolet. Conceptually, however, the statements are quite different, with the former being the picture advocated in \cite{Harlow:2015lma,Grimm:2018ohb}, and the latter in \cite{Heidenreich:2017sim,Heidenreich:2018kpg}.

The idea of the emergence proposal is that the species scale (\ref{spscb}), the (magnetic) Weak Gravity Conjecture (\ref{mwgc}) and the distance conjectures (\ref{sdc}) and (\ref{rsdc}), can all be thought of as statements relating the kinetic terms of fields in the infrared to tower of states which couple to those fields. Such a relation already exists in quantum field theory because integrating out the towers of states will lead to running for the kinetic terms into the infrared. The proposal is then that the two relations, Swampland constraints and quantum field theory renormalisation, are actually the same. For this to be true, the ultraviolet boundary conditions must be fixed in a certain way, and this way is precisely that they vanish (or are sufficiently sub-dominant to the running). A natural way to explain such boundary conditions is to assume that at this ultraviolet scale the fields are not dynamical. 

It is important to clearly state that the emergence proposal is speculative, it may or may not be correct, and it may or may not be the microscopic physics underlying the Swampland behaviour. Nonetheless, we will see that there are a number of striking relations which follow from it.

\subsection{A toy model for emergent gauge fields}
\label{sec:cpNmodel}

In order to first gain some intuition for emergent fields it is useful to consider a toy model. We will consider the so-called $\mathrm{CP}^{N-1}$-model of complex scalar fields, see for example \cite{DAdda:1978vbw,Witten:1978bc,Rabinovici:2011jj}. The idea that such a toy model could be related to the Weak Gravity Conjecture was first proposed in \cite{Harlow:2015lma}, and see \cite{Schadow} for a detailed review of this. The theory is often studied in two dimensions, where it is conformal in the ultraviolet. We will consider a four-dimensional version, which then has to have a cutoff $\Lambda_{CP}$ (for example by a lattice regularization as is often utilized in this context).  As discussed in \cite{Harlow:2015lma}, the part of interest for us of the analysis of the two-dimensional model then follows through for four dimensions with relatively minor modifications. The model consists of $N$ complex scalar fields $z_i$ which satisfy a constraint
\be
\sum_i z_i^* z_i = 1 \;.
\label{cpnconstraint}
\ee
They have the Lagrangian
\be
\cL = -\frac{N}{c^2} \left(D_{\mu} z_i \right)^* \left(D^{\mu} z^i \right) \;. 
\label{cpnstar}
\ee
Here the index on the $z_i$ is raised and lowered with a delta function, so is just used to denote sums. The parameter $c$ is introduced which acts as a type of coupling constant for the theory. The covariant derivative takes the form
\be
D_{\mu} = \partial_{\mu} - i A_{\mu} \;, 
\ee
where 
\be
A_{\mu} \equiv \frac{1}{2iN} \left( z_i^* \partial_{\mu} z^i - z^i \partial_{\mu} z_i^*\right) \;.
\label{Adef}
\ee
So the gauge field $A_{\mu}$ is not actually a dynamical fields in this theory. One can either just consider the theory in terms of scalars $z_i$, or write it in terms of a non-dynamical $A_{\mu}$ whose equations of motion are algebraic and lead to its definition (\ref{Adef}). Although there is no dynamical gauge field, the theory does have a gauge symmetry
\be
z_i \rightarrow e^{i\alpha\left(x^{\mu}\right)} z_i \;,
\ee
under which $A_{\mu}$ transforms as a gauge field would, thereby ensuring the gauge invariance of the theory. Similarly to $A_{\mu}$, we can introduce another field $\sigma$ which yields the constraint (\ref{cpnconstraint}) as its algebraic equations of motion, so we consider the Lagrangian
\be
\cL = -\frac{N}{c^2}\left[ \left(D_{\mu} z_i \right)^* \left(D^{\mu} z^i \right)  + \sigma \left( z_i^* z^i - 1\right) \right] \;. 
\label{cpnaction}
\ee
In four dimensions we then need to consider this as an effective theory with a cutoff $\Lambda_{CP}$.

The idea is now to treat $A_{\mu}$ as an independent field, so the action as quadratic in the $z_i$, and then integrate them out as a Gaussian in the path integral. So we write
\bea
{\cal Z} &=& \int \cD A \cD \sigma \cD z_i^* \cD z_i \;\mathrm{Exp}\left[-\frac{N}{c^2} \int d^4x \left( z_i^* \left(-D_{\mu} D^{\mu} + \sigma \right) z^i - \sigma\right) \right] \nn \\
&=& \int \cD A \cD \sigma \; \mathrm{Exp}\left[ -N \log \det \left( \frac{- D^2 + \sigma}{\Lambda_{CP^{N-1}}^2}\right) +\frac{N}{c^2} \int d^4x \sigma \right] \;.
\label{pieffact}
\eea
This yields an effective action for $A_{\mu}$ and $\sigma$. We first would like to determine the scalar potential for $\sigma$, for which we set $A_{\mu} =0$, take $\sigma$ to be constant, and go to momentum space
\bea
V\left(\sigma \right) = -N\left[ -\frac{\sigma}{c^2} + \int^{\Lambda_{CP}} d^4k \;\log \left( \frac{k^2+\sigma}{\Lambda^2_{CP}}\right) \right]\;.
\eea
The expectation value of $\sigma$, denoted $\sigma_0$, is then determined by the solution to
\be
\frac{1}{c^2} = \int^{\Lambda_{CP}} d^4k \; \frac{1}{k^2 + \sigma_0} \;.
\ee
Then by choosing the coupling constant $c$ sufficiently strong
\be
c > c_{\mathrm{crit}} = \left[ \int^{\Lambda_{CP}} d^4k \; \frac{1}{k^2} \right]^{-\frac12} \;,
\ee
we find that $\sigma_0 > 0$. The non-zero vacuum expectation value for $\sigma$ implies that the $z_i$ gain a mass
\be
m_z = \sqrt{\sigma_0}\;.
\ee
However, there will be one combination which will remain massless since it is protected by a gauge symmetry, this is $A_{\mu}$. We therefore obtain, at a scale below $m_z$, an effective theory for $A_{\mu}$.

The effective action for $A_{\mu}$ can be deduced by expanding (\ref{pieffact}) to second order in $A_{\mu}$ about the $\sigma=\sigma_0$ background. Performing the expansion then yields an effective action 
\be
\cL = -\frac{1}{4g^2} F_{\mu\nu} F^{\mu\nu} \;,
\label{effFact}
\ee
where $F_{\mu\nu}=\partial_{[\mu}A_{\nu]}$, the usual field strength. The key point is that the field $A_{\mu}$ has developed dynamics in the infrared, it has a kinetic term. It is therefore an example of an emergent dynamical field. As usual, the expansions of determinants in the path integral, as in (\ref{pieffact}), have an interpretation as Feynman diagrams. The relevant leading contributions to (\ref{effFact}) can be thought of as effective 1-loop diagrams. The result is that the gauge coupling for the emergent dynamical field takes the form
\be
\frac{1}{g^2} = \frac{N}{12 \pi^2} \log \left( \frac{\Lambda_{CP}}{m_z}\right) \;.
\label{cpnemrg}
\ee
We see that for large $N$ the emergent gauge field is weakly coupled, justifying the 1-loop type expansion of the determinant. 

The gauge coupling (\ref{cpnemrg}) almost matches exactly the form of the infrared gauge coupling of a fundamental $U(1)$ field coupled to $N$ charged scalars of mass $m_z$ in the infrared, which would look like
\be
\frac{1}{g^2} = \frac{1}{g^2_{\Lambda_{CP}}} + \frac{N}{12 \pi^2} \log \left( \frac{\Lambda_{CP}}{m_z}\right) \;.
\ee
The two agree by setting formally $g^2_{\Lambda_{CP}} \rightarrow \infty$. So it behaves as if the kinetic term was generated purely through integrating out the massive charged fields at 1-loop. 

It was pointed out in \cite{Harlow:2015lma} that the emergent gauge field of the $\mathrm{CP}^{N-1}$-model satisfies the electric Weak Gravity Conjecture in the following sense. We can identify $\Lambda_{CP}$ with the species scale (\ref{spscb}), in which case we obtain
\be
m_z \leq \Lambda_{CP} = \frac{M_p}{\sqrt{N}} < g M_p \;.
\ee
In fact, we see that there are many, though not an infinite number, of states appearing at the scale $m_z$. We also have in this case that the magnetic Weak Gravity Conjecture scale $g M_p$ is identified with the species scale. This is a little strange, we do not expect necessarily that gravity will become strongly-coupled at the Weak Gravity Conjecture scale. We will see that both of these effects are artifacts of this being a toy model, while in string theory settings, see section \ref{sec:evemstrl}, there will indeed be an infinite tower of states associated with the magnetic Weak Gravity Conjecture scale, and the species scale will be separated from it. Nonetheless, the key lesson to draw from this analysis is that emergent fields naturally behave as if their kinetic terms arise purely from integrating out many states at one-loop.

\subsection{The Swampland constraints from emergence}
\label{sec:emeswam}

In this section we will adopt the idea of emergence, in particular that the kinetic terms for emergent fields are induced at one-loop in the infrared from integrating out states, but apply it in a different way to the previous section which is more in line with how it is realized in string theory. This analysis then follows the results of \cite{Heidenreich:2017sim,Grimm:2018ohb,Heidenreich:2018kpg,Corvilain:2018lgw}. While the $\mathrm{CP}^{N-1}$-model is a toy model for an emergent gauge field, in this section we will consider {\it all} fields, including even gravity, to have emergent dynamics from some scale $\Lambda_s$. It is not clear what this really means at the ultraviolet scale, since we do not know of theories which can lead to emergent gravity in this sense. Indeed the Weinberg-Witten theorem \cite{WEINBERG198059} says that whatever such a theory might look like, it would not be some local quantum field theory. Nonetheless, at the level of analysis that we will perform, we only need to keep some track of the running of the kinetic terms into the infrared, which we will assume to be given by the prescription that we treat this running as if the field was fundamental but set the ultraviolet value of the kinetic terms to vanish (or in the picture of \cite{Heidenreich:2017sim,Heidenreich:2018kpg}, set it to be of order one).

We will consider a tower of particles, with an associated mass scale $m$. So the mass of the $\mathrm{n}^{\mathrm{th}}$ state is $m_n = m n$. The particles will be taken to have also increasing charge, so the quantized charge is $q_n = n$. Now let us impose the emergence of gravity, in this rather vague and mild sense where we just keep track of the order of magnitude of one-loop effects, so
\be
\left. M^2_p \right|_{\mathrm{IR}} \sim \cancelto{0}{\left. M^2_p \right|_{\Lambda_s}} \;\;\;\; + N \Lambda_s^2 \;.
\ee
We have set the kinetic term, so $M_p$, at the ultraviolet scale $\Lambda_s$ to vanish. We therefore recover the relation $\Lambda_s = \frac{M_p}{\sqrt{N}}$ which identifies the ultraviolet scale with the species scale (\ref{spscb}). This relation between one-loop effects and the species scale is well known, and naturally leads to the statement that $\Lambda_s$ is a scale where gravity becomes strongly coupled. Note that it is therefore a rather natural scale at which to expect that kinetic terms for fields may not make much sense as they involve explicit spatial derivatives, so it is a natural scale from which dynamics may be emergent. In any case, if we are unhappy with an abstract idea of emergent gravity, then we can simply take it as an input that the ultraviolet scale should be the species scale. 

Once the ultraviolet scale is fixed in this way, it is simple to calculate a relation between $N$ and $m$. Specifically, we have $N m \sim \Lambda_s$ and so
\be
N \sim \left(\frac{M_p}{m} \right)^{\frac23} \;.
\label{Nmrel}
\ee

Now let us consider an emergent gauge field in this scenario. Our prescription suggests that we need to take the gauge coupling to satisfy 
\be
\left.\frac{1}{g^2} \right|_{\mathrm{IR}}  \sim \cancelto{0}{\left. \frac{1}{g^2} \right|_{\Lambda_s}} \;\;\;\; + \sum_i^N q_i^2 \log \left( \frac{\Lambda_s}{m_i} \right) \;.
\ee
This yields for the infrared gauge coupling
\be
\frac{1}{g^2} \sim N^3 \sim \left(\frac{M_p}{m} \right)^{2} \;,
\ee
which is exactly the magnetic Weak Gravity Conjecture (\ref{mwgc}). So the scale $g M_p$ is associated with the mass scale of an infinite tower of states.

Next we consider emergent dynamics for a scalar field $\varphi$. We denote its kinetic term pre-factor as $g_{\varphi\varphi}$, and impose
\be
\left.g_{\varphi\varphi}\right|_{\mathrm{IR}}  \sim \cancelto{0}{\left.g_{\varphi\varphi}\right|_{\Lambda_s}} \;\;\;\; + \sum_i^N \left(\partial_{\varphi} m_i \right)^2 \log \left( \frac{\Lambda_s}{m_i} \right) \;.
\ee
Here we have used that the coupling of a scalar to a particle is given by the derivative of its mass, as discussed in section \ref{sec:wgcsca}. We therefore obtain for the infrared value
\be
g_{\varphi\varphi} \sim N^3 \left( \partial_{\varphi} m\right)^2 \sim \left(  \frac{M_p\partial_{\varphi} m}{m} \right)^2 \;.
\ee
Now consider the proper distance in the field space of $\varphi$, moving from $\varphi_i$ to $\varphi_f$, as measured by this emergent metric. It takes the form
\be
\Delta \phi = \int^{\varphi_f}_{\varphi_i}\sqrt{g_{\varphi\varphi} } \;d \varphi \sim M_p \int^{\varphi_f}_{\varphi_i} \left|\partial_{\varphi} \left( \log m \right) \right|\;d \varphi \sim M_p \log \left( \frac{m\left(\varphi_i\right)}{m\left(\varphi_f\right)} \right) \;.
\ee
We have taken, by definition of the path in field space, the mass $m\left(\varphi\right)$ to be a decreasing function along the path. Rearranging, we obtain
\be
m\left(\varphi_f\right) \sim m\left(\varphi_i\right) e^{- \alpha \frac{\Delta \phi}{M_p}} \;,
\ee
with $\alpha \sim \cO\left( 1\right)$, which is precisely the distance conjecture.

It is quite striking how the emergence proposal relates the species scale, the Weak Gravity Conjecture and the distance conjecture in this way. Indeed, we may say that the evidence for the Swampland program can be interpreted as evidence for the emergence proposal. However, we should keep in mind that this analysis is zeroth order, including assuming a particular structure for the tower of states.

Note that one might wonder how the emergence proposal could work for asymptotically free non-Abelian theories. The natural answer is that the tower of charged states corresponds to increasingly large representations, see for example \cite{Heidenreich:2017sim}, which turns the running into strong coupling in the ultraviolet. Another way to view this is to apply the proposal to the Cartan sub-algebra. 

\subsection{Evidence for emergence in String Theory}
\label{sec:evemstrl}

In this section we will consider evidence for emergence in string theory. This will follow the analysis and results of \cite{Grimm:2018ohb}. We focus on the setting of type IIB string theory compactified on a Calabi-Yau manifold, as discussed in section \ref{sec:cymod}. This setting is in many ways the simplest one to test the proposal because it maintains ${\cal N}=2$ supersymmetry which means that the corrections to the kinetic terms of the fields in the vector multiplets are  actually one-loop exact in perturbation theory. It is also a special setting in string theory because in any direction in moduli space the relevant tower of states are always particle-like, specifically D3 branes wrapping three-cycles. In other string theory settings, certain limits in moduli space will lead to extended objects becoming light, and it is not clear how to implement such a situation into the emergence proposal (though see section \ref{sec:empot}). 

The string theory setting is perfect for testing the emergence proposal because it has been known since \cite{Strominger:1995cz} that string theory automatically integrates out wrapped D3 branes. This is true both for the complex-structure moduli space but also for the gauge couplings of the closed-string $U(1)$ gauge fields in the vector multiplets. The most famous example of this being the conifold singularity. This is very much analogous to the understanding of Seiberg-Witten theory in terms of integrating out monopoles \cite{Seiberg:1994rs}. So when we study the moduli space of string theory, we know that already the tower of states has been integrated out. The test we can perform is then to compare the integrating out by hand analysis, as in section \ref{sec:emeswam}, with the properties of the moduli space. The emergence proposal would predict that they should give the same answer. This is opposed to say if the tower of states when integrated out only gave a sub-leading contribution to an already present moduli space geometry in the ultraviolet. Then the integrating out calculation should not match what the moduli space actually looks like in the infrared, which is what string theory gives us.\footnote{The moduli space of the Calabi-Yau has integrated out all the wrapped D3 branes down to zero energy. To see this we can note that the singularities in the moduli space only occur when wrapped D3 branes become massless.} 

Let us consider approaching an infinite distance locus in Calabi-Yau complex-structure moduli space, parameterized in section \ref{sec:cymod} as the limit $\mathrm{Im\;}t \rightarrow \infty$. The leading behaviour of the mass of BPS states takes the form (\ref{asycech}). We can write this in a simplified form by splitting the charge as
\be
{\bf q} = n {\bf q}_0 + m {\bf q}_1 \;,
\ee
where ${\bf q}_0$ and ${\bf q}_1$ are defined such that $S\left({\bf q}_0,{\bf a}_0 \right) \neq 0$ and $S\left({\bf q}_1,N{\bf a}_0 \right) \neq 0$, and all other inner products vanishing.\footnote{Note that in $S\left({\bf q}_1,N{\bf a}_0 \right) \neq 0$, and in section \ref{sec:cymod}, $N$ refers to the monodromy matrix not the number of states in the tower.} The $n$ and $m$ are integers which specify states in the tower. This decomposition picks out a (specific representative, not completely general) tower of states whose mass takes the form
\be
M_{(n,m)} \sim \frac{n}{\left(\mathrm{Im\;}t \right)^{\frac{d}{2}}} + \frac{m}{\left(\mathrm{Im\;}t \right)^{\frac{d-2}{2}}} \;.
\ee
For infinite distance loci where $d < 3$, we require $m=0$ in order for the state to become massless. In such cases we therefore find a tower of states with integer increasing masses. Integrating them out, as in the general analysis of section \ref{sec:emeswam}, leads to $g_{t\bar{t}} \sim \left(\mathrm{Im\;}t \right)^{-2}$ which is indeed the correct asymptotic behaviour of the metric on the moduli space. The case $d=3$ is more involved because there are three types of towers which can become massless
\be
M_{n,0} \sim \frac{n}{\left(\mathrm{Im\;}t \right)^{\frac{3}{2}}} \;,\;\; M_{n,1} \sim \frac{1}{\left(\mathrm{Im\;}t \right)^{\frac{1}{2}}}  \;, \;\; M_{0,m} \sim \frac{m}{\left(\mathrm{Im\;}t \right)^{\frac{1}{2}}} \;.
\ee
The first and third types, $M_{n,0}$ and $M_{0,m}$, have the standard integer increasing mass spectrum. The second type of tower $ M_{n,1}$ is a compressed spectrum. It is analogous to the toy $\mathrm{CP}^{N-1}$-model studied in section \ref{sec:emeswam}, where the masses of the massive states were degenerate. It still, nonetheless, leads to the correct behaviour for the scalar field kinetic term upon integrating out \cite{Grimm:2018ohb}. Such a tower is associated to a monodromy orbit as discussed in section \ref{sec:cymod}, but the other tower types may also be associated to more general monodromy orbits as studied in \cite{Grimm:2018cpv}. It is interesting to note that such a compressed spectrum may arise. However, the most likely scenario is the tower of type $M_{n,0}$ is populated by BPS states, and it is the one forming the lightest tower and leading behaviour.

Let us test now the emergence proposal for the gauge fields in the vector multiplets. First we can consider towers of type $M_{n,0}$, these have integer increasing charges and masses. They are charged under the gauge field with gauge coupling type $g_{(0)}$ according to the notation of section \ref{sec:cymod} \cite{Grimm:2018ohb}. Integrating out such a tower reproduces the general results of section \ref{sec:emeswam} leading to gauge coupling behaviour
\be
\frac{1}{g_{(0)}^2} \sim N^3 \sim\frac{1}{\left(\Delta m\right)^2}\sim \left(\mathrm{Im\;}t \right)^{d} \;,
\ee
where $\Delta m$ denotes the mass separation in the tower. 
Comparing with table \ref{tab:gauvan}, for $g_{(0)}$, we see that this reproduces precisely the behaviour of the gauge coupling, as predicted by the emergence proposal. For $d=3$ there is also another type G gauge field with gauge coupling $g_{(1)}$. The two possible towers charges under it are $M_{n,1}$ and $M_{0,m}$. The latter gives a tower of increasing charges and so goes as $N^3$, but now $\Delta m \sim \left(\mathrm{Im\;}t \right)^{-\frac12}$ and so we find
\be
\frac{1}{g_{(1)}^2} \sim \mathrm{Im\;}t  \;,
\label{g1Nrun}
\ee
which matches table  \ref{tab:gauvan}. The tower $M_{n,1}$ has $\Delta m \sim \left(\mathrm{Im\;}t \right)^{-\frac32}$ but the charges under the $U(1)$ do not increase up the tower. Integrating out such a tower leads to running as $N$ rather than $N^3$, again yielding (\ref{g1Nrun}) and matching the geometry result. 

The quantitative evidence for emergence in the string theory setting is striking. Further, similar results have been found for six-dimensional theories and ${\cal N}=1$ settings in \cite{Lee:2018urn,Lee:2018spm,Lee:2019tst}. However, it is important to state that we are only matching the leading scaling behaviour at weak coupling. It is possible that there could be an ultraviolet related scaling of the same type which is unrelated to integrating out the states. To rule this out would require matching the precise coefficient of the integrating out effects with the Calabi-Yau moduli space results, which has not been done.\footnote{Also we have not discussed type F $U(1)$ gauge fields, which are discussed in \cite{Grimm:2018ohb}. For these there are some potential mismatches between integrating out and the geometry, but such type F $U(1)$s are less well understood and so the results are uncertain.}

The string theory setting also clarifies another subtlety with the emergence proposal. The question is at what scale should we evaluate the masses of the states that are integrated out. As is always the case, we should evaluate the masses on their pole masses which effectively means, since below this scale the mass essentially stops running, we should evaluate the masses according to their values in the infrared. The formula for the BPS states masses (\ref{asycech}) is indeed their mass in the infrared, and so is the correct expression to use for the masses of the states in the integrating out calculation. So string theory was very useful here because it provides the expression for the masses of the tower of states having already included the effects of their running down from the ultraviolet scales. 

Note that there is a large degeneracy of states at each mass level in string theory, which is not accounted for in this simple analysis. See \cite{Heidenreich:2017sim} for an analysis of some highly degenerate string spectra in this context.

\subsection{Emergence and dimensional reduction}

(Note that this section is not a review of published work, and is mostly new material.) If the emergence proposal is correct, then it suggests re-examining some of the canonical examples for the weak gravity and distance conjectures. In particular, we can consider the case of dimensional reduction on a circle as discussed in section \ref{sec:fcsdc}. There the Swampland conjectures were respected with the Kaluza-Klein tower becoming exponentially light as a function of the radion field. But from the emergence perspective this seems like a coincidence. It would be strange if classical dimensional reduction on a circle integrated out the Kaluza-Klein modes.\footnote{Indeed, already in \cite{Ooguri:2006in} the idea of emergence was hinted at, but was ruled out as general due to the simple Kaluza-Klein example. Note that also there are subtleties with how to interpret integrating down from the species scale for Kaluza-Klein towers, since a gauge symmetry relates the full tower. Also note that, in the interpretation of \cite{Heidenreich:2017sim,Heidenreich:2018kpg} in terms of reaching strong coupling rather than full emergence, it may be that there is no conceptual problem with classical dimensional reduction.} It is fair to say that we do not fully understand dimensional reduction within the emergence proposal, but in this section we will at least outline a proposal for understanding it. In general, for dimensional reduction, the natural interpretation we propose is that already in the higher dimensional theory the kinetic terms are emergent, and then one simply dimensionally reduces them. For the radion, the kinetic term is part of the higher-dimensional Einstein-Hilbert term, the kinetic term of the higher dimensional graviton. The nice thing is that the emergence proposal is consistent under dimensional reduction which means that one can recover the higher-dimensional emergence from the lower-dimensional picture. Indeed, this property would be rather crucial for explaining why we can apply the emergence proposal in a four-dimensional context within string theory which is a ten-dimensional theory. We will see that it will also shed light on why Kaluza-Klein modes happen to also satisfy the distance conjecture. This analysis is mostly based on unpublished work \cite{erandaniel}, and we refer also to \cite{Corvilain:2018lgw,Bonetti:2013cza,Grimm:2018weo} for relevant work on this topic. 

We will consider the case of a scalar field $\varphi$, rather than graviton-radion, for now. We consider dimensional reduction from a 5-dimensional theory on a circle of radius $R$. First we can check emergence in five dimensions. The 1-loop contribution to the scalar kinetic term behaves as
\be
g_{\varphi\varphi}^{\mathrm{1-loop}} \sim \Lambda_{5} N^3_5 \left( \partial_{\varphi} m \right)^2 \;.
\label{1loop5d}
\ee
Here we consider a tower with mass $m_{n_T}\left(\varphi\right) = n_T m\left(\varphi\right)$ and define the 5-dimensional species scale as 
\be
\Lambda_5 = \frac{M_5}{N_5^{\frac13}} \;,
\ee
where $M_5$ is the five-dimensional Planck mass, and $N_5$ is the number of states in the tower below the five-dimensional species scale $\Lambda_5$. Using $\Lambda_5\sim N_5 m$ we can solve
\be
\Lambda_5 \sim \left(m M_5^3 \right)^{\frac14} \;.
\ee
Then substituting into (\ref{1loop5d}) we have 
\be
g_{\varphi\varphi}^{\mathrm{1-loop}} \sim \frac{M_5^3 }{m^2}\left( \partial_{\varphi} m \right)^2 \;,
\ee
which, as in section \ref{sec:emeswam}, leads to the distance conjecture, so the exponential dependence in $m$ on the proper distance in $\varphi$. 

Now we can dimensionally reduce this on a circle, and recover the distance conjecture from a four-dimensional perspective. We will denote the four-dimensional zero-mode of $\varphi$ as $\varphi$ again. It will then couple to the tower of states and also their Kaluza-Klein modes.\footnote{Keeping the Kaluza-Klein modes of the tower of states is similar to the analysis in \cite{Dvali:2009ks,Klaewer:2018yxi} where Kaluza-Klein modes of oscillator modes were important.} So the relevant tower takes the form
\be
m^2_{n_T,n_{KK}} = n_T^2 m\left(\varphi\right) ^2 + n^2_{KK} m^2_{KK} \;,
\label{mtk}
\ee
where $m_{KK}=\frac{1}{R}$. We have not gone to the Einstein frame and so have the relation
\be
\frac{M_5^3}{m_{KK}} = M^2_p \;,
\ee
where $M_p$ is the four-dimensional Planck mass. Now we would like to determine the number of states $N_4$ below a four-dimensional species scale $\Lambda_4$. The mass (\ref{mtk}) determines $N$ as the number of lattice points inside the ellipsoid of appropriate radii. This ellipsoid can be approximated as a rectangle, so the number of points is approximately
\be
N_4 \sim \left( \frac{\Lambda_4}{m}\right) \left( \frac{\Lambda_4}{m_{KK}}\right) \;.
\ee
This yields for the species scale
\be
\Lambda_4 \sim \left(M_p^2 m m_{KK} \right)^{\frac14} \sim \Lambda_5 \;.
\ee
So the lower dimensional species scale is the same as the higher dimensional one, which makes the four-dimensional analysis consistent. If we now look at the four-dimensional 1-loop term
\be
g_{\varphi\varphi}^{\mathrm{1-loop}} \sim \left( \frac{\Lambda_4}{m}\right)^3 \left( \frac{\Lambda_4}{m_{KK}}\right) \left( \partial_{\varphi} m \right)^2 \sim \frac{M_p^2}{m^2} \left( \partial_{\varphi} m \right)^2  \;.
\label{1loop4d}
\ee
Here it is crucial to note that the derivative acting on the mass picks out only the dependence on the tower level $n_T$ and not $n_{KK}$. We therefore recover the emergence result in the lower dimensional theory. 

Let us now consider if it is possible to parametrically separate the tower mass scale $m$ from the Kaluza-Klein mass scale $m_{KK}$ such that the emergence picture in the lower dimensional theory comes from integrating out only of these types of states. The species scale for just a Kaluza-Klein tower is the higher dimensional Planck scale, and so it is not consistent to take the tower mass scale higher than it, which implies we cannot discard the tower states. The species scale for just the tower states is $\left(M_p m^2 \right)^{\frac13}$. Taking the Kaluza-Klein scale larger than this can be shown to imply that the Kaluza-Klein scale is also higher than the five-dimensional species scale $\Lambda_5$. But this would be inconsistent since it would involve a circle radius which is smaller than the scale at which gravity is strongly coupled. We therefore find that the two towers cannot be decoupled in a consistent setting. 

This non-decoupling of the towers then suggests a possible explanation as to why the Kaluza-Klein modes satisfy the distance conjecture for the radion. The tower of states which induce the kinetic term for the radion (the Einstein-Hilbert term) already in the five-dimensional theory should satisfy the distance conjecture for the four-dimensional radion. But the Kaluza-Klein tower cannot be consistently decoupled from the mass scale of this tower, and so it must also satisfy the distance conjecture. 

In the type IIB string theory on Calabi-Yau setting the kinetic terms for say the gauge fields in the vector multiplets came from dimensionally reducing the kinetic terms for the Ramond-Ramond four-form $C^{(4)}$. Then it is natural to expect that these would already be emergent from D3-branes in ten dimensions, and the dimensional reduction of the D3 towers lead to the tower of charged BPS states in four dimensions.\footnote{Note that due to the structure of Calabi-Yau spaces there is only a single tower of states, rather than a lattice structure as in (\ref{mtk}).} These are then integrated out to recover the dimensionally reduced kinetic terms (for the full vector multiplets).

\subsection{Emergent potentials}
\label{sec:empot}

(Note that this section is not a review of published work, and is mostly new material.) One way to think about the emergence proposal is that the tower of states offers some sort of a dual description of the physics. In the toy model case this was clear, since the relation (\ref{Adef}) can be thought of as a duality. In string theory indeed we often interpret infinite distances in moduli space as regions where a description in terms of light states is useful, and is dual to a description in terms of other degrees of freedom. Indeed, one could even say that duality is a necessary consequence of the emergence proposal, since the existence of the direction in scalar field space is only due to the existence of the tower of states, and so the dual description, in the first place.

A duality would suggest that perhaps not only kinetic terms for fields are captured by properties of the tower of states, but also potentials for them. This was proposed as a possibility in \cite{Ooguri:2018wrx}, and also mentioned in \cite{Grimm:2018cpv}. Let us see a motivation from ${\cal N}=2$ supersymmetry. Consider a compactification of type IIB string theory on a Calabi-Yau with Ramond-Ramond fluxes turned on, so that a scalar potential is induced for the complex-structure moduli. Then this potential will be related to the gauge kinetic matrix for the gauge field in the vector multiplets by (\ref{VRRQ2}). Since we recovered the leading weak-coupling behaviour for the gauge kinetic matrix from integrating out the tower of charged BPS states, we see that the information on the potential should be recovered in the same way.

It is informative to think about the origin of the supersymmetry relation (\ref{VRRQ2}) in string theory. The right-hand-side is associated to charged particles, which in string theory are D3 branes wrapping 3-cycles, and gauge fields which are coming from expanding the Ramond-Ramond four-form $C^{(4)}$ in a three-form basis. The left-hand-side is a potential induced by turning on three-form Ramond-Ramond flux, which is threading three-cycles again, and so the three-cycle geometry captures both the structure of the potential and the gauge kinetic matrix. In the gauge field case we associated a tower of charged states to the gauge kinetic matrix, but the potential should instead have associated to it a tower of D-brane states which are linked to the Ramond-Ramond two-form $C^{(2)}$, so D1 branes. This would therefore suggest that the relevant tower of states to consider for the potential should be D1 branes, or more naturally their magnetic dual D5 branes which are wrapping three-cycles and forming domain walls in the external spacetime. Indeed, it is well known that domain walls are related to potentials, so this is a natural assignment. This leads to the striking direction that perhaps potentials should be thought of as emergent from integrating out towers of light domain walls. It is not clear exactly what this means, but we can make the analogy with the emergence of kinetic terms for fields more precise as follows. 

The D5 branes couple naturally to a Ramond-Ramond six-form $C^{(6)}$, and expanding this in an internal three-form basis leads to three-forms with legs in the external spacetime. The relation of these three-forms to the flux potential is an example of a general relation between three-forms and potentials. At zeroth level, the relation can be understood as the statement that a four-form flux filling all of spacetime is indistinguishable from a cosmological constant. Such a relation has been studied in early works \cite{Duff:1980qv,Hawking:1984hk,Duff:1989ah}, and more recently played an important role in the development of the relation between the string theory landscape and the cosmological constant \cite{Bousso:2000xa}. In particular, the relation between axion potentials and three-forms as been developed in \cite{Dvali:2004tma,Dvali:2005an,Dvali:2005zk,Dvali:2013cpa} and utilised in the context of inflation in \cite{Kaloper:2008fb,Kaloper:2011jz}. This relation has been extensively studied in string theory in a recent program of work \cite{Dudas:2014pva,Bielleman:2015ina,Escobar:2015ckf,Ibanez:2015fcv,Garcia-Valdecasas:2016voz,Carta:2016ynn,Herraez:2016dxn,Valenzuela:2016yny,Bielleman:2016olv,Blumenhagen:2017cxt,Tenreiro:2017fon,Valenzuela:2017bvg,Farakos:2017jme,LandeteMarcos:2017wky,Farakos:2017ocw,Herraez:2018vae,Bandos:2018gjp,Nitta:2018yzb,Escobar:2018tiu,Escobar:2018rna}. 

Let us introduce the idea through a simple toy model, following the notation of \cite{Kaloper:2008fb}.  We consider the action
\be
S = \int d^4x \sqrt{-g} \left[ \frac{R}{2} - \frac12 \left(\partial a\right)^2 - \frac{1}{48} F_{\mu\nu\lambda\sigma}F^{\mu\nu\lambda\sigma} + \frac{\mu}{24} a \frac{\epsilon^{\mu\nu\lambda\sigma}}{\sqrt{-g}}F_{\mu\nu\lambda\sigma} + \frac{q}{24} \frac{\epsilon^{\mu\nu\lambda\sigma}}{\sqrt{-g}} \left(F_{\mu\nu\lambda\sigma} - 4 \partial_{\mu} \left(c_3\right)_{\nu\lambda\sigma} \right) \right] \;.
\label{toyaxpot}
\ee
Here $a$ is an axion (pseudo-)scalar field, $F_{\mu\nu\lambda\sigma}$ is a spacetime filling four-form which, using a Lagrange multiplier $q$, is taken as the field strength of a three-form $c_3$. There is a coupling between the axion and the field-strength parameterized by $\mu$, and $\epsilon^{\mu\nu\lambda\sigma}$ is the Levi-Civita tensor density. Then integrating out $F$ leads to the action \cite{Kaloper:2008fb}
\be
S = \int d^4x \sqrt{-g} \left[ \frac{R}{2} - \frac12 \left(\partial a\right)^2 - \frac12 \left(q+ \mu a \right)^2 \right] \;.
\ee
We therefore see that a scalar potential has been induced for the axion. The potential is invariant under a transformation $a \rightarrow a + 1$ and $q \rightarrow q - \mu$. This transformation is analogous to (\ref{exaxperflux}). Indeed, all flux-induced potentials in string theory can be induced by a more complicated version of (\ref{toyaxpot}), in the notation of \cite{Herraez:2018vae} we can write the flux potential through four-forms as 
\be
S = \int d^4 x \left[ \frac{R}{2} \star 1- \frac{1}{8} Z_{AB} F^A \wedge \star F^B + \frac{1}{4} F^A \rho_A + ... \right] \;. 
\label{fffpst}
\ee
Here $F^A$ are four-form field strength where the index $A$ runs over all the fluxes of the compactification. The $\rho_A$ are functions of only the axion fields of the theory and the fluxes, while $Z_{AB}$ is a function of only the geometric moduli (and dilaton) fields.\footnote{This factorisation between moduli and axions is strictly only true at large distances in moduli space where the axion shift symmetry becomes manifest, and is broken by exponentially small corrections.} Integrating out the four-forms yields the scalar potential
\be
V = \frac18 Z^{AB} \rho_A \rho_B \;.
\ee
Writing scalar potentials in string theory in the form (\ref{fffpst}) suggests a natural interpretation in terms of emergence. The kinetic terms for the three-forms could be induced by integrating out towers of light domain walls. It would be very interesting to understand better how to actually make such an integrating out procedure quantitative. Particularly so since the ideas presented in \cite{Ooguri:2018wrx} would suggest that the microstates of the tower may account for the entropy of de Sitter space at large distances in field space.

\section{More Swampland conjectures}
\label{sec:moreswamp}

In this section we discuss a number of Swampland conjectures which are more than variations on the weak gravity or distance conjectures. Nonetheless, we will see that they will still be related in certain ways to the same underlying ideas. 

Let us first discuss a number of conjectures briefly, before studying some other ones in more detail. In \cite{Vafa:2005ui} it was conjectured that the rank of gauge groups, or more generally the number of matter fields, is bounded in string theory to be finite. Another conjecture is that all gauge groups are compact \cite{Banks:2010zn}. In \cite{Montero:2017yja} it was conjectured that appropriate Chern-Simons terms are always present to break any generalized global symmetries, as discussed in section \ref{sec:complconj}. More precisely, in certain compactifications to low-dimensional theories, the charged objects do not break all the global symmetries unless the conjectured Chern-Simons terms are present. In \cite{Aldazabal:2018nsj} it was conjectured that the spectrum of charged states in compactifications of the Heterotic string is such that there are always charged fields present which, through obtaining expectation values, allow to cancel any Fayet-Iliopoulos terms. In \cite{Cecotti:2018ufg} it was conjectured that in ${\cal N}=2$ supergravity with no vector multiplets, the $\theta$-angle for the graviphoton is fixed to be $0$ or $\pi$. Finally, an obviously central conjecture, was made in \cite{DouglasStrings2005} which is that the number of string vacua is finite. 

In section \ref{sec:complconj} we discussed the completeness conjecture which states that all matter representations are realised in quantum gravity at some mass scale. An interesting question is what types of massless matter can be realised in string theory. In particular, one can compare the possible consistent matter spectra which are anomaly free, with known constructions in string theory and see if all consistent possibilities can be realised. Early work on this program was comparing anomaly-free six-dimensional supergravity theories with possible constructions from F-theory, see for example \cite{Kumar:2009ae,Seiberg:2011dr,Kumar:2009us,Kumar:2009ac,Kumar:2010ru,Taylor:2018khc}. There is also some work on understanding the possible massless matter spectra for four-dimensional F-theory compactifications, see in particular \cite{Lawrie:2015hia,Baume:2015wia,Cvetic:2017epq} for systematic attempts at classification, and \cite{Maharana:2012tu,Weigand:2018rez} for reviews of F-theory constructions. The program is particularly interesting for very high-dimensional theories, since there are not so many such consistent theories, see for example \cite{Garcia-Etxebarria:2017crf} for such studies.\footnote{In some sense an opposite, but related, idea is whether string theory can lead to constructions of new theories that were not known from the quantum field theory perspective. An example are ${\cal N}=3$ superconformal field theories in four dimensions \cite{Garcia-Etxebarria:2015wns}, and even more exotic ${\cal N}=7$ supergravities \cite{Ferrara:2018iko}.}

An interesting way to possibly think about the Swampland is to look for consistency constraints on quantum field theories which arise from placing them on particular curved backgrounds. This may be thought of as related to the Swampland in the sense that coupling the theories to gravity can naturally be interpreted as the statement that they should be valid on any spacetime background. This leads then to additional constraints on the theory \cite{Kapustin:2014dxa,Hsieh:2015xaa,Witten:2015aba}. Implications of these constraints, termed Dai-Freed anomalies, were studied for possible particle physics models in \cite{Wang:2018cai,Wang:2018qoy,Garcia-Etxebarria:2018ajm}.

Another Swampland-related program of study is what type of superconformal theories can be constructed in string theory. See \cite{Heckman:2018jxk} for an extensive review of this. In some sense this program is somewhat orthogonal to the Swampland because such conformal field theories are engineered precisely in the decoupling limit of gravity, so effectively taking the Calabi-Yau to be non-compact. However, it may be that some remnant of the gravitational constraints still manifests \cite{Heckman:2018jxk}.

\subsection{Moduli space conjectures}
\label{sec:modsp}

In this section we will consider moduli spaces in string theory, defined as field spaces spanned by exactly flat directions. There are a number of conjectures about properties of moduli spaces in string theory that are related to, but not equivalent to the distance conjecture. Moduli spaces are known to be closely tied to the existence of extended ${\cal N}=2$ supersymmetry (at least 8 supercharges for general dimensions). One may then consider a possible zeroth order conjecture, that in quantum gravity exact moduli spaces with finite volume only exist for 8 supercharges or more. We are not aware of any counter-examples to this, but clearly a significant investigation would be required to make this a reasonable proposal.  

It was conjectured in \cite{DouglasStrings2005} that the volume of moduli spaces in string theory is finite (for earlier work along these lines see  for example \cite{Horne:1994mi}). We can think of the moduli space as defining a sigma model with a target space (in string theory this would be the Calabi-Yau). Then the volume is measured with respect to the metric on the subspace of Ricci-flat target-space metrics, which means we look at deformations of the target space which keep it Ricci-flat. It was noted that the moduli space composed purely from the radius of a circle has infinite volume, but in the presence of supersymmetry such a one real-dimensional moduli space cannot exist.\footnote{For compactifications on a circle there will also be a pseudo-scalar associated to the Wilson line of the graviphoton which will have a vanishing kinetic term in the infinite distance limit, thereby preserving the finiteness of the moduli space volume.} In the case of more than 8 supercharges, it is possible to relate the finite volume conjecture to the conjecture (in \cite{Banks:2010zn}) that all gauge groups are compact \cite{Cecotti:2015wqa}. 

While moduli spaces are conjectured to have finite volume, they are also conjectured to always be non-compact \cite{Ooguri:2006in}, and therefore it must be that near infinite distance loci some orthogonal direction has to shrink to zero size so as to maintain a finite volume.\footnote{This direction is related to the monodromy action discussed in section \ref{sec:cymod}.} This cone-type structure is related to two other conjectures made in \cite{Ooguri:2006in}. First that the scalar curvature of the moduli space approaching infinite distance loci is always negative. Second, that there are no finite minimal length one-cycles in the moduli space within a given homotopy class. Roughly speaking, this means that if there is an axion in the moduli space, then one can always make its decay constant vanish at infinite distance. Again, with more than 8 supercharges, these properties can be related to properties of gauge symmetries, such as the completeness conjecture, as shown in \cite{Cecotti:2015wqa}. 

\subsection{Non-SUSY AdS conjecture}

In \cite{Ooguri:2017njy} a slightly sharpened version of the electric Weak Gravity Conjecture (\ref{ewgc}) was proposed. It was argued that in the absence of supersymmetry the inequality should be replaced by a strict inequality
\be
m < \sqrt{2} g q M_p \;.
\label{rwgc}
\ee
This is certainly reasonable, the left hand side is related to Poincar\'e symmetry, while the right hand side is an internal symmetry, and by the Coleman-Mandula theorem the only symmetry that can relate the two is supersymmetry. In the absence of an exact local symmetry, it is difficult to see how an equality between the two sides could be maintained at all loop order. The refined version of the Weak Gravity Conjecture (\ref{rwgc}) becomes particularly interesting when generalized to higher dimensional objects as in (\ref{wgcpd}). This is because there is a process called fragmentation of anti-de Sitter space \cite{Maldacena:1998uz}. Specifically, in anti-de Sitter space supported by flux, so we can consider $AdS_{p+2}$ which has a spacetime filling $\left(p+2\right)$-form flux through it, there exists an instanton solution corresponding to a $p$-brane nucleating and expanding out do the boundary of the AdS space. The end result is the AdS space with one less unit of flux and a $p$-brane. This process is illustrated in figure \ref{fig:adsfrag}. It is also possible to consider many such branes in which case the AdS space splits into two AdS spaces with fluxes on them that sum to the original flux. The question of whether the process occurs dynamically relates to the tension and charge of the $p$-brane. On one hand the tension of the brane will want it to contract, but the charge will drive an expansion. If the charge is greater than the tension the expansion, and AdS fragmentation, will occur. The refinement of the Weak Gravity Conjecture states that in the absence of supersymmetry there always exists such a brane, and so AdS space supported by flux carries an intrinsic instability \cite{Ooguri:2017njy}. 
\begin{figure}[t]
\centering
 \includegraphics[width=0.8\textwidth]{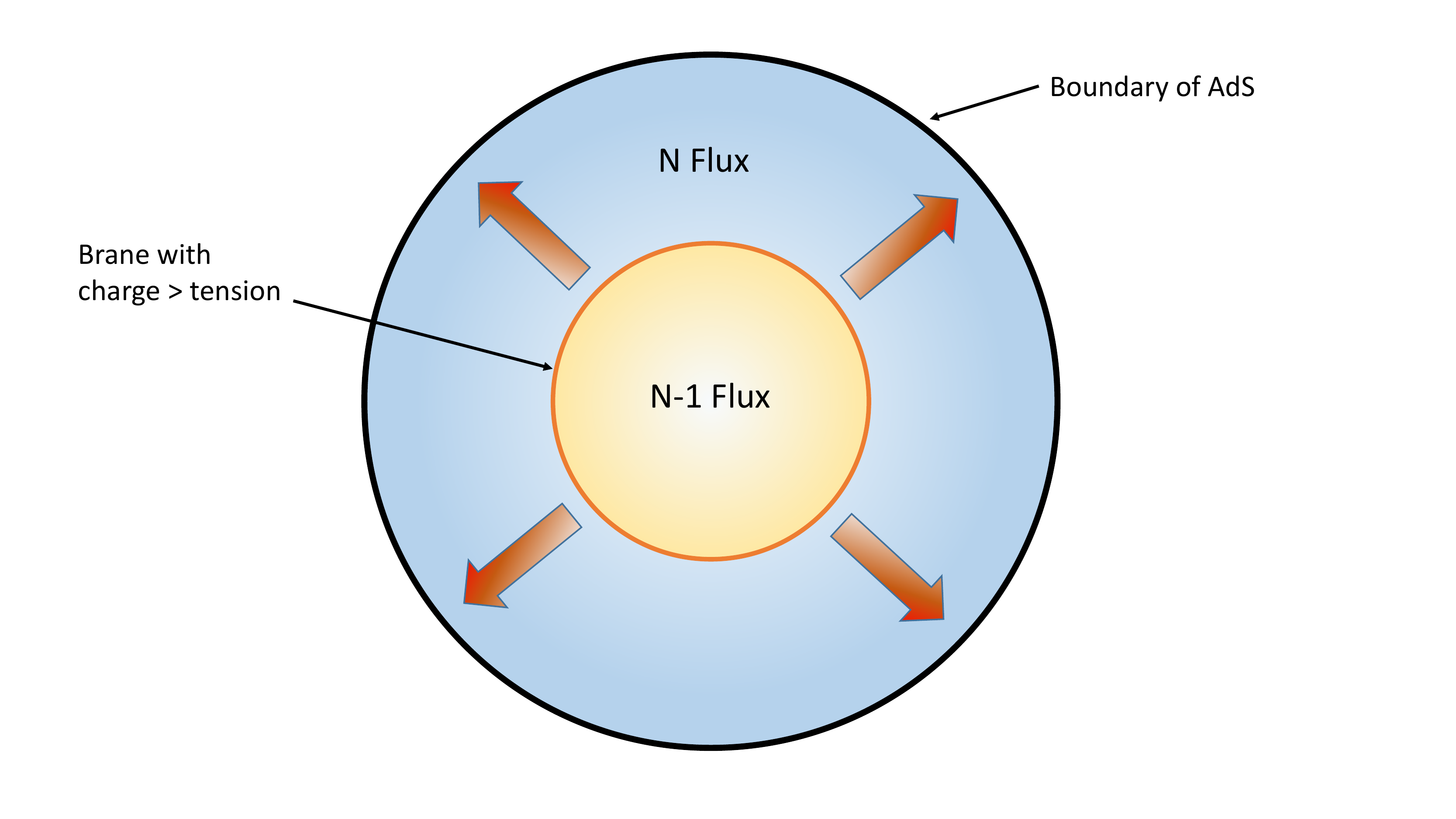}
\caption{Figure illustrating the fragmentation of anti-de Sitter space with spacetime-filling flux. A brane nucleates and, as long as its charge is larger than its tension, expands to the boundary of AdS reducing the flux by one unit in its interior.}
\label{fig:adsfrag}
\end{figure}

While this shows that AdS space supported by flux is unstable, it was conjectured in \cite{Ooguri:2016pdq} that this instability is part of a more general phenomenon and that any non-supersymmetric AdS space is unstable. 
\begin{tcolorbox}
{\bf Non-Supersymmetric AdS Instability Conjecture} \;\cite{Ooguri:2016pdq}
{\it 
\newline
\newline
Any realization of non-supersymmetric anti-de Sitter space must exhibit some instability. 
}
\end{tcolorbox}

A similar suggestion was also made in \cite{Freivogel:2016qwc}. This conjecture was studied and tested further in \cite{Danielsson:2016mtx,Banks:2016xpo,Ooguri:2017njy,Danielsson:2017max,Giombi:2017mxl,Aalsma:2018qwy,Antonelli:2018qwz}. In \cite{Buratti:2018onj} it was further argued that geometries which look locally like AdS also manifest such an instability. 

The instability of AdS space is due to a tunneling phenomenon and so takes a long time to manifest. However, it has rather crucial implications for holography in the AdS/CFT correspondence. Because from the boundary of AdS the decay would appear instantaneous, and so would manifest as such an instability of the dual CFT \cite{Horowitz:2007pr}. This would then imply that non-supersymmetric CFTs with gravitational duals should always be unstable.  

\subsubsection{Implications for Particle Physics}

An interesting application of the AdS instability conjecture relates to compactifications of the Standard Model on a circle. In \cite{ArkaniHamed:2007gg} compactifications of the Standard Model, with unfixed neutrino masses, on a circle to three dimensions were considered and the resulting effective potential for the radius of the circle was studied. It was shown that there are two competing effects, the magnitude of the cosmological constant, and the Casimir energy associated to light degrees of freedom. In the Standard Model these are the photon, the graviton, and the neutrinos. It was shown that for sufficiently large neutrino masses, larger than the magnitude of the cosmological constant, an AdS minimum forms for the radion field.  An interesting possibility arises which is that the AdS instability conjecture may forbid the existence of this stable AdS vacuum, which translates into bounds on properties of neutrinos. The idea was first raised in \cite{Ooguri:2016pdq}, and studied in detail in \cite{Ibanez:2017kvh,Ibanez:2017oqr,Hamada:2017yji,Gonzalo:2018tpb,Gonzalo:2018dxi}. A particularly striking possibility is that it bounds the neutrino masses by the cosmological constant
\be
m_{\nu} < \Lambda^{\frac14} \;.
\label{neutrinomassbound}
\ee
It was then pointed out in \cite{Ibanez:2017oqr} that this could lead to insights into the hierarchy problem since the neutrino masses are naturally tied to the scale of electroweak symmetry breaking.

An important difficulty with this chain of reasoning is that even if the AdS instability conjecture holds it is difficult to argue that there is no instability induced by physics at higher energy scales than the neutrinos. In particular, in \cite{Hamada:2017yji} it was shown that one must add the Wilson line of the electromagnetic gauge field on the circle to the analysis, and this leads to an instability associated with the electron.  In \cite{Gonzalo:2018tpb,Gonzalo:2018dxi} a possible way to remove the instability was considered by taking an orbifold compactification. It was then argued that the question of stability of the AdS vacuum would be related to physics of supersymmetry breaking. 

It is worth noting at this point that in \cite{Lust:2017wrl} another possible argument for an inequality like (\ref{neutrinomassbound}) was suggested. By modeling the cosmological constant as a linear repulsive force, it was argued that neutrinos would form a tower of (meta-)stable bound states if placed at a distance of order their Compton wavelength unless the inequality is satisfied. Of course, the question is whether this tower of bound states is problematic or not for quantum gravity.

\subsection{St\"uckelberg conjectures}
\label{sec:stu1con}

Massive fields can often propagate a different number of degrees of freedom to massless fields. The most well-known case is a massive photon which propagates three, rather than two, degrees of freedom. Such fields admit a helicity decomposition which picks out the internal additional degrees of freedom of lower spin, the St\"uckelberg fields, and the kinetic terms for those degrees of freedom then manifest explicitly form the mass term of the higher spin field. An interesting possibility is to apply the Weak Gravity Conjecture to the St\"uckelberg fields and thereby constrain some properties of the higher spin fields. This idea was studied for massive spin-1 fields in \cite{Reece:2018zvv} and for massive spin-2 fields in \cite{Klaewer:2018yxi,deRham:2018dqm}.

\subsubsection{Spin-1 conjecture}

A massive spin-1 field has a St\"uckelberg field which is a pseudo-scalar. We can decompose the massive gauge field $A_{\mu}$ into a transverse gauge field $A^{\perp}_{\mu}$ and a longitudinal St\"uckelberg axion $\theta$ as
\be
A_{\mu} = A_{\mu}^{\perp} + \partial_{\mu} \theta \;.
\ee
Then the mass term for the gauge field induces a kinetic term for $\theta$ making it dynamical
\be
m^2 A_{\mu} A^{\mu} \supset m^2 \left(\partial_{\mu} \theta \right)\left(\partial^{\mu} \theta \right) \;.
\ee 
The proposal of \cite{Reece:2018zvv} is that we may consider the mass term $m$ as an axion decay constant $f$, analogously to (\ref{actaxin}), and then apply the axion version of the magnetic Weak Gravity Conjecture (\ref{magf}) to deduce a condition
\be
\Lambda < \left(\frac{mM_p}{g} \right)^{\frac12} \;,
\label{spin1con}
\ee
where $g$ is the massive gauge field coupling. 

The proposal (\ref{spin1con}) has a number of subtleties discussed in \cite{Reece:2018zvv}. The most crucial one is that it only applies to gauge fields which gain a mass through a St\"uckelberg mechanism in string theory, rather than just a simple Higgs mechanism. The idea is to differentiate between the two possibilities by the statement that the $m \rightarrow 0$ limit is at infinite distance in field space for a St\"uckelberg massive field. Then the cutoff can be related to the distance conjecture. In this sense, the conjecture is more of an application of the distance conjecture than a completely general statement about massive gauge fields. 

It was argued in \cite{Craig:2018yld} that one can apply a similar reasoning to that discussed in section \ref{sec:strongwgc}, to show that the conjecture can be violated by a low-energy effective theory while respected in the ultraviolet. Similarly to the discussion in section \ref{sec:strongwgc}, it remains to see though if such setups can be realized in string theory. 

\subsubsection{Spin-2 conjecture}
\label{sec:sp2conj}

The application of the Weak Gravity Conjecture to the Spin-2 case was first studied in \cite{Klaewer:2018yxi}. In some sense it is actually simpler than the spin-1 case. Firstly because there is no known Higgs mechanism for massive spin-2 fields, and so there is no ambiguity regarding a stringy St\"uckelberg versus a Higgs mass. And secondly because the St\"uckelberg fields include a gauge field, which means that the gauge field version of the magnetic Weak Gravity Conjecture can be applied which is better understood than the axion version. 

A massive spin-2 field can be written as a symmetric tensor $w_{\mu\nu}$. It has 10 degrees of freedom, but only 5 are propagating. These can be decomposed into 2 propagating degrees of freedom of a massless spin-2 field, 2 propagating degrees of freedom of a massless gauge field, and 1 pseudo-scalar degree of freedom.\footnote{In analogy with the Higgs mechanism, we can think of the gauge field eating the scalar to become massive, and the massless spin-2 field eating the massive gauge field to become massive.}
This helicity decomposition reads (see, for example, \cite{Dvali:2006su})
\begin{equation} 
\label{decompo}
w_{\mu\nu} = h_{\mu\nu} + 2 \partial_{(\mu}\chi_{\nu)} + \Pi^L_{\mu\nu} \pi \;.
\end{equation} 
Here, $h_{\mu\nu}$ is the helicity-2 mode, $\chi_{\mu}$ the helicity-1 mode, and $\pi$ the scalar longitudinal mode, with $\Pi^L_{\mu\nu}$ being the associated projector. 
The action for a massive spin-2 field coupled to gravity is
\begin{eqnarray}
S_G &=& \int d^4x \; \sqrt{-g} \left[ M^2_p R  - \frac14 w^{\mu\nu} L_{\mu\nu}^{\rho\sigma} w_{\rho \sigma} -\frac18 m^2 \left( w_{\mu\nu}w^{\mu\nu} - w^2 \right) +... \right] \;. 
\label{ehgen}
\end{eqnarray}
The Lichnerowicz operator takes the form (in flat space)
\begin{eqnarray}
L_{\mu\nu}^{\rho\sigma} w_{\rho\sigma} &=& -\frac12 \left[ \Box w_{\mu\nu} - 2\partial_{(\mu}\partial_{\alpha} w^{\alpha}_{\nu)} + \partial_{\mu}\partial_{\nu} w - \eta_{\mu\nu} \left( \Box w - \partial_{\alpha} \partial_{\beta} w^{\alpha\beta}\right) \right] \;.
\end{eqnarray}
The last term in (\ref{ehgen}) is the Fierz-Pauli mass, with $w \equiv \eta^{\mu\nu}w_{\mu\nu}$. This mass term, after integration by parts, contains the kinetic term for $\chi_{\mu}$
\begin{equation}
\label{lfpfs}
{\cal L}_{\mathrm{FP}} \supset -\frac{1}{8}m^2 F_{\mu\nu} F^{\mu\nu} \;, 
\end{equation}
where $F_{\mu\nu} \equiv \partial_{\mu} \chi_{\nu} - \partial_{\nu}\chi_{\mu}$.

We would like to associate a gauge coupling to the gauge field $\chi_{\mu}$. To do that we need to specify its interactions, since this is what the gauge coupling really measures. This is exactly the same procedure as for the Weak Gravity Conjecture. We therefore specify the coupling of the gauge field to a canonically normalised matter current as
\begin{equation}
\label{h1int}
{\cal L}^{\chi}_{\mathrm{int}} =  \frac{m^2}{M_w} \chi_{\mu} J^{\mu} \;,
\end{equation}
where we have introduced a mass scale $M_w$ which specifies this coupling. With this definition we can read off the gauge coupling for $\chi_{\mu}$
\begin{equation}
\label{gdef}
g_m = \frac{m}{\sqrt{2}\;M_w} \;.
\end{equation}

The Spin-2 Conjecture is then nothing but the magnetic Weak Gravity Conjecture (\ref{mwgc}) applied to the gauge coupling (\ref{gdef}).
\begin{tcolorbox}
{\bf Spin-2 Conjecture} \;\cite{Klaewer:2018yxi}
{\it 
\newline
\newline
A theory with a massive spin-2 field, coupled to gravity, with mass $m$, and an interaction scale of the helicity-1 mode $M_w$ as in (\ref{h1int}), has a cut-off scale $\Lambda_m$ with
\begin{equation}
\label{s2c}
\Lambda_{m} \sim \frac{m \;M_p}{M_w} \;.
\end{equation}
This cutoff is associated with the mass scale of an infinite tower of states.
}
\end{tcolorbox}

The definition of $M_w$ through the explicit form of the current coupling (\ref{h1int}) is chosen such that it naturally coincides with the interaction scale of the massive spin-2 field itself. This is not an assumption underlying the conjecture, but rather a natural way to present it. To see this consider an interaction of $w_{\mu\nu}$ with some associated tensor (not necessarily the energy-momentum tensor)
\be
{\cal L}^{w}_{\mathrm{int}} = \frac{1}{M'_w} w_{\mu\nu} T_w^{\mu\nu}\;.
\ee
Then after integration by parts we can write this as
\be
{\cal L}^{w}_{\mathrm{int}} \sim \frac{1}{M'_w} \chi_{\mu} \left( \partial_{\nu} T_w^{\mu\nu}\right) \equiv  \frac{\left(m'\right)^2}{M'_w} \chi_{\mu} J^{\mu} \;.
\label{curscal}
\ee
We can now compare this with the definition (\ref{h1int}). We see that if $m' \sim m$ then $M'_w \sim M_w$. In \cite{Klaewer:2018yxi} it was argued that $m' \sim m$ is a natural expectation since we expect this interactions of the helicity-1 mode $\chi_{\mu}$ to vanish when $m \rightarrow 0$. The latter is due to the fact that if the mass was vanishing, a gauge symmetry would ensure the conservation of the tensor $T_w^{\mu\nu}$, this is true at least at the linear level to which we are working. While this argument suggests identifying $m \sim m'$, it is important to work towards proving this identification rigorously.  

It is informative to look at the canonical example of a massive spin-2 field, Kaluza-Klein gravitons, to see how these general expressions manifest. We consider a compactification of Einstein gravity from five dimensions to four dimensions on a circle. This is a special case of the general analysis in sections \ref{sec:compft} and \ref{sec:fecompga}, which we will follow. We first Fourier expand the metric components in (\ref{metana}) as
\bea
g_{\mu\nu} = \sum_n g_{\mu\nu}^n\left(x^{\mu} \right) e^{2 \pi i n X^5} \;,\;\; A_{\mu} = \sum_n A_{\mu}^n\left(x^{\mu} \right) e^{2 \pi i n X^5}\;,\;\; \Psi =  \sum_n \psi^n\left(x^{\mu} \right) e^{2 \pi i n X^5}\;.
\eea
Here we also introduced a 5-dimensional scalar field $\Psi$, and its Fourier expansion. Now we consider the kinetic term for $\Psi$
\bea
\frac12 \int d^5X \sqrt{-G} G^{MN} \partial_M \Psi \partial_N \Psi &\supset& \frac12 \int d^5X \sqrt{-G} G^{\mu5} \partial_{\mu} \Psi \partial_5 \Psi \nn \\
 &=& \frac12 \int d^4x \sqrt{-g} \sum_n A_{\mu}^n J_n^{\mu} \;,
 \label{kkAint}
\eea
where
\be
J_n^{\mu} = \sum_m \left(n+m\right) \psi^{n+m} \partial^{\mu} \psi^m \;. 
\ee
The gauge coupling for all the Kaluza-Klein gauge fields $A^n_{\mu}$ is the same as the one for the zero mode (\ref{zmgc}), which we can write as
\be
g_n \sim \frac{m_{KK}}{M_p} \;.
\label{gnkkmp}
\ee
The Kaluza-Klein gauge fields are nothing by the helicity-1 modes of the Kaluza-Klein gravitons, which are massive spin-2 fields. The interaction (\ref{kkAint}) is canonically normalised, which means we can read off the gauge coupling as (\ref{gnkkmp}), matching the proposal (\ref{gdef}) with $M_w=M_p$, as appropriate for gravitationally interacting massive spin-2 fields. One also seems that, relating to (\ref{curscal}), indeed $m' \sim m$. The Spin-2 conjecture then implies $\Lambda_m \sim m_{KK}$, indeed coinciding with an infinite tower of states. 

The application of the magnetic Weak Gravity Conjecture, to arrive at the Spin-2 conjecture, is here very much along the lines of its relation to the distance conjecture and the emergence proposal discussed in section \ref{sec:emergence}. These are related to the kinetic terms of the fields, and so naturally extends to the case when the graviton itself has a mass. This lead to the proposal for a strong version of the spin-2 conjecture \cite{Klaewer:2018yxi}. The {\bf Strong Spin-2 Conjecture} states that in the case when the graviton has a mass $m_{\mathrm{grav}}$, the cut-off scale is set my this mass
\be
\Lambda_m \sim m_{\mathrm{grav}} \;.
\label{Sspin2}
\ee
This has a rather strong implication since the mass of the graviton is experimentally constrained, for example by gravitational waves, as $m_{\mathrm{grav}} < 10^{-22} eV$ \cite{TheLIGOScientific:2016src}. Therefore, if the strong Spin-2 conjecture holds then it predicts that the mass of the graviton must be exactly zero.

In \cite{Klaewer:2018yxi}, it was proposed that the Spin-2 conjecture can be formulated in terms of Weyl gravity.
\begin{equation}
\label{wsgw}
S_G =   \int d^4x \sqrt{-g} \left[ M_p^2 R + \frac{1}{2g_W^2}W_{\mu\nu\rho\sigma}W^{\mu\nu\rho\sigma} + \dots \right] \;,
\end{equation}
where  $W_{\mu\nu\rho\sigma}$ is the Weyl tensor
\begin{equation}
W_{\mu\nu\rho\sigma} = R_{\mu\nu\rho\sigma} + g_{\mu[\sigma}R_{\rho]\nu} + g_{\nu[\rho}R_{\sigma]\mu} + \frac13 R g_{\mu[\rho}g_{\sigma]\nu} \;. \nonumber
\end{equation} 
Specifically, the action (\ref{wsgw}) has a massive spin-2 ghost, with mass $m \sim g_W M_p$. Following \cite{Gording:2018not,Ferrara:2018wqd,Ferrara:2018wlb}, it was proposed that this mass is related to the mass of a physical massive spin-2 field whose kinetic terms are made positive by an infinite series of higher derivative terms completing (\ref{wsgw}). In this sense, the spin-2 conjecture can be thought of as a Weak Gravity Conjecture for $g_W$, the coefficient controlling the $W^2$ term. 

The Spin-2 conjecture is satisfied in all fully understood string constructions since massive spin-2 fields are Kaluza-Klein gravitons, or string oscillator states, which are indeed associated with an infinite tower at the mass scale (\ref{s2c}). Interestingly, also the strong spin-2 conjecture (\ref{Sspin2}) can be tested in string theory in terms of S-fold W-superstring constructions  \cite{Ferrara:2018iko}. There is no massless spin-2 graviton and the spectrum starts right away with massive higher spin fields, which again set the mass scale of an infinite tower. Note however that in \cite{Kiritsis:2006hy,Aharony:2006hz} massive spin-2 constructions were studied for AdS backgrounds holographically, and possible embeddings of these in string theory were proposed in \cite{Bachas:2017rch,Bachas:2018zmb}. It would be interesting to study if these constructions satisfy the Spin-2 conjecture.

In \cite{deRham:2018dqm} various objections to the Spin-2 conjecture were raised. In particular, it was argued that the Weak Gravity Conjecture should not apply to St\"uckelberg gauge fields, and that the scaling $m' \sim m$ is incorrect. The latter objection does not really apply to the formulation of the conjecture, since $M_w$ is defined through the interaction of the helicity-1 mode. Also $m' \sim m$  can be seen to not be generally incorrect since it applies to the canonical case of Kaluza-Klein gravitons. It is of course true that both the scaling $m' \sim m$  and the application of the Weak Gravity Conjecture to St\"uckelberg fields are assumptions, with natural arguments discussed above, but are unproven. Another objection raised in \cite{deRham:2018dqm} was that the Kaluza-Klein case behaves the way it does due to the Weak Gravity Conjecture applying to the massless zero-mode gauge field. However, this can be seen to not be the case by generalising the reduction from a circle to an orbifold $\frac{S^1}{\mathbb{Z}_N}$, in which case the zero-mode gauge field is projected out, but the analysis of the massive modes remains largely unchanged. 

\section{The de Sitter Conjecture}
\label{sec:dsconj}

The idea of a String Theory Landscape dates back as far as the realization of string theory as a consistent theory. For example, in \cite{Lerche:173303} already appeared numbers such as $10^{1500}$ models. However, the Landscape became a particularly powerful idea in relation to the Cosmological Constant problem by forming a framework that can potentially realize the idea of anthropic selection of the cosmological constant as proposed in \cite{PhysRevLett.59.2607}. This role for the Landscape was first emphasized in \cite{Bousso:2000xa}, building on earlier similar ideas in \cite{Brown:1987dd,Brown:1988kg,Banks:1984cw}. The anthropic solution to the cosmological constant problem within string theory is that the Landscape provides a very large, but discrete, number of vacua whose cosmological constant can take any positive or negative value. Anthropic selection, in particular the formation of galaxies, then selects a small positive cosmological constant. An important part of this proposal is that the cosmological constant is not a free parameter in the theory. In 11-dimensional supergravity, the low-energy limit of M-theory, there is no cosmological constant. Rather, what we view as a cosmological constant is a minimum of a scalar potential induced by the compactification of the supersymmetric higher dimensional theory. Henceforth, when we refer to a de Sitter vacuum of string theory, we mean a vacuum which is a local minimum of the scalar potential where the potential takes a positive value. 

In practice, it has proven very difficult to construct de Sitter vacua in string theory. This may be due to the fact that the starting point is a supersymmetric theory, while de Sitter space is non-supersymmetric. Another difficulty is a practical one, a de Sitter vacuum requires stabilizing all the moduli, this is in general a difficult problem and also there are not so many different well-understood mechanisms for doing so. A reasonable perspective on these difficulties is then that simply we are not technically able enough to construct complete exact de Sitter solutions to string theory. From this perspective we may then aim only at a reasonable proposal on how a de Sitter vacuum could be constructed, without requiring a proof that this is a solution to string theory. The prototypical example of an extremely reasonable de Sitter prescription is the KKLT scenario \cite{Kachru:2003aw}. We will discuss de Sitter vacua in string theory in more detail in section \ref{sec:testsdS}. 

It is also reasonable, however, to take the technical difficulties more seriously as a hint that there is a deep obstruction to constructing de Sitter vacua in string theory. So in the spirit of this review, we may consider the possibility that de Sitter vacua are in the Swampland. This may at first appear inconsistent with observations since we have good evidence that the universe is entering a phase of late-time acceleration which is most likely driven by a positive cosmological constant. However, inflation is also a phase of acceleration that the universe most likely has passed through, and it was not due to a cosmological constant. Indeed, inflation was most likely driven by a scalar field rolling down a potential. It is therefore not completely unreasonable, given that our best ideas for past acceleration are based on scalar fields rolling, to consider that the late-time acceleration is also due to such a mechanism. Such a scenario is termed dynamical dark energy, the most well-known version of which is quintessence \cite{WETTERICH1988668,PhysRevD.37.3406}, and we refer to \cite{Copeland:2006wr} for a review. The possibility that de Sitter vacua are in the Swampland is therefore not ruled out by observation. It also does not mean that the anthropic selection solution to the cosmological constant problem is lost. Rather, we must consider that string theory may allow for a Landscape of potentials which have flat enough regions to lead to accelerated expansion of the universe, and anthropic arguments then limit the magnitude of the potential in those regions. 

The possibility that string theory does not allow for de Sitter vacua is not new, see \cite{Danielsson:2018ztv} for a recent review of the history of the field, and section \ref{sec:testsdS} for more detailed discussion. However, it has recently gained impetus by a quantitative proposal for a more detailed constraint on properties of potentials that are not in the Swampland. The de Sitter conjecture proposes a lower bound on the gradient of potentials in positive regions \cite{Obied:2018sgi}.  
\begin{tcolorbox}
{\bf The de Sitter Conjecture} \;\cite{Obied:2018sgi}
{\it 
\newline
\newline
The scalar potential of a theory coupled to gravity must satisfy a bound on its derivatives with respect to scalar fields
\begin{equation}
\label{dsc}
\left|\nabla V \right| \geq \frac{c}{M_p} V\;,
\end{equation}
where $c>0$ is a constant of order one. 
}
\end{tcolorbox}
The $\nabla$ in (\ref{dsc}) means one must consider the vector of derivatives with respect to all the scalar fields in the theory, such that the bound is on the norm of that vector. The constant $c$ is not specified sharply, and the possibility that it is of order say $0.01$ was not explicitly ruled out. Indeed, it was proposed in \cite{Obied:2018sgi} that it may be possible to make it quite small, but at the cost of having a cutoff due to an infinite tower of states become exponentially light, as in the distance conjectures (\ref{sdc}) and (\ref{rsdc}). It was shown in \cite{Agrawal:2018own} that the present acceleration constrains $c < 0.6$.

The original proposal of the de Sitter conjecture \cite{Obied:2018sgi} was based almost completely on evidence from examples in string theory. In \cite{Ooguri:2018wrx} a more general argument was presented for the conjecture based on the entropy of de Sitter space and the distance conjecture.\footnote{More precisely, as discussed in section \ref{sec:dsreldis}, the argument only held in parametrically large distances in scalar field space.} We will discuss this in more detail in section \ref{sec:dsreldis}. The relation to the entropy of de Sitter space lead to a natural refinement of the original conjecture. 
\begin{tcolorbox}
{\bf The Refined de Sitter Conjecture} \;\cite{Ooguri:2018wrx,Garg:2018reu} 
{\it 
\newline
\newline
The scalar potential of a theory coupled to gravity must satisfy either 
\begin{equation}
\label{rdsc1}
\left|\nabla V \right| \geq \frac{c}{M_p} V\;,
\end{equation}
or
\begin{equation}
\label{rdsc2}
\mathrm{min}\left(\nabla_i \nabla_j V \right) \leq -\frac{c'}{M^2_p} V\;.
\end{equation}
}
Here $c,c'>0$ are constants of order one, and the left hand side of (\ref{rdsc2}) is the minimum eigenvalue of the Hessian in an orthonormal frame.
\end{tcolorbox}
The second condition (\ref{rdsc2}) arises because, as explained below, if it is satisfied then there is an instability which implies that the entropy-based argument for the first condition (\ref{rdsc1}) breaks down. There are also other proposed refinements of the de Sitter conjecture \cite{Andriot:2018wzk,Garg:2018reu,Ben-Dayan:2018mhe,Dvali:2018jhn,Garg:2018zdg,Andriot:2018mav}.
They typically involve also a condition on the second derivative. 

In \cite{Ooguri:2018wrx} the refinement was motivated largely by the relation to the entropy of de Sitter space. Another good reason to consider such a refinement is to avoid some well motivated counter-examples in the Standard Model and extensions of it. In particular, in \cite{Denef:2018etk} it was pointed out that the top of the Higgs potential would strongly violate the original de Sitter conjecture (\ref{dsc}) since we have there
\be
\frac{\left|\nabla V \right|  }{V} \sim \frac{10^{-55}}{M_p} \;,\;\; \frac{\mathrm{min}\left(\nabla_i \nabla_j V \right) }{V} \sim - \frac{10^{35}}{M_p^2}\;.
\ee
It may be possible to satisfy the de Sitter conjecture (\ref{dsc}) through some difficult to realize scenarios \cite{Denef:2018etk}. But we see that the Refined de Sitter conjecture is certainly satisfied. Similarly, it was argued that the QCD axion and the QCD phase transition would be associated to violations of the original conjecture \cite{Murayama:2018lie,Choi:2018rze}.\footnote{Interestingly, the conjecture could actually be utilised to argue for the existence of the QCD axion in the first place \cite{Dvali:2018dce}.} It was shown in \cite{Ooguri:2018wrx} that the refined de Sitter conjecture avoids these counter examples. More generally, the relation to the top of any axion potential is interesting because the second derivative of such a potential is bounded as
\be
\frac{\mathrm{min}\left(\nabla_i \nabla_j V \right) }{V} \leq -\frac{1}{f^2} \;.
\ee
The refined de Sitter conjecture therefore is satisfied for any such axion as long as $f \leq M_p$, which is implied by the Weak Gravity Conjecture for axions (\ref{wgcax}). 

The rest of this section is devoted to reviewing various aspects of the de Sitter conjecture. In section \ref{sec:qaods} we will introduce some important concepts related to de Sitter space, and also some general arguments which closely relate to the de Sitter conjecture but are not necessarily set within string theory. In section \ref{sec:dsreldis} we will review a connection between the refined de Sitter conjecture and the distance conjecture that was pointed out in \cite{Ooguri:2018wrx}. In section we will present a (brief) review of attempts at constructing de Sitter vacua in string theory, the various no-go theorems and the best candidates for de Sitter constructions. There has been a significant program of study of the cosmological consequences of the de Sitter conjecture, and these will be reviewed in section \ref{sec:imswcos}.

\subsection{Some quantum aspects of de Sitter space}
\label{sec:qaods}

In this section we will review some properties of de Sitter space, following \cite{Witten:2001kn,Spradlin:2001pw}. We can describe $d$-dimensional de Sitter space $dS_d$ as a hypersurface, embedded in $d+1$ dimensional Minkowski space $\cM_{d+1}$, defined by the equation
\be
-X_0^2 + X_1^2 + ... + X_d^2 = R^2 \;,
\ee
where $R$ is the radius of the de Sitter space. This is related to the cosmological constant on the space $\Lambda$ as
\be
\Lambda = \frac{\left(d-2\right)\left(d-1 \right)}{2 R^2} \;. 
\ee 
In global coordinates we can write the metric on $dS_d$ as
\be
ds^2 = -dt^2 + R^2 \cosh^2\left(\frac{t}{R}\right) d\Omega^2_{d-1}\;,
\ee
where $d\Omega^2_{d-1}$ is the metric on the unit Euclidean $\left(d-1\right)-$sphere. This shows that de Sitter space can be thought of as a sphere whose radius evolves in time, starting from infinite size, shrinking to radius R, and then becoming infinitely large again. We can also write the metric in Static coordinates which reads
\be
ds^2 = - \left(1-\frac{r^2}{R^2} \right) dt^2 + \left(1-\frac{r^2}{R^2} \right)^{-1} dr^2 + r^2 d\Omega^2_{d-2} \;.
\ee
This shows that de Sitter space has a horizon of radius $R$. This horizon is a bit different to that of a black hole, it is associated to an observer. It is defined by the causal patch available to a given observer, so an observer cannot probe a distance further than R. Even though the horizon is different to a black hole, it may still be associated a temperature, the Gibbons-Hawking temperature $T_{dS}$ \cite{PhysRevD.15.2738},
\be
T_{dS} = \frac{1}{2 \pi R} \;.
\ee
From the temperature we can associate an entropy $S_{dS}$. We will restrict henceforth to four-dimensional spacetime $d=4$, so the entropy reads
\be
S_{dS} = 8\pi^2 \left(R M_p\right)^2 \;.
\label{dsentropy}
\ee
For black holes we interpret the Bekenstein-Hawking entropy as the logarithm of the number of microstates of the black hole. This has been made explicit for certain black holes in string theory following the seminal work \cite{Strominger:1996sh}.  It was proposed in \cite{Banks:2000fe,Witten:2001kn} that a similar interpretation in terms of microstates should be assigned to de Sitter entropy. So we define asymptotically de Sitter microstates as those which asymptote to de Sitter space in the future, and assign the entropy to the logarithm of the dimension of the Hilbert space associated to those microstates
\be
S_{dS} = \log \mathrm{dim} \;\cH \;.
\label{dsSH}
\ee
It is not completely clear how to define this Hilbert space, see for example \cite{Banks:2000fe,Witten:2001kn,Banks:2003cg} for discussions. Indeed, there are many aspects of de Sitter space which are difficult to make precise which are due to the absence of a spatial infinity to define asymptotics. For example, there is no definition of an S-matrix in de Sitter space, which makes it difficult to define string theory on it.\footnote{In anti-de Sitter space the S-matrix is replaced by correlation functions on the boundary, leading to a precise definition of physics in the bulk.} There is no conserved notion of energy, and there is no supersymmetry.

It is expected that in quantum gravity de Sitter space is at best meta-stable. This is consistent with the definition discussed above for string theory of de Sitter vacua being minima of a scalar potential, since one always expects a decompactification limit leading to a supersymmetric higher dimensional theory with zero energy to exist. It is then always possible to tunnel to this zero energy limit, though the tunneling time can easily exceed the age of the universe.\footnote{It is not clear to the author how to actually define the de Sitter horizon in quantum gravity, since to fully probe the causal patch of an observer would take an infinite amount of time, by which the de Sitter space would have decayed.} This may also be related to an obstruction to reaching the Poincar\'e recurrence time of de Sitter $t_R \sim e^{S_{dS}}$. Poincar\'e recurrence, the system returning to its original state, occurs in any system which has a finite dimensional Hilbert space, like de Sitter space, and which evolves for a sufficiently long time. Another possibly important time scale for de Sitter space is a possible analogue to the Page time of black holes \cite{Page:1993wv,Page:1993df}.

Some arguments, prior to the de Sitter conjecture, have been made that de Sitter space must be not only meta-stable but actually unstable. In \cite{PhysRevD.31.754}, see \cite{Antoniadis_2007} for a review, and also similarly in \cite{Polyakov:2007mm,Polyakov:2009nq,Polyakov:2012uc}, it was argued that de Sitter space is unstable when coupled to an interacting quantum field theory.\footnote{Some arguments were presented also for a non-interacting quantum field theory, for example in \cite{Mottola:1985qt,Padmanabhan:2002ji,Markkanen:2017abw}.} The idea is that the (Bunch-Davies) vacuum is such that there is particle production from the vacuum, and the interactions of these produced particles lead to perturbations which break the de Sitter isometry group. This leads to a time-dependence which decreases the spacetime curvature. This is a type of discharge of the cosmological constant into matter and radiation. In \cite{Dvali:2014gua,Dvali:2017eba,Dvali:2018jhn} a different argument for an instability of de Sitter space was presented. The idea is to consider de Sitter space as a coherent state of gravitons on top of a Minkowski background. The graviton fields within the coherent state can interact with each other, and over a sufficiently long time these interactions will cause the coherent state to become a fully quantum state with no semi-classical description. This time scale was denoted the quantum break time of de Sitter space, and is in some sense analogous to the Page time of black holes. It was argued that quantum breaking of de Sitter should not occur and instead the space must exhibit an instability. The instability must be sufficiently strong to censor quantum breaking, which lead to an expression similar to the refined de Sitter conjecture.

\subsection{Relation to the distance conjecture}
\label{sec:dsreldis}

In \cite{Ooguri:2018wrx} a relation between the distance conjecture (\ref{sdc},\ref{rsdc}) and the de Sitter conjecture was studied. It was argued that the refined de Sitter conjecture (\ref{rdsc1}-\ref{rdsc2}) follows from the distance conjecture, or more precisely from an interpretation of it in terms of duality, at parametrically large distances in field space. The main idea is that at large distances in field space towers of states become light thereby increasing the number of states in the theory and so the dimension of the Hilbert space. The interpretation of de Sitter entropy in terms of the dimension of a Hilbert space (\ref{dsSH}) suggests then that the entropy should increase with field distance. The proposal of \cite{Ooguri:2018wrx} is that at parametrically large distances in field space the exponentially large number of light states dominate the Hilbert space and so determine the behaviour of the de Sitter entropy. The monotonic increase in the number of such states, according to the distance conjecture, then implies a monotonic increase in entropy. Noting that the entropy of de Sitter space is inversely proportional to the cosmological constant this implies that the constant would need to decrease at large distances in field space. In other words, it cannot be a constant but must be a scalar potential which has a non-vanishing derivative. The exponential nature of the tower of states in the distance conjecture is then mapped to the property of the potential that its derivative should be proportional to itself, reproducing the first condition of the refined de Sitter conjecture (\ref{rdsc1}). This argument obviously relies on some semi-classical notion of de Sitter entropy, and the second condition of the refined de Sitter conjecture (\ref{rdsc2}) states that if such a notion does not exist, due to a sufficiently strong instability of the potential, then the conjecture may not hold. This summarizes the essential points, and the rest of this section will review some more details.

The first thing to establish is how to assign an entropy to a field rolling on a potential rather than a cosmological constant. The suggestion in \cite{Ooguri:2018wrx} is to utilize the covariant entropy bound \cite{Fischler:1998st,Bousso:1999xy} and apply it to the largest possible area. The point is that if the scalar field is rolling slowly enough such that there is an accelerated expansion, then an apparent horizon (as opposed to a cosmological horizon) will exist. This requires
\be
\left| \nabla V \right| \leq \sqrt{2} V \;.
\label{accex}
\ee
If (\ref{accex}) is violated then clearly the de Sitter conjecture is satisfied, so we are only interested in the case where it is satisfied and an apparent horizon exists. The covariant entropy bound is defined in terms of the entropy on converging light sheets, and the apparent horizon is the largest radius for which we have such converging sheets. This is illustrated in figure \ref{fig:ls}. 
\begin{figure}[t]
\centering
 \includegraphics[width=0.8\textwidth]{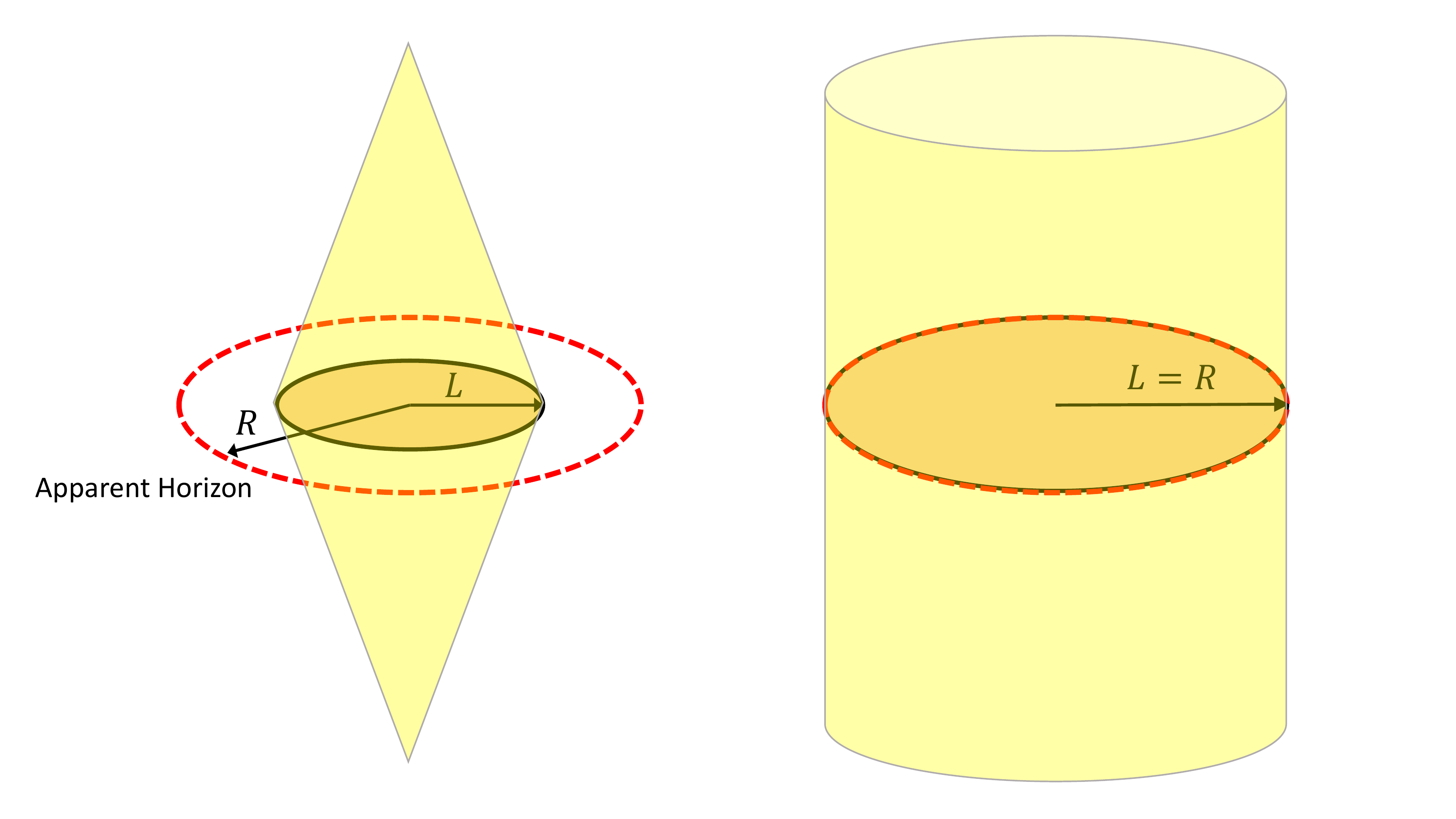}
\caption{Figure illustrating the light sheets associated to spheres with radii $L$ smaller than, or equal to, the apparent horizon of an accelerating universe of radius $R$. The sheets converge to a caustic at finite distance for $L<R$. The covariant entropy bound relates to the entropy of states on the light sheets, thereby naturally associating a maximum entropy to the apparent horizon. }
\label{fig:ls}
\end{figure}
In this sense the radius of the apparent horizon $R$, where
\be
R \sim \frac{1}{\sqrt{V}} \;,
\ee
gives a natural measure of the entropy of a quasi-de Sitter space, so with a rolling scalar field.

The entropy associated to the states in the tower is not so easy to determine since it amounts to counting the associated microstates. Indeed, the emergence proposal, see section \ref{sec:emergence}, would suggest that the appropriate tower is actually a tower of domain walls. But we can expect some coarse properties of the entropy, it should increase as some order one power of the number of states in the tower $N$, and possibly as some order one power of the apparent horizon radius $R$. So we can write
\be
S_{\mathrm{tower}} \sim N^{\gamma} R^{\delta} \;,
\ee
with $\gamma$ and $\delta$ some undetermined order one numbers. Now the crucial point is to assume that the tower dominates the Hilbert space, and so we can associate the entropy of the quasi-de Sitter setting equal to its entropy
\be
R^2 \sim N^{\gamma} R^{\delta} \;,
\label{entsim}
\ee
which yields
\be
V\left( \phi \right) \sim R^{-2} \sim N^{-\frac{2\gamma}{2-\delta}}\;.
\label{Vexp}
\ee
The number of light states in the tower is taken to be
\be
N \sim n\left( \phi \right) e^{b \phi} \;,
\label{Nstates}
\ee
where $n\left( \phi \right)$ is accounting for the number of different towers becoming light, and is taken to increase monotonically with $\phi$. $b$ is some order one number. Together (\ref{Vexp}) and (\ref{Nstates}) yield the de Sitter conjecture with 
\be
c \sim \frac{2b\gamma}{2-\delta} \;.
\ee
The crucial step of setting the entropies equal (\ref{entsim}) can be motivated in two ways. First, it is natural to expect that with an exponentially large number of states the tower would eventually dominate the Hilbert space. Second, the interpretation of the distance conjecture in terms of duality, specifically that the light states are offering a dual description of the physics, implies that the entropies should be equated. 

It is worth noting that it is not the case that one expects the covariant entropy bound to be violated unless $R$ increases with $N$ to satisfy (\ref{entsim}). Rather it is expected that it will be more and more closely saturated. A violation would be censored, for example through the species scale becoming lighter than the Hubble scale, as discussed in section \ref{sec:cebss}. Indeed, another way to think about this censorship is that the decreasing species scale due to the tower of states induces a consistency bound on the magnitude of the potential. This agrees with the ideas in \cite{Grimm:2018ohb,Heidenreich:2018kpg,talk-madrid,Hebecker:2018vxz}. 

The second condition of the refined de Sitter conjecture (\ref{rdsc2}) can be understood by noting that the finite temperature of de Sitter space induces a positive mass to scalar fields, at horizon scales, of order the potential. Therefore, the second derivative of the potential may be negative and still no instability would manifest, but if its magnitude is more than the potential there will be an instability on horizon scales spoiling an entropic interpretation of the horizon. Another way to think of it is as the condition that the apparent horizon should exist for at least one Hubble time.

The relation between the distance conjecture and de Sitter conjecture is manifest by this argument at parametrically large distances in field space. Since in string theory all coupling constants are field dependent, such that weak coupling is at large distance in field space, the argument holds at any parametrically controlled regime of string theory. In this sense it is a type of generalization of the Dine-Seiberg problem \cite{Dine:1985he}, which states that the string coupling induces a runaway direction in field space in any sufficiently perturbative regime of string theory. However, it is significantly stronger since it applies at any point where the Hilbert space is dominated by any of the many possible light towers of states. 

Perhaps more importantly, it connects microscopic and quantum aspects of de Sitter space with properties of potentials in effective field theories that can arise from string theory. This forms a contribution towards the goal that, eventually, we would like string theory to shed some light on quantum and microscopic aspects of de Sitter space, or if it does not exist within the theory then an accelerating universe with an apparent horizon, similarly to how string theory in anti-de Sitter space yielded deep insights into quantum gravity. It would certainly be disappointing if all we learned from string theory about de Sitter space is how to construct it in an effective low energy theory. With this in mind, we go on to review attempts at constructing de sitter vacua in string theory.

\subsection{de Sitter vacua in string theory}
\label{sec:testsdS}

String theory has proven rather resistant to realizing de Sitter vacua. This is in stark contrast to anti-de Sitter vacua, which are common and very well understood. The difference could well be due to the fact that de Sitter space cannot be supersymmetric and, at least the best understood, string theories start supersymmetric in higher dimensions. More generally, the apparent obstructions to de Sitter vacua may just be related to technical difficulties. However, it is certainly the case that there were plenty of opportunities for string theory to realize simple de Sitter vacua which it seems not to have taken. As we will review below, there are certain no-go theorems which from the perspective of the low-energy four-dimensional effective theory appear very surprising. Potentials of immense complexity, with many turning points, can be shown to never exhibit a de Sitter vacuum. 

It is therefore an open question whether string theory admits de Sitter vacua. The question manifests typical aspects of the Swampland program. On one hand there are various no-go theorems, general ideas, and various simple string theory examples supporting the possibility that string theory does not exhibit de Sitter vacua. On the other hand there are multiple proposals for counter-examples to this possibility. Prime among these is the KKLT construction \cite{Kachru:2003aw}, which we introduced in section \ref{sec:4dnoscale}, and will review further in section \ref{sec:dskklt}. In this section we present a brief review of some of the key results in the program of de Sitter vacua in string theory. We refer to \cite{Danielsson:2018ztv}, and also \cite{Obied:2018sgi,Andriot:2018ept}, for more detailed reviews of this field.

With respect to the Swampland, the existence of de Sitter vacua is critical as a test of the de Sitter conjecture (\ref{dsc}). However, it may be that de Sitter vacua cannot exist in string theory and yet the de Sitter, or more interestingly, the refined de Sitter conjecture  (\ref{rdsc1}-\ref{rdsc2}) may be violated. We will mostly consider de Sitter vacua, but will also discuss direct tests of the refined de Sitter conjecture.

\subsubsection{Classical vacua}
\label{sec:classIIA}

The first set of vacua we will consider are classical in nature, so where quantum corrections are neglected. We can further initially restrict to vacua which can be embedded in supergravity. That means we consider the (two-derivative) supergravities that are the low-energy limits of string theories and M-theory. We will allow the presence of positive tension objects, so any D-branes but no orientifolds. These assumptions appear rather mild, and indeed there is a Landscape of possible potentials that can arise from such compactifications. However, the Maldacena-Nunez no-go theorem states that there are no de Sitter vacua in such settings \cite{Maldacena:2000mw}. For the case of 11-dimensional supergravity, this no-go theorem was further extended in \cite{Obied:2018sgi} to show that there is a lower bound on the derivative of the potential for any compactification to $d$ dimensions
\be
\frac{\left|\nabla V \right|}{V} \geq \frac{6}{\sqrt{\left(d-2\right)\left( 11-d\right)}}\;.
\ee
This is consistent with the de Sitter conjecture. 

The no-go theorems do not hold in the presence of orientifold planes, and so they motivate studying the string theory setting. The only string theory setting where classical vacua with all the moduli stabilized are known is type IIA string theory. To be more precise, this is the only setting which admits a geometric interpretation as in Einstein gravity. String theory allows for non-geometric backgrounds as solutions for the world-sheet target space. These actually most likely form the bulk of string theory solutions. However, they are not sufficiently well understood yet to yield a definite answer regarding whether de Sitter vacua can be generated through such compactifications. We refer to \cite{Plauschinn:2018wbo} for a review of non-geometric compactifications. Within geometric type IIA compactifications, there are some general no-go theorems. An interesting such theorem applies to compactifications of type IIA on Calabi-Yau manifolds \cite{Hertzberg:2007wc}. For compactifications of type IIA string theory with fluxes and also orientifolds there is a bound for any $V>0$, 
\be
\frac{\left|\nabla V \right|}{V} \geq \sqrt{\frac{54}{13}} \;.
\label{CYiianogo}
\ee

Moving away from Calabi-Yau compactifications it is possible to consider manifolds with scalar curvature.\footnote{In \cite{Palti:2008mg} it was pointed out that already the leading $\left(\alpha'\right)^3$ corrections to the Calabi-Yau setting violate the assumptions leading to the no-go theorem (\ref{CYiianogo}). Note that $\alpha'$ corrections are classical effects, and in type IIA also correct the superpotential.}  In such cases, there is no general no-go theorem against de Sitter vacua, and there have been many searches for such vacua \cite{Silverstein:2007ac,Covi:2008ea,Haque:2008jz,Caviezel:2008tf,Flauger:2008ad,Danielsson:2009ff,deCarlos:2009fq,Caviezel:2009tu,Dibitetto:2010rg,Wrase:2010ew,Danielsson:2010bc,Blaback:2010sj,Danielsson:2011au,Shiu:2011zt,Burgess:2011rv,VanRiet:2011yc,Danielsson:2012by,Danielsson:2012et,Gautason:2013zw,Kallosh:2014oja,Junghans:2016uvg,Andriot:2016xvq,Junghans:2016abx,Andriot:2017jhf,Andriot:2018ept,Roupec:2018mbn,Junghans:2018gdb,Banlaki:2018ayh,Cordova:2018dbb,Cribiori:2019clo,Berglund:2019pxr,Blaback:2019zig,Blaback:2018hdo,Andriot:2019wrs}. We refer to \cite{Danielsson:2018ztv}, for example, for a detailed description of this research program. Here we will give an overview by first outlining the main challenges that face an attempt to realize fully understood de Sitter vacua this way:
\begin{itemize}
\item {\bf Flux quantization:} Studying vacua without quantizing fluxes is technically simpler and also leads to a richer set of possibilities. These are valid supergravity vacua, but their embedding in string theory where fluxes are quantized is unclear. It is therefore a challenge to study the potentials in a systematic way while respecting flux quantization.
\item {\bf Consistent truncations:} Typically it is difficult to induce a mass hierarchy between the moduli fields whose potential is studied, and higher modes such as Kaluza-Klein states. Then a consistent truncation must be utilised which relies on a group structure in the internal manifold. Once various ingredients are added to the compactification it is less clear if such truncations remain valid. Further, it is not clear that the sector of fields not within the truncation does not manifest instabilities. This is all part of the difficult challenge of understanding the geometry, and its deformations, for non Calabi-Yau spaces.
\item {\bf Smeared sources:}  Almost all constructions rely on cancelling tadpoles between fluxes and localized objects, thereby employing a smeared-source approximation (see section \ref{sec:simpleiia}). Some of these sources are intersecting orientifold planes which are particularly difficult to understand from a ten-dimensional uplift perspective. It is therefore a challenge to determine how uplifting to fully local solutions may modify the conclusions.
\item {\bf Coupling expansions:} It is typically difficult to find vacua where coupling constants controlling the effective theory are under parametric control. Sometimes a vacuum may push the validity of the coupling expansion to the limit, for example by having a string coupling of order one or string-scale volumes of cycles. Ideally we would like parametrically controlled couplings, but this seems particularly challenging to realize, see for example \cite{Banlaki:2018ayh}.
\end{itemize}
There is no known violation of the de Sitter conjecture, or the refined de Sitter conjecture, within a construction where all the challenges above are fully addressed. There are some proposed counter examples, see for example \cite{Caviezel:2008tf,Flauger:2008ad,Danielsson:2009ff,Caviezel:2009tu,Danielsson:2010bc}, to the de Sitter conjecture, which respect the refined de Sitter conjecture, but which still do not fully address some elements listed above. The refined de Sitter conjecture is particularly difficult to find a counter example to, see \cite{Blaback:2018hdo,Blaback:2019zig} for proposals, requiring leaving almost all of the challenges above unanswered. The results of the classical de Sitter vacua program are therefore as yet unclear. It seems that there is no strong evidence against the refined de Sitter conjecture, and the best understood cases rather offer evidence for it. However, in the absence of a full no-go theorem, this status may well be due to just the technical difficulty of finding completely understood solutions.

An interesting approach to testing potentials even away from weak-coupling limits was adopted in \cite{Gonzalo:2018guu} which utilised duality properties of the potentials. It was found that the refined de Sitter conjecture was satisfied for those potentials. This was also the case for heterotic studies in \cite{Parameswaran:2010ec,Olguin-Tejo:2018pfq}. However, in \cite{Blaback:2013qza} duality invariant potentials were found to support de Sitter minima, which suggests that duality invariance by itself is not sufficiently strong to rule out such minima.

\subsubsection{Quantum vacua}
\label{sec:dskklt}

By quantum vacua we refer to vacua which crucially rely on quantum effects for moduli stabilisation. Typically these would be non-perturbative in nature, as in the KKLT proposal discussed in section \ref{sec:4dnoscale}, or even more radically the racetrack proposal discussed in section \ref{sec:reacetrack}.\footnote{For some recent general discussions regarding the computation of the necessary non-perturbative effects in such settings see \cite{Sethi:2017phn,Kachru:2018aqn}.} The main body of work with regards to de Sitter vacua constructions is built upon the anti-de Sitter vacuum of KKLT \cite{Kachru:2003aw}, or possibly the Large Volume scenario \cite{Balasubramanian:2005zx}, where some positive energy contribution is added to the potential. This is often termed an uplift of the vacuum to de Sitter. There are a number of different type of uplifts proposed in the literature, for example \cite{Burgess:2003ic,Saltman:2004sn,Lebedev:2006qq,Cremades:2007ig,Achucarro:2006zf,Villadoro:2005yq,Parameswaran:2006jh,Parameswaran:2007kf,Westphal:2006tn,Rummel:2011cd,Louis:2012nb,Cicoli:2012fh,Cicoli:2013cha,Aparicio:2014wxa,Cicoli:2015ylx,Kallosh:2018nrk,Kane:2019nod,Weissenbacher:2019mef}.

One of the earliest, and in some sense simplest, proposal is to uplift the KKLT vacuum to de Sitter by using the positive energy due to D-terms \cite{Burgess:2003ic}. We may consider adding some hidden sector with a $U(1)$ gauge symmetry which has a D-term that leads to a positive energy contribution, this is then added on top of the KKLT anti-de Sitter vacuum. There is a general reason, emphasized in \cite{deAlwis:2005tf}, why such a simple construction would not work, which is that in supergravity as long as the superpotential is non-vanishing $W \neq 0$, vanishing F-terms imply vanishing D-terms. Since the original anti-de Sitter KKLT vacuum was supersymmetric with vanishing F-terms one cannot simply add to it some non-vanishing D-terms. This is not a serious obstruction, for example it may be possible to start from a non-supersymmetric vacuum as in the LVS scenario, and uplift that using D-terms \cite{Cremades:2007ig,Cicoli:2012fh,Cicoli:2013cha,Aparicio:2014wxa,Cicoli:2015ylx}. However, it does yield a valuable general lesson which is that whatever uplifts the anti-de Sitter vacuum to a de Sitter one will most likely modify the stabilisation of the moduli fields that occurred in the anti-de Sitter vacuum. This modification may not be important, yielding only small corrections, or it may be crucial and induce a new instability. Much of the contemporary discussion about uplifts is regarding this point. 

The most studied, and probably most robust, proposal for uplifts of quantum vacua is by utilising anti-D3 branes. This was proposed in \cite{Kachru:2003aw}, building on the work in \cite{Kachru:2002gs}. The idea is to consider the warped throat which develops in type IIB Calabi-Yau compactifications with fluxes, as discussed in section \ref{sec:iib10d}, and illustrated in figure \ref{fig:wcyt}. The throat is described, at least near the tip, by the Klebanov-Strassler (KS) solution \cite{Klebanov:2000hb}. Locally, near the tip, the metric takes the form \cite{Herzog:2001xk,Kachru:2002gs}
\be
ds^2 = a_0^2 ds^2_{4d} + g_s M b_0^2 \left( \frac12 dr^2 + d\Omega_3^2 + r^2 d\tilde{\Omega}_2^2 \right) \;.
\ee
Here $a_0$ is a constant given in \cite{Kachru:2002gs} and $b^2_0\simeq 0.9$, $g_s$ is the string coupling, and for this analysis we will work in units where $\alpha'=1$, rather than Planck units. The coordinate along the length of the throat is $r$, and there is a 2-sphere, with metric denoted $d\tilde{\Omega}_2^2$, which shrinks to zero size at the tip. There is also a 3-sphere with metric denoted $d\Omega_3^2$. There are $M$ units of Ramond-Ramond flux threading this 3-sphere
\be
\frac{1}{4 \pi^2}\int_{S^3} F_3 =  M \;.
\ee
To this background now are added $p$ anti-D3 branes, where $p \ll M$ so that they can be treated as probes only. Due to the background flux, these anti-D3 branes will puff up to an NS5 brane wrapping a two-sphere, denoted $S^2$, inside the three-sphere $S^3$. More precisely they can be described as such an NS5 brane supporting $p$ units of magnetic flux $\cF_2$ threading the $S^2$ 
\be
\frac{1}{2\pi} \int_{S^2} \cF_2 = p \;.
\ee
The flux will want to make the two-sphere that the NS5 brane warps expand to minimize the energy, but due to the background fluxes in the solution there will also be other forces acting on the NS5 brane. If we parameterize the radius of the $S^2$ that the NS5 brane wraps by the polar angle $\psi$ of its embedding in the $S^3$, the NS5 will feel a potential of the form \cite{Kachru:2002gs}
\be
V\left( \psi \right) = A_0 \left[ \frac{1}{\pi} \left( b_0^4 \sin^4 \psi + \left(\pi \frac{p}{M}-\psi+\frac12 \sin 2\psi \right)^2\right)^{\frac12} - \frac{1}{2\pi} \left(2 \psi - \sin 2 \psi \right) \right]\;,
\label{kpvpot}
\ee
where $A_0$ is a constant given in \cite{Kachru:2002gs}. We plot this potential for three values of $\frac{p}{M}$ in figure \ref{fig:KPV}. 
\begin{figure}[t]
\centering
 \includegraphics[width=0.8\textwidth]{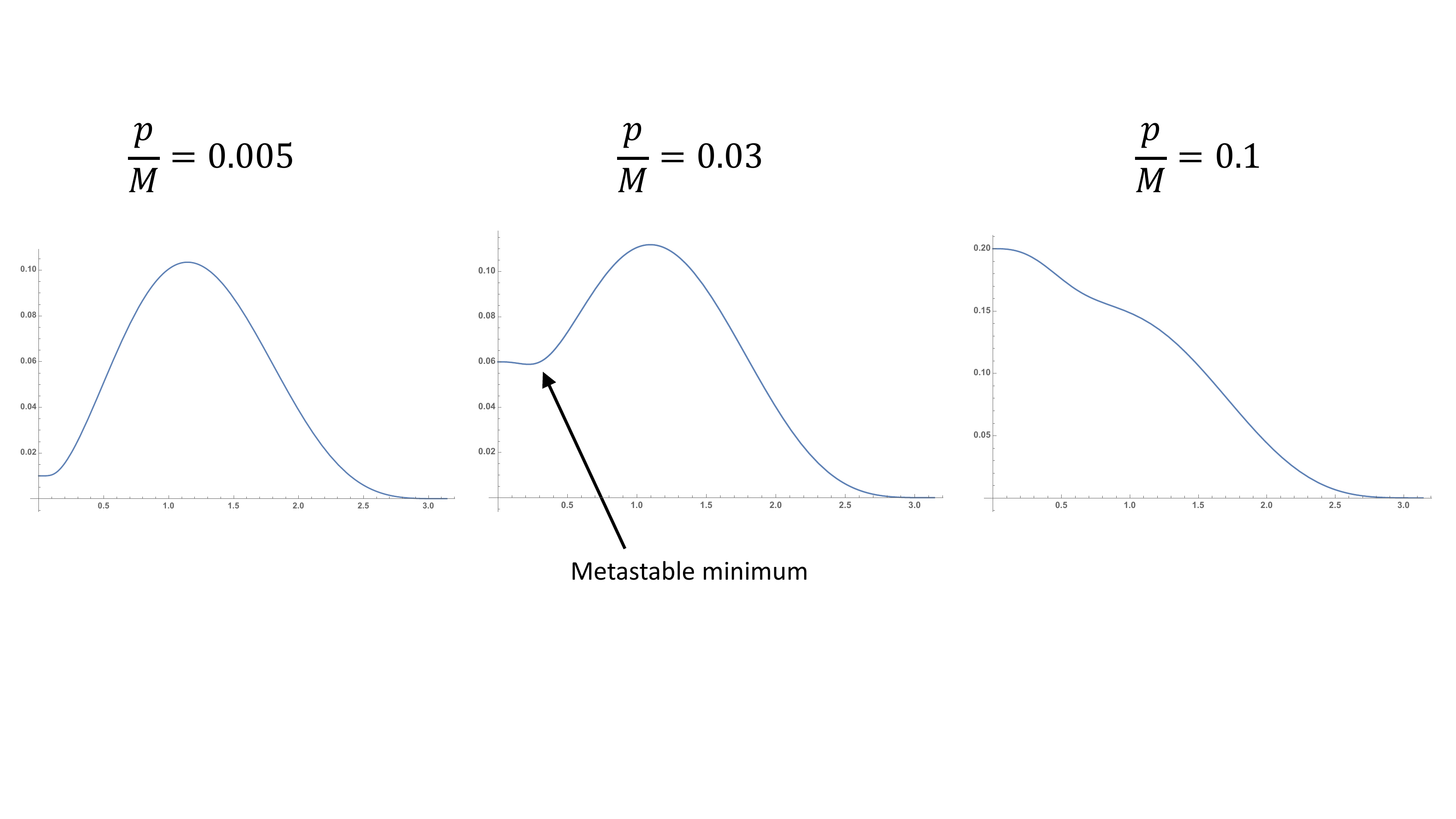}
\caption{Figure showing the potential (\ref{kpvpot}) for different values of $\frac{p}{M}$, where $p$ is the number of anti-D3 branes and $M$ is the amount of Ramond-Ramond flux in a KS throat. For appropriately chosen values of $p$ and $M$ there develops a meta-stable positive energy local minimum. Note that the overall value of the potential has been lifted to asymptote to zero energy.}
\label{fig:KPV}
\end{figure}
For appropriately chosen values of $p$ and $M$ there develops a meta-stable positive energy local minimum. The full minimal energy configuration at $\psi = \pi$ corresponds to the NS5 brane wrapping a vanishing volume sphere and in terms of the D3 picture this corresponds to $M-p$ D3 branes with no anti-D3 branes or Ramond-Ramond flux, and one less unit of NS flux.

In \cite{Kachru:2003aw} it was argued that the positive energy meta-stable minimum can then be implemented into the four-dimensional effective potential by adding an appropriate term to the potential (\ref{kkltwef}). Let us denote $T+\bar{T}=2t$, then the potential takes the form \cite{Kachru:2003aw,Kachru:2003sx}\footnote{The power of $t$ in the uplift term was corrected in \cite{Kachru:2003sx} to be $\frac{1}{t^2}$ from the $\frac{1}{t^3}$ proposed in \cite{Kachru:2003aw}. We have also set the axion in $T$ to vanish.}
\be
V\left(t\right) = \frac{2 \pi a A e^{-2 \pi a t}}{2 t^2} \left( \frac13 2 \pi a t A e^{-2 \pi a t} + W_0 + A e^{-2 \pi a t} \right) + \frac{D}{t^2} \;.
\label{potkklt}
\ee
The last term in (\ref{potkklt}) is added to account for the positive energy contribution from the anti-D3 branes, and the parameter $D$ may be naturally exponentially small since it is suppressed by the warping of the throat. For appropriately chosen parameters, the potential (\ref{potkklt}) exhibits positive energy de Sitter minima. For example, taking $A=1$, $2 \pi a = 0.1$, $W_0 = -10^{-4}$ and $D=2.65\times10^{-11}$ yields such a minimum. This is shown in figure \ref{fig:kklt}.
\begin{figure}[t]
\centering
 \includegraphics[width=0.8\textwidth]{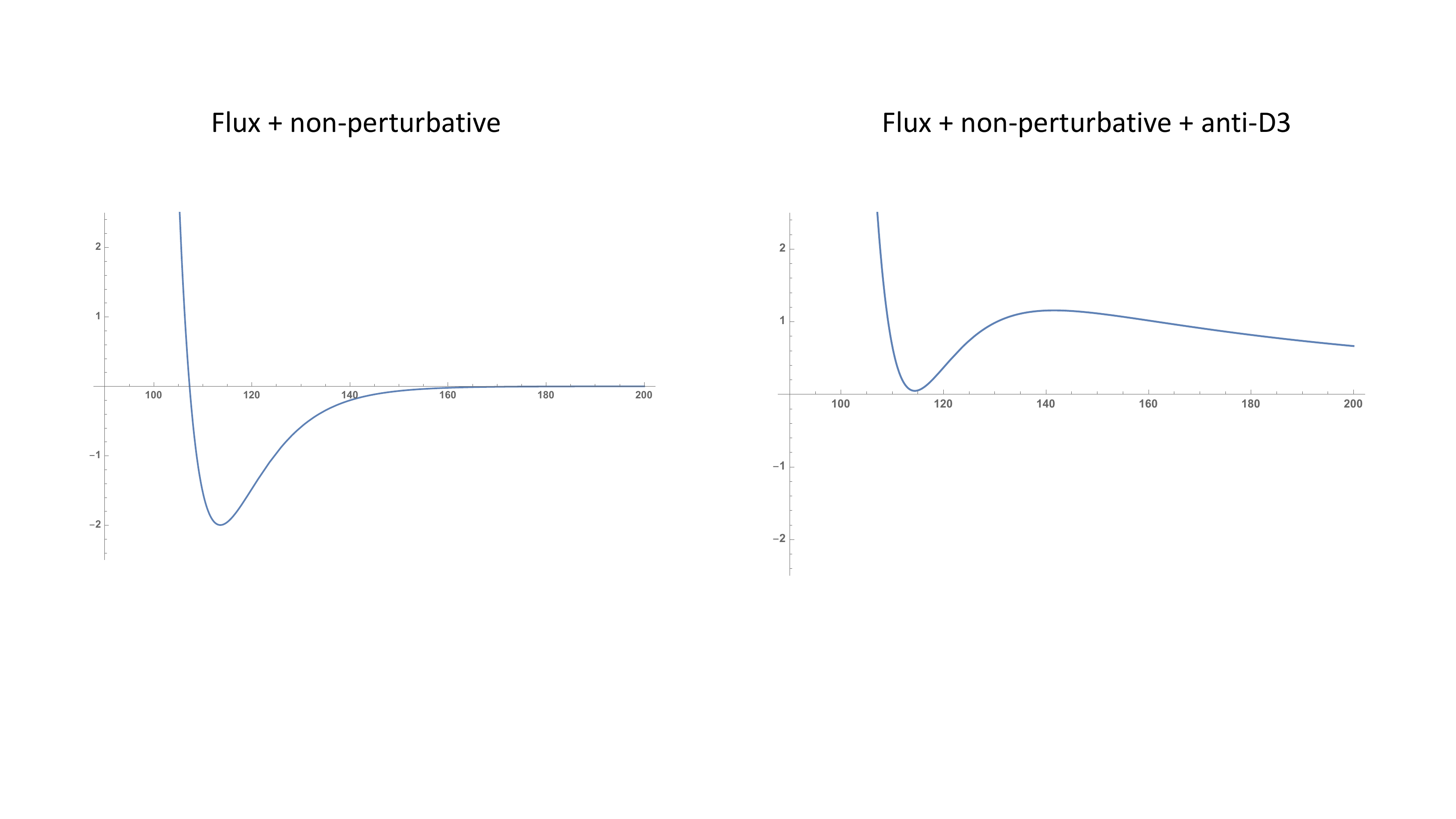}
\caption{Figure showing the potential (\ref{potkklt}) for $A=1$, $2 \pi a = 0.1$ and $W_0 = -10^{-4}$ (multiplied by $10^{15}$). The left hand side shows the case with $D=0$, so with no anti-D3 branes, and the right hand side has $D=2.65\times10^{-11}$, so with anti-D3 branes down a sufficiently warped throat. The latter setup provides an uplift to positive energy and leads to the proposal of \cite{Kachru:2003aw} for de Sitter vacua in string theory.}
\label{fig:kklt}
\end{figure}
This therefore is the KKLT proposal for de Sitter minima in string theory. 

There has been significant work studying in detail various aspects of the anti-D3 uplift proposal. One program focussed on understanding the supergravity solution of the KS throat with an anti-D3, and in particular its singular behaviour \cite{Bena:2009xk,Bena:2012ek,Bena:2014bxa,Bena:2015kia,Bena:2016fqp,Blaback:2012nf,Gautason:2013zw,Danielsson:2014yga,Michel:2014lva,Danielsson:2016cit,Cohen-Maldonado:2015ssa,Cohen-Maldonado:2016cjh}. One of the issues was whether there is an instability or inconsistency associated to singularities in the supergravity equations, or whether those singularities are not problematic and are resolved in string theory. The conclusions are not completely clear, but it appears that, in particular see \cite{Michel:2014lva}, at least for a single anti-D3 there is no inconsistency. Another aspect which was studied is the effect of the anti-D3 uplift on the stabilisation of the K{\"a}hler moduli. In \cite{Moritz:2017xto} it was argued that a simple part of the ten-dimensional equations, coming from integrating over the internal space, suggests that the K{\"a}hler moduli are destabilised by the uplift. Further investigations however argued that at least this particular part of the ten-dimensional equations does not imply such an instability \cite{Hamada:2018qef,Kallosh:2019axr,Kallosh:2019oxv,Hamada:2019ack,Gautason:2019jwq,Carta:2019rhx}. In a similar direction, the backreaction of the uplift on complex-structure moduli was studied in \cite{Blumenhagen:2016bfp,Wolf:2017wmu,Bena:2018fqc,Blumenhagen:2019qcg}. In particular, in \cite{Bena:2018fqc} it was argued that keeping the complex-structure modulus stable requires a minimum amount of flux, which is not allowed in type IIB but can be realized in the context of F-theory (so including strongly coupled physics). In \cite{Blumenhagen:2019qcg} it was argued that the effective theory of KKLT may be problematic in the sense that there are Kaluza-Klein modes in the throat which are lighter than the moduli which are in the KKLT effective theory. It remained inconclusive whether this leads to a fundamental problem or instability, or whether it only affects the vacuum in a mild way. 

Overall, the question of whether the KKLT scenario leads to true de Sitter vacua of string theory remains somewhat subjective and controversial. Certainly it is a very reasonable prescription, and it has passed many tests. To the authors' knowledge, there is no explicit calculation which shows that the proposal is incorrect. On the other hand, it is very much a prescription, rather than an explicit ten-dimensional solution, and it combines many different elements, some of which are only really understood in the four-dimensional effective theory. Many of the elements are also implicit rather than explicitly calculated, even in the anti-de Sitter construction, such as the required small $W_0$. In the absence of a ten-dimensional understanding of the KKLT de Sitter solutions, it is not clear, at least to the author, that we can decide with certainty either way regarding their validity.

As mentioned earlier, we can also consider quantum vacua which do not have a de Sitter vacuum yet still violate the de Sitter conjecture. An interesting example was proposed in \cite{Conlon:2018eyr}, based on a simple KKLT construction, where the existence of a positive energy maximum was argued for on general grounds. Such a maximum would violate the de Sitter conjecture (\ref{dsc}), but its existence could naturally be compatible with the refined de Sitter conjecture due to the condition (\ref{rdsc2}). 

\subsubsection{Relation to scale separation in AdS}
\label{sec:scsepcon}

The discussion of the KKLT proposal in the previous section utilised quite precisely chosen values for the parameters. For example, the de Sitter minimum in figure \ref{fig:kklt} utilised $D=2.65 \times 10^{-11}$, but if we had taken $D= 10^{-10}$ say, then the minimum would have been washed away. And if we had taken $D=10^{-11}$ then the minimum would be still anti-de Sitter. This type of delicate tuning is universal to attempts at constructing de Sitter vacua by starting from an anti-de Sitter vacuum and uplifting it, as long as the amount of uplifting energy is comparable to the barrier which separated the original anti-de Sitter minimum from the Minkowski decompactification limit.  

This suggests that the most robust realizations of de Sitter vacua could be reached by uplifting an anti-de Sitter vacuum which has a large separation of scales between the magnitude of the negative cosmological constant and the height of the barrier to decompactification. This hierarchy is naturally tied to a hierarchy between the anti-de Sitter length scale $R_{AdS}$, and the extra-dimensional Kaluza-Klein scale and moduli masses $M_{\mathrm{Mod/KK}}$, 
\be
M_{\mathrm{Mod/KK}}R_{AdS} \gg 1 \;.
\label{kadshier}
\ee
This allows for potentials in the effective theory with energy scales much larger than the AdS scale. One might consider then that the existence of de Sitter vacua may be related to whether a parametric hierarchy of the type (\ref{kadshier}) can be reached. This has in some sense been appreciated for a while, see for example \cite{Kallosh:2004yh,Conlon:2008cj}. But has become more prominent in recent studies. In \cite{Gautason:2018gln} it was even conjectured that vacua which support a parametric hierarchy as in (\ref{kadshier}) are in the Swampland. 

In fact, we have encountered an obstruction to a hierarchy (\ref{kadshier}) already in section \ref{sec:wgciios}, where we discussed how the Weak Gravity Conjecture for open-string $U(1)$s may forbid such a hierarchy, at least in certain situations \cite{ArkaniHamed:2006dz}. A relation between (\ref{kadshier}) and the Weak Gravity Conjecture was also proposed in \cite{Moritz:2018sui}, as an obstruction to realizing the hierarchy (\ref{kadshier}) within the scenario of \cite{Kallosh:2004yh}, although this application was criticized in \cite{Kallosh:2019axr}.

The limiting version of the hierarchy (\ref{kadshier}) can be reached within a supersymmetric Minkowski vacuum where all the moduli are stabilized. Since then $R_{AdS} = \infty$ while $M_{\mathrm{Mod/KK}}$ remains finite. The first way we may consider constructing such a vacuum is through a classical compactification, in which case we are restricted to type IIA string theory. However, a no-go theorem was presented in \cite{Micu:2007rd} which states that it is not possible to fix all the moduli in a supersymmetric Minkowski vacuum for any geometric compactification of type IIA string theory.\footnote{This holds at the classical and also two-derivative level, so without $\alpha'$ corrections.} This is an interesting piece of evidence towards an obstruction to such vacua. Therefore, classical vacua would have to come from non-geometric compactifications, see \cite{Palti:2007pm} and \cite{Becker:2006ks} for example directions towards this. Quantum vacua could in principle lead to supersymmetric Minkowski solutions, see for example \cite{Kallosh:2004yh}, though this would require some unnatural tuning of parameters. Indeed, in \cite{Dine:2005gz} it was generally argued that a natural way to reach supersymmetric Minkowski vacua is through R-symmetries, which would be easier to implement in a perturbative setting.
 
A class of vacua which allowed for a parametric hierarchy (\ref{kadshier}) are type IIA compactifications on Calabi-Yau manifolds as studied in \cite{DeWolfe:2005uu}. A simple version of these vacua, replacing the Calabi-Yau with a torus, were discussed in section \ref{sec:iiat6flux}. However, as discussed in section \ref{sec:iias3s3}, these vacua do not solve the ten-dimensional equations of motion locally, but rather utilised the smeared source approximation. In the case of the torus, the associated local ten-dimensional solution was $S^3 \times S^3$, in which case there was no separation of scales (\ref{kadshier}). It is therefore unclear if completing the vacua of \cite{DeWolfe:2005uu} to local ten-dimensional solutions would preserve this hierarchy. 

One way to be sure of the consistency of an anti-de Sitter construction with a hierarchy (\ref{kadshier}) is to construct its Conformal Field Theory (CFT) dual. For example, the CFT duals to the type IIA vacua of \cite{DeWolfe:2005uu} were studied in \cite{Aharony:2008wz}. For the KKLT setup, work on the CFT duals was done in \cite{Silverstein:2003jp,Fabinger:2003gp,deAlwis:2014wia}, and for the LVS scenario the CFT duals were studied in \cite{deAlwis:2014wia,Conlon:2018vov} (see also \cite{Banks:2004xh,Polchinski:2009ch} for general studies). In all situations, there were no explicit constructions of the CFTs, but rather only certain properties of them were deduced. A particularly interesting property of these CFTs, assuming they exist, would be that the hierarchy (\ref{kadshier}) would imply that all single-trace operators must have large conformal dimension $\Delta$. This is because of the relation between this and the mass of the states in the bulk $m_{\Delta}$
\be
m_{\Delta}^2 R_{AdS}^2 = \Delta\left(\Delta - 3 \right) \;.
\ee
Having such a spectrum of operators is unusual, for example, perturbative CFTs posses operators of $\cO\left(1\right)$ conformal dimensions.\footnote{Note that, in this context, it is interesting to also consider the sole constraint on the Kaluza-Klein masses of having an effective four-dimensional theory $M_{\mathrm{KK}}R_{AdS} \gg 1$. This would mean that one can allow for a few low-dimensional operators corresponding to the moduli. Finding such CFT duals would help better establish the moduli stabilisation models studied in section \ref{sec:stringcomp}.}

\subsection{Implications for Cosmology}
\label{sec:imswcos}

In this section we present a brief overview of the cosmological implications of the de Sitter conjecture (\ref{dsc}) and its refinement (\ref{rdsc1}-\ref{rdsc2}). The most immediate implication is for the current universe with regards to the nature of dark energy. The observation that the universe is entering a late-time phase of accelerated expansion suggests that the overall scalar potential for the universe should be at some positive value $V>0$. The de Sitter conjecture therefore implies that it cannot be at a minimum, where we would have $\nabla V=0$. Instead it would imply that the universe should be currently rolling down a potential, sufficiently slowly that the potential energy dominates over the kinetic energy so as to drive an accelerated expansion. This is illustrated in figure \ref{fig:dspotunv}. 
\begin{figure}[t]
\centering
 \includegraphics[width=0.8\textwidth]{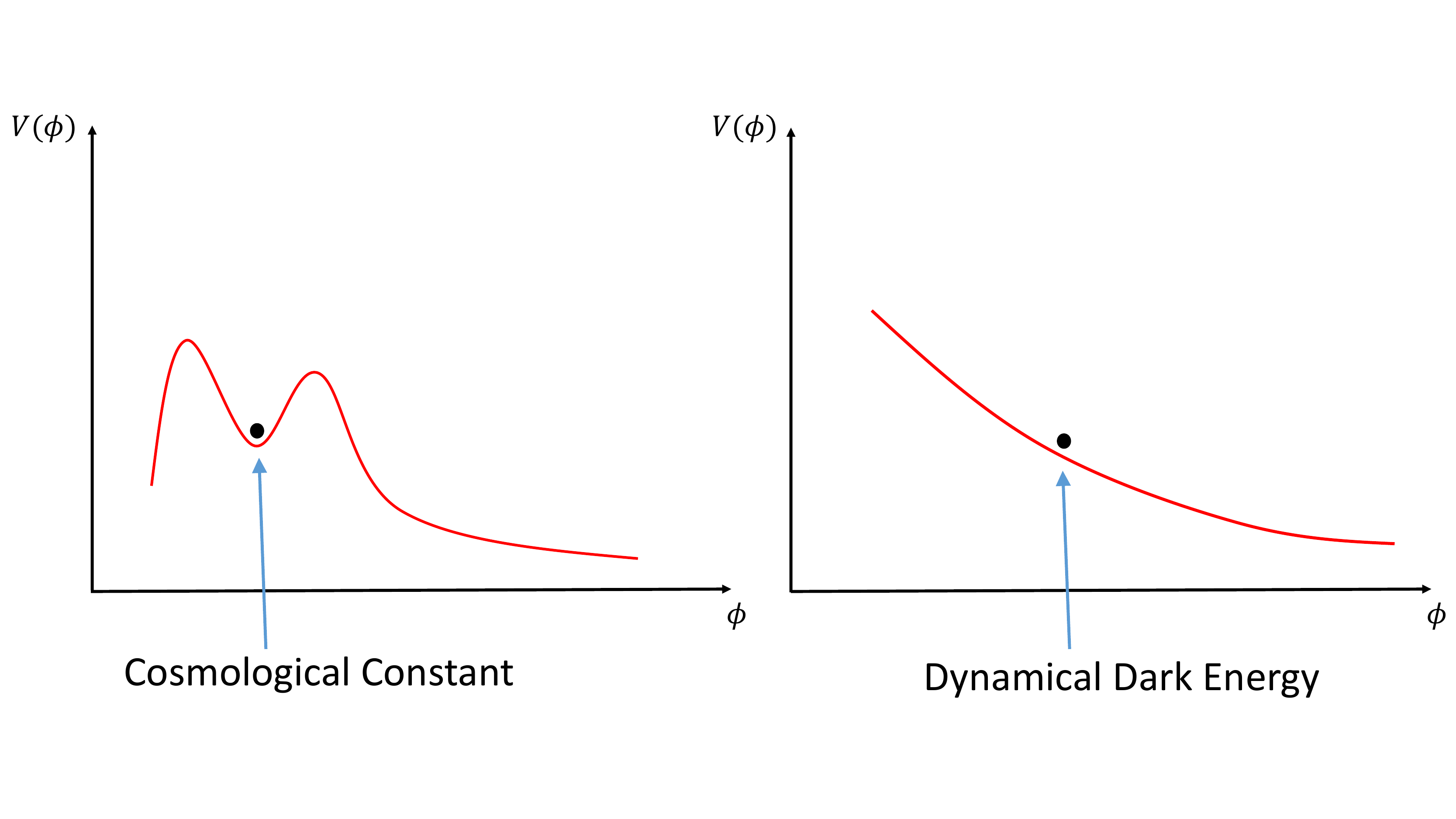}
\caption{Figure showing the scalar potential of the universe along a particular scalar direction denoted by $\phi$, with the current state of the universe corresponding to the black dot. The possibility on the left corresponds to a Cosmological Constant driving the present day accelerated expansion, and violates the de Sitter conjecture. The potential on the right corresponds to a dynamical dark energy scenario where the accelerated expansion is driven by a rolling scalar field, and is compatible with the de Sitter conjecture. }
\label{fig:dspotunv}
\end{figure}
Such a scenario is termed dynamical dark energy, the most well-known version of which is quintessence \cite{WETTERICH1988668,PhysRevD.37.3406}, and we refer to \cite{Copeland:2006wr} for a review. 

A prediction of dynamical dark energy scenarios is that the dark energy equation of state must vary in time. So taking dark energy to be describable by a fluid with pressure $p$ and energy density $\rho$, the equation of state is parametrized by $\omega$ as
\be
p = \omega \rho \;.
\ee
For a scalar field rolling down a potential, this corresponds to
\be
\omega \simeq \frac{\frac12 \dot{\phi}^2 - V\left( \phi \right)}{\frac12 \dot{\phi}^2 + V\left( \phi \right)} \;,
\ee
so that a cosmological constant scenario is the limit $\omega = -1$. Current observation of the dark energy equation of state $\omega$ place bounds on its possible deviation from a cosmological constant. These bounds were first analyzed in \cite{Agrawal:2018own} where they were found to constrain the parameter $c$ in the de Sitter conjecture (\ref{dsc}) as 
\be
c < 0.6 \;,
\ee
which is compatible with $c \sim \cO\left(1\right)$. 

The conjecture also has interactions with inflationary models, since $c$ is related directly to the slow-roll parameter during inflation. Perhaps the most direct bounds from inflation come from the non-observation of tensor modes and restrict \cite{Agrawal:2018own}
\be
c < 0.09 \;.
\ee
This can be sharpened to $c < 0.035$ if no detection is made down to $r \sim 0.01$. Specific models require even smaller values of $c$, for example plateau-type models require $c < 0.02$. This is in some tension with the conjecture, though that depends on how strictly one interprets $c \sim \cO\left( 1\right)$.

It was also noted in \cite{Agrawal:2018own} that a cosmological implication of the conjecture is that the universe must correspond to a scalar field that is rolling to increasingly large expectation values. It is possible that eventually this would lead to an effective negative cosmological constant, causing a phase transition in the universe. Or the expectation value of the scalar field could become so large that the light states of the distance conjecture (\ref{sdc}-\ref{rsdc}) start to affect the universe, possibly causing also a phase transition. The point where this would happen though is uncertain, since in principle the tower of states could be very light before its effects become important. See also \cite{Blumenhagen:2018hsh} for a similar discussion.

Following the initial analysis of the cosmological implications of the de Sitter conjecture in \cite{Agrawal:2018own}, there has been significant further work. The cosmological bounds from dark energy on the de Sitter conjecture were further developed in \cite{Heisenberg:2018yae,Cicoli:2018kdo,Akrami:2018ylq,Heisenberg:2018rdu,Wang:2018duq,Raveri:2018ddi}, and on the refined de Sitter conjecture in \cite{Agrawal:2018rcg,Chiang:2018lqx,Thompson:2018ifr,Elizalde:2018dvw,Tosone:2018qei}. In \cite{Dvali:2018fqu,Dvali:2018jhn} the relation to cosmological implications from quantum break time were studied. In \cite{Achucarro:2018vey} it was argued that the de Sitter and distance conjectures favour multi-field inflation. Multi-field scenarios, particularly with non-geodesic inflaton motion, were further studied in the Swampland context in \cite{Brown:2017osf,Damian:2018tlf,Landete:2018kqf,Schimmrigk:2018gch,Bjorkmo:2019aev,Fumagalli:2019noh,Bjorkmo:2019fls,Achucarro:2019pux}. In \cite{Garg:2018reu} tension with a large number of e-folds in inflation was studied. The interaction of the de Sitter conjecture with tensor modes in inflation were studied in \cite{Kehagias:2018uem,Dias:2018ngv}, and with the spectral tilt in \cite{Ben-Dayan:2018mhe,Kinney:2018nny}. Studies more specific to the refined de Sitter conjecture and inflation were performed in \cite{Fukuda:2018haz,Lin:2018rnx,Cheong:2018udx,Holman:2018inr,Kinney:2018kew,Seo:2018abc,Kobakhidze:2019ppv}. In \cite{Colgain:2018wgk} a relation between the de Sitter conjecture and the tension in different measurements of the Hubble constant was studied. The interaction of the conjecture with warm inflation models was studied in \cite{Rasouli:2018kvy,Das:2018hqy,Motaharfar:2018zyb,Das:2018rpg,Kamali:2019hgv}, with eternal inflation in \cite{Matsui:2018bsy,Dimopoulos:2018upl}, with $\alpha$-attractor models in \cite{Scalisi:2018eaz} and with inflation models which do not utilise the Bunch-Davies vacuum in \cite{Brahma:2018hrd,Ashoorioon:2018sqb}. The conjectures were also utilised as motivation to study in more detail quintessence model building \cite{Chiang:2018jdg,Emelin:2018igk,Han:2018yrk,DAmico:2018mnx,Olguin-Tejo:2018pfq}, and to study the general cosmological bounds and theoretical constraints and challenges of quintessence models \cite{Marsh:2018kub,Hamaguchi:2018vtv,Acharya:2018deu,Hertzberg:2018suv}. The implications for isocurvature perturbations and dark sectors were studied in \cite{Matsui:2018xwa,Haque:2019prw}.

The conjectures were also considered as motivating cosmological scenarios qualitatively different from the usual $\Lambda$CDM or quintessence ideas \cite{Banerjee:2018qey,Lin:2018kjm,Lehners:2018vgi,Lin:2018edm,Cai:2018ebs,Bramberger:2019zez,Heckman:2018mxl,Heckman:2019dsj}. The interaction of theories of modified gravity with the de Sitter conjectures was studied in \cite{Artymowski:2019vfy,Heisenberg:2019qxz,Brahma:2019kch,Yi:2018dhl}. The conjectures were applied to properties of compact objects such as boson stars \cite{Herdeiro:2018hfp} and primordial black holes \cite{Kawasaki:2018daf}. The conjectures were also studied in the context of the possible singularity structures in cosmology in \cite{Odintsov:2018zai}.

\section{Summary and outlook}
\label{sec:summary}

In this article we introduced and reviewed the Swampland program. We began with a quick introduction to string theory, which led to a first encounter with the Swampland in the form of the Weak Gravity Conjecture and the Swampland Distance Conjecture. We then reviewed the overall structure of the different Swampland proposals, the motivating arguments for them, as well as their tests in string theory. We introduced the Emergence Proposal, which aims to explain the Swampland constraints as consequences of the emergent nature of fields within quantum gravity. Finally, we reviewed the de Sitter conjecture which, while more speculative than the other conjectures, would have remarkable consequences if correct. 

Our understanding of the Swampland is still very much in flux. It is not yet a mature field where there are some underlying principles from which the results are rigorously derived. Nonetheless, the past few years have seen significant interest and work on the Swampland and from this attention the ideas and proposals have come out only stronger and framed within a more coherent structure. This is encouraging, and suggests that we are on the right track towards uncovering some interesting and possibly impactful new physics. Of course, it is quite likely that in time the various proposed conjectures will be modified, perhaps some will be ruled out by counter-examples, and others will be refined. It also increasingly appears that there are strong links between the different conjectures suggesting that we may possibly reach a picture where they are all consequences of a single underlying new principle. 


While string theory has played, and will continue to play, a central role in the Swampland program, ideally we would like to be able to formulate the underlying physics as general properties of quantum gravity. This is in some sense in the spirit of AdS/CFT, which started life as a specific example in string theory, but is now studied in a much more general way. It would be fantastic if the ideas underlying the Swampland program eventually reached such a mature stage. 

The Swampland program covers a very wide range of topics, from string theory, through black holes and quantum gravity, to particle physics and cosmology. This may make entering the field as a researcher a somewhat daunting task. The aim of this introduction and review is to form a starting point where the wide range of ideas are gathered together, and where simple examples are worked through so as to form a base from which the more complicated constructions can be explored. The hope is that it will therefore act as a resource encouraging researchers to interact with this fascinating field. 

{\bf Acknowledgements:} I would like to thank Gia Dvali, Thomas Grimm, Daniel Kl\"awer, Dieter L\"ust, Fernando Marchesano, Hirosi Ooguri, Gary Shiu, Cumrun Vafa, Irene Valenzuela and Timo Weigand for useful discussions and explanations. I also thank Joe Conlon, Thomas Grimm, Arthur Hebecker, Daniel Junghans, Daniel Kl\"awer, Dieter L\"ust, Gary Shiu and Irene Valenzuela for extremely helpful comments on the manuscript. Like all reviews, there is bias by the author towards material they are more familiar with, and so I apologise for not dedicating enough space to topics that probably deserved more attention. I also apologise for any mistakes or misunderstandings I may have made in interpreting the work of others. In any such cases, the fault is, of course, completely mine. 

\appendix
	
\bibliographystyle{jhep}
\bibliography{swampland.bib}  

\end{document}